\documentclass[twoside]{article}
\usepackage{rotating}

\usepackage{graphicx}
\usepackage{subfigure}
\usepackage{latexsym}
\usepackage{amsfonts}
\usepackage{amsmath}
\usepackage{amssymb}
\usepackage{url}
\usepackage{subfigure}
\usepackage[dvips]{color}

\newlength{\figurewidth}
\renewcommand{\sectionmark}[1]{\markboth{\textsc{Leskovec, Lang, Dasgupta, and Mahoney}}
{\textsc{Community structure in large networks}}}
\renewcommand{\subsectionmark}[1]{\markboth{\textsc{Leskovec, Lang, Dasgupta, and Mahoney}}
{\textsc{Community structure in large networks}}}

\newtheorem{definition}{Definion}
\newtheorem{theorem}{Theorem}
\newtheorem{lemma}{Lemma}

\newtheorem{observation}{Observation}
\newenvironment{proof}{{\bf Proof:}}{\hfill{\rule{2mm}{2mm}}}

\newcommand{\hide}[1]{}

\def\Pr{{\mathbf{Pr}}}
\def\Ex{{\mathbf{E}}}

\def\BAS{\begin{eqnarray*}}
\def\EAS{\end{eqnarray*}}

\newcommand{\net}[1]{{\textsc{#1}}}

\newcommand{\captionfonts}{\small}
\makeatletter
\long\def\@makecaption#1#2{%
  \vskip\abovecaptionskip
  \sbox\@tempboxa{{\captionfonts #1: #2}}%
  \ifdim \wd\@tempboxa >\hsize
    {\captionfonts #1: #2\par}
  \else
    \hbox to\hsize{\hfil\box\@tempboxa\hfil}%
  \fi
  \vskip\belowcaptionskip}
\makeatother

\begin{document}


\title{
Community Structure in Large Networks:
Natural Cluster Sizes and the Absence of Large Well-Defined Clusters
\thanks{
A conference proceedings version of this paper appeared 
in WWW 2008 as~\cite{LLDM08_communities_CONF}.
In addition, this paper is substantially the same as our 
manuscript of March 7, 2008~\cite{LLDM08_communities_DRAFT}.
}
}

\author{
Jure Leskovec
\thanks{
Carnegie Mellon University,
Pittsburgh, PA 15213.
Email: jure@cs.cmu.edu
}
\and Kevin J. Lang
\thanks{
Yahoo! Research, Sunnyvale, CA 94089.
Email: \{langk,
anirban\}@yahoo-inc.com 
}
\and
Anirban Dasgupta \footnotemark[2]
\and
Michael W. Mahoney
\thanks{
Stanford University, 
Stanford, CA 94305. 
Email: mmahoney@cs.stanford.edu
}
}

\date{}
\maketitle


\begin{abstract}
A large body of work has been devoted to defining and identifying clusters or
communities in social and information networks, \emph{i.e.}, in graphs in 
which the nodes represent underlying social entities and the edges represent 
some sort of interaction between pairs of nodes.  Most such research begins
with the premise that a community or a cluster should be thought of as a
set of nodes that has more and/or better connections between its members 
than to the remainder of the network. In this paper, we explore from a novel
perspective several questions related to identifying meaningful
communities in large social and information networks, and we come to
several striking conclusions.

Rather than defining a procedure to extract sets of nodes from a graph and
then attempt to interpret these sets as a ``real'' communities, we employ
approximation algorithms for the graph partitioning problem to characterize 
as a function of size the
statistical and structural properties of partitions of graphs that could
plausibly be interpreted as communities. In particular, we define the
\emph{network community profile plot}, which characterizes the ``best''
possible community---according to the conductance measure---over a wide
range of size scales. We study over $100$ large real-world networks,
ranging from traditional and on-line social networks, to technological and
information networks and web graphs, and ranging in size from thousands up
to tens of millions of nodes.

Our results suggest a significantly more refined picture of community
structure in large networks than has been appreciated previously. Our
observations agree with previous work on small networks, but we show that 
large networks have a very different structure.  In particular, we observe tight
communities that are barely connected to the rest of the network at
very small size scales (up to $\approx 100$ nodes); and communities of
size scale beyond $\approx 100$ nodes gradually ``blend into'' the
expander-like core of the network and thus become less ``community-like,''
with a roughly inverse relationship between community size and optimal
community quality. This observation agrees well with the so-called Dunbar number 
which gives a limit to the size of a well-functioning community.

However, this behavior is not explained, even at a qualitative level, by
any of the commonly-used network generation models. Moreover, it is
exactly the opposite of what one would expect based on intuition from
expander graphs, low-dimensional or manifold-like graphs, and from small social networks that
have served as testbeds of community detection algorithms. The relatively
gradual increase of the network community profile plot as a function of
increasing community size depends in a subtle manner on the way in which
local clustering information is propagated from smaller to larger size
scales in the network. We have found that a generative graph model, in
which new edges are added via an iterative ``forest fire'' burning
process, is able to produce graphs exhibiting a network community profile
plot similar to what we observe in our network datasets.
\end{abstract}

\newpage
\rule{6in}{.1mm}
\tableofcontents
\rule{6in}{.1mm}
\newpage

\newpage

\section{Introduction}
\label{sxn:intro}

A large amount of research has been devoted to the task of defining and
identifying communities in social and information networks, {\em i.e.},
in graphs in which the nodes represent underlying social entities and the
edges represent interactions between pairs of nodes. Most recent papers on
the subject of community detection in large networks begin by noting that
it is a matter of common experience that communities exist in such
networks. These papers then note that, although there is no agreed-upon
definition for a community, a community should be thought of as a set of
nodes that has more and/or better connections between its members than
between its members and the remainder of the network. These papers then
apply a range of algorithmic techniques and intuitions to extract subsets
of nodes and then interpret these subsets as meaningful communities
corresponding to some underlying ``true'' real-world communities. In this
paper, we explore from a novel perspective several questions related to
identifying meaningful communities in large sparse networks, and we come
to several striking conclusions that have implications for community
detection and graph partitioning in such networks. 
We emphasize that, in contrast to most of the previous work on this subject, we look at very 
large networks of up to millions of nodes, and we observe very different 
phenomena than is seen in small commonly-analyzed networks.

\subsection{Overview of our approach}
\label{sxn:intro:approach}

At the risk of oversimplifying the large and often intricate body of work
on community detection in complex networks, the following five-part story
describes the general methodology:
\begin{enumerate}
\item[(1)] Data are modeled by an ``interaction graph.'' In
    particular, part of the world gets mapped to a graph in which
    nodes represent entities and edges represent some type of
    interaction between pairs of those entities. For example, in a
    social network, nodes may represent individual people and edges
    may represent friendships, interactions or communication between
    pairs of those people.
\item[(2)] The hypothesis is made that the world contains groups of
    entities that interact more strongly amongst themselves than with
    the outside world, and hence the interaction graph should contain
    sets of nodes, {\em i.e.}, communities, that have more and/or
    better-connected ``internal edges'' connecting members of the set
    than ``cut edges'' connecting the set to the rest of the world.
\item[(3)] A objective function or metric is chosen to formalize this
    idea of groups with more intra-group than inter-group
    connectivity.
\item[(4)] An algorithm is then selected to find sets of nodes that
    exactly or approximately optimize this or some other related
    metric. Sets of nodes that the algorithm finds are then called
    ``clusters,'' ``communities,'' ``groups,'' ``classes,'' or
    ``modules''.
\item[(5)] The clusters or communities or modules are evaluated in
    some way. For example, one may map the sets of nodes back to the
    real world to see whether they appear to make intuitive sense as a
    plausible ``real'' community. Alternatively, one may attempt to
    acquire some form of ``ground truth,'' in which case the set of
    nodes output by the algorithm may be compared with it.
\end{enumerate}
\setcounter{equation}{5}

\begin{figure}[t]
	\begin{center}
		\includegraphics[width=0.50\textwidth,height=6cm]{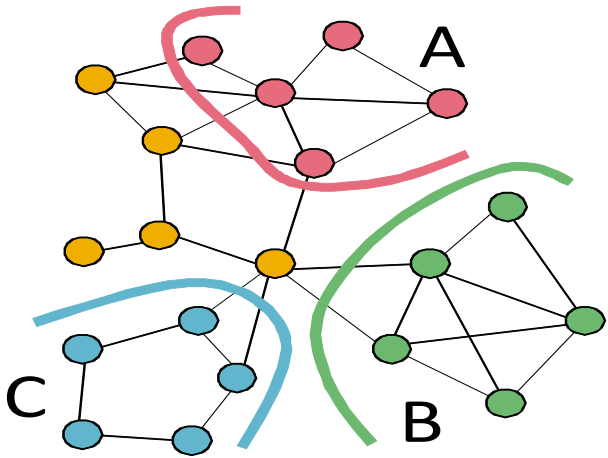}
	\end{center}
  \caption{Network communities. Of the three $5$-nodes sets that have been 
  marked, $B$ has the best (\emph{i.e.}, the lowest) conductance, as it has 
  the lowest ratio between the number of edges cut and the number of edges
  inside. So, set $B$ is the best $5$-node community or the most 
  community-like set of $5$ nodes in this particular network.}
\label{fig:conduct2}
\end{figure}

With respect to points (1)--(4), we follow the usual path. In particular,
we adopt points (1) and (2), and we then explore the consequence of making
such a choice, {\em i.e.}, of making such an hypothesis and modeling
assumption. For point (3), we choose a natural and widely-adopted notion
of community goodness (community quality score) called {\em conductance}, which is also
known as the normalized cut
metric~\cite{Chung:1997,ShiMalik00_NCut,kannan04_gbs}. Informally, the
conductance of a set of nodes (defined and discussed in more detail in
Section~\ref{sxn:related:LowCondAlgs}) is the ratio of the number of
``cut'' edges between that set and its complement divided by the number of
``internal'' edges inside that set. Thus, to be a good community, a set of
nodes should have small conductance, \emph{i.e.}, it should have many
internal edges and few edges pointing to the rest of the network. 
Conductance is widely used to capture the intuition of a good community;
it is a fundamental combinatorial quantity; and it has a very natural
interpretation in terms of random walks on the interaction graph.
Moreover, since there exist a rich suite of both theoretical and practical
algorithms~\cite{hendrickson95,spielman96_spectral,Leighton:1988,Leighton:1999,Arora:2004,karypis98_metis,karypis98metis,zhao04cluto,dhillon07graclus},
we can for point (4) compare and contrast several methods to approximately
optimize it.
To illustrate conductance, note that of the three $5$-node sets $A$, $B$, 
and $C$ illustrated in the graph in Figure~\ref{fig:conduct2}, $B$ has the 
best (the lowest) conductance and is thus the most community-like.


However, it is in point (5) that we deviate from previous work. Instead of
focusing on individual groups of nodes and trying to interpret them as
``real'' communities, we investigate statistical properties of a large
number of communities over a wide range of size scales in over
$100$ large sparse real-world social and information networks.
We take a step back and ask questions such as: How well do real graphs
split into communities? What is a good way to measure and characterize
presence or absence of community structure in networks? What are typical
community sizes and typical community qualities?

To address these and related questions, we introduce the concept of a
\emph{network community profile (NCP) plot} that we define and describe in
more detail in Section~\ref{sxn:ncpp:def}. Intuitively, the network
community profile plot measures the score of ``best'' community as a
function of community size in a network. Formally, we define it as the
conductance value of the minimum conductance set of cardinality $k$ in the
network, as a function of $k$. As defined, the NCP plot will be NP-hard to 
compute exactly,
so operationally we will use several natural approximation algorithms for
solving the Minimum Conductance Cut Problem in order to compute different
approximations to it. By comparing and
contrasting these plots for a large number of networks, and by computing
other related structural properties, we obtain results that suggest a
significantly more refined picture of the community structure in large
real-world networks than has been appreciated previously.


We have gone to a great deal of effort to be confident that we are
computing quantities fundamental to the networks we are considering,
rather than artifacts of the approximation algorithms we employ. 
In particular:
\begin{itemize}
\item
We
use several classes of graph partitioning algorithms to probe the networks
for sets of nodes that could plausibly be interpreted as communities.
These algorithms, including flow-based methods, spectral methods, and
hierarchical methods, have complementary strengths and weaknesses that are
well understood both in theory and in practice. For example, flow-based
methods are known to have difficulties with
expanders~\cite{Leighton:1988,Leighton:1999}, and flow-based
post-processing of other methods are known in practice to yield cuts with
extremely good conductance values~\cite{Lang04_BalanceTR,kevin04mqi}. On
the other hand, spectral methods are known to have difficulties when they
confuse long paths with deep
cuts~\cite{spielman96_spectral,guatterymiller98}, a consequence of which
is that they may be viewed as computing a ``regularized'' approximation to
the network community profile plot. (See Section~\ref{algo-notes-section}
for a more detailed discussion of these and related issues.) 
\item 
We
compute spectral-based lower bounds and also
semidefinite-programming-based lower bounds for the conductance of our
network datasets. 
\item
We compute a wide range of other structural
properties of the networks, {\em e.g.}, sizes, degree distributions,
maximum and average diameters of the purported communities, internal
versus external conductance values of the purported communities, etc.
\item
We recompute statistics on versions of the networks that have been
modified in well-understood ways, {\em e.g.}, by removing small
barely-connected sets of nodes or by randomizing the edges. 
\item
We
compare our results across not only over $100$ large social and
information networks, but also numerous commonly-studied small social
networks, expanders, and low-dimensional manifold-like objects, and we
compare our results on each network with what is known from the field from
which the network is drawn. To our knowledge, this makes ours the most
extensive such analysis of the community structure in large real-world
social and information networks. 
\item
We compare results with
analytical and/or simulational results on a wide range of commonly and
not-so-commonly used network generation
models~\cite{newman2003_review,bollobas03_review,barabasi99emergence,kumar00stochastic,ravasz03_hierarchical,jure05dpl,FFV04_geometric1,FFV07_geometric2}.
\end{itemize}

\subsection{Summary of our results}
\label{sxn:intro:results}

{\bf Main Empirical Findings:} Taken as a whole, the results we present in
this paper suggest a rather detailed and somewhat counterintuitive picture
of the community structure in large social and information networks.
Several qualitative properties of community structure, as revealed by the
network community profile plot, are nearly universal:
\begin{itemize}
\item Up to a size scale, which empirically is roughly $100$ nodes,
    there not only exist cuts with relatively good conductance, {\em
    i.e.}, good communities, but also the slope of the network
    community profile plot is generally sloping downward. This latter
    point suggests that smaller communities can be combined into
    meaningful larger communities, a phenomenon that we empirically
    observe in many cases.
\item At the size scale of roughly $100$ nodes, we often observe the
    global minimum of the network community profile plot; these are
    the ``best'' communities, according to the conductance measure, in
    the entire graph. These are, however, rather interestingly
    connected to the rest of the network; for example, in most cases,
    we observe empirically that they are a small set of nodes barely
    connected to the remainder of the network by just a {\em single}
    edge.
\item 
Above the size scale of roughly $100$ nodes, the network community profile 
plot gradually increases, and thus there is a nearly inverse relationship 
between community size and community quality.  As a
function of increasing size, the best possible communities become more and 
more ``blended into'' the remainder of the network.
Intuitively, communities blend in with one another and gradually disappear 
as they grow larger. 
In particular, in many cases, larger communities can be broken into smaller and 
smaller pieces, often recursively, each of which is more community-like than 
the original supposed community.
\item Even up to the largest size scales, we observe significantly
    more structure than would be seen, for example, in an
    expander-like random graph on the same degree sequence.
\end{itemize}

\begin{figure}
	\begin{center}
		\subfigure[Typical NCP plot]{
			\includegraphics[width=0.40\textwidth]{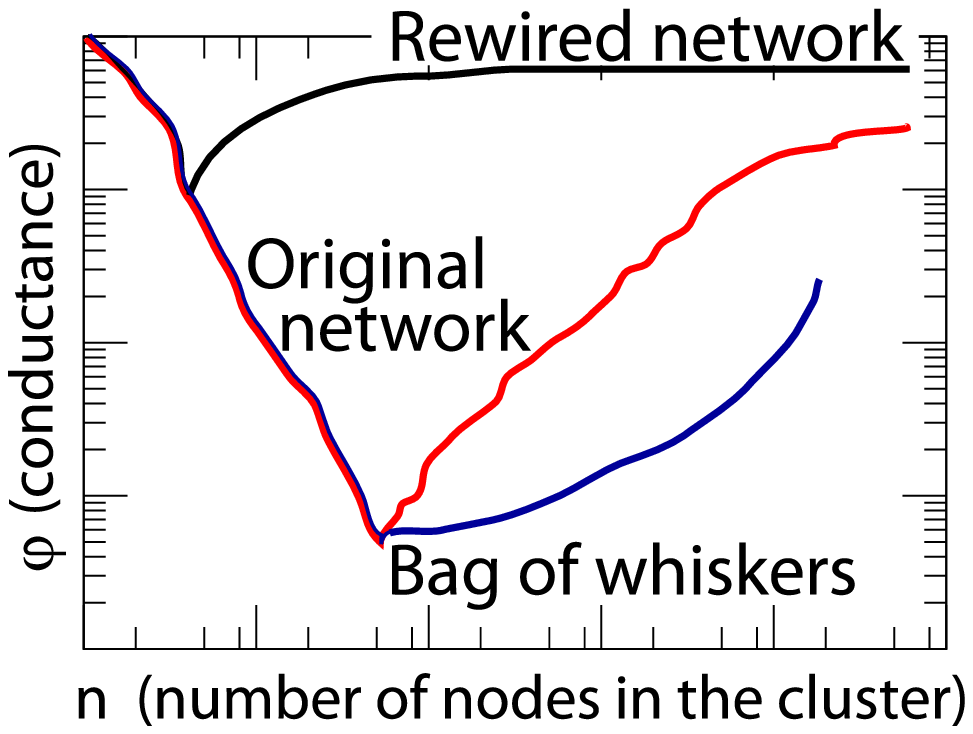} \quad
			\label{fig:intro_ncpp}
		}
		\subfigure[Caricature of network structure]{
    			\includegraphics[width=0.45\textwidth]{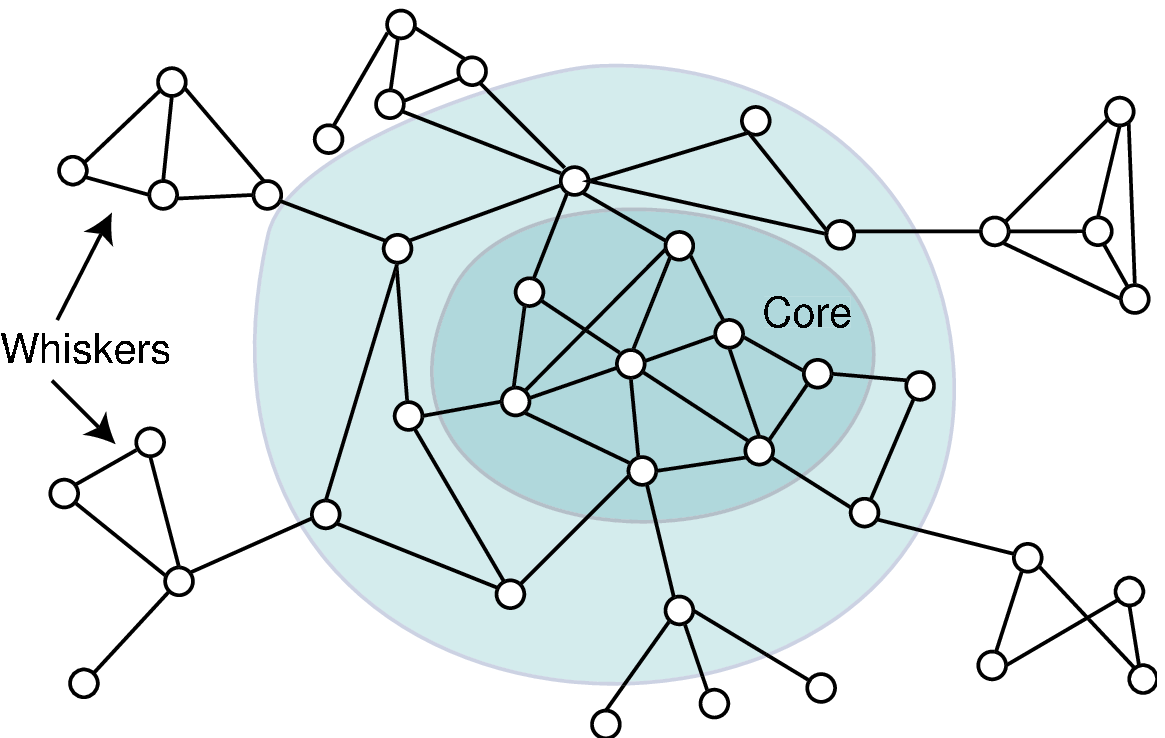} \quad
			\label{fig:intro_graph}
		}
	\end{center}
\caption{ (a) Typical network community profile plot for a large social or
information network: networks have better and better communities up to 
a size scale of $\approx 100$ nodes, and after that size scale communities 
``blend-in'' with
the rest of the network (red curve). However, real networks still have
more structure than their randomized (conditioned on the same degree
distribution) counterparts (black curve). Even more surprisingly, if one
allows for disconnected communities (blue curve), the community quality 
scores often get even
better (even though such communities have no intuitive meaning).
(b) Network structure for a large social or information network,
as suggested by our empirical evaluations.
See the text for more information on the ``core'' and ``whiskers,'' and note
that the core in our real-world networks is actually extremely sparse.
}
\label{fig:intro}
\end{figure}

A schematic picture of a typical network community profile plot is
illustrated in Figure~\ref{fig:intro_ncpp}. In red (labeled as ``original
network''), we plot community size vs. community quality score for the sets of
nodes extracted from the original network.  In black (rewired network), we plot
the scores of communities extracted from a random network conditioned on
the same degree distribution as the original network.
This illustrates not only tight communities at very small scales, but also that 
at larger and larger size scales (the precise cutoff point for which is 
difficult to specify precisely) the best possible communities gradually 
``blend in'' more and more with the rest of the network and thus gradually 
become less and less community-like.
Eventually, even the existence of large well-defined communities is quite
questionable if one models the world with an interaction graph, as in point~(1) 
above, and if one also defines good communities as densely linked clusters that 
are weakly-connected to the outside, as 
in hypothesis~(2) above.
Finally, in blue (bag of whiskers), we also plot the scores of communities that
are composed of disconnected pieces (found according to a procedure we describe 
in Section~\ref{sxn:obs_struct}).
This blue curve shows, perhaps somewhat surprisingly, that one can often 
obtain better community quality scores by combining unrelated disconnected pieces.

To understand the properties of generative models sufficient to reproduce
the phenomena we have observed, we have examined in detail the structure
of our social and information networks. Although nearly every
network is an exception to any simple rule, we have observed that an
``octopus'' or ``jellyfish''
model~\cite{ChungLu:2006,tauro01topology,siganos06_jellyfish} provides a
rough first approximation to structure of many of the networks we have
examined. That is, most networks may be viewed as having a ``core,'' with
no obvious underlying geometry and which contains a constant fraction of
the nodes, and then there is a periphery consisting of a large number of relatively small
``whiskers'' that are only tenuously connected to the core.
Figure~\ref{fig:intro_graph} presents a caricature of this network
structure. Of course, our network datasets are far from random in numerous
ways---{\em e.g.}, they have higher edge density in the core; the small
barely-connected whisker-like pieces are generally larger, denser, and
more common than in corresponding random graphs; they have higher local
clustering coefficients; and this local clustering information gets
propagated globally into larger clusters or communities in a subtle and
location-specific manner.
More interestingly, as shown in Figure~\ref{fig:phiCore} in
Section~\ref{sxn:obs_struct:remove_whiskers}, the core itself consists of a
nested core-periphery structure.

{\bf Main Modeling Results:} 
The behavior that we observe is not reproduced,
at even a qualitative level, by any of the commonly-used network
generation models we have examined, including but not limited to
preferential attachment models, copying models, small-world models, and
hierarchical network models. Moreover, this behavior is qualitatively
different than what is observed in networks with an underlying mesh-like
or manifold-like geometry (which may not be surprising, but is significant
insofar as these structures are often used as a scaffolding upon which to
build other models), in networks that are good expanders (which may be
surprising, since it is often observed that large social networks are
expander-like), and in small social networks such as those used as
testbeds for community detection algorithms (which may have implications
for the applicability of these methods to detect large community-like
structures in these networks). For the commonly-used network generation
models, as well as for expander-like, low-dimensional, and small social
networks, the network community profile plots are generally downward
sloping or relatively flat.

Although it is well understood at a qualitative level that nodes that are
``far apart'' or ``less alike'' (in some sense) should be less likely to
be connected in a generative model, understanding this point
quantitatively so as to reproduce the empirically-observed relationship
between small-scale and large-scale community structure turns out to be
rather subtle. We can make the following observations:
\begin{itemize}

\item Very sparse random graph models with no underlying geometry have
    relatively deep cuts at small size scales, the best cuts at large
    size scales are very shallow, and there is a relatively abrupt
    transition in between. (This is shown pictorially in
    Figure~\ref{fig:intro_ncpp} for a randomly rewired version of the
    original network.) This is a consequence of the extreme sparsity
    of the data: sufficiently dense random graphs do not have these
    small deep cuts; and the relatively deep cuts in sparse graphs are
    due to small tree-like pieces that are connected by a single edge
    to a core which is an extremely good expander.
\item A Forest Fire generative
    model~\cite{jure05dpl,jure07evolution}, in which edges are added
    in a manner that imitates a fire-spreading process, reproduces not
    only the deep cuts at small size scales and the absence of deep
    cuts at large size scales but other properties as well: the small
    barely connected pieces are significantly larger and denser than
    random; and for appropriate parameter settings the network
    community profile plot increases relatively gradually as the size
    of the communities increases.
\item The details of the ``forest fire'' burning mechanism are crucial
    for reproducing how local clustering information gets propagated
    to larger size scales in the network, and those details shed light
    on the failures of commonly-used network generation models. In the
    Forest Fire Model, a new node selects a ``seed'' node and links to
    it. Then with some probability it ``burns'' or adds an edge to the
    each of the seed's neighbors, and so on, recursively. Although
    there is a ``preferential attachment'' and also a ``copying''
    flavor to this mechanism, two factors are particularly important:
    first is the local (in a graph sense, as there is no underlying 
    geometry in the model) manner in which the edges
    are added; and second is that the number of edges that a new node
    can add can vary widely, depending on the local structure around
    the seed node. Depending on the neighborhood structure around the
    seed, small fires will keep the community well-separated from the
    network, but occasional large fires will connect the community to
    the rest of the network and make it blend into the network core.
\end{itemize}
Thus, intuitively, the structure of the whiskers (components connected to
the rest of the graph via a single edge) are responsible for the downward
part of the network community profile plot, while the core of the network
and the manner in which the whiskers root themselves to the core helps to
determine the upward part of the network community profile plot. Due to
local clustering effects, whiskers in real networks are larger and give
deeper cuts than whiskers in corresponding randomized graphs, fluctuations
in the core are larger and deeper than in corresponding randomized graphs,
and thus the network community profile plot increases more gradually and
levels off to a conductance value well below the value for a corresponding
rewired network.

{\bf Main Methodological Contributions:} To obtain these and other
conclusions, we have employed approximation algorithms for graph
partitioning to investigate structural properties of our network datasets.
Briefly, we have done the following:

\begin{itemize}
\item We have used Metis+MQI, which consists of using the popular
    graph partitioning package Metis~\cite{karypis98_metis} followed
    by a flow-based MQI post-processing~\cite{kevin04mqi}. With this
    procedure, we obtain sets of nodes that have very good
    conductance scores. At very small size scales, these sets of nodes
    could plausibly be interpreted as good communities, but at larger
    size scales, we often obtain tenuously-connected (and in some
    cases unions of disconnected) pieces, which perhaps do not
    correspond to intuitive communities.
\item Thus, we have also used the Local Spectral method of Anderson,
    Chung, and Lang~\cite{andersen06local} to obtain sets of nodes
    with good conductance value that that are ``compact'' or more
    ``regularized'' than those pieces returned by Metis+MQI. Since
    spectral methods confuse long paths with deep
    cuts~\cite{spielman96_spectral,guatterymiller98}, empirically we
    obtain sets of nodes that have worse conductance scores than sets
    returned by Metis+MQI, but which are ``tighter'' and more
    ``community-like.'' For example, at small size scales the sets of
    nodes returned by the Local Spectral Algorithm agrees with the
    output of Metis+MQI, but at larger scales this algorithm returns
    sets of nodes with substantially smaller diameter and average
    diameter, which seem plausibly more community-like.
\end{itemize}
We have also used what we call the Bag-of-Whiskers Heuristic to identify
small barely connected sets of nodes that exert a surprisingly large
influence on the network community profile plot.

Both Metis+MQI and the Local Spectral Algorithm scale well and thus either
may be used to obtain sets of nodes from very large graphs. For many of
the small to medium-sized networks, we have checked our results by
applying one or more other spectral, flow-based, or heuristic algorithms,
although these do not scale as well to very large graphs. Finally, for
some of our smaller network datasets, we have computed spectral-based and
semidefinite-programming-based lower bounds, and the results are
consistent with the conclusions we have drawn.

{\bf Broader implications: } 
Our observation that, independently of the network size, compact 
communities exist only up to a size scale of around $100$ nodes agrees 
well with the ``Dunbar number''~\cite{dunbar98gossip}, which predicts 
that roughly $150$ individuals is the upper limit on the size of a well-functioning human 
community. 
Moreover, we should emphasize that our results do not disagree with the 
literature at small sizes scales.  
One reason for the difference in our findings is that previous studies 
mainly focused on small networks, which are simply not large enough for 
the clusters to gradually blend into one another as one looks at larger 
size scales. 
In order to make our observations, one needs to look at large number (due 
to the complex noise properties of real graphs) of large networks. 
It is only when Dunbar's limit is exceeded by several orders of magnitude 
that it is relatively easy to observe large communities blurring together 
and eventually vanishing. 
A second reason for the difference is that previous work did not measure 
and examine the {\em network community profile} of cluster size vs. cluster 
quality.
Finally, we should note that our explanation also aligns well with the 
{\em common bond} vs. {\em
common identity} theory of group attachment~\cite{ren07bond} from social
psychology, where it has been noted that bond communities tend to be
smaller and more cohesive~\cite{back51influence}, as they are based on
interpersonal ties, while identity communities are focused around common
theme or interest.
We discuss these implications and connections further in
Section~\ref{sxn:discussion}.

\subsection{Outline of the paper}
\label{sxn:intro:outline}

The rest of the paper is organized as follows. In
Section~\ref{sxn:related} we describe some useful background, including a
brief description of the network datasets we have analyzed. Then, in
Section~\ref{sxn:ncpp} we present our main results on the properties of
the network community profile plot for our network datasets. We place an
emphasis on how the phenomena we observe in large social and information
networks are qualitatively different than what one would expect based on
intuition from and experience with expander-like graphs, low-dimensional
networks, and commonly-studied small social networks. Then, in
Sections~\ref{sxn:obs_struct} and~\ref{algo-notes-section}, we summarize
the results of additional empirical evaluations. In particular, in
Section~\ref{sxn:obs_struct}, we describe some of the observations we have
made in an effort to understand what structural properties of these large
networks are responsible for the phenomena we observe; and in
Section~\ref{algo-notes-section}, we describe some of the results of
probing the networks with different approximation algorithms in an effort
to be confident that the phenomena we observed really are properties of
the networks we study, rather than artifactual properties of the
algorithms we chose to use to study those networks. We follow this in
Section~\ref{sxn:models} with a discussion of complex network generation
models. We observe that the commonly-used network generation models fail
to reproduce the counterintuitive phenomena we observe. We also notice
that very sparse random networks reproduce certain aspects of the
phenomena, and that a generative model based on an iterative ``forest
fire'' burning mechanism reproduces very well the qualitative properties
of the phenomena we observe. Finally, in Section~\ref{sxn:discussion} we
provide a discussion of our results in a broader context, and in
Section~\ref{sxn:conclusion} we present a brief conclusion.
\section{Background on communities and overview of our methods}
\label{sxn:related}

In this section, we will provide background on our data and methods. We
start in Section~\ref{sxn:related:NetworkData} with a description of the
network datasets we will analyze. Then, in
Section~\ref{sxn:related:clusters}, we review related community detection
and graph clustering ideas. Finally, in
Section~\ref{sxn:related:LowCondAlgs}, we provide a brief description of
approximation algorithms that we will use. There exist a large number of
reviews on topics related to those discussed in this paper. For example,
see the reviews on community
identification~\cite{newman2004_detect,danon05community}, data
clustering~\cite{jain99data}, graph and spectral
clustering~\cite{gaertler05_clustering,luxburg05_survey,Schaeffer07_survey},
graph and heavy-tailed data
analysis~\cite{newman2005_zipf,CF06_survey,CSN07_powerlaw}, surveys on
various aspects of complex
networks~\cite{barabasi02_reviewRMP,dorogovtsev02_reviewAIP,newman2003_review,bollobas03_review,CRTV05_survey,LDW06_survey,BLMCW06_survey},
the monographs on spectral graph theory and complex
networks~\cite{Chung:1997,ChungLu:2006}, and the book on social network
analysis~\cite{WassermanFaust94}. See Section~\ref{sxn:discussion} for a
more detailed discussion of the relationship of our work with some of this
prior work.

\subsection{Social and information network datasets we analyze}
\label{sxn:related:NetworkData}

We have examined a large number of real-world complex networks.
See Tables~\ref{tab:data_StatsDesc_1}, \ref{tab:data_StatsDesc_2},
and~\ref{tab:data_StatsDesc_3} for a summary.
For convenience, we have organized the networks into the following categories:
Social networks;
Information/citation networks;
Collaboration networks;
Web graphs;
Internet networks;
Bipartite affiliation networks;
Biological networks;
Low-dimensional networks;
IMDB networks; and
Amazon networks.
We have also examined numerous small social networks that have been used as a
testbed for community detection algorithms ({\em e.g.}, Zachary's karate
club~\cite{zachary77karate,newman_netdata}, interactions between
dolphins~\cite{lusseau03dolphins,newman_netdata}, interactions between
monks~\cite{sampson68monks,newman_netdata}, Newman's network science
network~\cite{newman2006finding,newman_netdata}, etc.),
numerous simple network models in which by design there is an underlying
geometry ({\em e.g.}, power grid and road networks~\cite{watts98collective},
simple meshes, low-dimensional manifolds including graphs corresponding to the
well-studied ``swiss roll'' data set~\cite{isomap00_science}, a geometric
preferential attachment model~\cite{FFV04_geometric1,FFV07_geometric2}, etc.),
several networks that are very good expanders, and many simulated networks
generated by commonly-used network generation models({\em e.g.}, preferential
attachment models~\cite{newman2003_review}, copying
models~\cite{kumar00stochastic}, hierarchical
models~\cite{ravasz03_hierarchical}, etc.).


\begin{sidewaystable}
\begin{center}
{\footnotesize
\begin{tabular}{l|r|r|r|r|r|r|r|r|r|l}
Network & $N$ & $E$ & $N_b$ & $E_b$ & $\bar{d}$ & $\tilde{d}$ & $\bar{C}$ & $D$ & $\bar{D}$  & Description \\
\hline \hline
\multicolumn{11}{l}{Social networks} \\ 
\hline \hline
\net{Delicious} & 147,567 & 301,921 & 0.40 & 0.65 & 4.09 & 48.44 & 0.30 & 24 & 6.28  &  \url{del.icio.us} collaborative tagging social network  \\
\net{Epinions} & 75,877 & 405,739 & 0.48 & 0.90 & 10.69 & 183.88 & 0.26 & 15 & 4.27  &  Who-trusts-whom network from epinions.com~\cite{richardson03trust}  \\
\net{Flickr} & 404,733 & 2,110,078 & 0.33 & 0.86 & 10.43 & 442.75 & 0.40 & 18 & 5.42  &  Flickr photo sharing social network~\cite{ravi06structure}  \\
\net{LinkedIn} & 6,946,668 & 30,507,070 & 0.47 & 0.88 & 8.78 & 351.66 & 0.23 & 23 & 5.43  &  Social network of professional contacts  \\
\net{LiveJournal01} & 3,766,521 & 30,629,297 & 0.78 & 0.97 & 16.26 & 111.24 & 0.36 & 23 & 5.55  &  Friendship network of a blogging community~\cite{lars06groups}  \\
\net{LiveJournal11} & 4,145,160 & 34,469,135 & 0.77 & 0.97 & 16.63 & 122.44 & 0.36 & 23 & 5.61  &  Friendship network of a blogging community~\cite{lars06groups}  \\
\net{LiveJournal12} & 4,843,953 & 42,845,684 & 0.76 & 0.97 & 17.69 & 170.66 & 0.35 & 20 & 5.53  &  Friendship network of a blogging community~\cite{lars06groups}  \\
\net{Messenger} & 1,878,736 & 4,079,161 & 0.53 & 0.78 & 4.34 & 15.40 & 0.09 & 26 & 7.42  &  Instant messenger social network  \\
\net{Email-All} & 234,352 & 383,111 & 0.18 & 0.50 & 3.27 & 576.87 & 0.50 & 14 & 4.07  &  Research organization email network (all addresses)~\cite{jure07evolution}  \\
\net{Email-InOut} & 37,803 & 114,199 & 0.47 & 0.82 & 6.04 & 165.73 & 0.58 & 8 & 3.74  &  (all addresses but email has to be sent both ways)~\cite{jure07evolution}  \\
\net{Email-Inside} & 986 &	16,064 &	0.90 &	0.99 &	32.58 &	74.66 &	0.45 &	7 &	2.60 &  (only emails inside the research organization)~\cite{jure07evolution}  \\
\net{Email-Enron} & 33,696 & 180,811 & 0.61 & 0.90 & 10.73 & 142.36 & 0.71 & 13 & 3.99  &  Enron email dataset~\cite{klimt04enron}  \\
\net{Answers} & 488,484 & 1,240,189 & 0.45 & 0.78 & 5.08 & 251.78 & 0.11 & 22 & 5.72  &  Yahoo Answers social network  \\
\net{Answers-1} & 26,971 & 91,812 & 0.56 & 0.87 & 6.81 & 59.17 & 0.08 & 16 & 4.49  &  Cluster 1 from Yahoo Answers  \\
\net{Answers-2} & 25,431 & 65,551 & 0.48 & 0.80 & 5.16 & 56.57 & 0.10 & 15 & 4.76  &  Cluster 2 from Yahoo Answers  \\
\net{Answers-3} & 45,122 & 165,648 & 0.53 & 0.87 & 7.34 & 417.83 & 0.21 & 15 & 3.94  &  Cluster 3 from Yahoo Answers  \\
\net{Answers-4} & 93,971 & 266,199 & 0.49 & 0.82 & 5.67 & 94.48 & 0.08 & 16 & 4.91  &  Cluster 4 from Yahoo Answers  \\
\net{Answers-5} & 5,313 & 11,528 & 0.41 & 0.73 & 4.34 & 29.55 & 0.12 & 14 & 4.75  &  Cluster 5 from Yahoo Answers  \\
\net{Answers-6} & 290,351 & 613,237 & 0.40 & 0.71 & 4.22 & 57.16 & 0.09 & 22 & 5.92  &  Cluster 6 from Yahoo Answers  \\
\hline \hline
\multicolumn{11}{l}{Information (citation) networks} \\ 
\hline \hline
\net{Cit-Patents} & 3,764,105 & 16,511,682 & 0.82 & 0.96 & 8.77 & 21.34 & 0.09 & 26 & 8.15  &  Citation network of all US patents~\cite{jure05dpl}  \\
\net{Cit-hep-ph} & 34,401 & 420,784 & 0.96 & 1.00 & 24.46 & 63.50 & 0.30 & 14 & 4.33  &  Citations between physics (arxiv {\tt hep-th}) papers~\cite{gehrke03kddcup}  \\
\net{Cit-hep-th} & 27,400 & 352,021 & 0.94 & 0.99 & 25.69 & 106.40 & 0.33 & 15 & 4.20  &  Citations between physics (arxiv {\tt hep-ph}) papers~\cite{gehrke03kddcup}  \\
\net{Blog-nat05-6m} & 29,150 & 182,212 & 0.74 & 0.96 & 12.50 & 342.51 & 0.24 & 10 & 3.40  &  Blog citation network (6 months of data)~\cite{jure07cascades}  \\
\net{Blog-nat06all} & 32,384 & 315,713 & 0.87 & 0.99 & 19.50 & 153.08 & 0.20 & 18 & 3.94  &  Blog citation network (1 year of data)~\cite{jure07cascades}  \\
\net{Post-nat05-6m} & 238,305 & 297,338 & 0.21 & 0.34 & 2.50 & 39.51 & 0.13 & 45 & 10.34  &  Blog post citation network (6 months)~\cite{jure07cascades}  \\
\net{Post-nat06all} & 437,305 & 565,072 & 0.22 & 0.38 & 2.58 & 35.54 & 0.11 & 54 & 10.48  &  Blog post citation network (1 year)~\cite{jure07cascades}  \\
\hline \hline
\multicolumn{11}{l}{Collaboration networks} \\ 
\hline \hline
\net{AtA-IMDB} & 883,963 & 27,473,042 & 0.87 & 0.99 & 62.16 & 517.40 & 0.79 & 15 & 3.48  &  IMDB actor collaboration network from Dec 2007 \\
\net{CA-astro-ph} & 17,903 & 196,972 & 0.89 & 0.98 & 22.00 & 65.70 & 0.67 & 14 & 4.21  &  Co-authorship in {\tt astro-ph} of arxiv.org~\cite{jure05dpl}  \\
\net{CA-cond-mat} & 21,363 & 91,286 & 0.81 & 0.93 & 8.55 & 22.47 & 0.70 & 15 & 5.36  &  Co-authorship in {\tt cond-mat} category~\cite{jure05dpl}  \\
\net{CA-gr-qc} & 4,158 & 13,422 & 0.64 & 0.78 & 6.46 & 17.98 & 0.66 & 17 & 6.10  &  Co-authorship in {\tt gr-qc} category~\cite{jure05dpl}  \\
\net{CA-hep-ph} & 11,204 & 117,619 & 0.81 & 0.97 & 21.00 & 130.88 & 0.69 & 13 & 4.71  &  Co-authorship in {\tt hep-ph} category~\cite{jure05dpl}  \\
\net{CA-hep-th} & 8,638 & 24,806 & 0.68 & 0.85 & 5.74 & 12.99 & 0.58 & 18 & 5.96  &  Co-authorship in {\tt hep-th} category~\cite{jure05dpl}  \\
\net{CA-DBLP} & 317,080 & 1,049,866 & 0.67 & 0.84 & 6.62 & 21.75 & 0.73 & 23 & 6.75  &  DBLP co-authorship network~\cite{lars06groups}  \\
\end{tabular}
}
\end{center}
\caption{
Network datasets we analyzed. 
Statistics of networks we consider: 
number of nodes $N$; 
number of edges $E$; 
fraction nodes not in whiskers (size of largest biconnected component) $N_b/N$; 
fraction of edges in biconnected component $E_b/E$; 
average degree $\bar{d}=2E/N$; 
second order average degree $\tilde{d}$; 
average clustering coefficient $\bar{C}$; 
diameter $D$; and
average path length $\bar{D}$. 
}
\label{tab:data_StatsDesc_1}
\end{sidewaystable}


\begin{sidewaystable}
\begin{center}
{\footnotesize
\begin{tabular}{l|r|r|r|r|r|r|r|r|r|l}
Network & $N$ & $E$ & $N_b$ & $E_b$ & $\bar{d}$ & $\tilde{d}$ & $\bar{C}$ & $D$ & $\bar{D}$  & Description \\
\hline \hline
\multicolumn{11}{l}{Web graphs} \\ 
\hline \hline
\net{Web-BerkStan} & 319,717 & 1,542,940 & 0.57 & 0.88 & 9.65 & 1,067.55 & 0.32 & 35 & 5.66  &  Web graph of Stanford and UC Berkeley~\cite{ashraf04pagerank}  \\
\net{Web-Google} & 855,802 & 4,291,352 & 0.75 & 0.92 & 10.03 & 170.35 & 0.62 & 24 & 6.27  &  Web graph Google released in 2002~\cite{google02}  \\
\net{Web-Notredame} & 325,729 & 1,090,108 & 0.41 & 0.76 & 6.69 & 280.68 & 0.47 & 46 & 7.22  &  Web graph of University of Notre Dame~\cite{albert1999dww}  \\
\net{Web-Trec} & 1,458,316 & 6,225,033 & 0.59 & 0.78 & 8.54 & 682.89 & 0.68 & 112 & 8.58  &  Web graph of TREC WT10G web corpus~\cite{trec00wt10g}  \\
\hline \hline 
\multicolumn{11}{l}{Internet networks} \\ 
\hline \hline
\net{As-RouteViews} & 6,474 & 12,572 & 0.62 & 0.80 & 3.88 & 164.81 & 0.40 & 9 & 3.72  &  AS from Oregon Exchange BGP Route View~\cite{jure05dpl}  \\
\net{As-Caida} & 26,389 & 52,861 & 0.61 & 0.81 & 4.01 & 281.93 & 0.33 & 17 & 3.86  &  CAIDA AS Relationships Dataset  \\
\net{As-Skitter} & 1,719,037 & 12,814,089 & 0.99 & 1.00 & 14.91 & 9,934.01 & 0.17 & 5 & 3.44  &  AS from traceroutes run daily in 2005 by Skitter  \\
\net{As-Newman} & 22,963 & 48,436 & 0.65 & 0.83 & 4.22 & 261.46 & 0.35 & 11 & 3.83  &  AS graph from Newman~\cite{newman_netdata}  \\
\net{As-Oregon} & 13,579 & 37,448 & 0.72 & 0.90 & 5.52 & 235.97 & 0.46 & 9 & 3.58  &  Autonomous systems~\cite{AsOregon}  \\
\net{Gnutella-25} & 22,663 & 54,693 & 0.59 & 0.83 & 4.83 & 10.75 & 0.01 & 11 & 5.57  &  Gnutella network on March 25 2000~\cite{ripeanu02gnutella}  \\
\net{Gnutella-30} & 36,646 & 88,303 & 0.55 & 0.81 & 4.82 & 11.46 & 0.01 & 11 & 5.75  &  Gnutella P2P network on March 30 2000~\cite{ripeanu02gnutella}  \\
\net{Gnutella-31} & 62,561 & 147,878 & 0.54 & 0.81 & 4.73 & 11.60 & 0.01 & 11 & 5.94  &  Gnutella network on March 31 2000~\cite{ripeanu02gnutella}  \\
\net{eDonkey} & 5,792,297 & 147,829,887 & 0.93 & 1.00 & 51.04 & 6,139.99 & 0.08 & 5 & 3.66  &  P2P eDonkey graph for a period of 47 hours in 2004  \\
\hline \hline 
\multicolumn{11}{l}{Bi-partite networks} \\ 
\hline \hline
\net{IpTraffic} & 2,250,498 & 21,643,497 & 1.00 & 1.00 & 19.23 & 94,889.05 & 0.00 & 5 & 2.53  &  IP traffic graph a single router for 24 hours \\
\net{AtP-astro-ph} & 54,498 & 131,123 & 0.70 & 0.87 & 4.81 & 16.67 & 0.00 & 28 & 7.78  &  Authors-to-papers network of {\tt astro-ph}~\cite{jure07cascades}  \\
\net{AtP-cond-mat} & 57,552 & 104,179 & 0.65 & 0.79 & 3.62 & 10.54 & 0.00 & 31 & 9.96  &  Authors-to-papers network of {\tt cond-mat}~\cite{jure07cascades}  \\
\net{AtP-gr-qc} & 14,832 & 22,266 & 0.47 & 0.60 & 3.00 & 9.72 & 0.00 & 35 & 11.08  &  Authors-to-papers network of {\tt gr-qc}~\cite{jure07cascades}  \\
\net{AtP-hep-ph} & 47,832 & 86,434 & 0.60 & 0.76 & 3.61 & 16.80 & 0.00 & 27 & 8.55  &  Authors-to-papers network of {\tt hep-ph}~\cite{jure07cascades}  \\
\net{AtP-hep-th} & 39,986 & 64,154 & 0.53 & 0.68 & 3.21 & 13.07 & 0.00 & 36 & 10.74  &  Authors-to-papers network of {\tt hep-th}~\cite{jure07cascades}  \\
\net{AtP-DBLP} & 615,678 & 944,456 & 0.49 & 0.64 & 3.07 & 13.61 & 0.00 & 48 & 12.69  &  DBLP authors-to-papers bipartite network  \\
\net{Spending} & 1,831,540 & 2,918,920 & 0.34 & 0.58 & 3.19 & 1,536.35 & 0.00 & 26 & 5.62  &  Users-to-keywords they bid  \\
\net{Hw7} & 653,260 & 2,278,448 & 0.99 & 0.99 & 6.98 & 346.85 & 0.00 & 24 & 6.26  &  Downsampled advertiser-query bid graph \\
\net{Netflix} & 497,959 & 100,480,507 & 1.00 & 1.00 & 403.57 & 28,432.89 & 0.00 & 5 & 2.31  &  Users-to-movies they rated. From Netflix prize~\cite{netflixprize}  \\
\net{QueryTerms} & 13,805,808 & 17,498,668 & 0.28 & 0.41 & 2.53 & 14.92 & 0.00 & 86 & 19.81  &  Users-to-queries they submit to a search engine  \\
\net{Clickstream} & 199,308 & 951,649 & 0.39 & 0.87 & 9.55 & 430.74 & 0.00 & 7 & 3.83  &  Users-to-URLs they visited~\cite{montgomery01clickstream}  \\
\hline \hline 
\multicolumn{11}{l}{Biological networks} \\ 
\hline \hline
\net{Bio-Proteins} & 4,626 & 14,801 & 0.72 & 0.91 & 6.40 & 24.25 & 0.12 & 12 & 4.24  &  Yeast protein interaction network~\cite{coliza05protein}  \\
\net{Bio-Yeast} & 1,458 & 1,948 & 0.37 & 0.51 & 2.67 & 7.13 & 0.14 & 19 & 6.89  &  Yeast protein interaction network data~\cite{jeong01protein}  \\
\net{Bio-YeastP0.001} & 353 & 1,517 & 0.73 & 0.93 & 8.59 & 20.18 & 0.57 & 11 & 4.33  &  Yeast protein-protein interaction map~\cite{qi05ppi}  \\
\net{Bio-YeastP0.01} & 1,266 & 8,511 & 0.79 & 0.97 & 13.45 & 47.73 & 0.44 & 12 & 3.87  &  Yeast protein-protein interaction map~\cite{qi05ppi}  \\
\end{tabular}
}
\end{center}
\caption{
Network datasets we analyzed. 
Statistics of networks we consider: 
number of nodes $N$; 
number of edges $E$; 
fraction nodes not in whiskers (size of largest biconnected component) $N_b/N$; 
fraction of edges in biconnected component $E_b/E$; 
average degree $\bar{d}=2E/N$; 
second order average degree $\tilde{d}$; 
average clustering coefficient $\bar{C}$; 
diameter $D$; and
average path length $\bar{D}$. 
}
\label{tab:data_StatsDesc_2}
\end{sidewaystable}


\begin{sidewaystable}
\begin{center}
{\footnotesize
\begin{tabular}{l|r|r|r|r|r|r|r|r|r|l}
Network & $N$ & $E$ & $N_b$ & $E_b$ & $\bar{d}$ & $\tilde{d}$ & $\bar{C}$ & $D$ & $\bar{D}$  & Description \\
\hline \hline 
\multicolumn{11}{l}{Nearly low-dimensional networks} \\ 
\hline \hline
\net{Road-CA} & 1,957,027 & 2,760,388 & 0.80 & 0.85 & 2.82 & 3.17 & 0.06 & 865 & 310.97  &  California road network  \\
\net{Road-USA} & 126,146 & 161,950 & 0.97 & 0.98 & 2.57 & 2.81 & 0.03 & 617 & 218.55  &  USA road network (only main roads)  \\
\net{Road-PA} & 1,087,562 & 1,541,514 & 0.79 & 0.85 & 2.83 & 3.20 & 0.06 & 794 & 306.89  &  Pennsylvania road network  \\
\net{Road-TX} & 1,351,137 & 1,879,201 & 0.78 & 0.84 & 2.78 & 3.15 & 0.06 & 1,064 & 418.73  &  Texas road network  \\
\net{PowerGrid} & 4,941 & 6,594 & 0.62 & 0.69 & 2.67 & 3.87 & 0.11 & 46 & 19.07  &  Power grid of Western States Power Grid~\cite{watts98collective}  \\
\net{Mani-faces7k} & 696 & 6,979 & 0.98 & 0.99 & 20.05 & 37.99 & 0.56 & 16 & 5.52  &  Faces (64x64 grayscale images) (connect 7k closest pairs)  \\
\net{Mani-faces4k} & 663 & 3,465 & 0.90 & 0.97 & 10.45 & 20.20 & 0.56 & 29 & 8.96  &  Faces (connect 4k closest pairs)  \\
\net{Mani-faces2k} & 551 & 1,981 & 0.84 & 0.94 & 7.19 & 12.77 & 0.54 & 32 & 11.07  &  Faces (connect 2k closest pairs)  \\
\net{Mani-facesK10} & 698 & 6,935 & 1.00 & 1.00 & 19.87 & 25.32 & 0.51 & 6 & 3.25  &  Faces (connect every to 10 nearest neighbors)  \\
\net{Mani-facesK3} & 698 & 2,091 & 1.00 & 1.00 & 5.99 & 7.98 & 0.45 & 9 & 4.89  &  Faces (connect every to 5 nearest neighbors)  \\
\net{Mani-facesK5} & 698 & 3,480 & 1.00 & 1.00 & 9.97 & 12.91 & 0.48 & 7 & 4.03  &  Faces (connect every to 3 nearest neighbors)  \\
\net{Mani-swiss200k} & 20,000 & 200,000 & 1.00 & 1.00 & 20.00 & 21.08 & 0.59 & 103 & 37.21  &  Swiss-roll (connect 200k nearest pairs of nodes)  \\
\net{Mani-swiss100k} & 19,990 & 99,979 & 1.00 & 1.00 & 10.00 & 11.02 & 0.59 & 162 & 58.32  &  Swiss-roll (connect 100k nearest pairs of nodes)  \\
\net{Mani-swiss60k} & 19,042 & 57,747 & 0.93 & 0.96 & 6.07 & 7.03 & 0.59 & 243 & 89.15  &  Swiss-roll (connect 60k nearest pairs of nodes)  \\
\net{Mani-swissK10} & 20,000 & 199,955 & 1.00 & 1.00 & 20.00 & 25.38 & 0.56 & 10 & 5.47  &  Swiss-roll (every node connects to 10 nearest neighbors)  \\
\net{Mani-swissK5} & 20,000 & 99,990 & 1.00 & 1.00 & 10.00 & 12.89 & 0.54 & 13 & 8.34  &  Swiss-roll (every node connects to 5 nearest neighbors)  \\
\net{Mani-swissK3} & 20,000 & 59,997 & 1.00 & 1.00 & 6.00 & 7.88 & 0.50 & 17 & 6.89  &  Swiss-roll (every node connects to 3 nearest neighbors)  \\
\hline \hline 
\multicolumn{11}{l}{IMDB Actor-to-Movie graphs} \\ 
\hline \hline
\net{AtM-IMDB} & 2,076,978 & 5,847,693 & 0.49 & 0.82 & 5.63 & 65.41 & 0.00 & 32 & 6.82  &  Actors-to-movies graph from IMDB (\url{imdb.com})  \\
\net{Imdb-top30} & 198,430 & 566,756 & 0.99 & 1.00 & 5.71 & 18.19 & 0.00 & 26 & 8.32  &  Actors-to-movies graph heavily preprocessed   \\
\net{Imdb-raw07} & 601,481 & 1,320,616 & 0.54 & 0.79 & 4.39 & 20.94 & 0.00 & 32 & 8.55  &  Country clusters were extracted from this graph  \\
\net{Imdb-France} & 35,827 & 74,201 & 0.51 & 0.76 & 4.14 & 14.62 & 0.00 & 20 & 6.57  &  Cluster of French movies  \\
\net{Imdb-Germany} & 21,258 & 42,197 & 0.56 & 0.78 & 3.97 & 13.69 & 0.00 & 34 & 7.47  &  German movies (to actors that played in them)  \\
\net{Imdb-India} & 12,999 & 25,836 & 0.57 & 0.78 & 3.98 & 31.55 & 0.00 & 19 & 6.00  &  Indian movies  \\
\net{Imdb-Italy} & 19,189 & 37,534 & 0.55 & 0.77 & 3.91 & 11.66 & 0.00 & 30 & 6.91  &  Italian movies  \\
\net{Imdb-Japan} & 15,042 & 34,131 & 0.60 & 0.82 & 4.54 & 16.98 & 0.00 & 19 & 6.81  &  Japanese movies  \\
\net{Imdb-Mexico} & 13,783 & 36,986 & 0.64 & 0.86 & 5.37 & 24.15 & 0.00 & 19 & 5.43  &  Mexican movies  \\
\net{Imdb-Spain} & 15,494 & 31,313 & 0.51 & 0.76 & 4.04 & 14.22 & 0.00 & 28 & 6.44  &  Spanish movies  \\
\net{Imdb-UK} & 42,133 & 82,915 & 0.52 & 0.76 & 3.94 & 15.14 & 0.00 & 23 & 7.04  &  UK movies  \\
\net{Imdb-USA} & 241,360 & 530,494 & 0.51 & 0.78 & 4.40 & 25.25 & 0.00 & 30 & 7.63  &  USA movies  \\
\net{Imdb-WGermany} & 12,120 & 24,117 & 0.56 & 0.78 & 3.98 & 11.73 & 0.00 & 22 & 6.26  &  West German movies  \\
\hline \hline 
\multicolumn{11}{l}{Amazon product co-purchasing networks} \\
\hline \hline
\net{Amazon0302} & 262,111 & 899,792 & 0.95 & 0.97 & 6.87 & 11.14 & 0.43 & 38 & 8.85  &  Amazon products from 2003 03 02~\cite{clauset04large}  \\
\net{Amazon0312} & 400,727 & 2,349,869 & 0.94 & 0.99 & 11.73 & 30.33 & 0.42 & 20 & 6.46  &  Amazon products from 2003 03 12~\cite{clauset04large}  \\
\net{Amazon0505} & 410,236 & 2,439,437 & 0.94 & 0.99 & 11.89 & 30.93 & 0.43 & 22 & 6.48  &  Amazon products from 2003 05 05~\cite{clauset04large}  \\
\net{Amazon0601} & 403,364 & 2,443,311 & 0.96 & 0.99 & 12.11 & 30.55 & 0.43 & 25 & 6.42  &  Amazon products from 2003 06 01~\cite{clauset04large}  \\
\net{AmazonAll} & 473,315 & 3,505,519 & 0.94 & 0.99 & 14.81 & 52.70 & 0.41 & 19 & 5.66  &  Amazon products (all 4 graphs merged)~\cite{clauset04large}  \\
\net{AmazonAllProd} & 524,371 & 1,491,793 & 0.80 & 0.91 & 5.69 & 11.75 & 0.35 & 42 & 11.18  &  Products (all products, source+target)~\cite{jure07viral}  \\
\net{AmazonSrcProd} & 334,863 & 925,872 & 0.84 & 0.91 & 5.53 & 11.53 & 0.43 & 47 & 12.11  &  Products (only source products)~\cite{jure07viral}  \\
\end{tabular}
}
\end{center}
\caption{
Network datasets we analyzed. 
Statistics of networks we consider: 
number of nodes $N$; 
number of edges $E$; 
fraction nodes not in whiskers (size of largest biconnected component) $N_b/N$; 
fraction of edges in biconnected component $E_b/E$; 
average degree $\bar{d}=2E/N$; 
second order average degree $\tilde{d}$; 
average clustering coefficient $\bar{C}$; 
diameter $D$; and
average path length $\bar{D}$. 
}
\label{tab:data_StatsDesc_3}
\end{sidewaystable}


\textbf{Social networks:} The class of social networks in
Table~\ref{tab:data_StatsDesc_1} is particularly diverse and interesting.
It includes several large on-line social networks: a network of
professional contacts from LinkedIn (\net{LinkedIn}); a friendship network
of a LiveJournal blogging community (\net{LiveJournal01}); and a
who-trusts-whom network of Epinions (\net{Epinions}). It also includes an
email network from Enron (\net{Email-Enron}) and from a large European
research organization. For the latter we generated three networks:
\net{Email-Inside} uses only the communication inside organization;
\net{Email-InOut} also adds external email addresses where email has been
sent both way; and \net{Email-All} adds all communication inside the
organization and to the outside world. Also included in the class of
social networks are networks that are not the central focus of the
websites from which they come, but which instead serve as a tool for
people to share information more easily. For example, we have: the
networks of a social bookmarking site Delicious (\net{Delicious}); a
Flickr photo sharing website (\net{Flickr}); and a network from Yahoo!
Answers question answering website (\net{Answers}). In all these networks,
a node refers to an individual and an edge is used to indicate that means
that one person has some sort of interaction with another person, {\em
e.g.}, one person subscribes to their neighbor's bookmarks or photos, or
answers their questions.

\textbf{Information and citation networks:} The class of
Information/citation networks contains several different citation
networks. It contains two citation networks of physics papers on
\url{arxiv.org}, (\net{Cit-hep-th} and \net{Cit-hep-ph}), and a network of
citations of US patents (\net{Cit-Patents}). (These paper-to-paper
citation networks are to be distinguished from scientific collaboration
networks and author-to-paper bipartite networks, as described below.) It
also contains two types of blog citation networks. In the so-called post
networks, nodes are posts and edges represent hyperlinks between blog
posts (\net{Post-nat05-6m} and \net{Post-nat06all}). On the other hand,
the so-called blog network is the blog-level-aggregation of the same data,
{\em i.e.}, there is a link between two blogs if there is a post in first
that links the post in a second blog (\net{Blog-nat05-6m} and
\net{Blog-nat06all}).

\textbf{Collaboration networks:} The class of collaboration networks
contain academic collaboration ({\em i.e.}, co-authorship) networks
between physicists from various categories in \url{arxiv.org}
(\net{CA-astro-ph}, etc.) and between authors in computer science
(\net{CA-DBLP}). It also contains a network of collaborations between
pairs of actors in IMDB (\net{AtA-IMDB}), {\em i.e.}, there is an edge
connecting a pair of actors if they appeared in the same movie. (Again,
this should be distinguished from actor-to-movie bipartite networks, as
described below.)

\textbf{Web graphs:} The class of Web graph networks includes four
different web-graphs in which nodes represent web-pages and edges
represent hyperlinks between those pages. Networks were obtained from
Google (\net{Web-Google}), the University of Notre Dame
(\net{Web-Notredame}), TREC (\net{Web-Trec}), and Stanford University
(\net{Web-BerkStan}).
 The class of Internet networks consists of various autonomous systems networks
obtained at different sources, as well as a Gnutella and eDonkey peer-to-peer
file sharing networks.

\textbf{Bipartite networks:} The class of Bipartite networks is
particularly diverse and includes: authors-to-papers graphs from both
computer science (\net{AtP-DBLP}) and physics (\net{AtP-astro-ph}, etc.);
a network representing users and the URLs they visited
(\net{Clickstream}); a network representing users and the movies they
rated (\net{Netflix}); and a users-to-queries network representing query
terms that users typed into a search engine (\net{QueryTerms}). (We also
have analyzed several bipartite actors-to-movies networks extracted from
the IMDB database, which we have listed separately below.)

\textbf{Biological networks:} The class of Biological networks include
protein-protein interaction networks of yeast obtained from various
sources.

\textbf{Low dimensional grid-like networks:} The class of Low-dimensional
networks consists of graphs constructed from road (\net{Road-CA}, etc.) or
power grid (\net{PowerGrid}) connections and as such might be expected to
``live'' on a two-dimensional surface in a way that all of the other
networks do not. We also added a ``swiss roll'' network, a $2$-dimensional
manifold embedded in $3$-dimensions, and a ``Faces'' dataset where each
point is an $64$ by $64$ gray-scale image of a face (embedded in $4,096$
dimensional space) and where we connected the faces that were most similar
(using the Euclidean distance).

\textbf{IMDB, Yahoo! Answers and Amazon networks:} Finally, we have networks
from IMDB, Amazon, and Yahoo! Answers, and for each of these we have
separately analyzed subnetworks. The IMDB networks consist of
actor-to-movie links, and we include the full network as well as
subnetworks associated with individual countries based on the country of
production. For the Amazon networks, recall that Amazon sells a variety of
products, and for each item $A$ one may compile the list the up to ten
other items most frequently purchased by buyers of $A$. This information
can be presented as a directed network in which vertices represent items
and there is a edge from item $A$ to another item $B$ if $B$ was
frequently purchased by buyers of $A$. We consider the network as
undirected. We use five networks from a study of Clauset {\em et
al.}~\cite{clauset04large}, and two networks from the viral marketing
study from Leskovec {\em et al.}~\cite{jure07viral}. Finally, for the
Yahoo! Answers networks, we observe several deep cuts at large size
scales, and so in addition the full network, we analyze the top six most
well-connected subnetworks.

In addition to providing a brief description of the network,
Tables~\ref{tab:data_StatsDesc_1}, \ref{tab:data_StatsDesc_2} and
\ref{tab:data_StatsDesc_3} show the number of nodes and edges in each
network, as well as other statistics which will be described in
Section~\ref{sxn:obs_struct:stats}. (In all cases, we consider the network
as undirected, and we extract and analyze the largest connected
component.) The sizes of these networks range from about $5,000$ nodes up
to nearly $14$ million nodes, and from about $6,000$ edges up to more than
$100$ million edges. All of the networks are quite sparse---their
densities range from an average degree of about $2.5$ for the blog post
network, up to an average degree of about $400$ in the network of movie
ratings from Netflix, and most of the other networks, including the purely
social networks, have average degree around $10$ (median average degree of
$6$). In many cases, we examined several versions of a given network. For
example, we considered the entire IMDB actor-to-movie network, as well as
sub-pieces of it corresponding to different language and country groups.
Detailed statistics for all these networks are presented in
Tables~\ref{tab:data_StatsDesc_1}, \ref{tab:data_StatsDesc_2}
and~\ref{tab:data_StatsDesc_3} and are described in
Section~\ref{sxn:obs_struct}. In total, we have examined over $100$ large 
real-world social and
information networks, making this, to our knowledge, the largest and most
comprehensive study of such networks.

\subsection{Clusters and communities in networks}
\label{sxn:related:clusters}

Hierarchical clustering is a common approach to community identification
in the social sciences~\cite{WassermanFaust94}, but it has also found
application more
generally~\cite{newman02community,hopcroft03_communitiesPNAS}. In this
procedure, one first defines a distance metric between pairs of nodes and
then produces a tree (in either a bottom-up or a top-down manner)
describing how nodes group into communities and how these group further
into super-communities. A quite different approach that has received a
great deal of attention (and that will be central to our analysis) is
based on ideas from \emph{graph
partitioning}~\cite{Schaeffer07_survey,bgw07_graphclustering}. In this
case, the network is a modeled as simple undirected graph, where nodes and
edges have no attributes, and a partition of the graph is determined by
optimizing a merit function. The graph partitioning problem is find some
number $k$ groups of nodes, generally with roughly equal size, such that
the number of edges between the groups, perhaps normalized in some way, is
minimized.

Let $G=(V,E)$ denote a graph, then the \emph{conductance} $\phi$ of a set
of nodes $S \subset V$, (where $S$ is assumed to contain no more than half
of all the nodes), is defined as follows. Let $v$ be the sum of degrees of
nodes in $S$, and let $s$ be the number of edges with one endpoint in $S$
and one endpoint in $\overline{S}$, where $\overline{S}$ denotes the
complement of $S$. Then, the conductance of $S$ is $\phi=s/v$, or
equivalently $\phi = s/(s+2e)$, where $e$ is the number of edges with both
endpoints is $S$. More formally:
\begin{definition}
Given a graph $G$ with adjacency matrix $A$ the {\em conductance of a set}
of nodes $S$ is defined as:
\begin{equation}
\label{eqn:conductance_set}
\phi(S) = \frac{ \sum_{i \in S, j \notin S} A_{ij} }
               { \min\{ A(S), A(\overline{S}) \} }   ,
\end{equation}
where $ A(S) = \sum_{i \in S} \sum_{j \in V} A_{ij}$, or equivalently $
A(S) = \sum_{i \in S} d(i)$, where $d(i)$ is a degree of node $i$ in $G$.
Moreover, in this case, the {\em conductance of the graph} $G$ is:
\begin{equation}
\label{eqn:conductance_graph}
\phi_G = \min_{S \subset V} \phi(S) .
\end{equation}
\label{def:conductance}
\end{definition}
Thus, the conductance of a set provides a measure for the quality of the
cut
$(S,\overline{S})$, or relatedly the goodness of a community $S$.%
\footnote{ Throughout this chapter we consistently use shorthand phrases
like ``this piece has good conductance'' to mean ``this piece is separated
from the rest of the graph by a low-conductance cut.''}

\begin{figure}
	\begin{center}
		\subfigure[Three communities]{
			\includegraphics[width=0.35\textwidth]{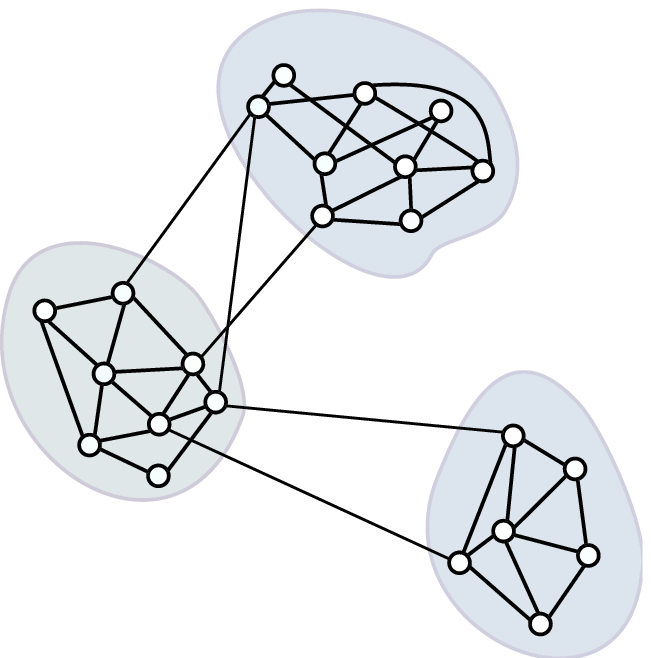} \quad
			\label{fig:conductance_comm}
		}
		\subfigure[Conductance bottleneck]{
    			\includegraphics[width=0.45\textwidth]{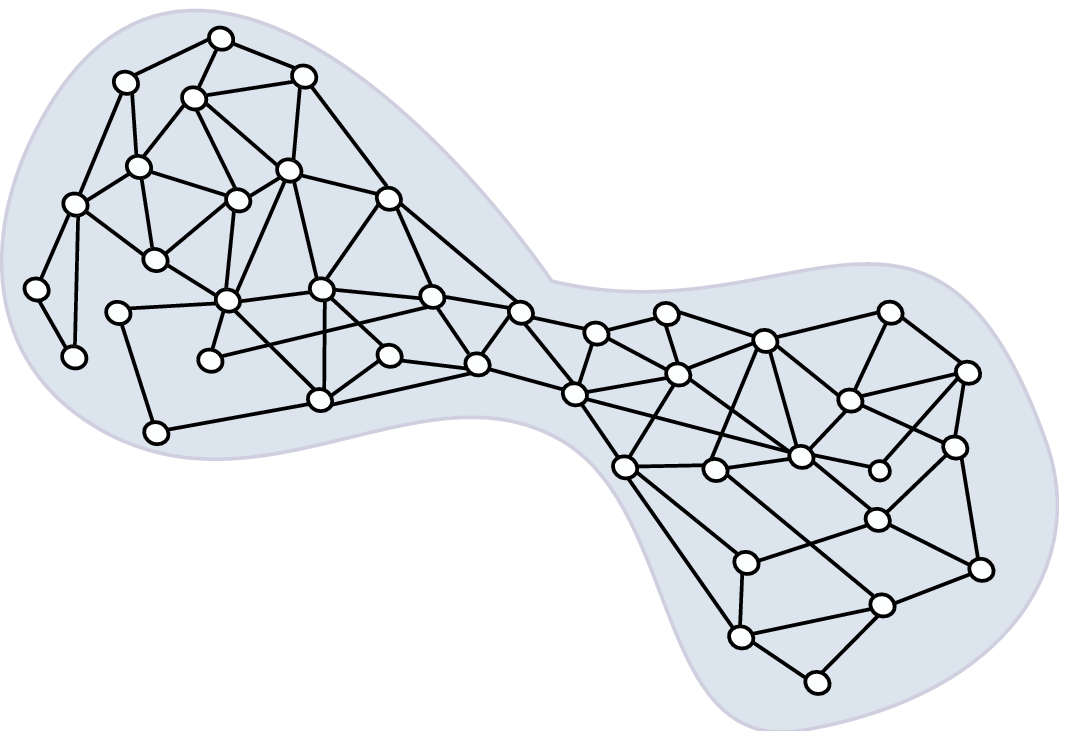} \quad
			\label{fig:conductance_cond}
		}
	\end{center}
\caption{
(a) Caricature of the traditional view of communities as being sets of nodes
with more and/or better intra-connections than inter-connections.
(b) A graph with its minimum conductance bottleneck illustrated.
}
\label{fig:conductance}
\end{figure}

Indeed, it is often noted that communities should be thought of as sets of
nodes with more and/or better intra-connections than inter-connections;
see Figure~\ref{fig:conductance} for an illustration.
When interested in detecting communities and evaluating their quality, we
prefer sets with small conductances, {\em i.e.}, sets that are densely linked
inside and sparsely linked to the outside.
Although numerous measures have been proposed for how community-like is a set
of nodes, it is commonly noted---{\em e.g.}, see Shi and
Malik~\cite{ShiMalik00_NCut} and Kannan, Vempala, and
Vetta~\cite{kannan04_gbs}---that conductance captures the ``gestalt'' notion
of clustering~\cite{zahn71_gestalt}, and as such it has been widely-used for
graph clustering and community
detection~\cite{gaertler05_clustering,luxburg05_survey,Schaeffer07_survey}.

There are many other density-based measures that have been used to
partition a graph into a set of
communities~\cite{gaertler05_clustering,luxburg05_survey,Schaeffer07_survey}.
One that deserves particular mention is
modularity~\cite{newman04community,newman2006_ModularityPNAS}. For a given
partition of a network into a set of communities, modularity measures the
number of within-community edges, relative to a null model that is usually
taken to be a random graph with the same degree distribution. Thus,
modularity was originally introduced and it typically used to measure the
strength or quality of a particular partition of a network. We, however,
are interested in a quite different question than those that motivated the
introduction of modularity. Rather than seeking to partition a graph into
the ``best'' possible partition of communities, we would like to know how good
is a particular element of that partition, \emph{i.e.}, how community-like
are the best possible communities that modularity or any other merit
function can hope to find, in particular as a function of the size of that partition.

\subsection{Approximation algorithms for finding low-conductance cuts}
\label{sxn:related:LowCondAlgs}

In addition to capturing very well our intuitive notion of what it means
for a set of nodes to be a good community, the use of conductance as an
objective function has an added benefit: there exists an extensive
theoretical and practical literature on methods for approximately
optimizing it. (Finding cuts with exactly minimal conductance is NP-hard.)
In particular, the theory literature contains several algorithms with
provable approximation performance guarantees.

First, there is the spectral method, which uses an eigenvector of the
graph's Laplacian matrix to find a cut whose conductance is no bigger than
$\phi$ if the graph actually contains a cut with conductance
$O(\phi^2)$~\cite{Cheeger69_bound,Donath:1972,fiedler73graphs,mohar91_survey,Chung:1997}.
The spectral method also produces lower bounds which can show that the
solution for a given graph is closer to optimal than promised by the
worst-case guarantee. Second, there is an algorithm that uses
multi-commodity flow to find a cut whose conductance is within an $O(\log
n)$ factor of optimal~\cite{Leighton:1988,Leighton:1999}. Spectral and
multi-commodity flow based methods are complementary in that the
worst-case $O(\log n)$ approximation factor is obtained for flow-based
methods on expander graphs~\cite{Leighton:1988,Leighton:1999}, a class of
graphs which does not cause problems for spectral methods, whereas
spectral methods can confuse long path with deep
cuts~\cite{guatterymiller98,spielman96_spectral}, a difference that does
not cause problems for flow-based methods. Third, and very recently, there
exists an algorithm that uses semidefinite programming to find a solution
that is within $O(\sqrt{\log n})$ of optimal~\cite{Arora:2004}. This paper
sparked a flurry of theoretical research on a family of closely related
algorithms including~\cite{AHK04,khandekar06_partitioning,Arora:2007}, all
of which can be informally described as combinations of spectral and
flow-based techniques which exploit their complementary strengths.
However, none of those algorithms are currently practical enough to use in
our study.

Of the above three theoretical algorithms, the spectral method is by far
the most practical. Also very common are recursive bisection heuristics:
recursively divide the graph into two groups, and then further subdivide
the new groups until the desired number of clusters groups is achieved.
This may be combined with local improvement methods like the Kernighan-Lin
and Fiduccia-Mattheyses procedures~\cite{Kernighan:1970,Fiduccia:1982},
which are fast and can climb out of some local minima. The latter was
combined with a multi-resolution framework to create
Metis~\cite{karypis98_metis,karypis98metis}, a very fast program intended
to split mesh-like graphs into equal sized pieces. The authors of Metis
later created Cluto~\cite{zhao04cluto}, which is better tuned for
clustering-type tasks. Finally we mention Graclus~\cite{dhillon07graclus},
which uses multi-resolution techniques and kernel $k$-means to optimize a
metric that is closely related to conductance.

While the preceding were all approximate algorithms for finding the lowest
conductance cut in a whole graph, we now mention
MQI~\cite{Gallo:1989,kevin04mqi}, an \emph{exact} algorithm for the
slightly different problem of finding the lowest conductance cut in
\emph{half} of a graph. This algorithm can be combined with a good method
for initially splitting the graph into two pieces (such as Metis or the
Spectral method) to obtain a surprisingly strong heuristic method for
finding low conductance cuts in the whole graph~\cite{kevin04mqi}. The
exactness of the second optimization step frequently results in cuts with
extremely low conductance scores, as will be visible in many of our plots.
MQI can be implemented by solving single parametric max flow problems, or
sequences of ordinary max flow problems. Parametric max flow (with MQI
described as one of the applications) was introduced by~\cite{Gallo:1989},
and recent empirical work is described in~\cite{goldberg07_parametric},
but currently there is no publicly available code that scales to the
sizes we need. Ordinary max flow is a very thoroughly studied problem.
Currently, the best theoretical time bounds are~\cite{goldberg98_beyond},
the most practical algorithm is~\cite{goldberg88_maxflow}, while the best
implementation is {\tt hi\_pr} by~\cite{goldberg95_push}. Since Metis+MQI
using the {\tt hi\_pr} code is very fast and scalable, while the method
empirically seems to usually find the lowest or nearly lowest conductance
cuts in a wide variety of graphs, we have used it extensively in this
study.

We will also extensively use Local Spectral Algorithm of Andersen, Chung,
and Lang~\cite{andersen06local} to find node sets of low conductance, {\em
i.e.}, good communities, around a seed node. This algorithm is also very
fast, and it can be successfully applied to very large graphs to obtain
more ``well-rounded'', ``compact,'' or ``evenly-connected'' 
communities than those returned by Meits+MQI. The latter
observation (described in more detail in Section~\ref{algo-notes-section})
is since local spectral methods also confuse long paths (which tend to
occur in our very sparse network datasets) with deep cuts. This algorithm
takes as input two parameters---the seed node and a parameter $\epsilon$
that intuitively controls the locality of the computation---and it outputs
a set of nodes. Local spectral methods were introduced by Spielman and
Teng~\cite{Spielman:2004,andersen06local}, and they have roughly the same
kind of quadratic approximation guarantees as the global spectral method,
but they have computational cost is proportional to the size of the
obtained
piece~\cite{chung07_fourproofs,Chung07_localcutsLAA,Chung07_heatkernelPNAS}.
\section{The Network Community Profile Plot (NCP plot)}
\label{sxn:ncpp}

In this section, we discuss the \emph{network community profile plot} (NCP
plot), which measures the quality of network communities at different size
scales. We start in Section~\ref{sxn:ncpp:def} by introducing it. Then, in
Section~\ref{sxn:ncpp:low_small}, we present the NCP plot for several
examples of networks which inform peoples' intuition and for which the NCP
plot behaves in a characteristic manner. Then, in
Sections~\ref{sxn:ncpp:large_sparse} and~\ref{sxn:ncpp:large_sparse_more}
we present the NCP plot for a wide range of large real world social and
information networks. We will see that in such networks the NCP plot
behaves in a qualitatively different manner.

\subsection{Definitions for the network community profile plot}
\label{sxn:ncpp:def}

In order to more finely resolve community structure in large networks, we
introduce the {\em network community profile plot} (NCP plot).
Intuitively, the NCP plot measures the quality of the best possible
community in a large network, as a function of the community size.
Formally, we may define it as the conductance value of the best
conductance set of cardinality $k$ in the entire network, as a function of
$k$.

\begin{definition}
Given a graph $G$ with adjacency matrix $A$, the {\em network community 
profile plot (NCP plot)} plots $\Phi(k)$ as a function of $k$, where
\begin{equation}
\label{eqn:conductance_k}
\Phi(k) = \min_{S \subset V, |S|=k} \phi(S) ,
\end{equation}
where $|S|$ denotes the cardinality of the set $S$, and where the conductance
$\phi(S)$ of $S$ is given by equation~(\ref{eqn:conductance_set}).
\end{definition}

Since this quantity is intractable to compute, we will employ well-studied 
approximation algorithms for the Minimum Conductance Cut Problem to 
approximate it.
In particular, operationally we will use several natural heuristics based on 
approximation algorithms to do graph partitioning in order to compute 
different approximations to the NCP plot.
Although other procedures will be described in 
Section~\ref{algo-notes-section}, we will primarily employ two procedures. 
First, Metis+MQI, {\em i.e.}, the graph partitioning package 
Metis~\cite{karypis98_metis} followed by the flow-based post-processing 
procedure MQI~\cite{kevin04mqi}; this procedure returns sets that have very 
good conductance values. 
Second, the Local Spectral Algorithm of Andersen, Chung, and 
Lang~\cite{andersen06local}; this procedure returns sets that are somewhat 
more ``compact'' or ``smoothed'' or ``regularized,'' but that often have
somewhat worse conductance values. 

Just as the conductance of a set of nodes provides a quality measure of
that set as a community, the shape of the NCP plot provides insight into
the community structure of a graph as a whole. 
For example, the magnitude of the conductance tells us how well clusters of 
different sizes are separated from the rest of the network. 
One might hope to obtain some sort of ``smoothed'' measure of the notion of 
the best community of size $k$ (\emph{e.g.}, by considering an average of 
the conductance value over all sets of a given size or by considering a 
smoothed extremal statistic such as a $95$-th percentile) rather than 
conductance of the best set of that size.  
We have not defined such a measure since there is no obvious way to average 
over all subsets of size $k$ and obtain a meaningful approximation to the 
minimum. 
On the other hand, our approximation algorithm methodology implicitly 
incorporates such an effect.
Although Metis+MQI finds sets of nodes with extremely good conductance value, 
empirically we observe that they often have little or no internal 
structure---they can even be disconnected.
On the other hand, since spectral methods in general tend to confuse long 
paths with deep cuts~\cite{spielman96_spectral,guatterymiller98}, the Local 
Spectral Algorithm finds sets that are ``tighter'' and more ``well-rounded'' 
and thus in many ways more community-like.
(See Sections~\ref{sxn:related:LowCondAlgs} and~\ref{algo-notes-section} for 
details on these algorithmic issues and interpretations.)

\subsection{Community profile plots for expander, low-dimensional, and small social networks}
\label{sxn:ncpp:low_small}

\begin{figure}
	\begin{center}
		\subfigure[Several low-dimensional meshes.]{
			\includegraphics[width=0.45\textwidth]{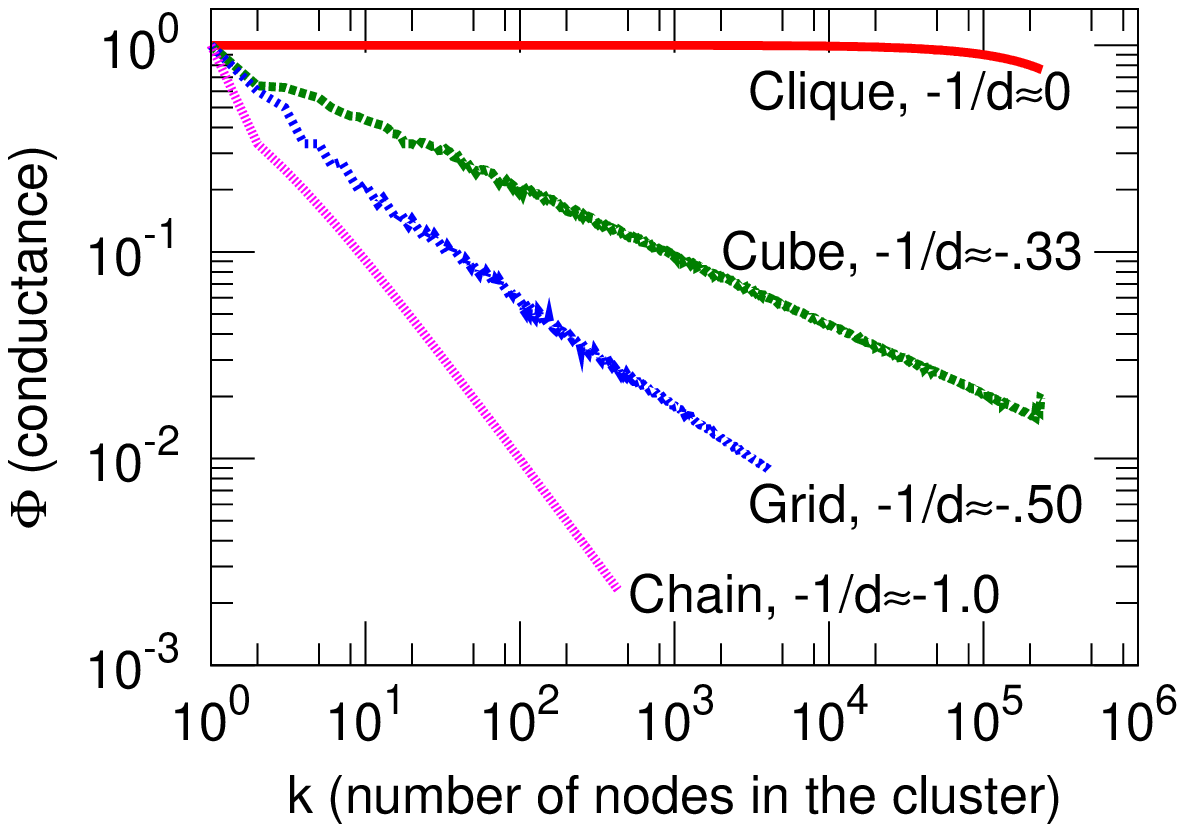} \quad
			\label{fig:ncpp_lowdim:toy}
		}
		\subfigure[\net{PowerGrid}]{
			\includegraphics[width=0.45\textwidth]{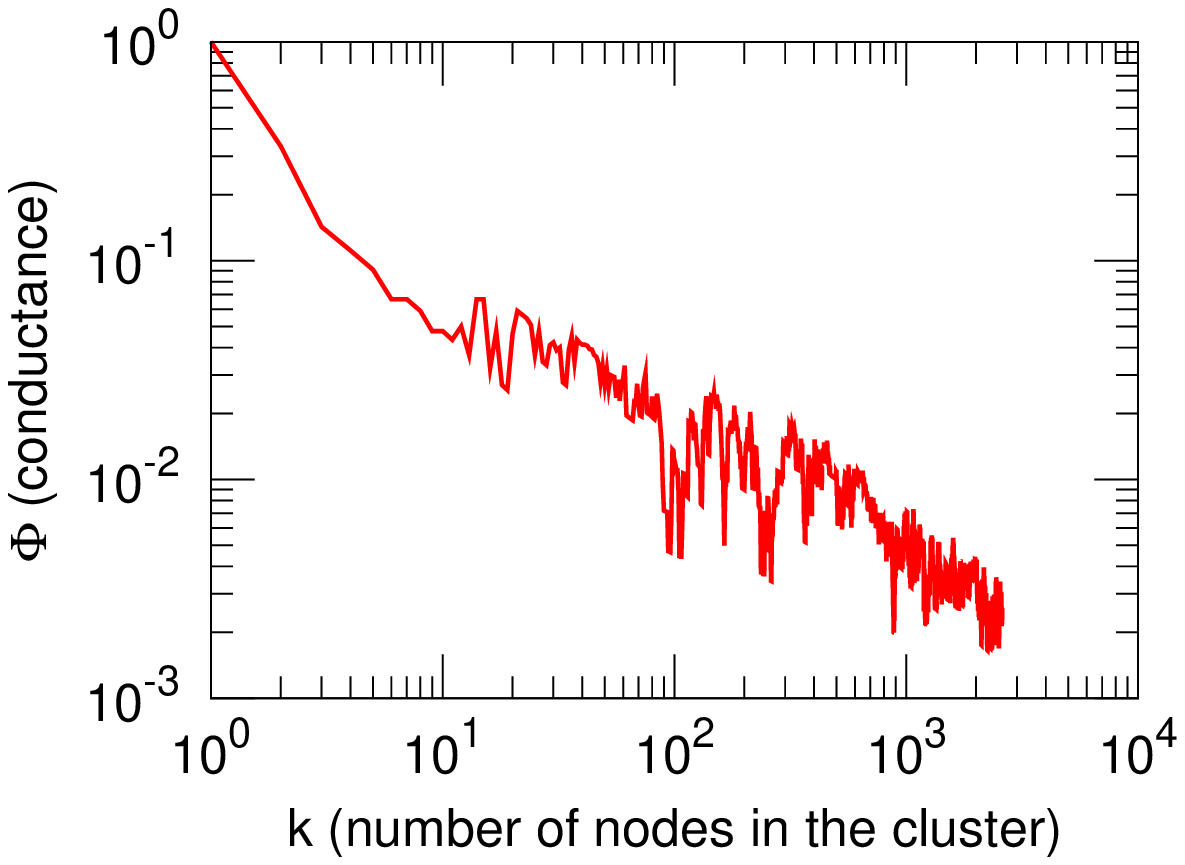} \quad
			\label{fig:ncpp_lowdim:power}
		}
		\subfigure[\net{Road-CA}]{
			\includegraphics[width=0.45\textwidth]{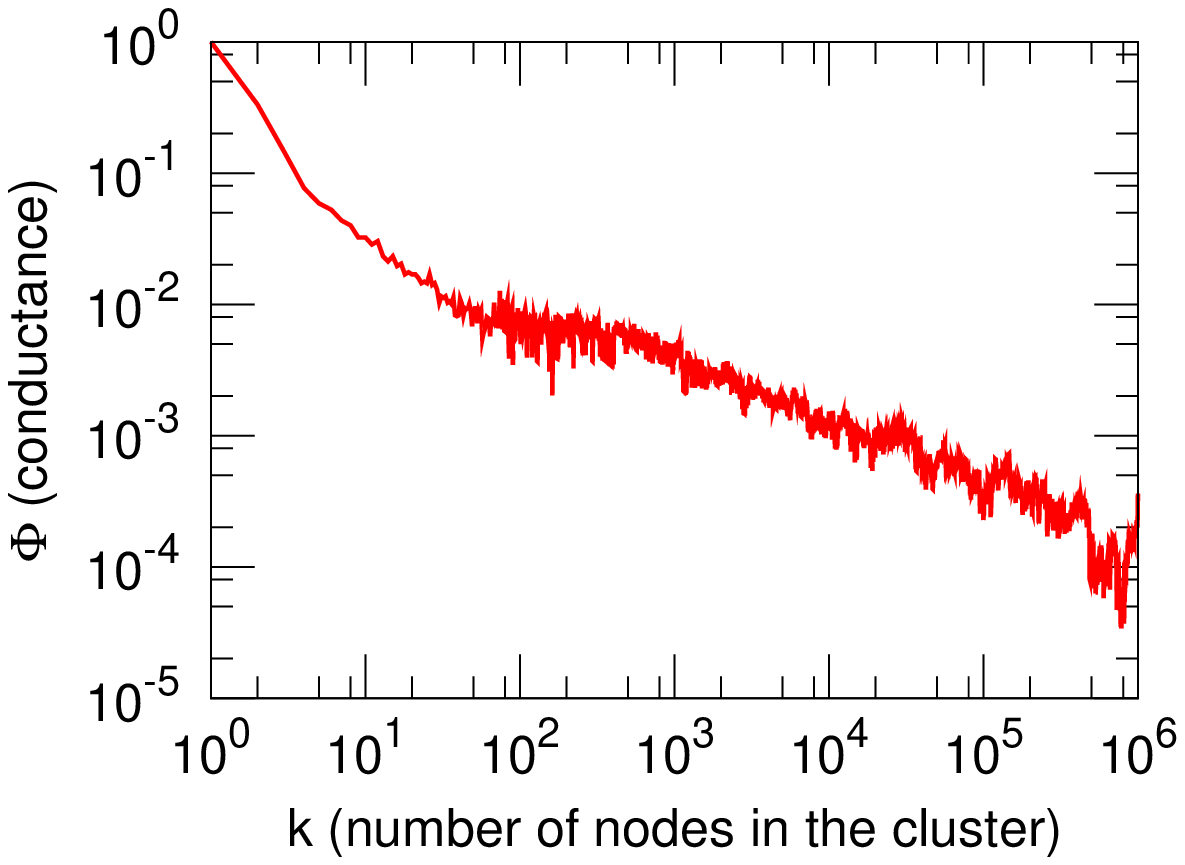}    \quad
			\label{fig:ncpp_lowdim:road}
		}
		\subfigure[Manifold]{
			\includegraphics[width=0.45\textwidth]{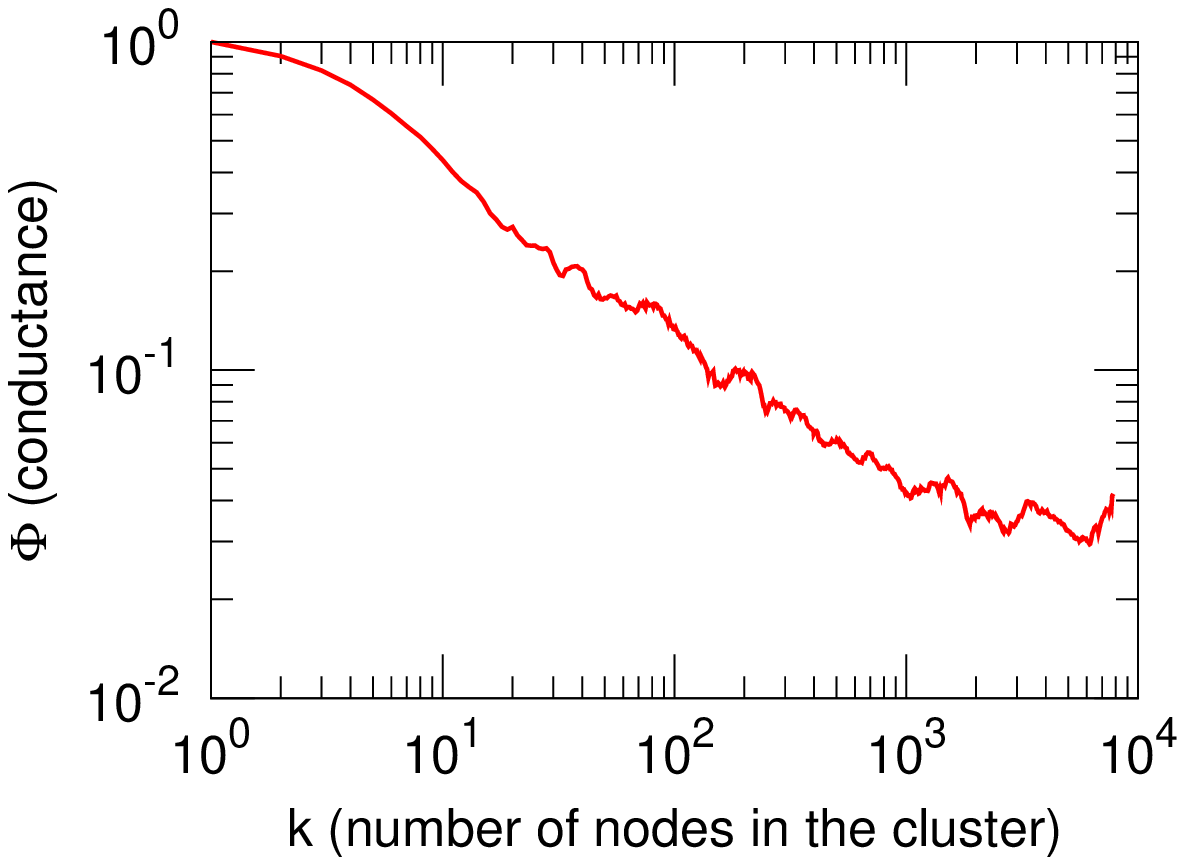} \quad
			\label{fig:phiModels:Manifold}
		}
		\subfigure[Expander: dense $G_{nm}$ graph]{
			\includegraphics[width=0.45\textwidth]{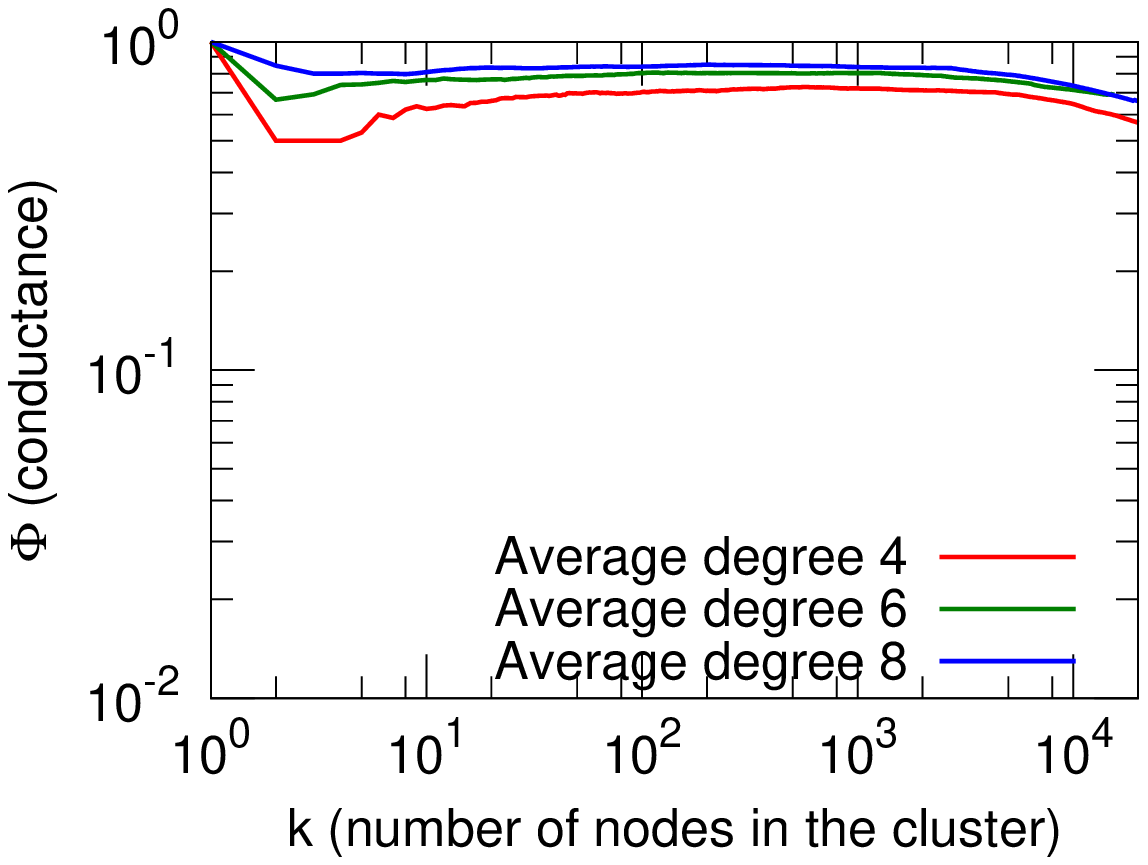}    \quad
			\label{fig:phiModels:expander1}
		}
		\subfigure[Expander: union of matchings]{
			\includegraphics[width=0.45\textwidth]{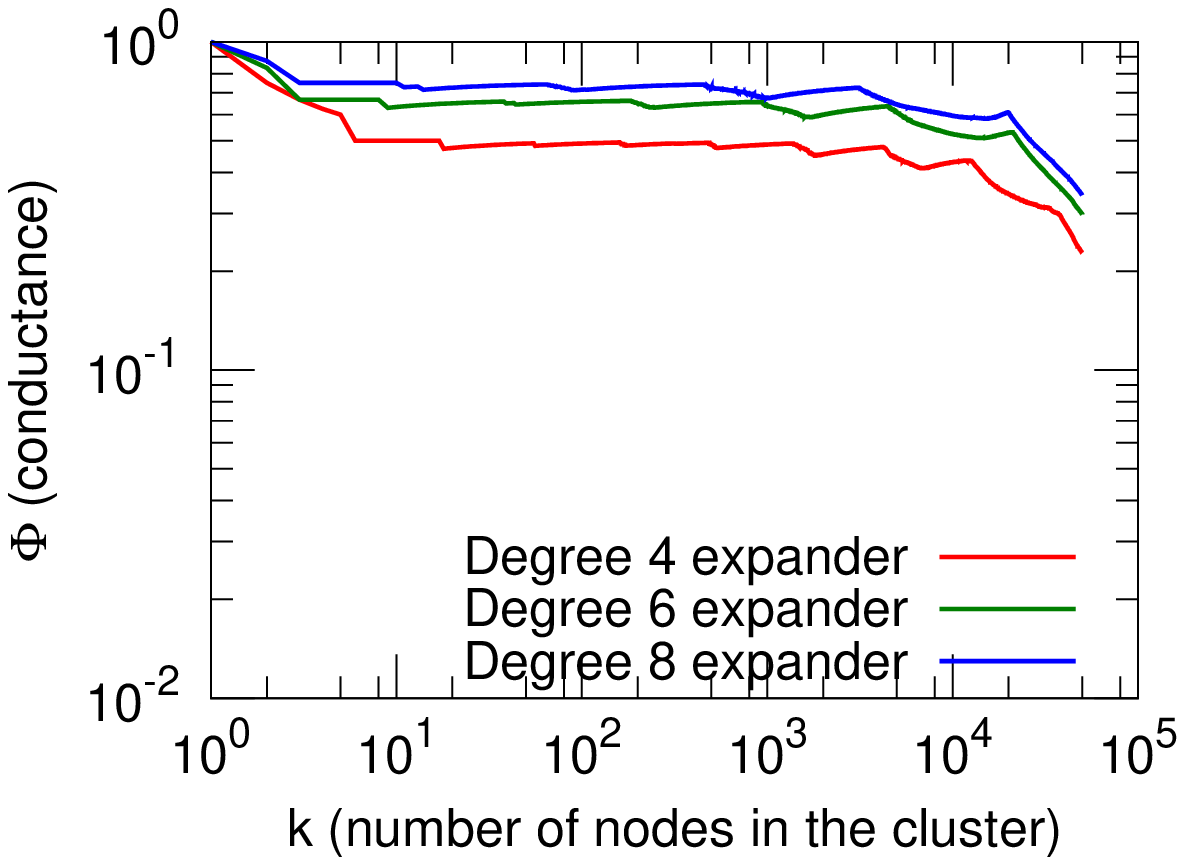} \quad
			\label{fig:phiModels:expander2}
		}
	\end{center}
\caption{
Network community profile plots for expander-like graphs and several networks 
that ``live'' in low-dimensional spaces.
(\ref{fig:ncpp_lowdim:toy})
A large clique graph, a cube (3d mesh), a grid (2d mesh) and a chain (line).
Note that the slope of community profile plot directly corresponds to
dimensionality of the graph.
(\ref{fig:ncpp_lowdim:power}) and (\ref{fig:ncpp_lowdim:road})
Two networks on the Earth's surface and thus that are reasonably
well-embeddable in two dimensions.
(\ref{fig:phiModels:Manifold})
A 2d ``swiss roll'' manifold embedded in $3$ dimensions, where every we 
connected every point to $10$ nearest neighbors.
(\ref{fig:phiModels:expander1}) and
(\ref{fig:phiModels:expander2})
Two networks that are very good expanders.
}
\label{fig:ncpp_lowdim}
\end{figure}

The NCP plot behaves in a characteristic manner for graphs that are
``well-embeddable'' into an underlying low-dimensional geometric
structure. To illustrate this, consider Figure~\ref{fig:ncpp_lowdim}. In
Figure~\ref{fig:ncpp_lowdim:toy}, we show the results for a
$1$-dimensional chain, a $2$-dimensional grid, and a $3$-dimensional cube.
In each case, the NCP plot is steadily downward sloping as a function of
the number of nodes in the smaller cluster. Moreover, the curves are
straight lines with a slope equal to $-1/d$, where $d$ is the
dimensionality of the underlying grids. In particular, as the underlying
dimension increases then the slope of the NCP plot gets less steep. Thus,
we observe:
\begin{observation}
  If the network under consideration corresponds to a $d$-dimensional grid,
  then the NCP plot shows that
  \begin{equation}
  \label{eqn:slope_dim}
  -\frac{1}{d} = \frac{\log(\phi(k))}{\log(k)}  .
  \end{equation}
\end{observation}
This is simply a manifestation of the isoperimetric (\emph{i.e.}, surface
area to volume) phenomenon: for a grid, the ``best'' cut is obtained by
cutting out a set of adjacent nodes, in which case the surface area
(number of edges cut) increases as $O(m^{d-1}$), while the volume (number
of vertices/edges inside the cluster) increases as $O(m^d)$.

This qualitative phenomenon of a steadily downward sloping NCP plot is
quite robust for networks that ``live'' in a low-dimensional structure,
\emph{e.g.}, on a manifold or the surface of the earth. For example,
Figure~\ref{fig:ncpp_lowdim:power} shows the NCP plot for a power grid
network of Western States Power Grid~\cite{watts98collective}, and
Figure~\ref{fig:ncpp_lowdim:road} shows the NCP plot for a road network of
California. These two networks have very different sizes---the power grid
network has $4,941$ nodes and $6,594$ edges, and the road network has
$1,957,027$ nodes and $2,760,388$ edges---and they arise in very different
application domains. In both cases, however, we see predominantly downward
sloping NCP plot, very much similar to the profile of a simple 
$2$-dimensional grid. Indeed, the ``best-fit'' line for power grid gives the
slope of $\approx-0.45$, which by~(\ref{eqn:slope_dim}) suggests that
$d\approx2.2$, which is not far from the ``true'' dimensionality of $2$.
Moreover, empirically we observe that minima in the NCP plot correspond to 
community-like sets, which are occasionally nested.
This corresponds to hierarchical community organization. 
For example, the nodes giving the dip at $k=19$ are included in the nodes
giving the dip at $k=883$, while dips at $k=94$ and $k=105$ are both included 
in the dip at $k=262$.

In a similar manner, Figure~\ref{fig:phiModels:Manifold} shows the profile 
plot for
a graph generated from a ``swiss roll'' dataset which is commonly examined
in the manifold and machine learning literature~\cite{isomap00_science}.
In this case, we still observe a downward sloping NCP plot that
corresponds to internal dimensionally of the manifold (2 in this case).
Finally, Figures~\ref{fig:phiModels:expander1}
and~\ref{fig:phiModels:expander2} show NCP plots for two graphs that are
very good expanders. The first is a $G_{nm}$ graph with $100,000$ nodes
and a number of edges such that the average degree is $4$, $6$, and $8$.
The second is a constant degree expander: to make one with degree $d$, we
take the union of $d$ disjoint but otherwise random complete matchings,
and we have plotted the results for $d=4,6,8$. In both of these cases, the
NCP plot is roughly flat, which we also observed in 
Figure~\ref{fig:ncpp_lowdim:toy} for a clique, which is to be expected
since the minimum conductance cut in the entire graph cannot be too small
for a good expander~\cite{HLW06_expanders}.

\begin{figure}
\begin{center}
  \begin{tabular}{cc}
		\subfigure[Zachary's karate club network \ldots]{
			\includegraphics[width=0.35\textwidth]{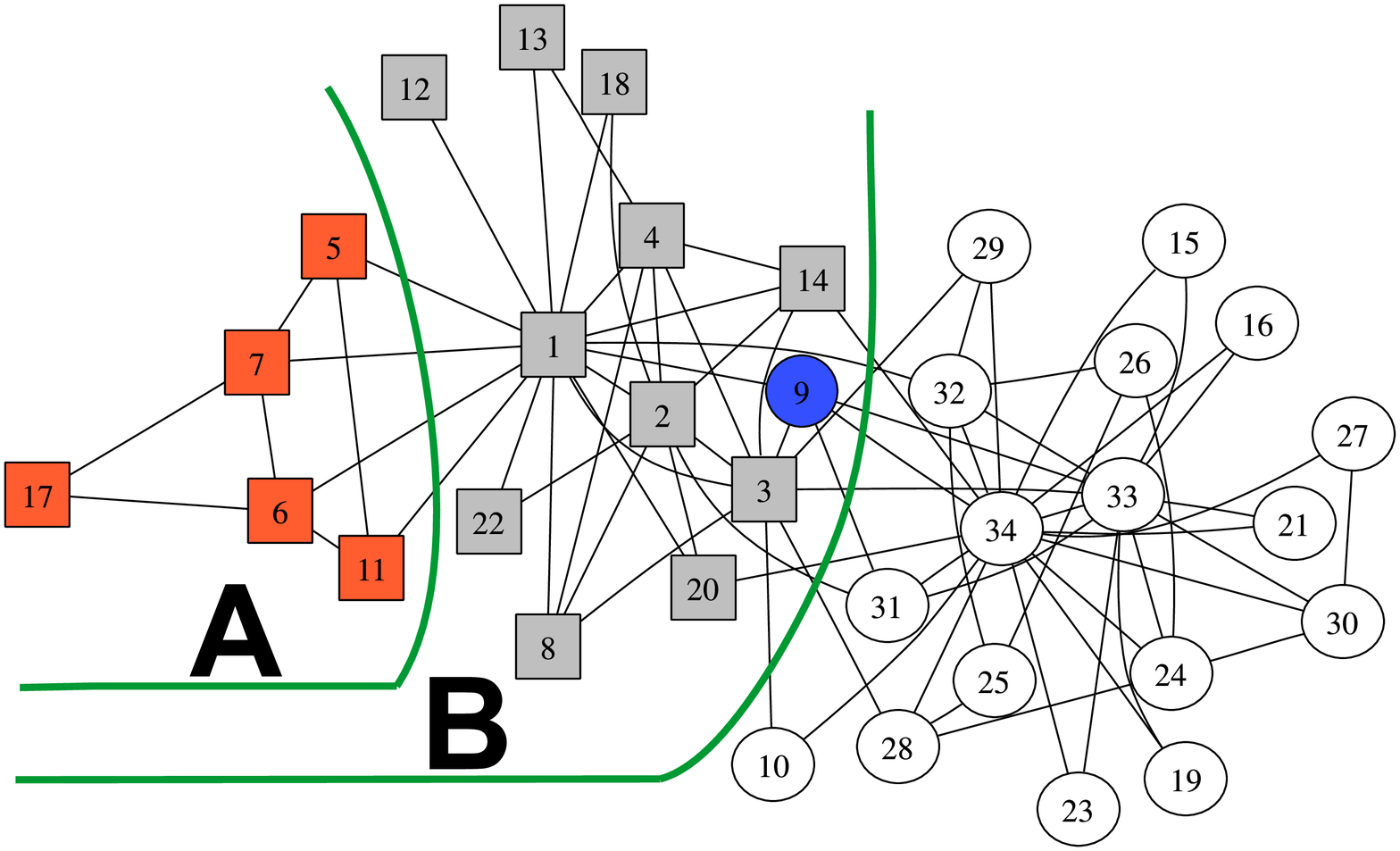}
			\label{fig:ncpp_small:karate_graph} }
  &
		\subfigure[\ldots and it's community profile plot]{
			\includegraphics[width=0.35\textwidth]{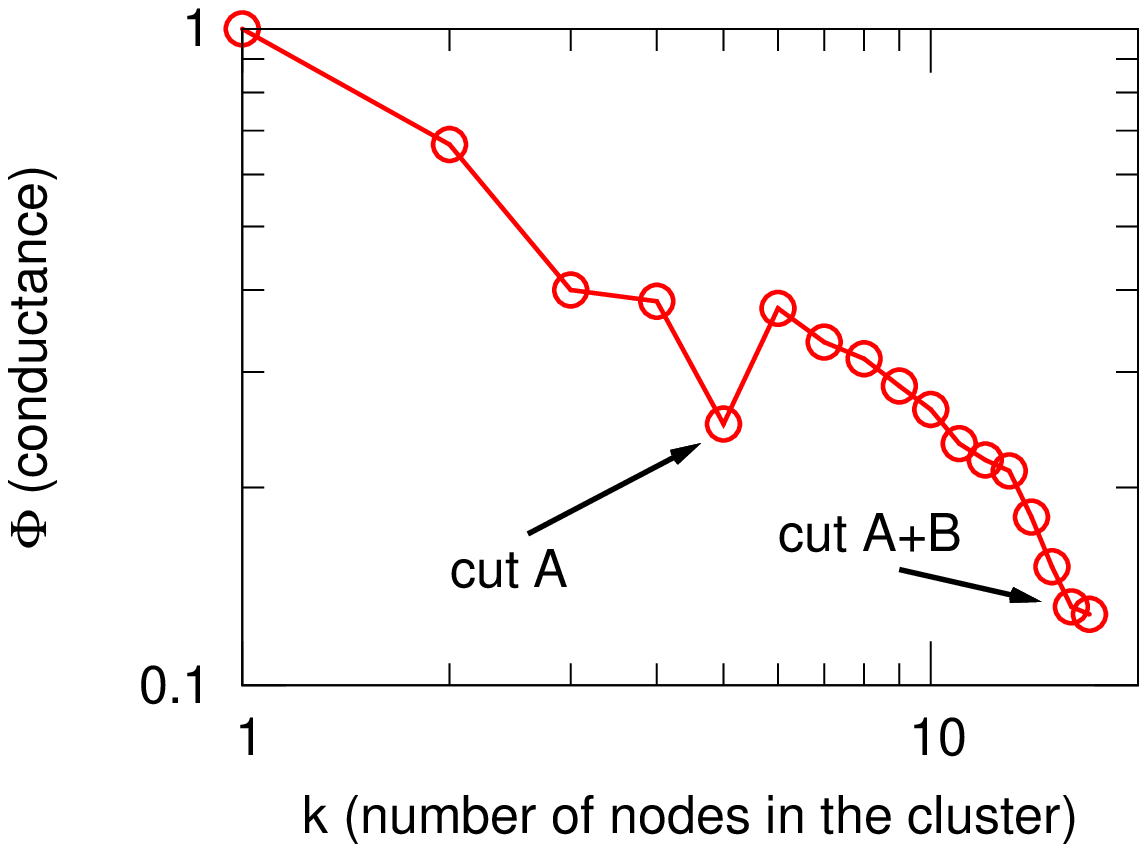}
			\label{fig:ncpp_small:karate_plot} }
  \\
		\subfigure[Dolphins social network \ldots]{
			\includegraphics[width=0.35\textwidth]{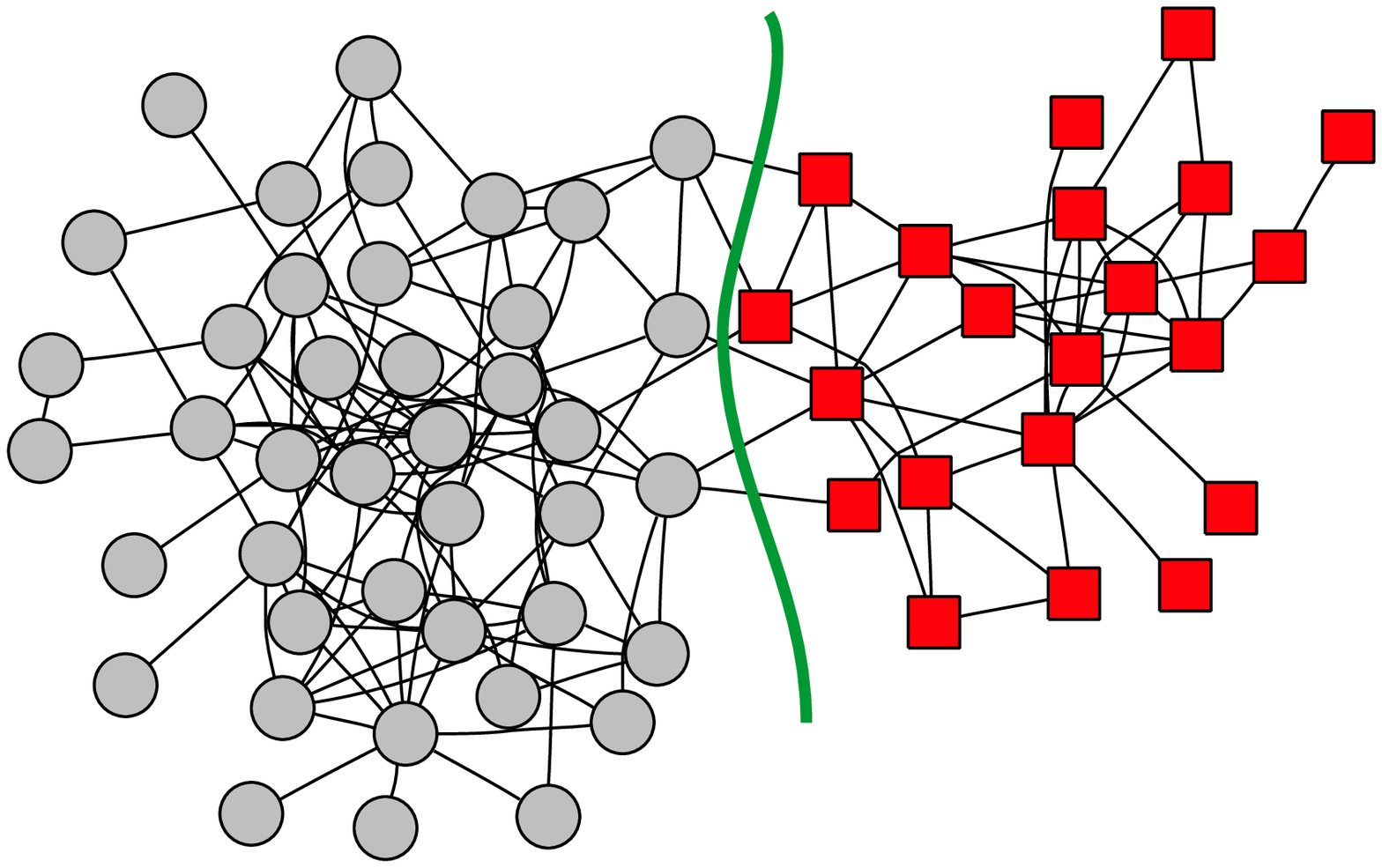}
			\label{fig:ncpp_small:dolphin_graph} }
	&
		\subfigure[\ldots and it's community profile plot]{
			\includegraphics[width=0.35\textwidth]{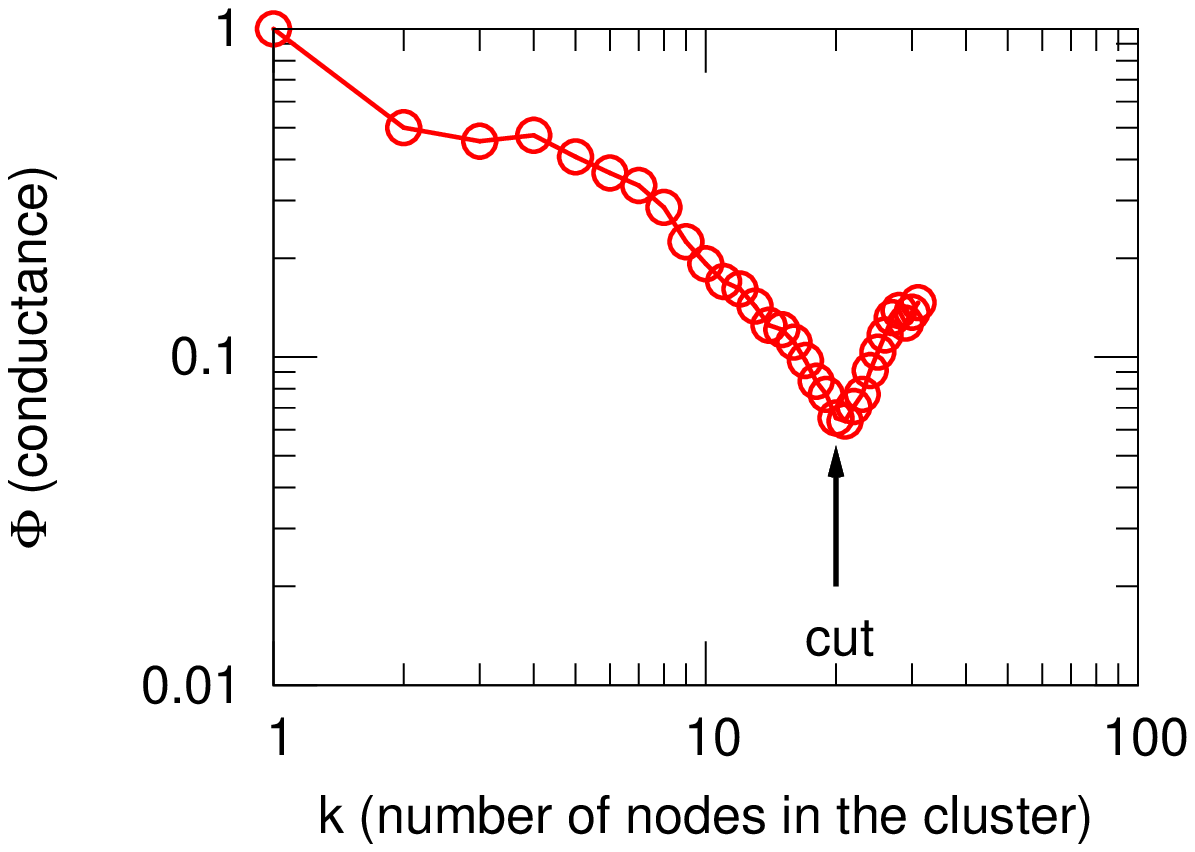}
			\label{fig:ncpp_small:dolphin_plot} }
	\\
		\subfigure[Monks social network \ldots]{
			\includegraphics[width=0.35\textwidth]{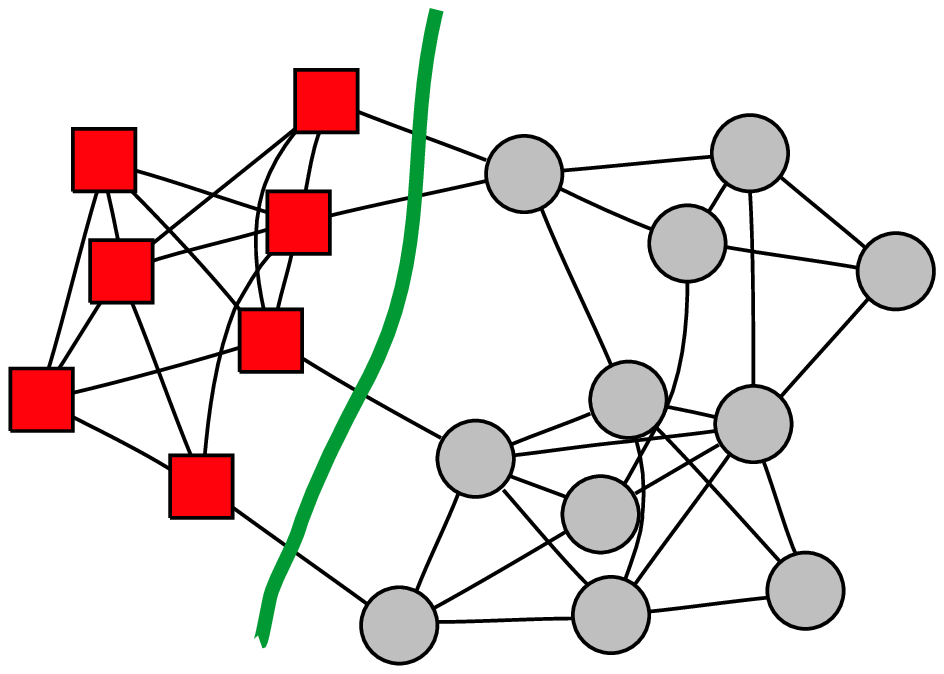}
			\label{fig:ncpp_small:monk_graph} }
		&
		\subfigure[\ldots and it's community profile plot]{
			\includegraphics[width=0.35\textwidth]{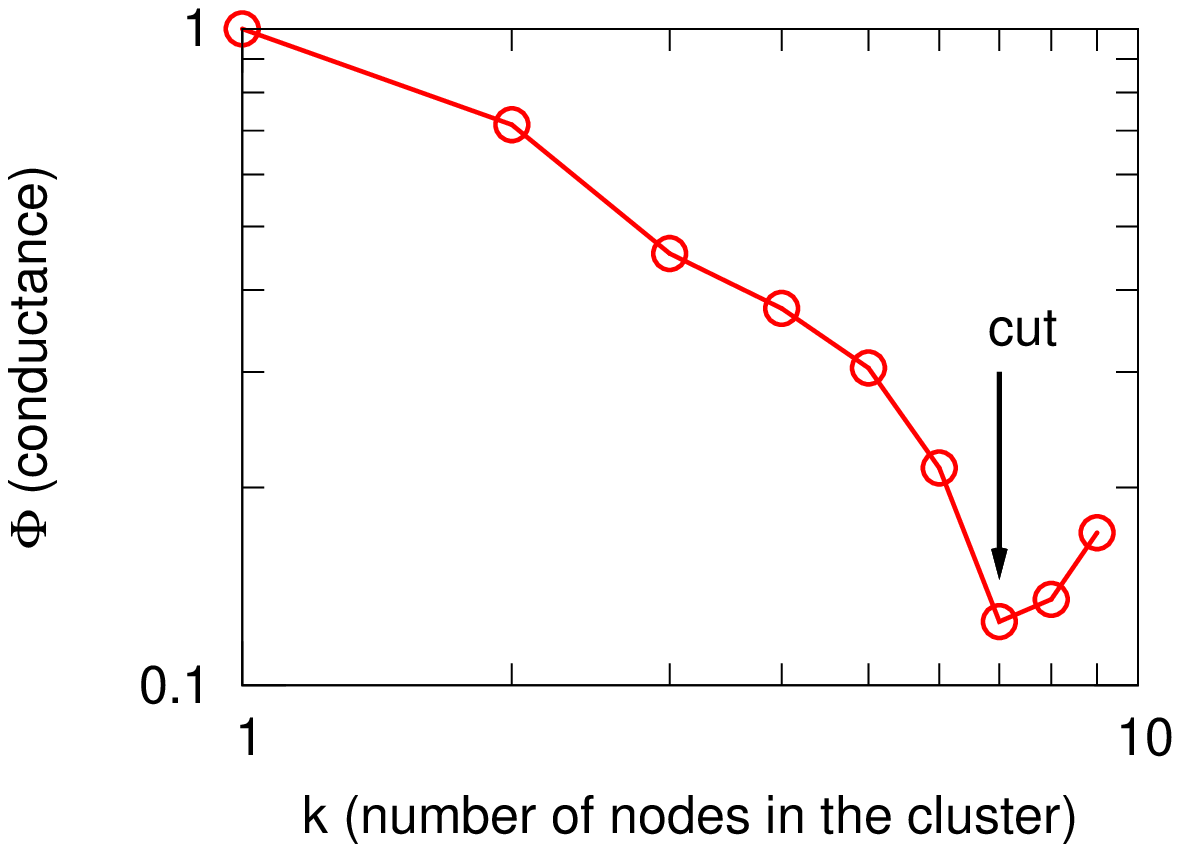}
			\label{fig:ncpp_small:monk_plot} }
  \\
		\subfigure[Network science network \ldots]{
			\includegraphics[width=0.30\textwidth]{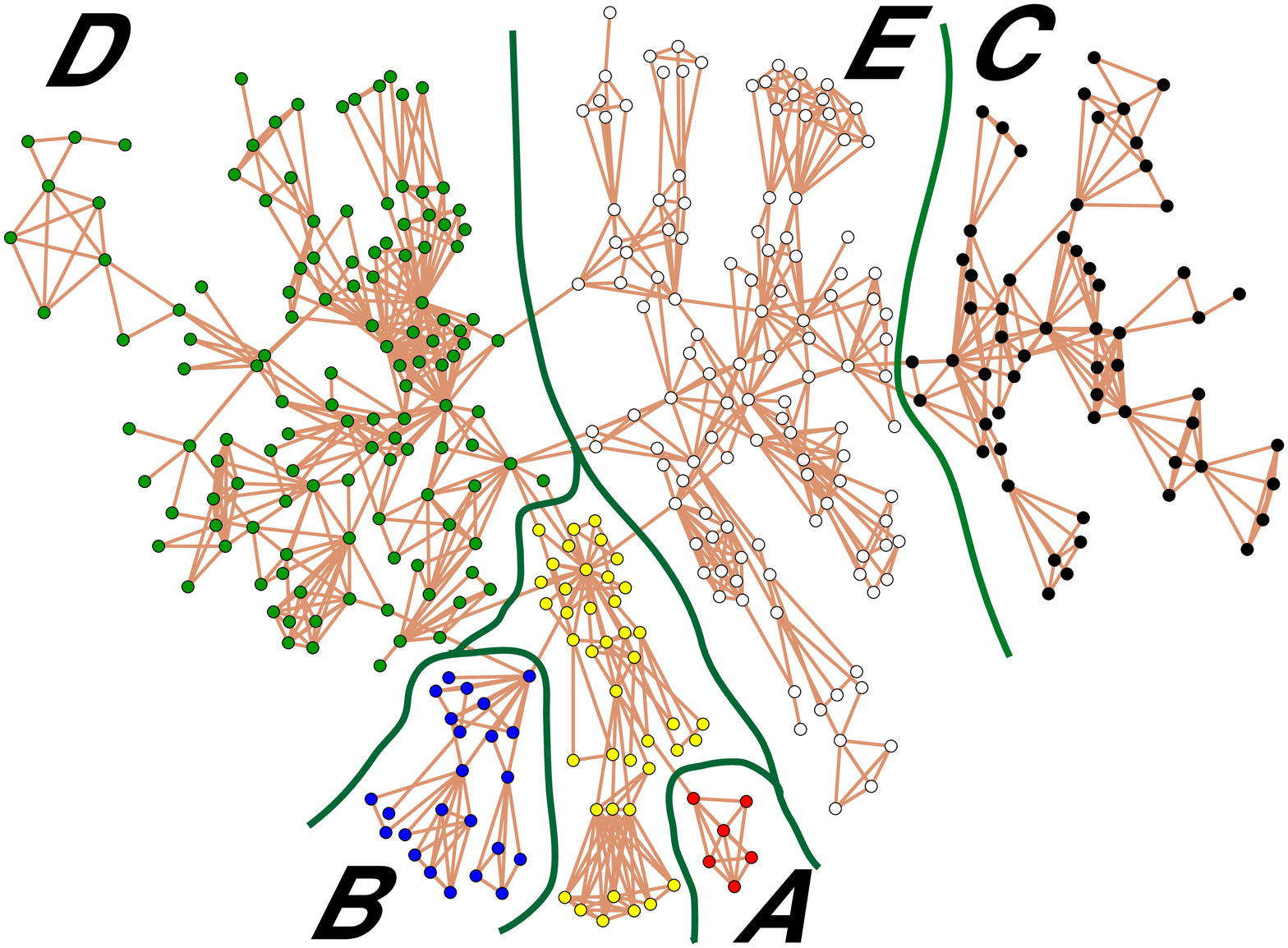}
			\label{fig:ncpp_small:netsci_graph} }
	&
		\subfigure[\ldots and it's community profile plot]{
			\includegraphics[width=0.35\textwidth]{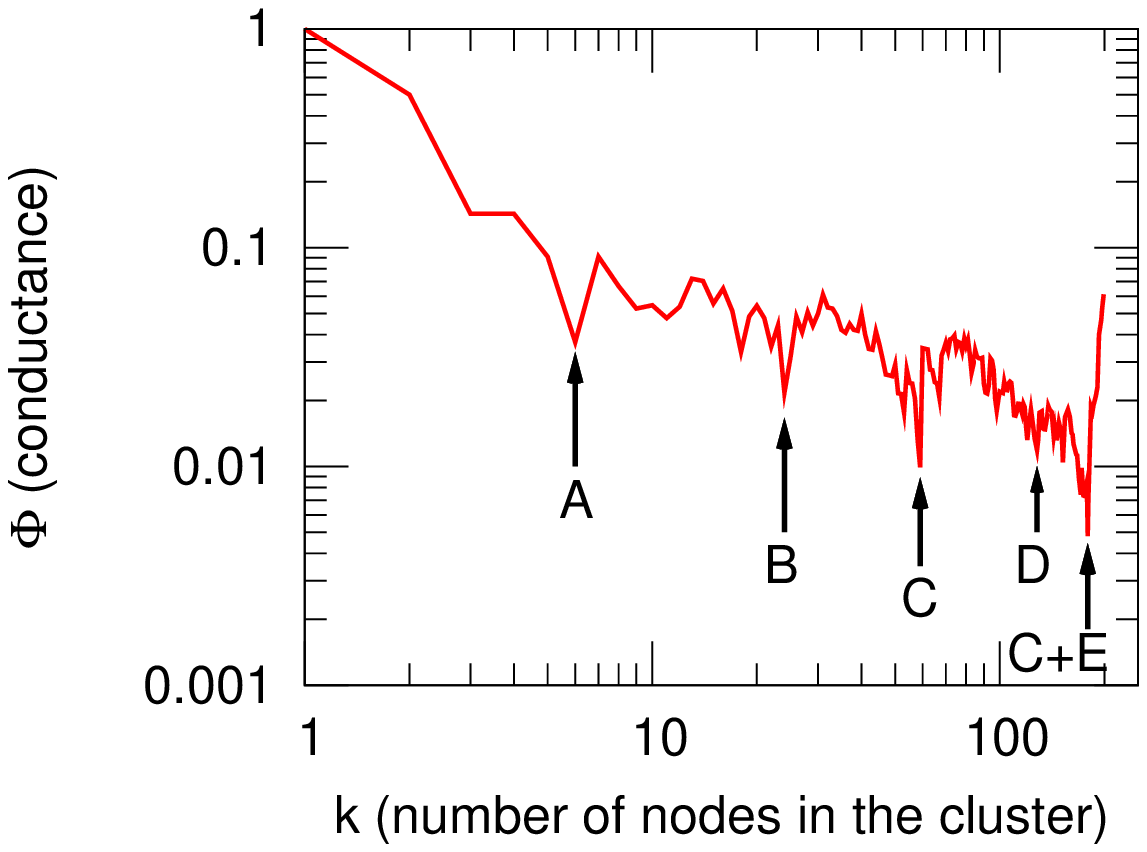}
			\label{fig:ncpp_small:netsci_plot} }
	\end{tabular}
\end{center}
\caption{
Depiction of several small social networks that are common test sets for
community detection algorithms and their network community profile plots.
(\ref{fig:ncpp_small:karate_graph}--\ref{fig:ncpp_small:karate_plot})
Zachary's karate club network.
(\ref{fig:ncpp_small:dolphin_graph}--\ref{fig:ncpp_small:dolphin_plot})
A network of dolphins.
(\ref{fig:ncpp_small:monk_graph}--\ref{fig:ncpp_small:monk_plot})
A network of monks.
(\ref{fig:ncpp_small:netsci_graph}--\ref{fig:ncpp_small:netsci_plot})
A network of researchers researching networks.
}
\label{fig:ncpp_small}
\end{figure}

Somewhat surprisingly (especially when compared with large networks in
Section~\ref{sxn:ncpp:large_sparse}), a steadily decreasing downward NCP
plot is seen for small social networks that have been extensively studied
in validating community detection algorithms. Several examples are shown
in Figures~\ref{fig:ncpp_small}. For these networks, the interpretation is
similar to that for the low-dimensional networks: the downward slope
indicates that as potential communities get larger and larger, there are
relatively more intra-edges than inter-edges; and empirically we observe
that local minima in the NCP plot correspond to sets of nodes that are
plausible communities. Consider, \emph{e.g.}, Zachary's karate
club~\cite{zachary77karate} network (\net{ZacharyKarate}), an
extensively-analyzed social
network~\cite{newman2004_detect,newman2006_ModularityPNAS,KLN07_robustness}.
The network has $34$ nodes, each of which represents a member of a karate
club, and $78$ edges, each of which represent a friendship tie between two
members.
Figure~\ref{fig:ncpp_small:karate_graph} depicts the karate club network,
and Figure~\ref{fig:ncpp_small:karate_plot} shows its NCP plot. There are
two local minima in the plot: the first dip at $k=5$ corresponds to the
Cut $A$, and the second dip at $k=17$ corresponds to Cut $B$. Note that
Cut $B$, which separates the graph roughly in half, has better conductance
value than Cut $A$. This corresponds with the intuition about the NCP plot
derived from studying low-dimensional graphs. Note also that the karate
network corresponds well with the intuitive notion of a community, where
nodes of the community are densely linked among themselves and there are
few edges between nodes of different communities.

In a similar manner: Figure~\ref{fig:ncpp_small:dolphin_graph} shows a
social network (with $62$ nodes and $159$ edges) of interactions within a
group of dolphins~\cite{lusseau03dolphins};
Figure~\ref{fig:ncpp_small:monk_graph} shows a social network of monks
(with $18$ nodes representing individual monks and $41$ edges representing
social ties between pairs of monks) in a cloister~\cite{sampson68monks};
and Figure~\ref{fig:ncpp_small:netsci_graph} depicts Newman's network
(with $914$ collaborations between $379$ researchers) of scientists who
conduct research on networks~\cite{newman04community}. For each network,
the NCP plot exhibits a downward trend, and it has local minima at cluster sizes
that correspond to good communities: the minimum for the
dolphins network~(Figure~\ref{fig:ncpp_small:dolphin_plot}) corresponds to
the separation of the network into two communities denoted with different
shape and color of the nodes (gray circles versus red squares); the minima
of the monk network~(Figure~\ref{fig:ncpp_small:monk_plot}) corresponds to
the split of $7$ Turks (red squares) and the so-called loyal opposition
(gray circles)~\cite{sampson68monks}; and empirically both local minima
and the global minimum in the network science
network~(Figure~\ref{fig:ncpp_small:netsci_plot}) correspond to plausible
communities. Note that in the last case, the figure also displays
hierarchical structure in which case the community defined by Cut $C$ is
included in a larger community that has better conductance value.

At this point, we can observe that the following two general observations
hold for networks that are well-embeddable in a low-dimensional space and
also for small social networks that have been extensively studied and used
to validate community detection algorithms. First, minima in the NCP
plots, \emph{i.e.}, the best low-conductance cuts of a given size,
correspond to communities-like sets of nodes. Second, the NCP plots are
generally relatively gradually sloping downwards, meaning that smaller
communities can be combined into larger sets of nodes that can also be
meaningfully interpreted as communities.

\subsection{Community profile plots for large social and information networks}
\label{sxn:ncpp:large_sparse}

\begin{figure}
   \begin{center}
   \begin{tabular}{cc}
      \subfigure[\net{LiveJournal01}]{
         \includegraphics[width=0.40\textwidth]{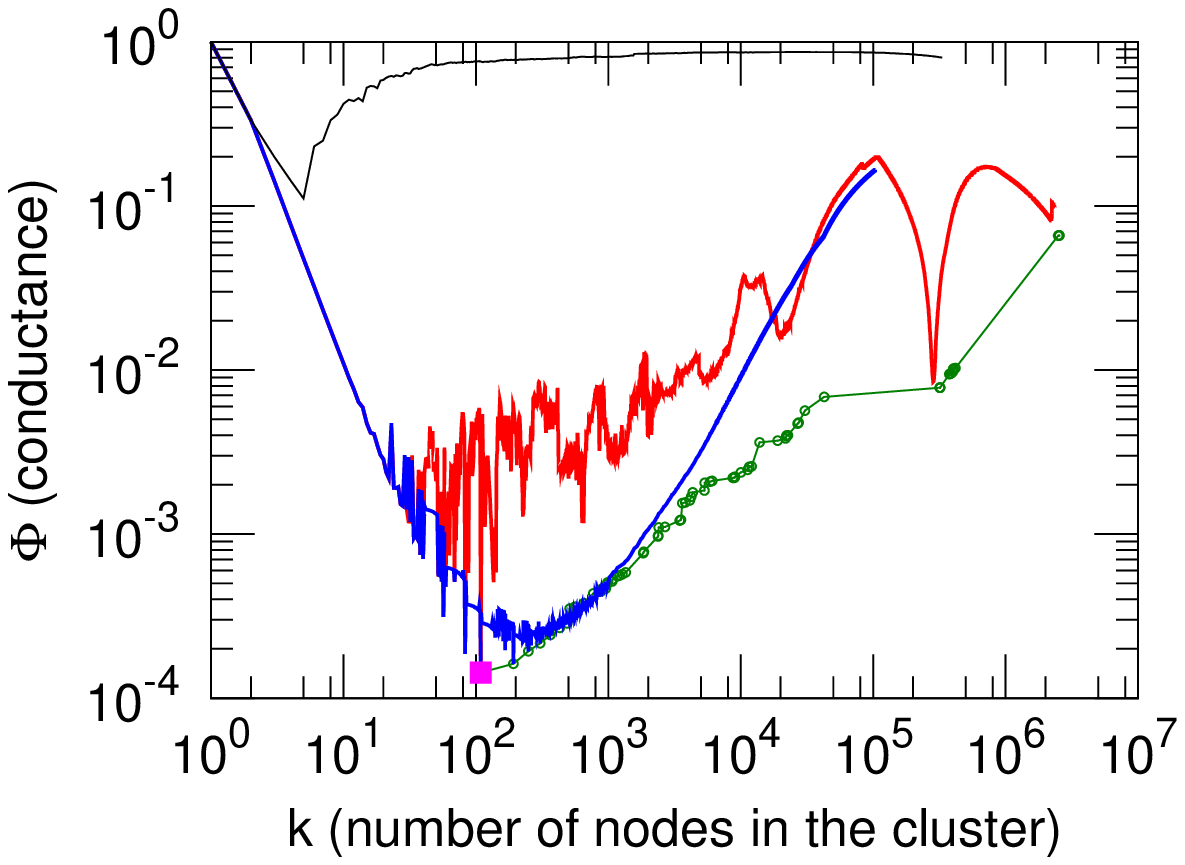}
         \label{fig:netPhiPlot:LiveJournal1}
      } &
      \subfigure[\net{Epinions}]{
         \includegraphics[width=0.40\textwidth]{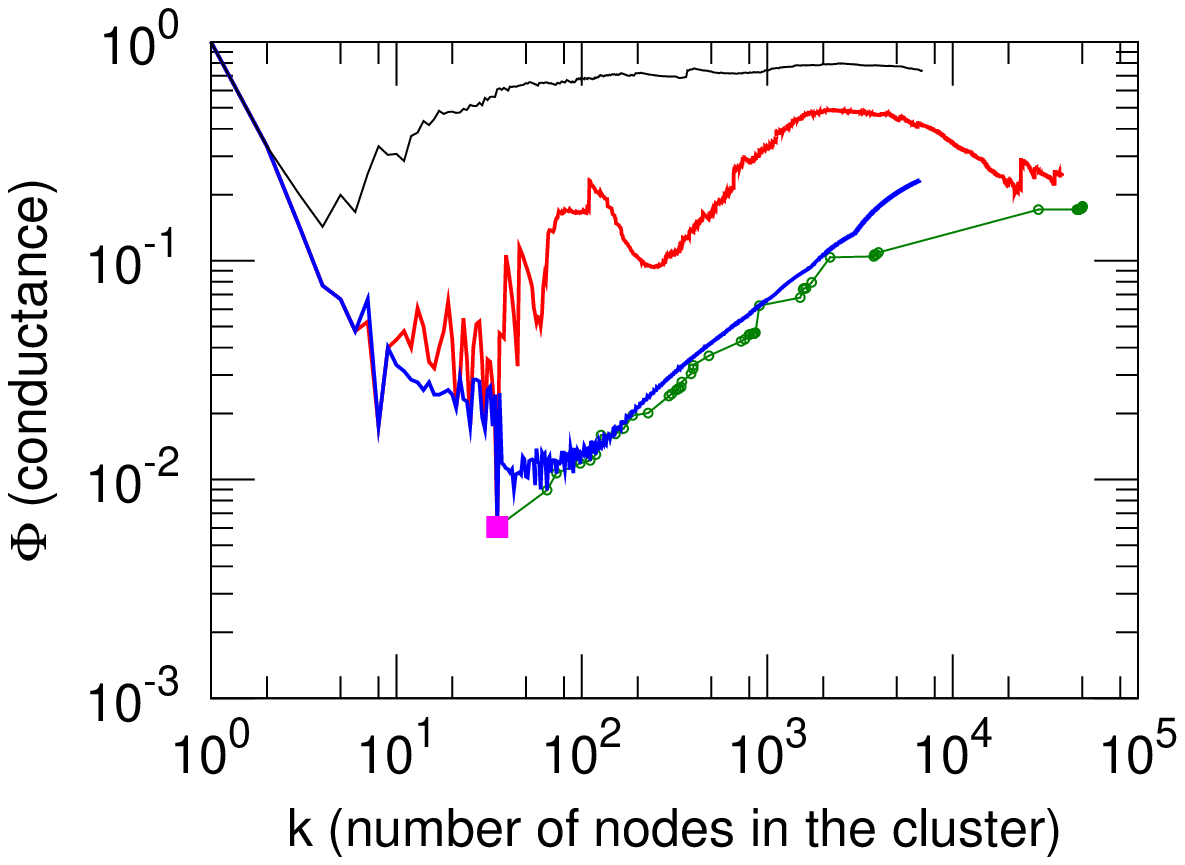}
         \label{fig:netPhiPlot:Epinions}
      } \\
      \subfigure[\net{Cit-hep-th}]{
         \includegraphics[width=0.40\textwidth]{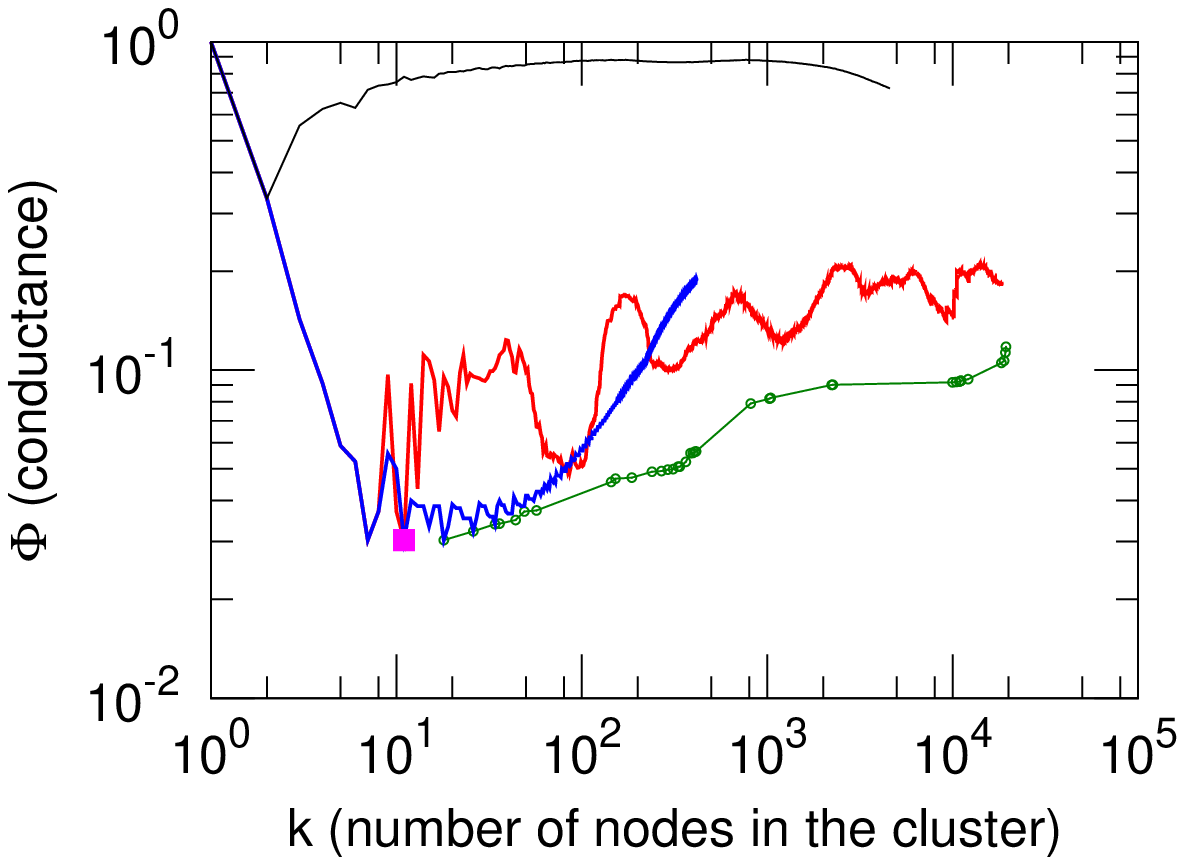}
         \label{fig:netPhiPlot:Cit-hep-th}
      } &
      \subfigure[\net{Web-Google}]{
         \includegraphics[width=0.40\textwidth]{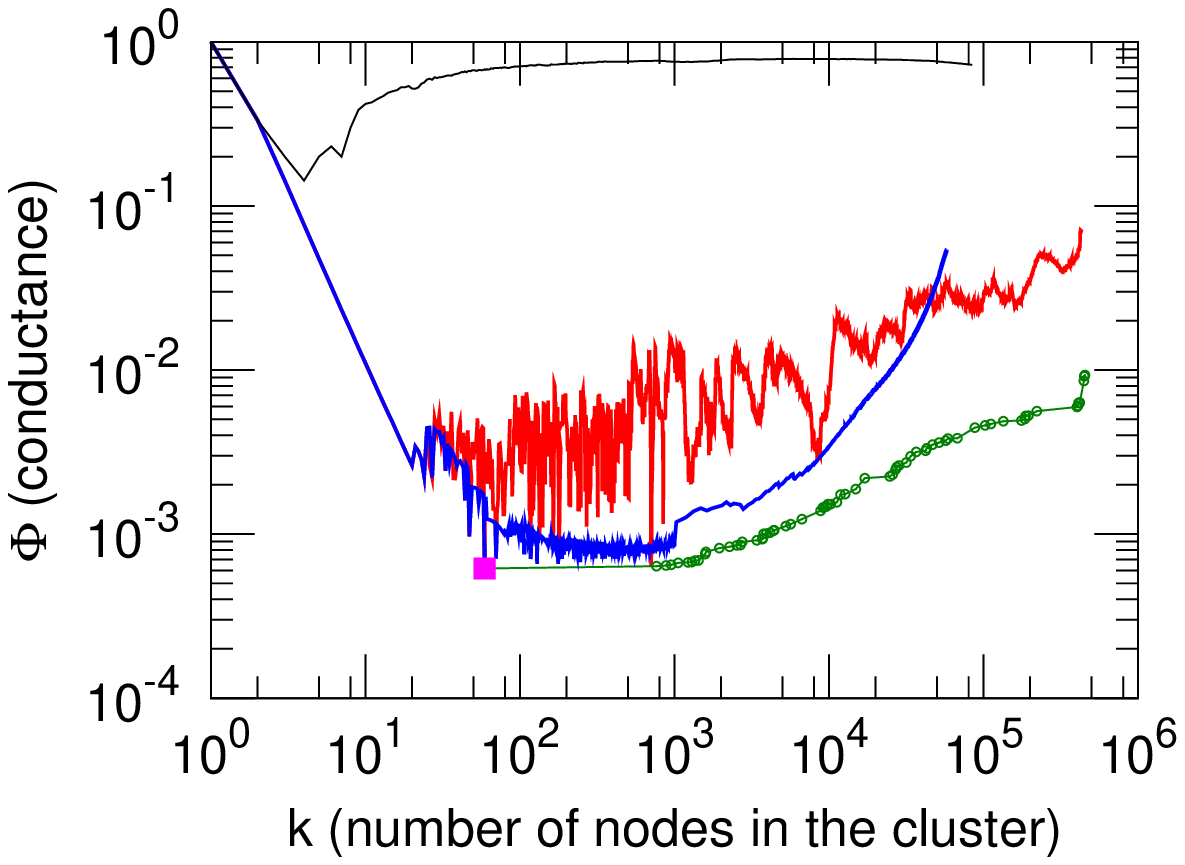}
         \label{fig:netPhiPlot:Web-Google}
      } \\
      \subfigure[\net{Atp-DBLP}]{
         \includegraphics[width=0.40\textwidth]{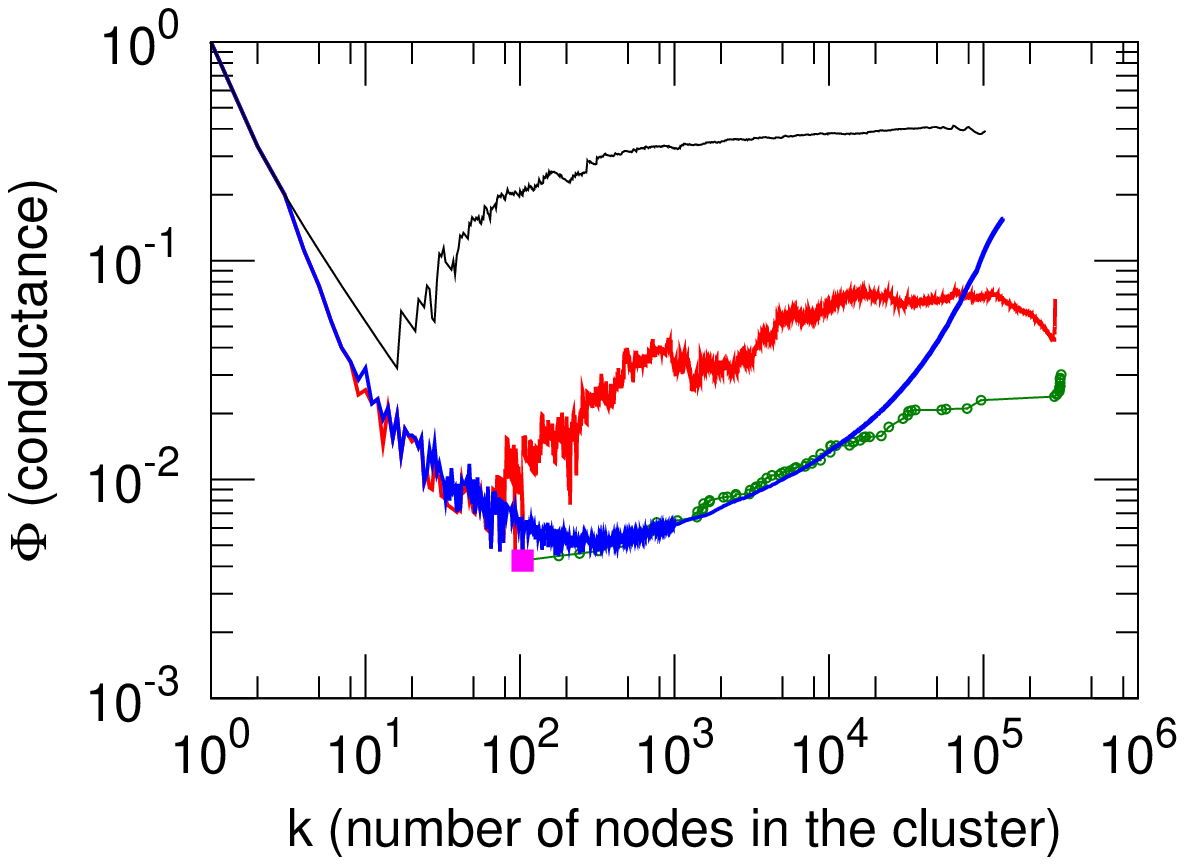}
         \label{fig:netPhiPlot:Atp-DBLP}
      } &
      \subfigure[\net{Gnutella-31}]{
         \includegraphics[width=0.40\textwidth]{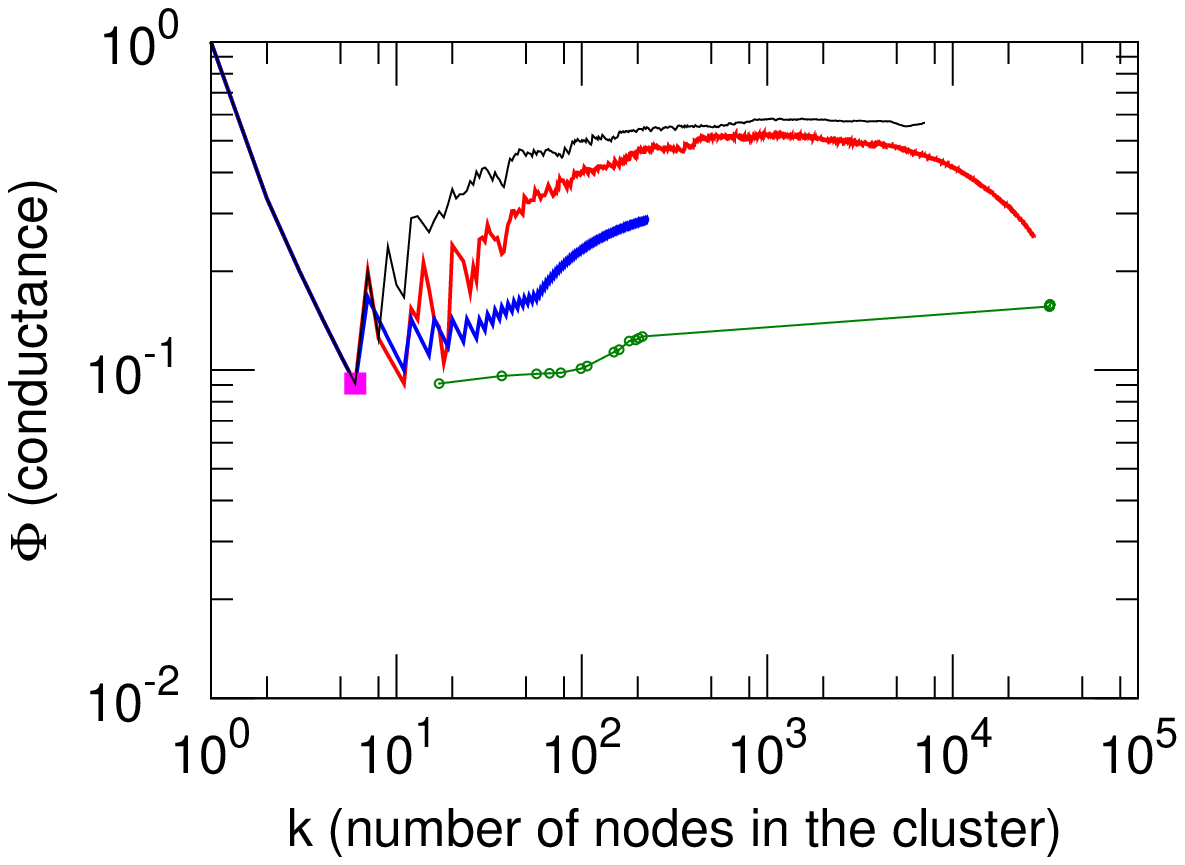}
         \label{fig:netPhiPlot:Gnutella}
      } \\
      \multicolumn{2}{c}{\includegraphics[width=0.6\textwidth]{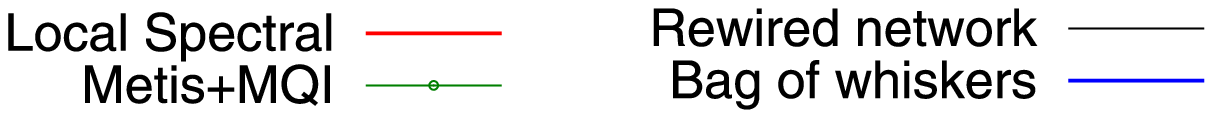}}
   \end{tabular}
   \end{center}
\caption{
[Best viewed in color.]
Network community profile plots for a representative sample of large networks
listed in
Tables~\ref{tab:data_StatsDesc_1}, \ref{tab:data_StatsDesc_2}
and~\ref{tab:data_StatsDesc_3}.
The red curves plot the results of the Local Spectral Algorithm on the
specified network; green curves plot the results of Metis+MQI; blue curves plot
the results of the Bag-of-Whiskers Heuristic; and black curves plot the results
of the Local Spectral Algorithm applied to a randomly rewired version of the
same network.
Notice that in all cases the ``best'' communities are quite small (typically
between $10$ and $100$ nodes) and that the network community profile plot
steadily increases for nearly its entire range.
See Figures~\ref{fig:phiDatasets1}, \ref{fig:phiDatasets2}, and
\ref{fig:phiDatasets3} for the NCP plots of other networks.
}
\label{fig:netPhiPlot}
\end{figure}

We have examined NCP plots for each of the networks listed in
Tables~\ref{tab:data_StatsDesc_1}, \ref{tab:data_StatsDesc_2}
and~\ref{tab:data_StatsDesc_3}. In Figure~\ref{fig:netPhiPlot}, we present
NCP plots for six of these networks. (These particular networks were
chosen to be representative of the wide range of networks we have
examined, and for ease of comparison we will compute other properties for
them in future sections. See Figures~\ref{fig:phiDatasets1},
\ref{fig:phiDatasets2}, and \ref{fig:phiDatasets3} in
Section~\ref{sxn:ncpp:large_sparse_more} for the NCP plots of other
networks listed in Tables~\ref{tab:data_StatsDesc_1},
\ref{tab:data_StatsDesc_2} and~\ref{tab:data_StatsDesc_3}, and for a
discussion of them.) The most striking feature of these plots is that the
NCP plot is steadily increasing for nearly its entire range.

Consider, first, the NCP plot for the \net{LiveJournal01} social network,
as shown in Figure~\ref{fig:netPhiPlot:LiveJournal1}, and focus first on
the red curve, which presents the results of applying the Local Spectral
Algorithm.%
\footnote{ The algorithm takes as input two parameters---the seed node and
the parameter $\epsilon$ that intuitively controls the locality of the
computation---and it outputs a set of nodes. For a given seed node and
resolution parameter $\epsilon$ we obtain a local community profile plot,
which tells us about conductance of cuts in vicinity of the seed node. By
taking the lower-envelope over community profiles of different seed nodes
and $\epsilon$ values we obtain the global network community profile plot.
For our experiments, we typically considered $100$ different values of
$\epsilon$. Since very local random walks discover small clusters, in this
case we considered every node as a seed node. As we examine larger
clusters, the random walk computation spreads farther away from the seed
node, in which case the exact choice of seed node becomes less important.
Thus, in this case, we sampled fewer seed nodes. Additionally, in our
experiments, for each value of $\epsilon$ we randomly sampled nodes until
each node in the network was visited by random walks starting from, $10$
different seed nodes on average. }
We make the following observations:
\begin{itemize}
\item Up to a size scale, which empirically is roughly $100$ nodes,
    the slope of the NCP plot is generally sloping downward.
\item At that size scale, we observe the global minimum of the NCP
    plot. This set of nodes as well as others achieving local minima
    of the NCP plot in the same size range are the ``best''
    communities, according to the conductance measure, in the entire
    graph.
\item These best communities (the best denoted by a square) are barely
    connected to the rest of the graph, \emph{e.g.}, they are
    typically connected to the rest of the nodes by a {\em single}
    edge.
\item Above the size scale of roughly $100$ nodes, the NCP plot
    gradually increases over several orders of magnitude. The ``best''
    communities in the entire graph are quite good (in that they have
    size roughly $10^{2}$ nodes and conductance scores less than
    $10^{-3}$) whereas the ``best'' communities of size $10^{5}$ or
    $10^{6}$ have conductance scores of about $10^{-1}$. In between
    these two size extremes, the conductance scores get gradually
    worse, although there are numerous local dips and even one
    relatively large dip between $10^{5}$ and $10^{6}$ nodes.
\end{itemize}
Note that both axes in Figure~\ref{fig:netPhiPlot} are logarithmic, and
thus the upward trend of the NCP plot is over a wide range of size scales.
Note also that the green curve plots the results of Metis+MQI (that returns 
disconnected clusters), and the blue curve plots the results of applying 
the Bag-of-Whiskers Heuristic, as described in 
Section~\ref{sxn:obs_struct:bags_of_whiskers}. These procedures will be
discussed in detail in Sections~\ref{sxn:obs_struct}
and~\ref{algo-notes-section}.

The black curve in Figure~\ref{fig:netPhiPlot:LiveJournal1} plots the
results of the Local Spectral Algorithm applied to a \emph{rewired
version} of the \net{LiveJournal01} network, \emph{i.e.}, to a random
graph conditioned on the same degree distribution as the original network.
(We obtain such random graph by starting with the original network and
then randomly selecting pairs of edges and rewiring the endpoints. By
doing the rewiring long enough, we obtain a random graph that has the same
degree sequence as the original network~\cite{milo04graphs}.)

Interestingly, the NCP of a rewired network first slightly decreases but
then increases and flattens out. 
Several things should be noted:
\begin{itemize}
\item The original \net{LiveJournal01} network has considerably more
    structure, \emph{i.e.}, deeper/better cuts, than its rewired
    version, even up to the largest size scales. That is, we observe
    significantly more structure than would be seen, for example, in
    an random graph on the same degree sequence.
\item Relative to the original network, the ``best'' community in the
    rewired graph, \emph{i.e.}, the global minimum of the conductance
    curve, shifts upward and towards the left. This means that in
    rewired networks the best conductance clusters get smaller and
    have worse conductance scores.
\item Sets at and near the minimum are small trees that are connected
    to the core of the random graph by a single edge.
\item After the small dip at a very small size scale ($\approx 10$
    nodes), the NCP plot increases to a high level rather quickly.
    This is due to the absence of structure in the core.
\end{itemize}
Finally, also note that the variance in the rewired version of the NCP
plot (data not shown) is not much larger than the width of the curve in
the figure.

We have observed qualitatively similar results in nearly every large
social and information network we have examined. For example, several
additional examples are presented in Figure~\ref{fig:netPhiPlot}: another
network from the class of social networks (\net{Epinions}, in
Figure~\ref{fig:netPhiPlot:Epinions}); an information/citation network
(\net{Cit-hep-th}, in Figure~\ref{fig:netPhiPlot:Cit-hep-th}); a Web graph
(\net{Web-Google}, in Figure~\ref{fig:netPhiPlot:Web-Google}); a Bipartite
affiliation network (\net{AtP-DBLP}, in
Figure~\ref{fig:netPhiPlot:Atp-DBLP}); and an Internet network
(\net{Gnutella-31}, in Figure~\ref{fig:netPhiPlot:Gnutella}).

Qualitative observations are consistent across the range of network sizes,
densities, and different domains from which the networks are drawn. Of
course, these six networks are very different than each other---some of
these differences are hidden due to the definition of the NCP plot,
whereas others are evident. Perhaps the most obvious example of the latter
is that even the best cuts in \net{Gnutella-31} are not significantly
smaller or deeper than in the corresponding rewired network, whereas for
\net{Web-Google} we observe cuts that are orders of magnitude deeper.

Intuitively, the upward trend in the NCP plot means that separating large
clusters from the rest of the network is especially expensive. It suggests
that larger and larger clusters are ``blended in'' more and more with the
rest of the network. The interpretation we draw, based on these data and
data presented in subsequent sections is that, if a density-based concept
such as conductance captures our intuitive notion of community goodness
and if we model large networks with interaction graphs, then the best
possible communities get less and less community-like as they grow in
size.

\subsection{More community profile plots for large social and information networks}
\label{sxn:ncpp:large_sparse_more}

\begin{figure*}
	\begin{center}
	\begin{tabular}{ccc}
    \multicolumn{3}{c}{\bf Social networks}\\
	\includegraphics[width=0.3\textwidth]{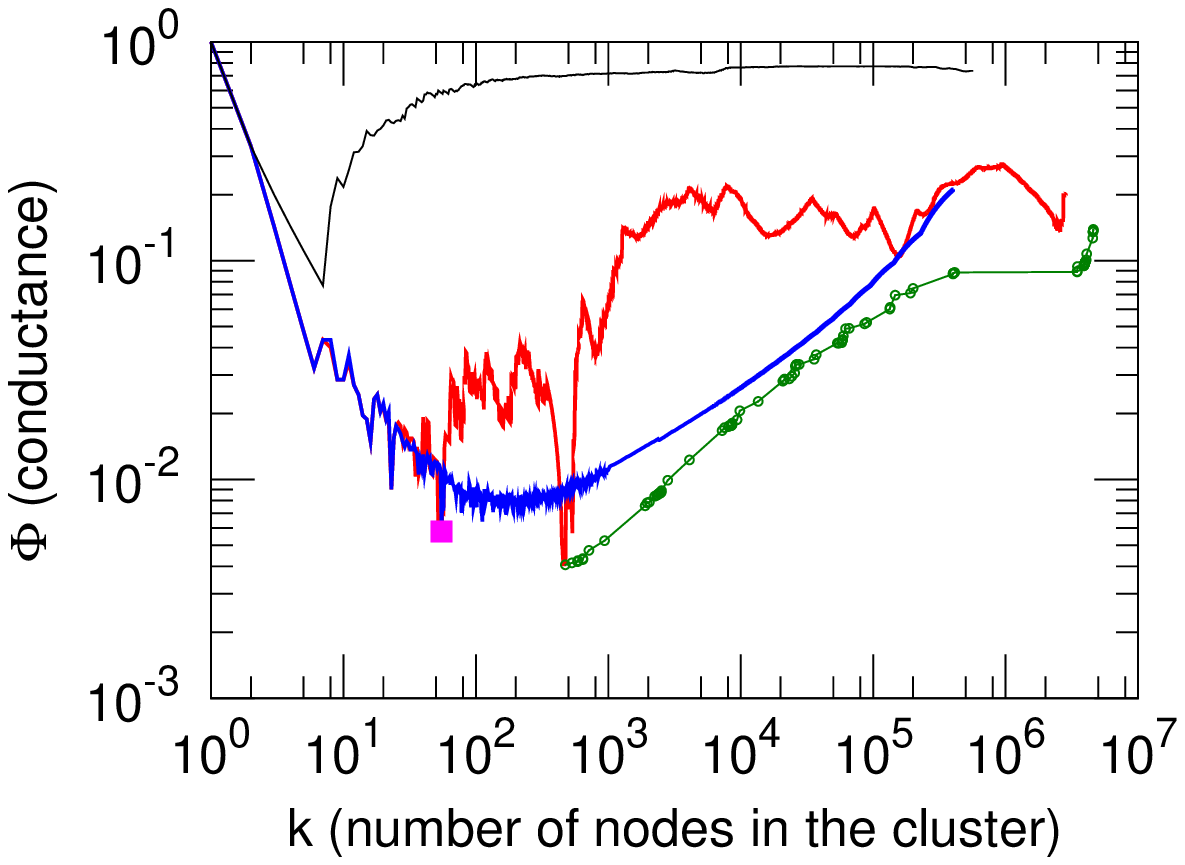} &
  \includegraphics[width=0.3\textwidth]{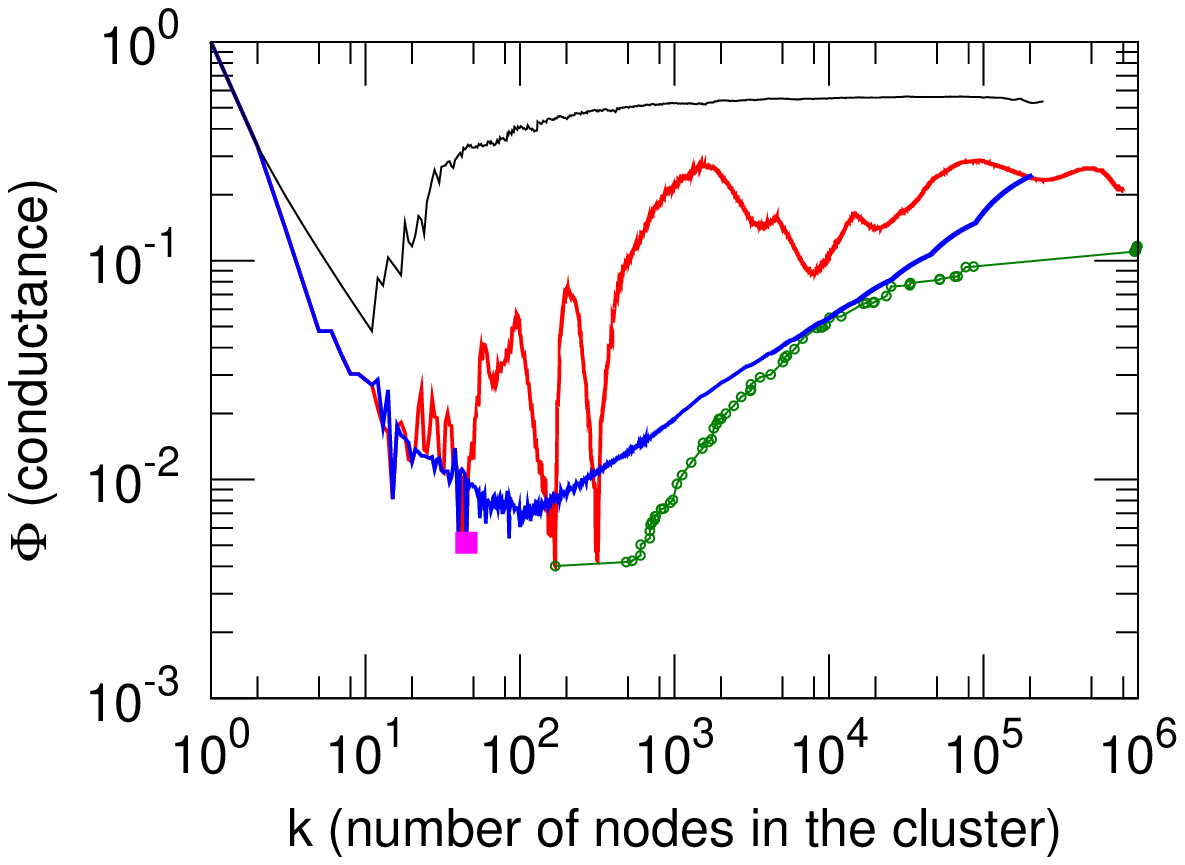} &
  \includegraphics[width=0.3\textwidth]{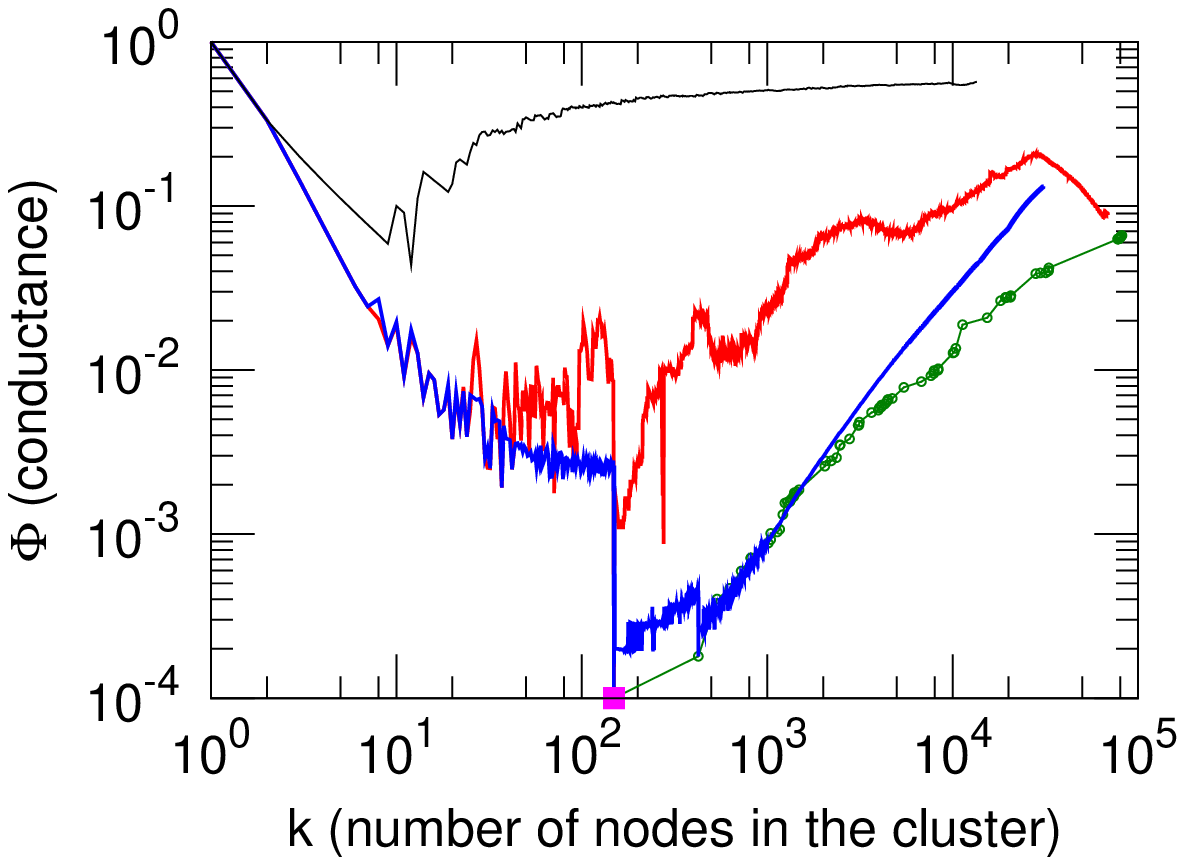} \\
	\net{LinkedIn} & \net{Messenger}  & \net{Delicious}\\
	\includegraphics[width=0.3\textwidth]{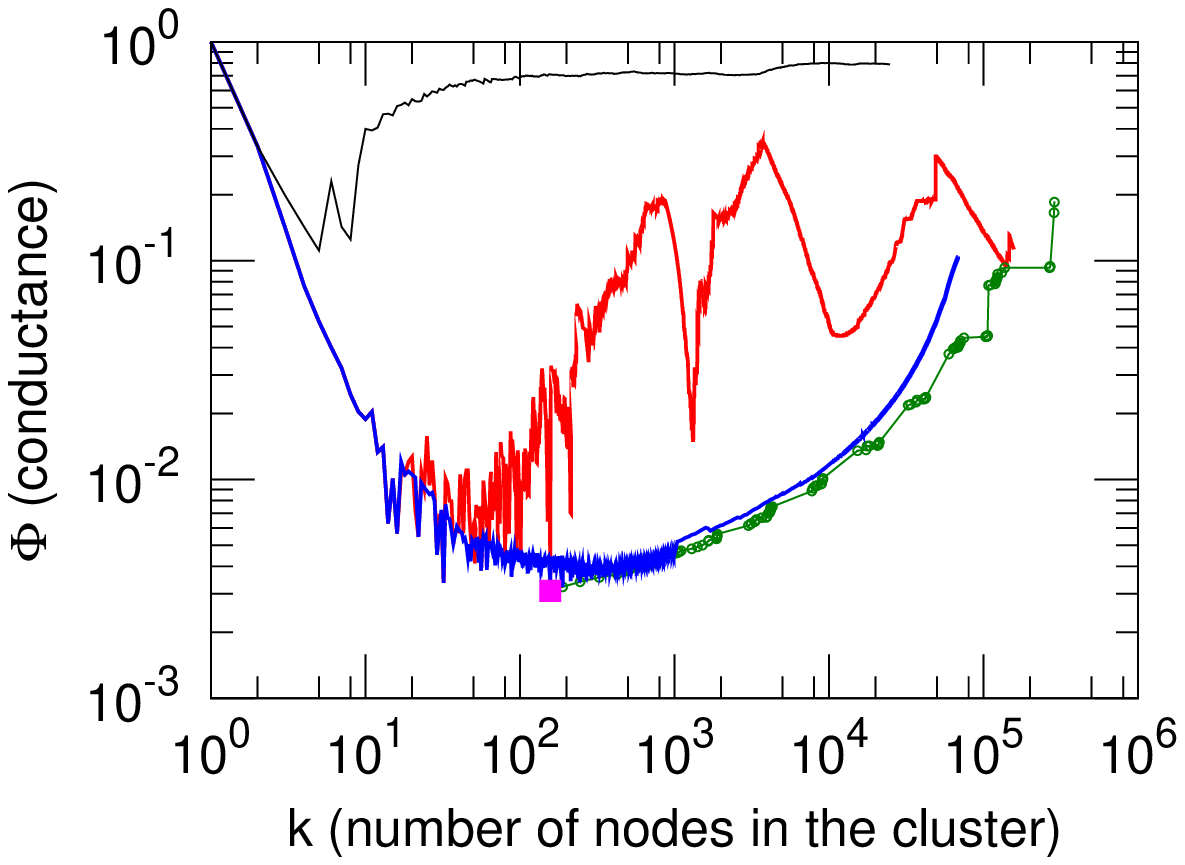} &
	\includegraphics[width=0.3\textwidth]{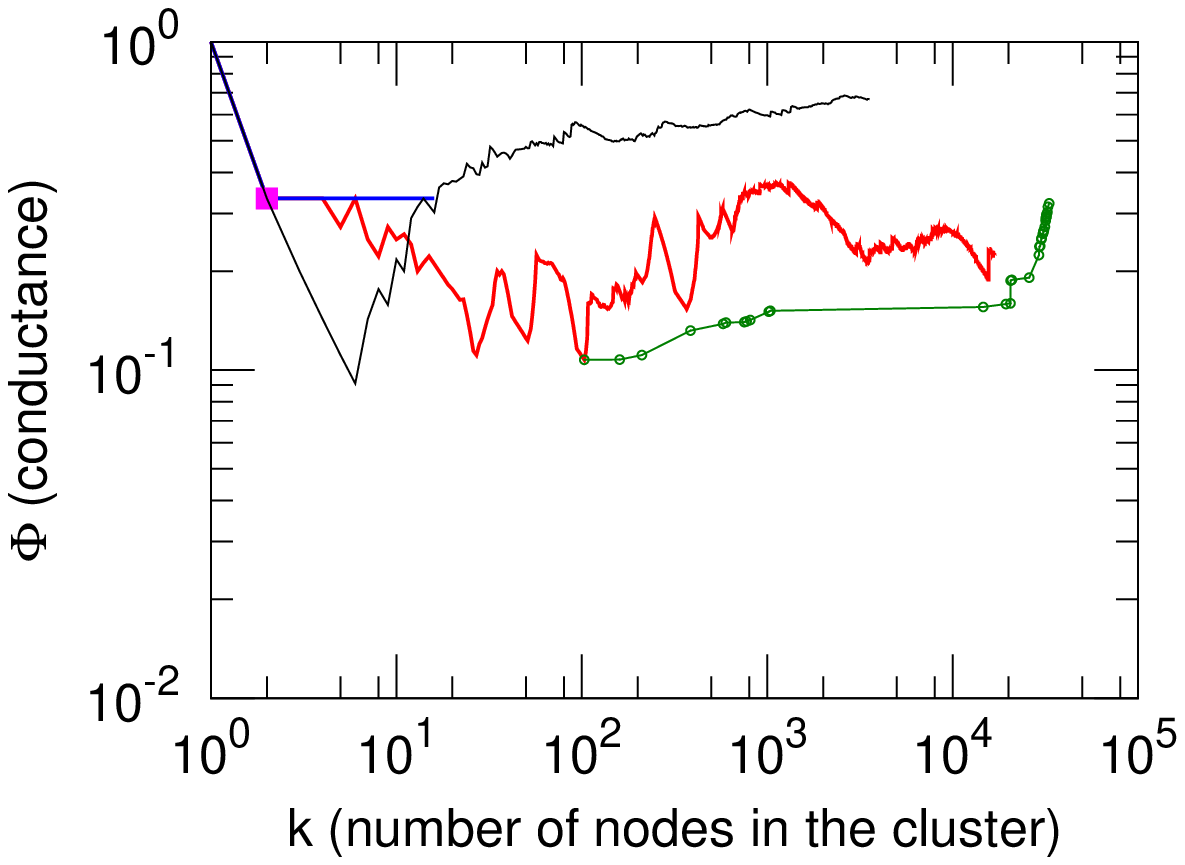} &
	\includegraphics[width=0.3\textwidth]{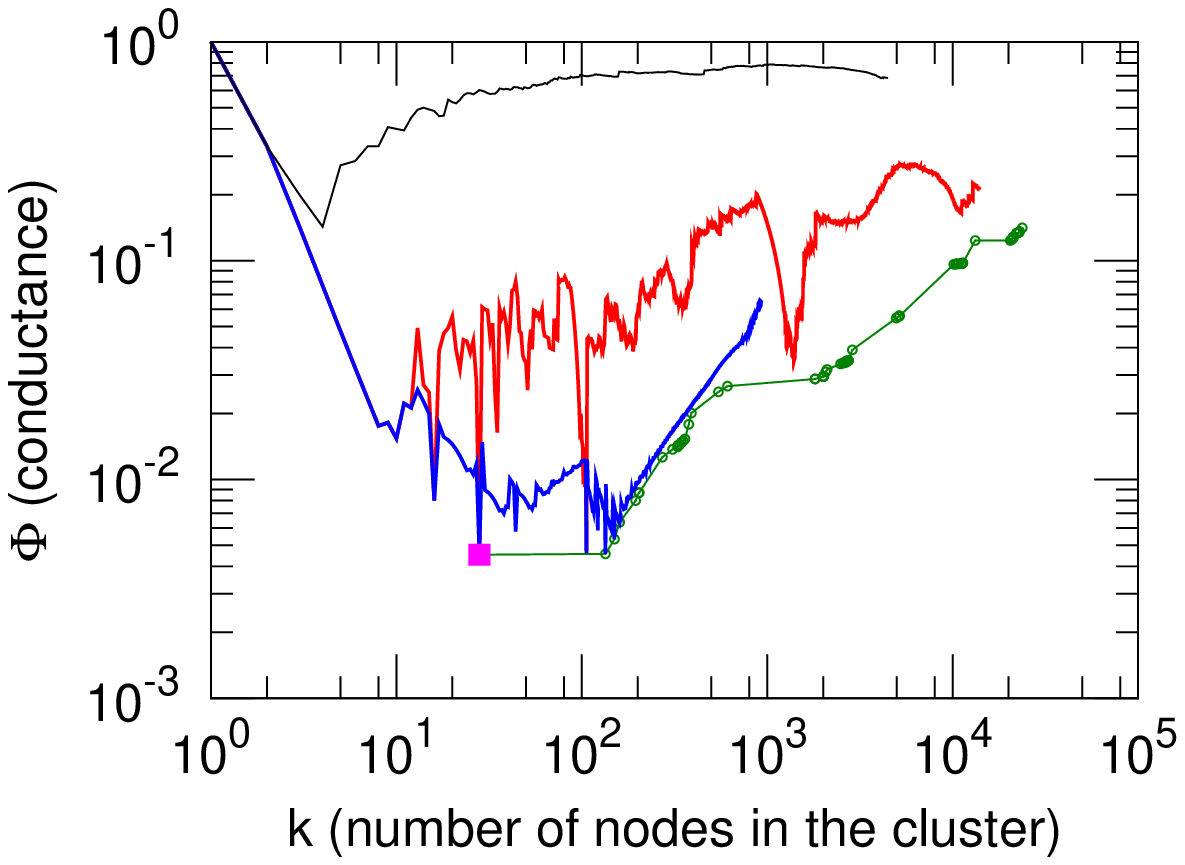} \\
	\net{Flickr} & \net{Email-InOut}  & \net{Email-enron}\\
    & & \\
    \multicolumn{3}{c}{\bf Information networks (citation and blog networks)}\\
	\includegraphics[width=0.3\textwidth]{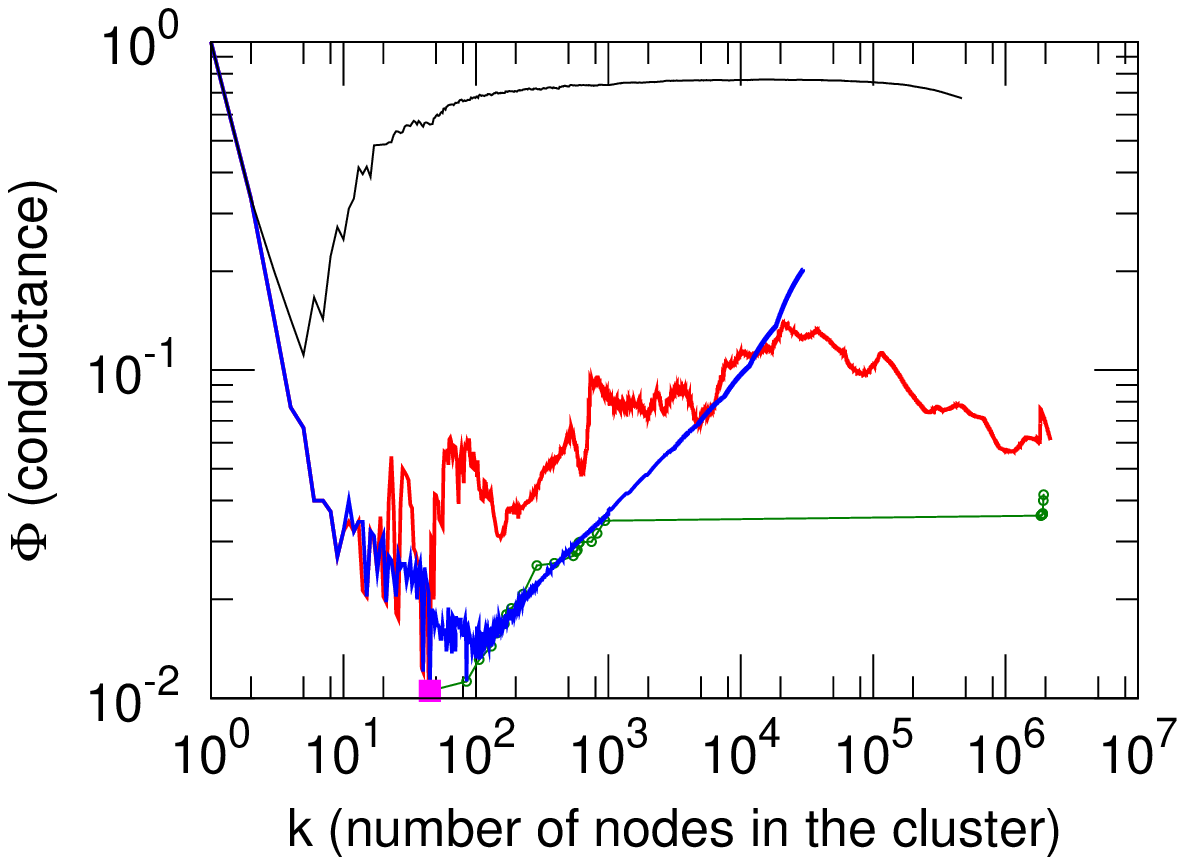} &
  \includegraphics[width=0.3\textwidth]{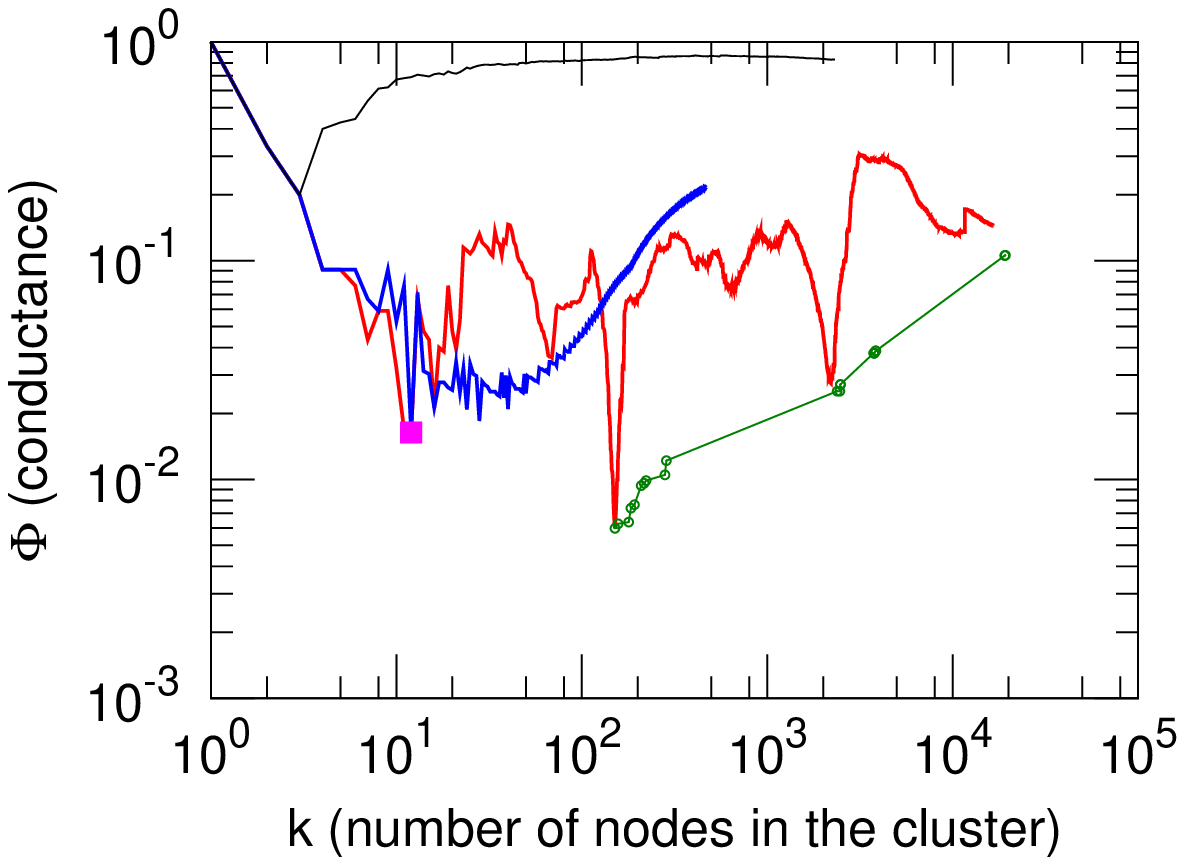} &
	\includegraphics[width=0.3\textwidth]{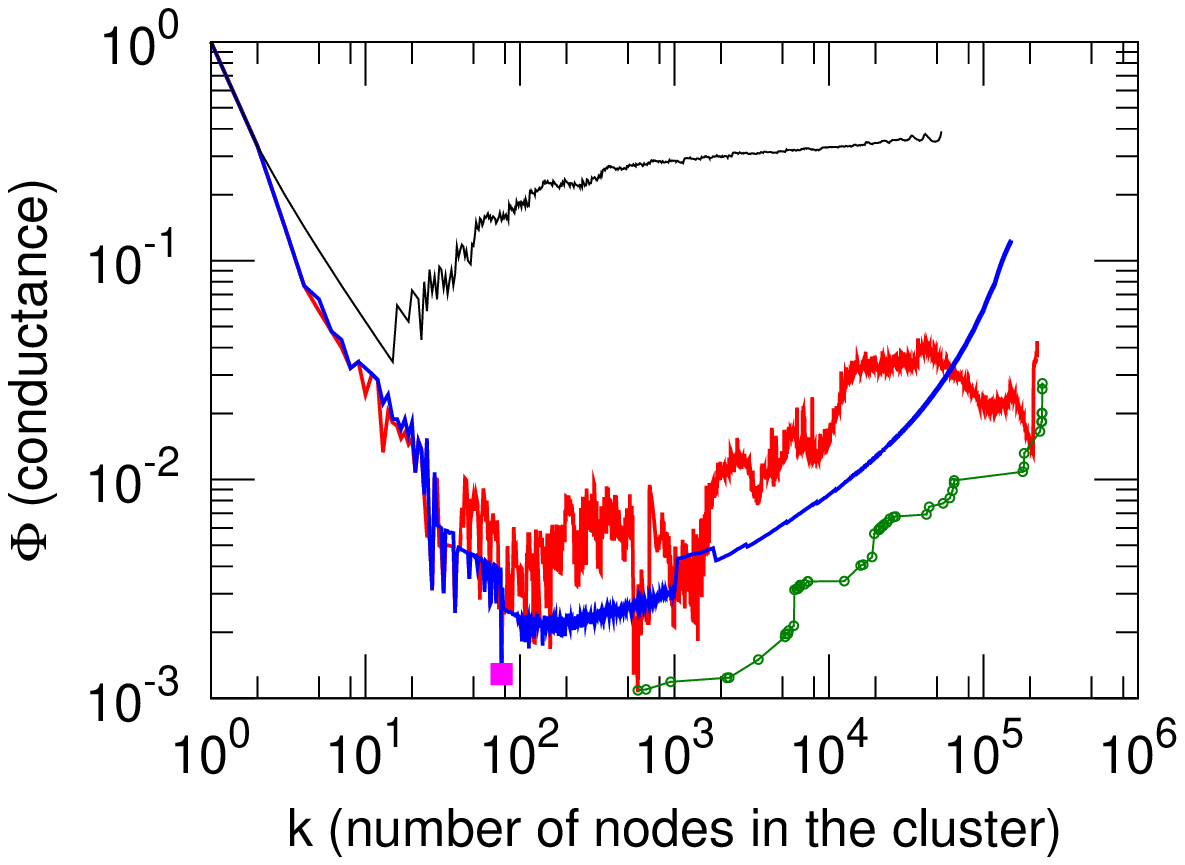} \\
	\net{Patents} & \net{Blog-nat06all}  & \net{Post-nat06all}\\
    & & \\
    \multicolumn{3}{c}{\bf Collaboration networks}\\
	\includegraphics[width=0.3\textwidth]{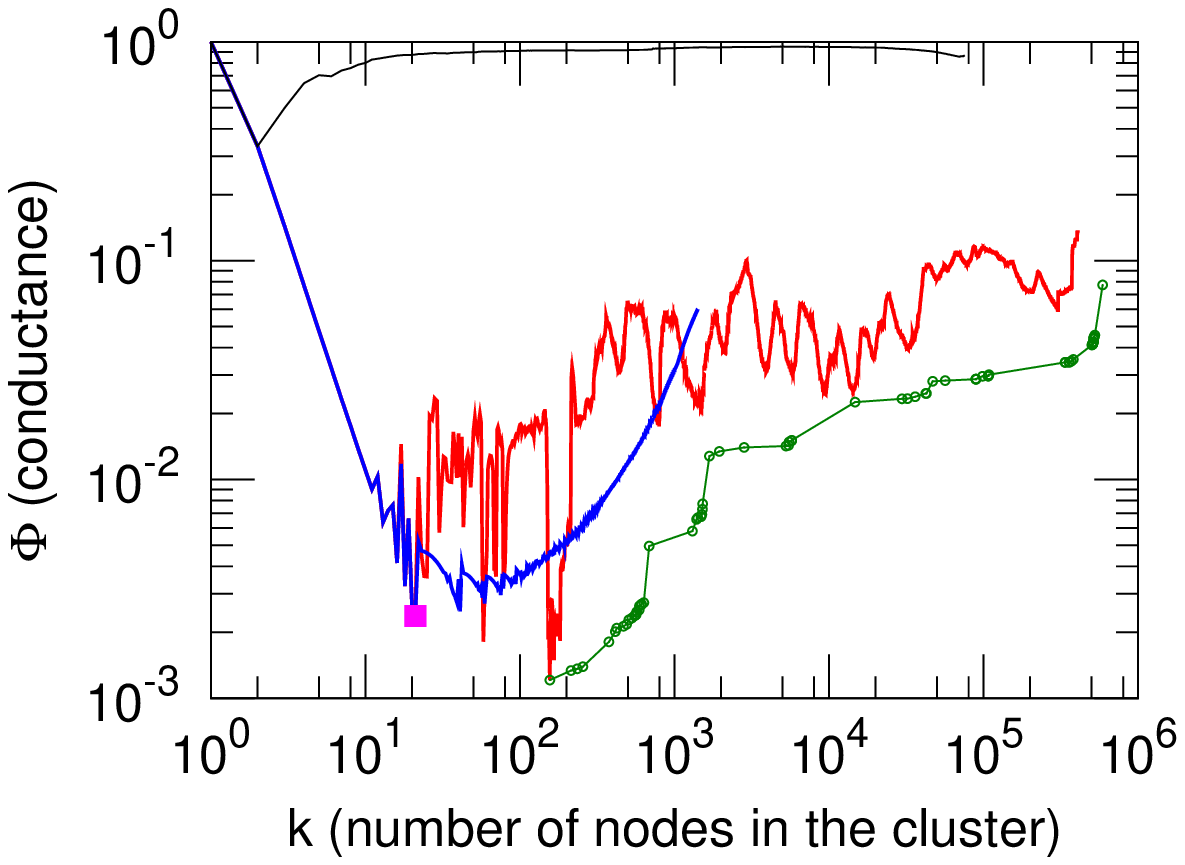} &
  \includegraphics[width=0.3\textwidth]{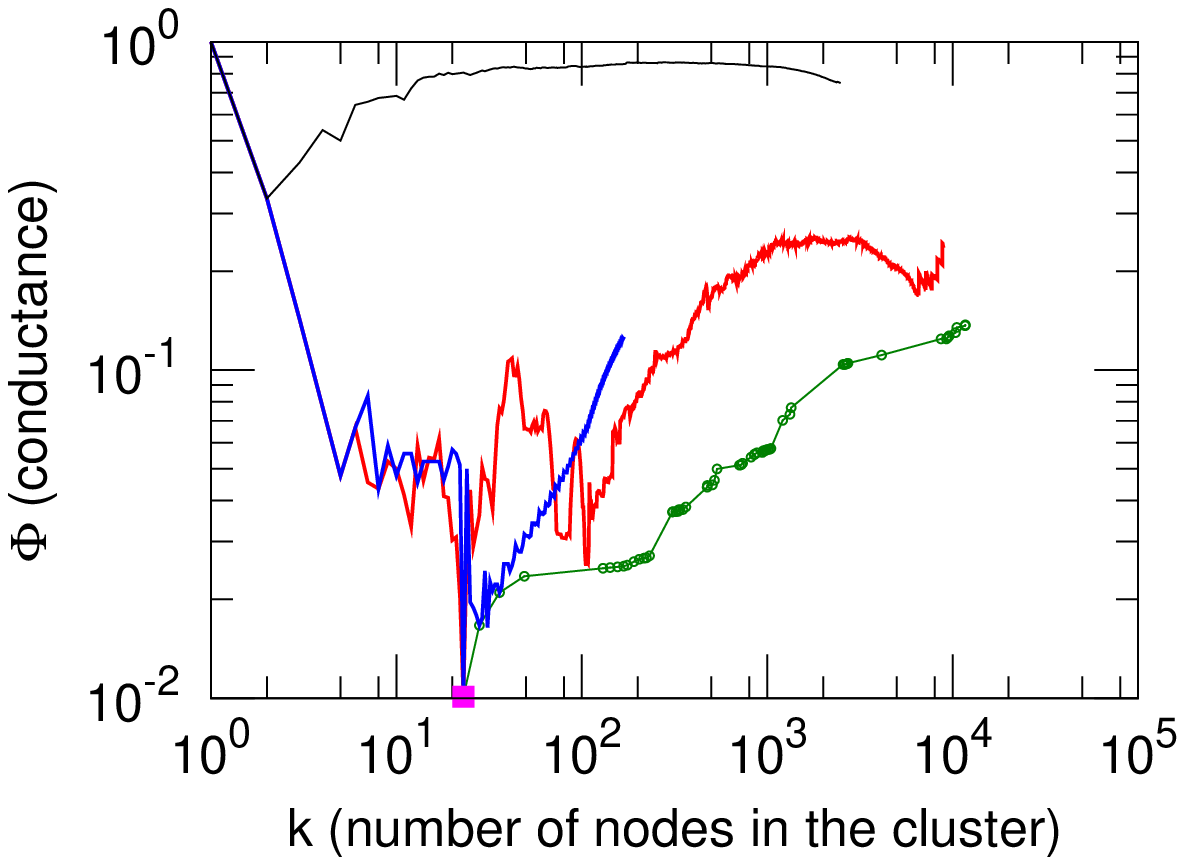} &
	\includegraphics[width=0.3\textwidth]{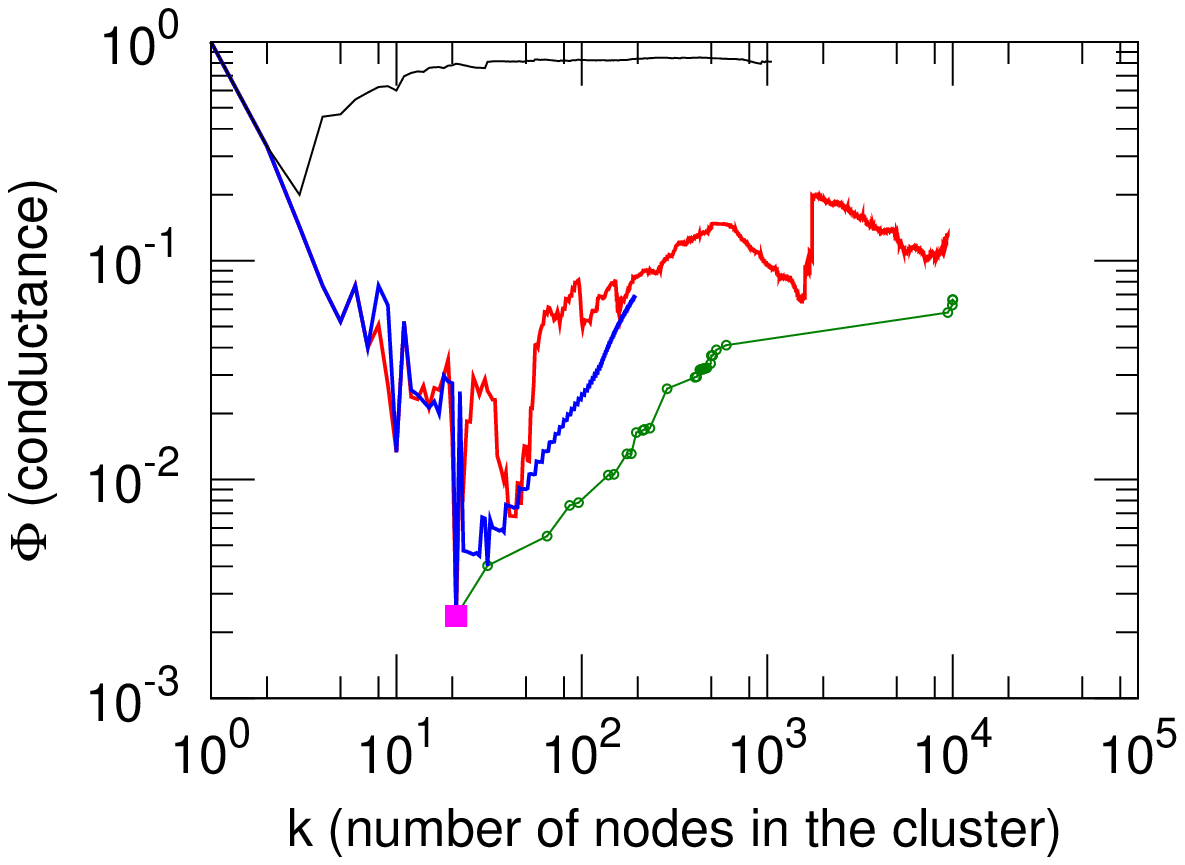} \\
	\net{AtA-IMDB} & \net{CA-astro-ph}  & \net{CA-hep-ph}\\
  \multicolumn{3}{c}{\includegraphics[width=0.6\textwidth]{phiTR-legend2}} \\
	\end{tabular}
	\end{center}

\caption{ [Best viewed in color.] Community profile plots of networks from
Table~\ref{tab:data_StatsDesc_1}. } \label{fig:phiDatasets1}
\end{figure*}

\begin{figure*}
	\begin{center}
	\begin{tabular}{ccc}
    \multicolumn{3}{c}{\bf Web graphs}\\
	\includegraphics[width=0.3\textwidth]{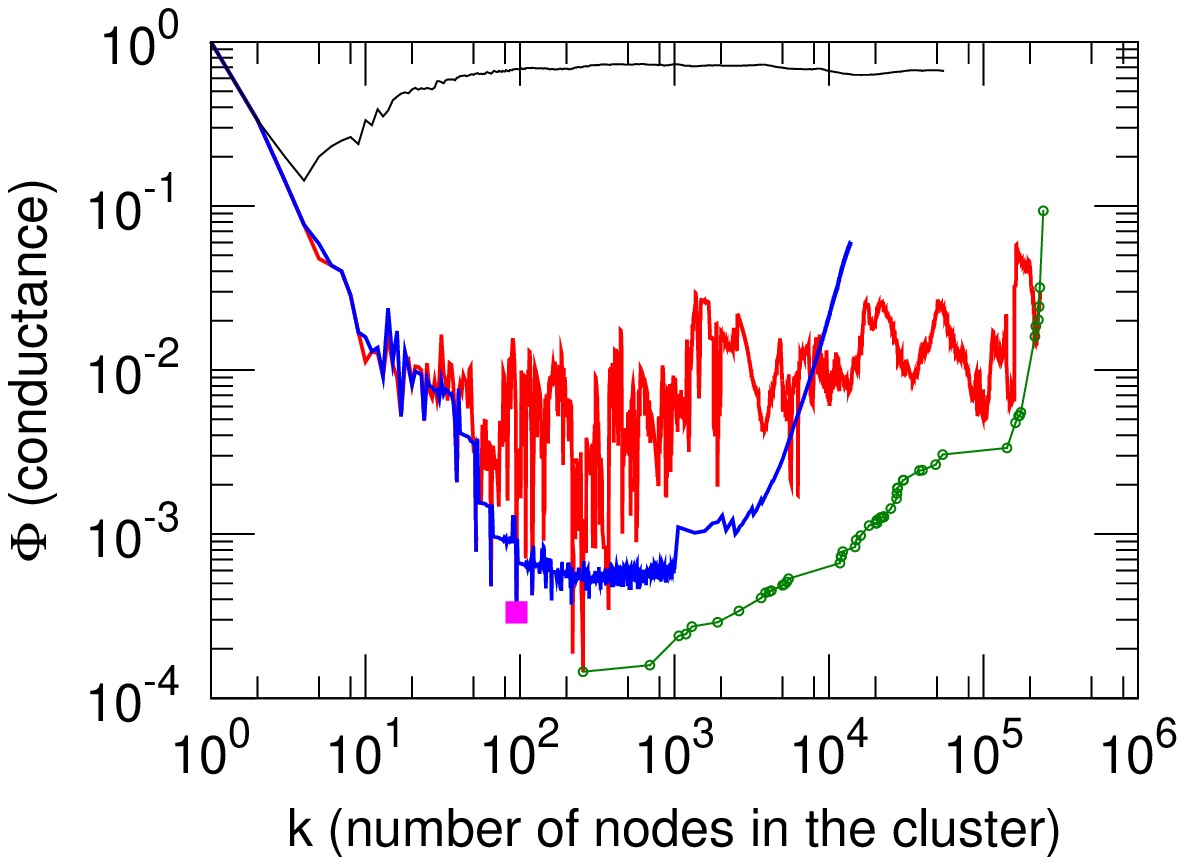} &
  \includegraphics[width=0.3\textwidth]{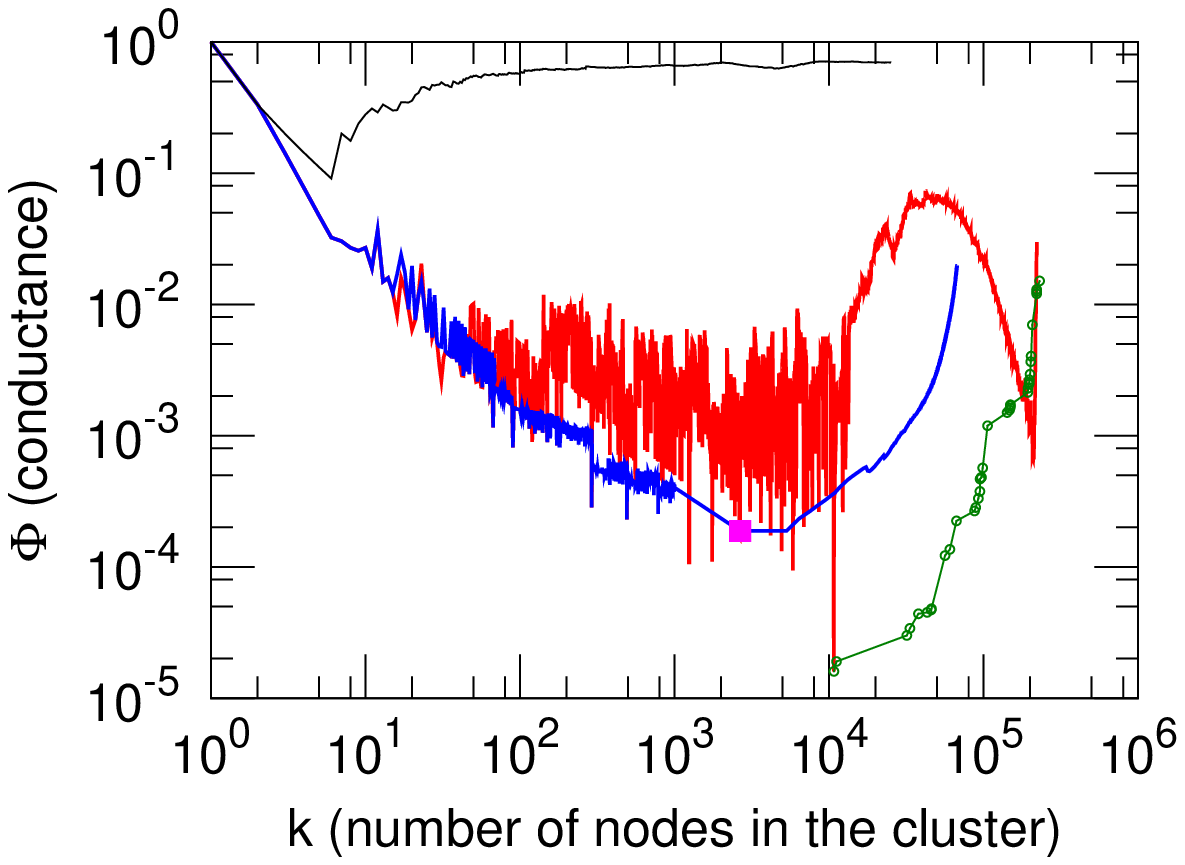} &
	\includegraphics[width=0.3\textwidth]{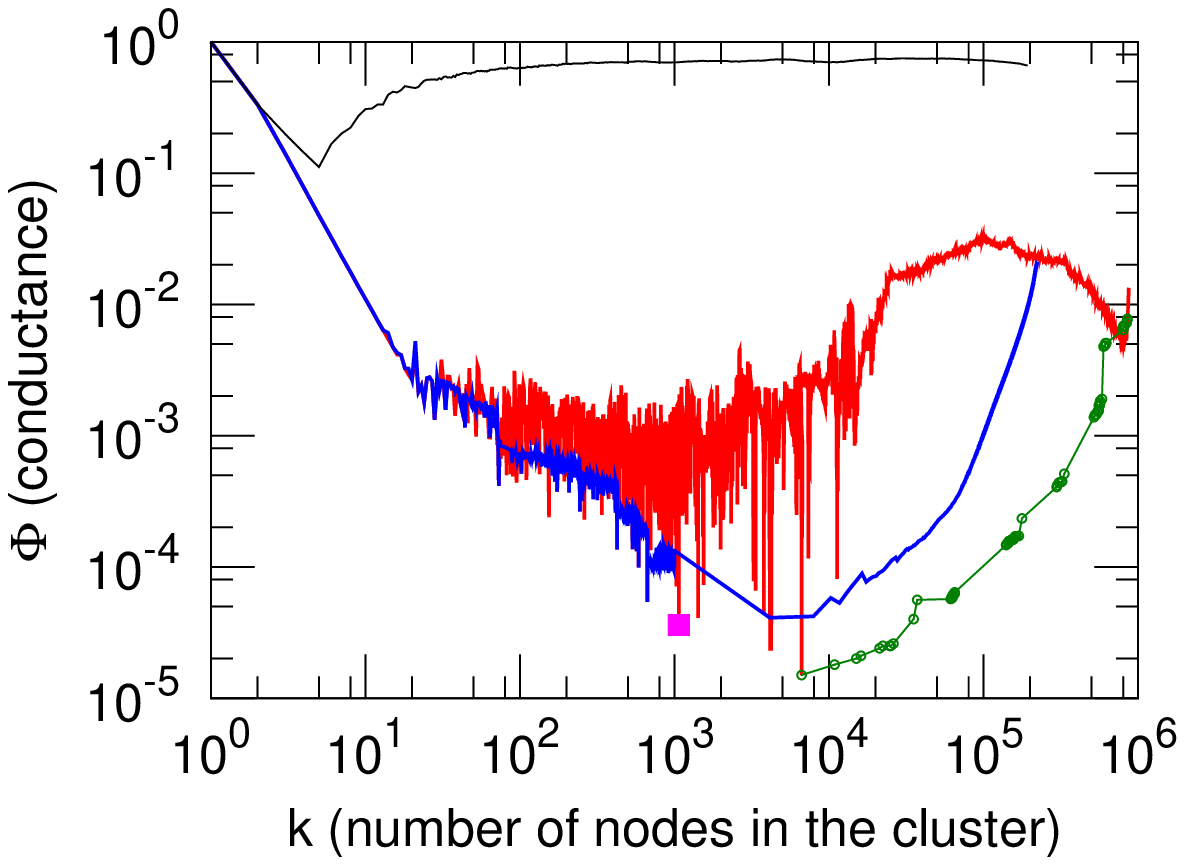} \\
	\net{Web-BerkStan} & \net{Web-Notredame}  & \net{Web-Trec}\\
    & & \\
    \multicolumn{3}{c}{\bf Internet networks}\\
	\includegraphics[width=0.3\textwidth]{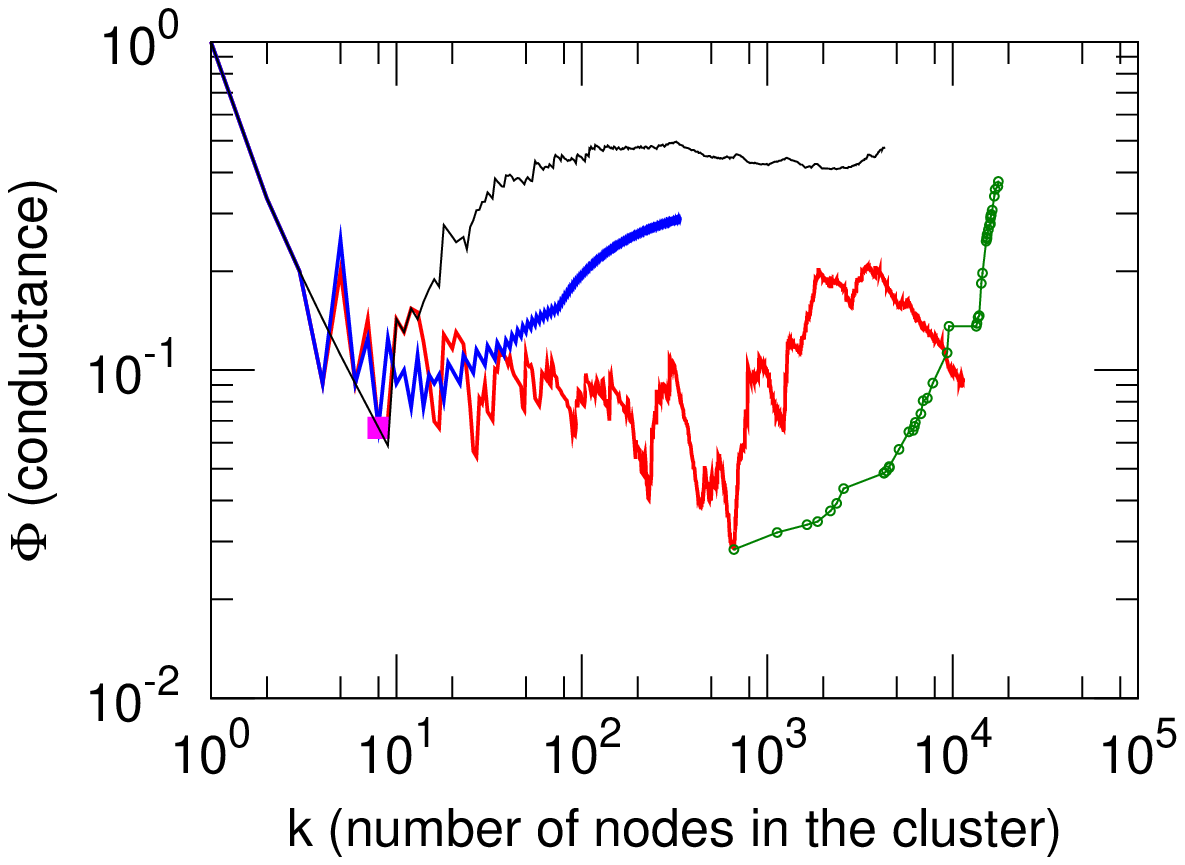} &
  \includegraphics[width=0.3\textwidth]{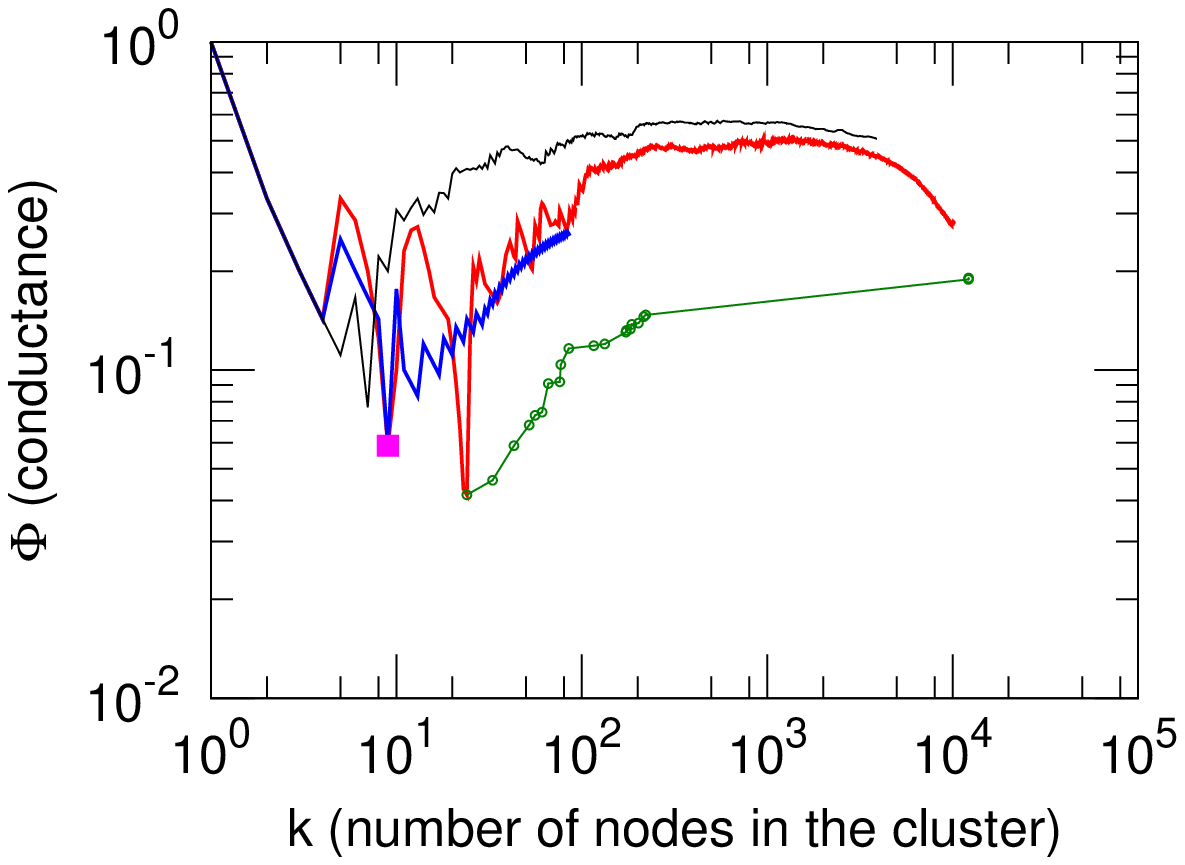} &
	\includegraphics[width=0.3\textwidth]{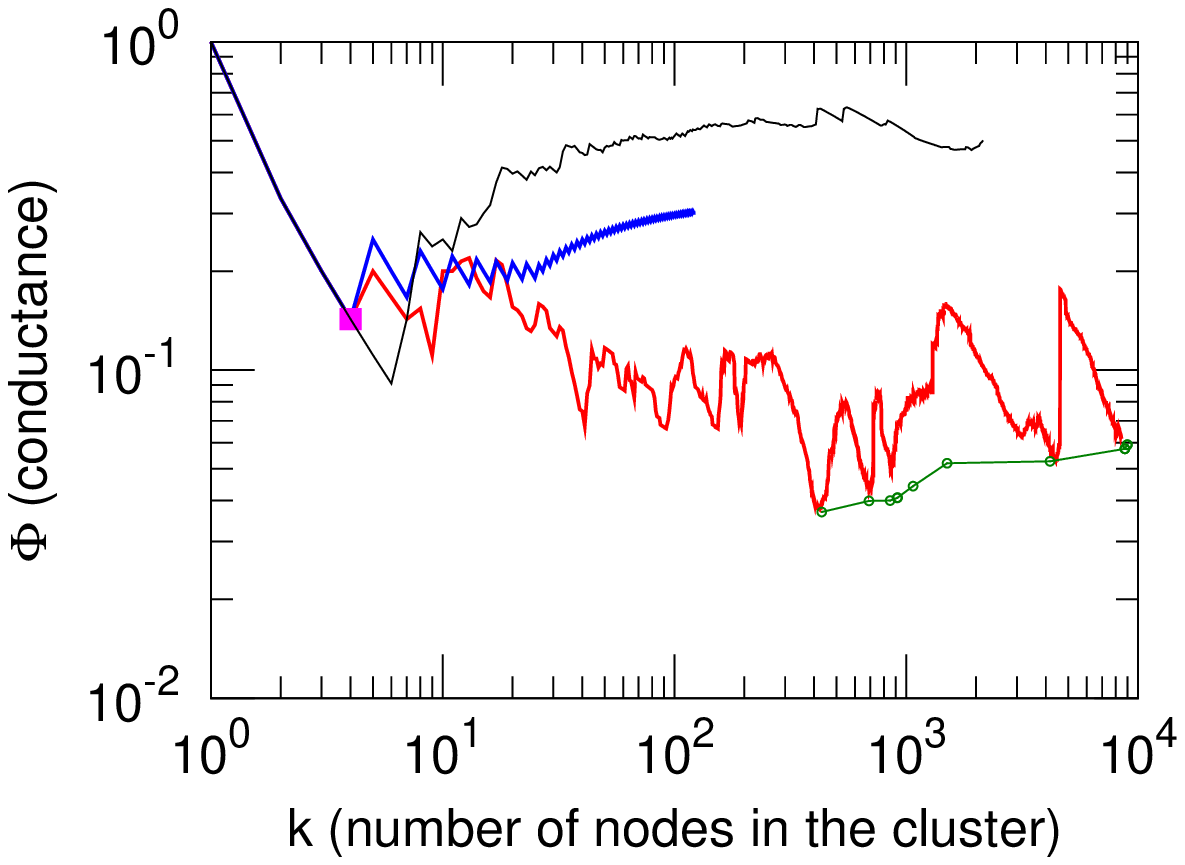} \\
	\net{As-Newman} & \net{Gnutella-25}  & \net{As-Oregon}\\
    & & \\
    \multicolumn{3}{c}{\bf Bipartite affiliation networks}\\
	\includegraphics[width=0.3\textwidth]{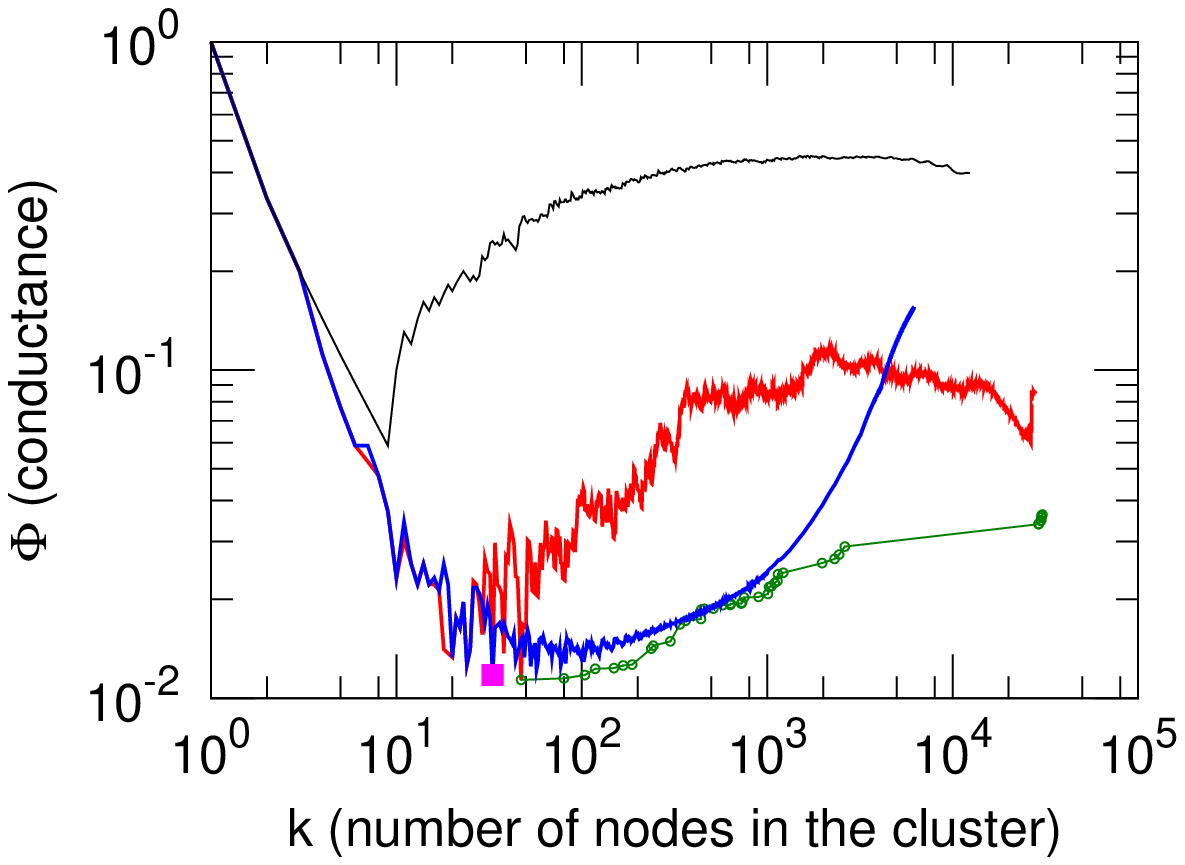} &
  \includegraphics[width=0.3\textwidth]{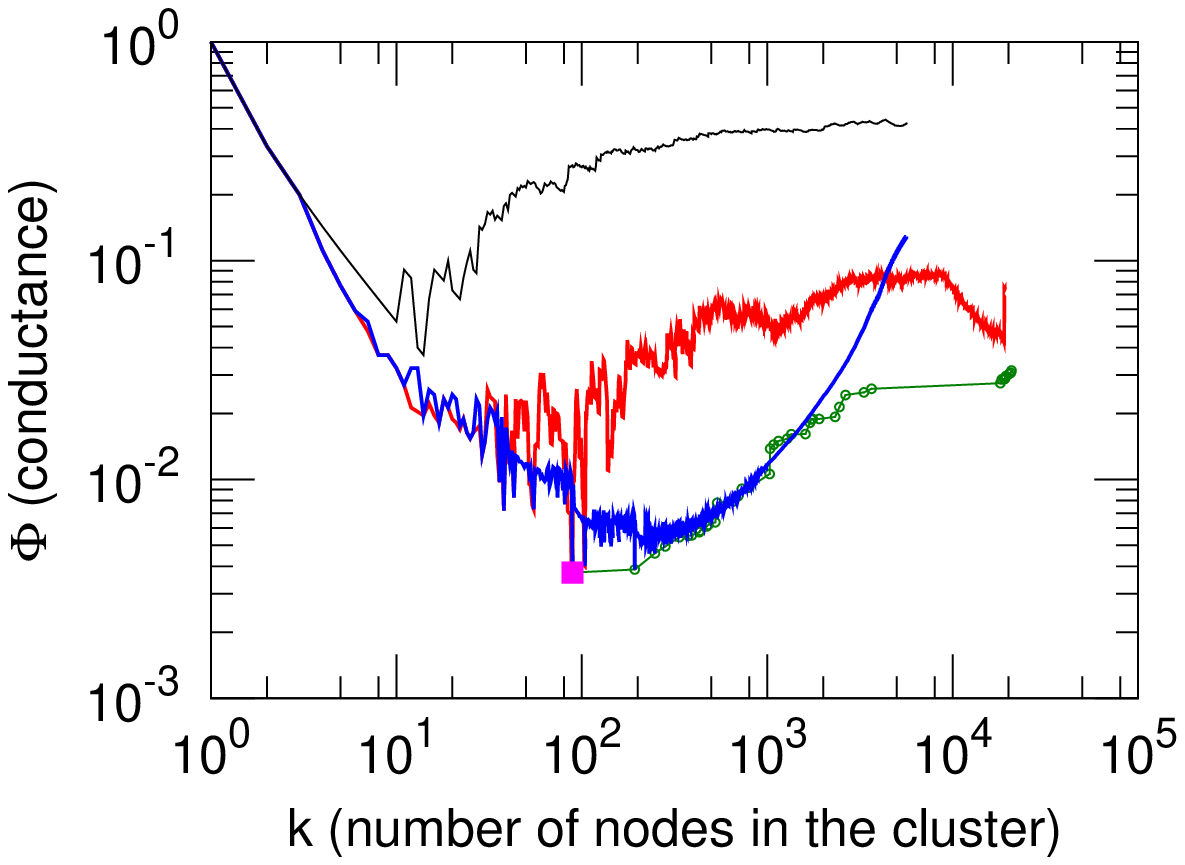} &
	\includegraphics[width=0.3\textwidth]{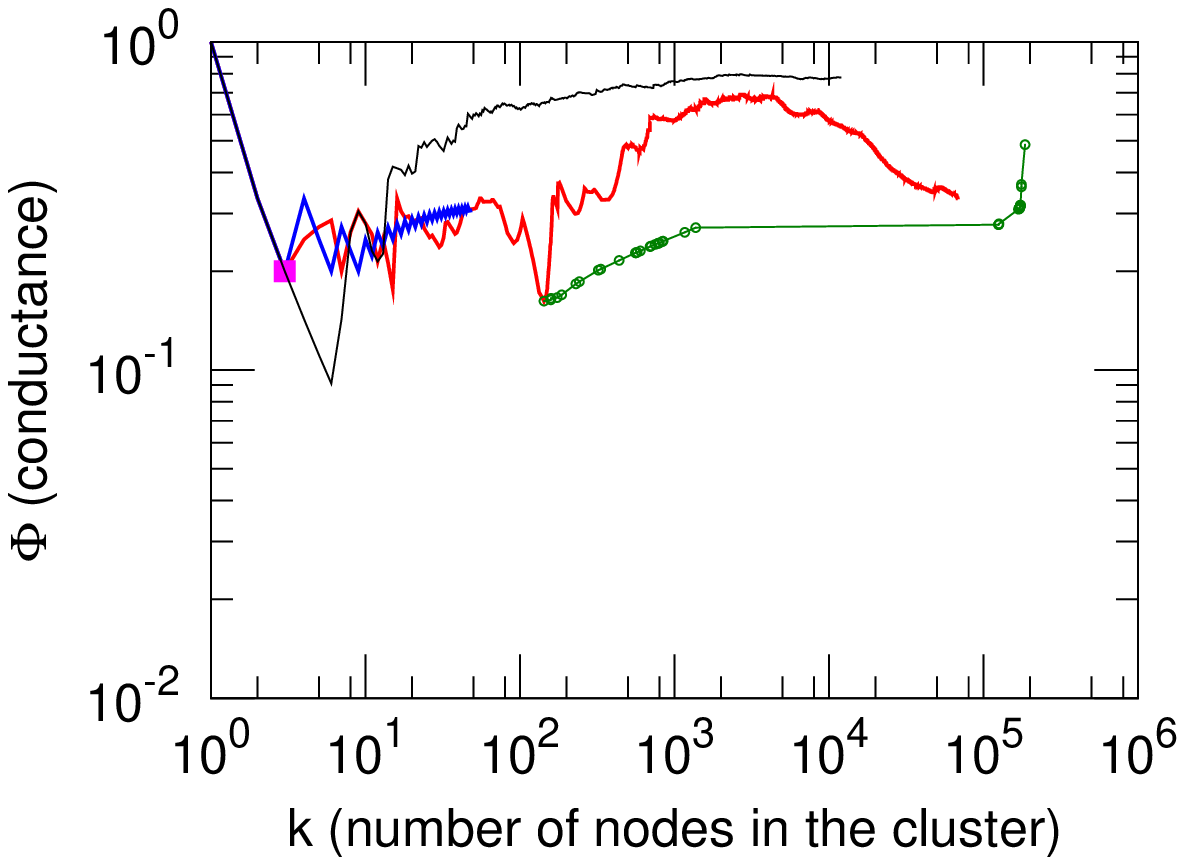} \\
	\net{AtP-cond-mat} & \net{AtP-hep-th}  & \net{Clickstream}\\
    & & \\
    \multicolumn{3}{c}{\bf Biological networks}\\
	\includegraphics[width=0.3\textwidth]{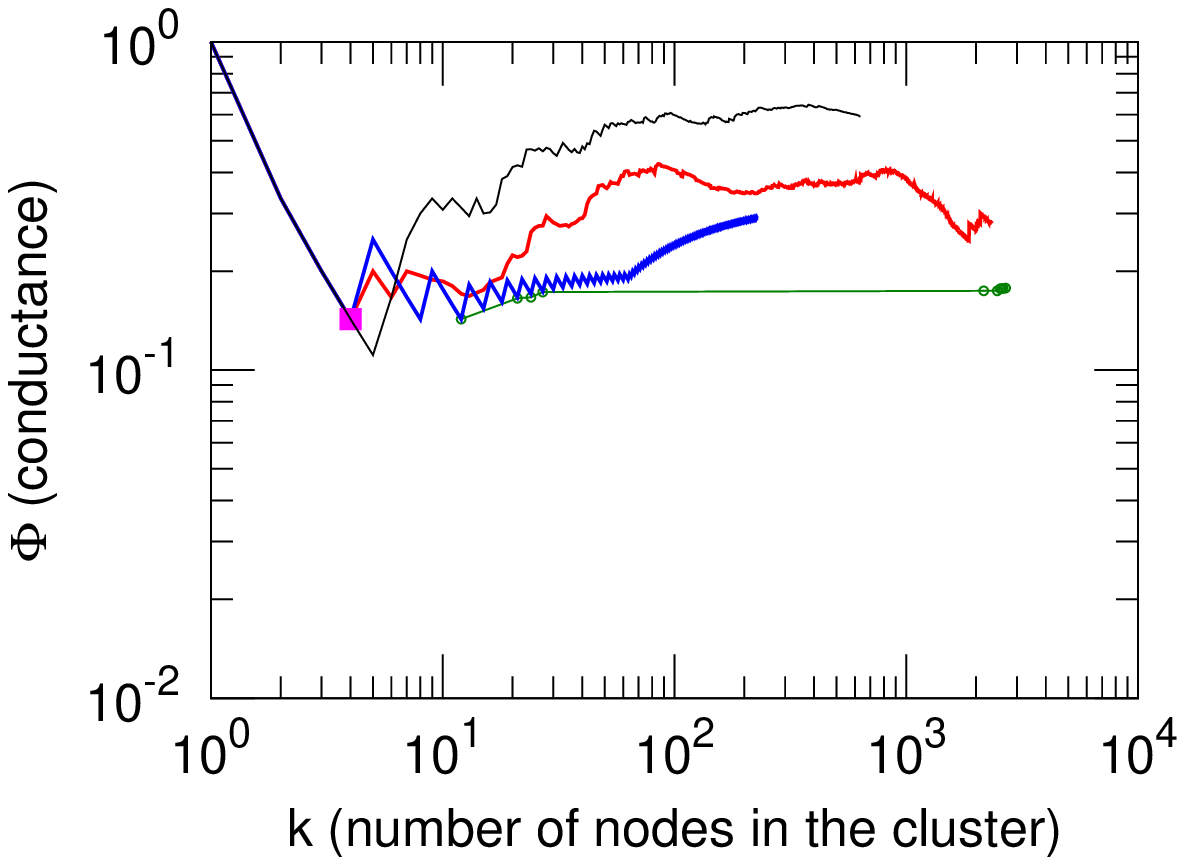} &
  \includegraphics[width=0.3\textwidth]{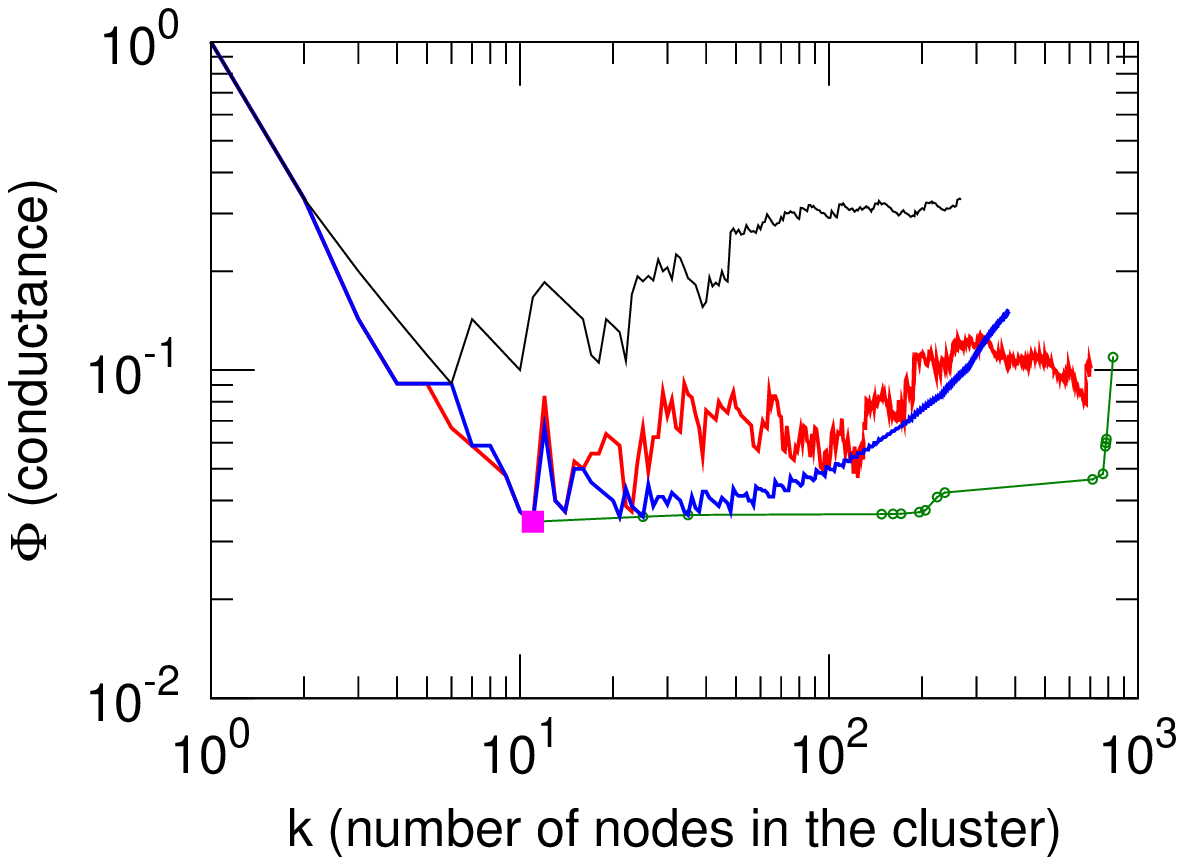} &
	\includegraphics[width=0.3\textwidth]{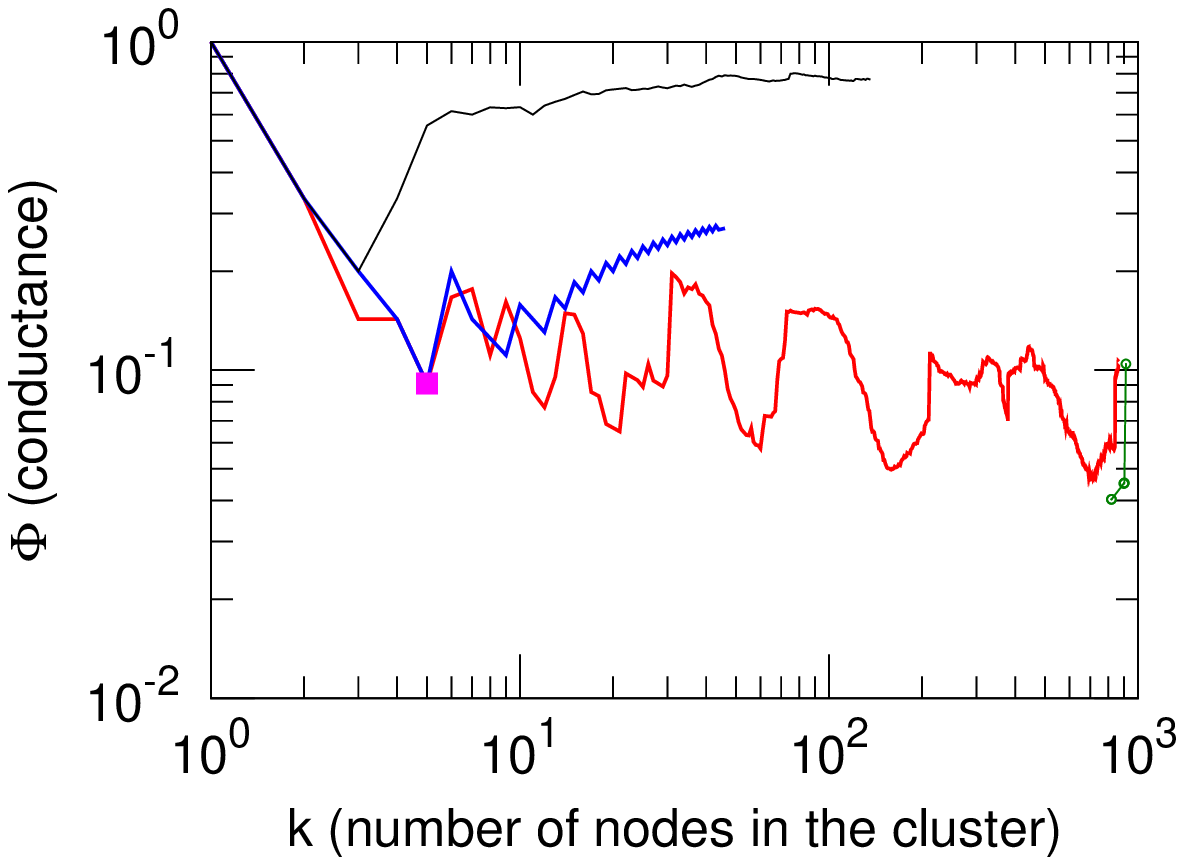} \\
	\net{Bio-Proteins} & \net{Bio-Yeast}  & \net{Bio-YeastP0.01}\\
  \multicolumn{3}{c}{\includegraphics[width=0.6\textwidth]{phiTR-legend2}}\\
	\end{tabular}
	\end{center} \caption{ [Best viewed in color.] Community profile plots
of networks from Table~\ref{tab:data_StatsDesc_2}. 
} \label{fig:phiDatasets2}
\end{figure*}

\begin{figure*}
	\begin{center}
	\begin{tabular}{ccc}
    \multicolumn{3}{c}{\bf Low-dimensional networks}\\
	\includegraphics[width=0.3\textwidth]{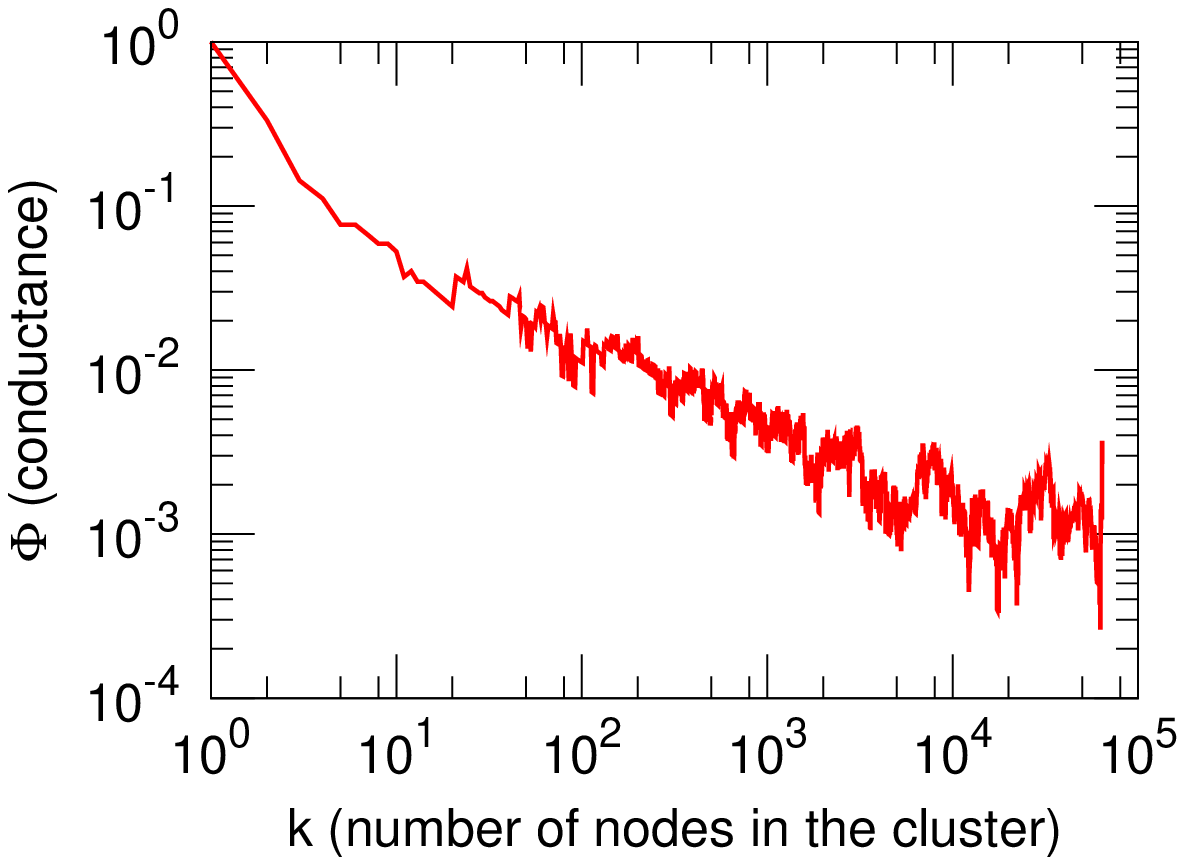} &
	\includegraphics[width=0.3\textwidth]{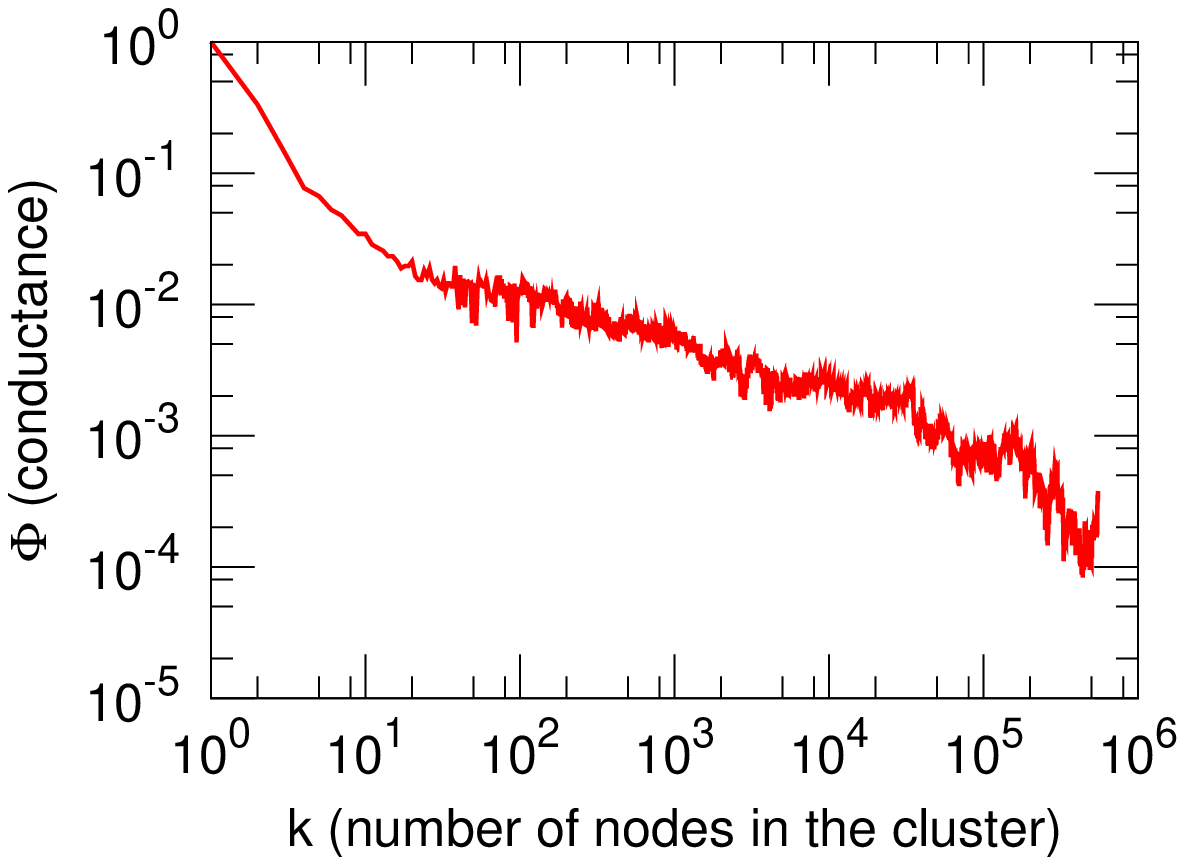} &
	\includegraphics[width=0.3\textwidth]{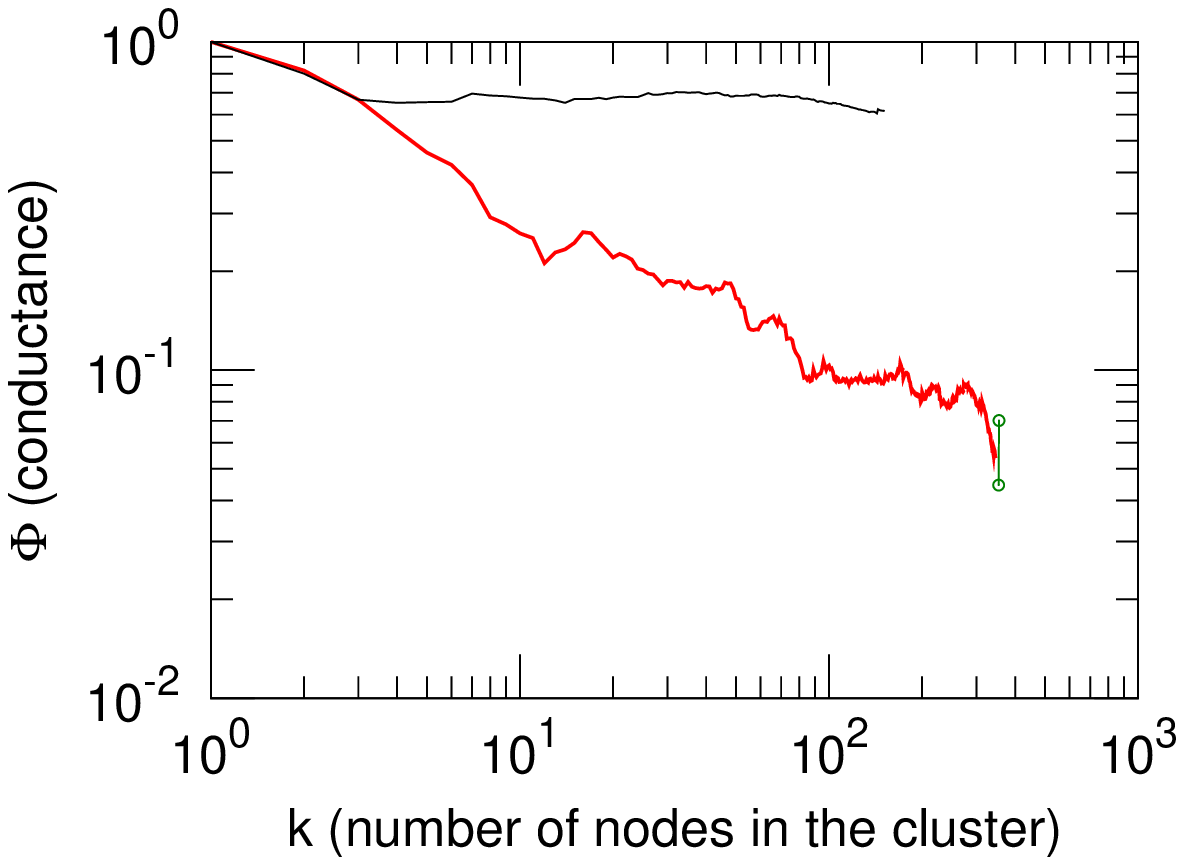} \\
	\net{Road-USA} & \net{Road-PA}  & \net{Mani-facesK5}\\
    & & \\
  \multicolumn{3}{c}{\bf IMDB Actor-to-Movie graphs}\\
	\includegraphics[width=0.3\textwidth]{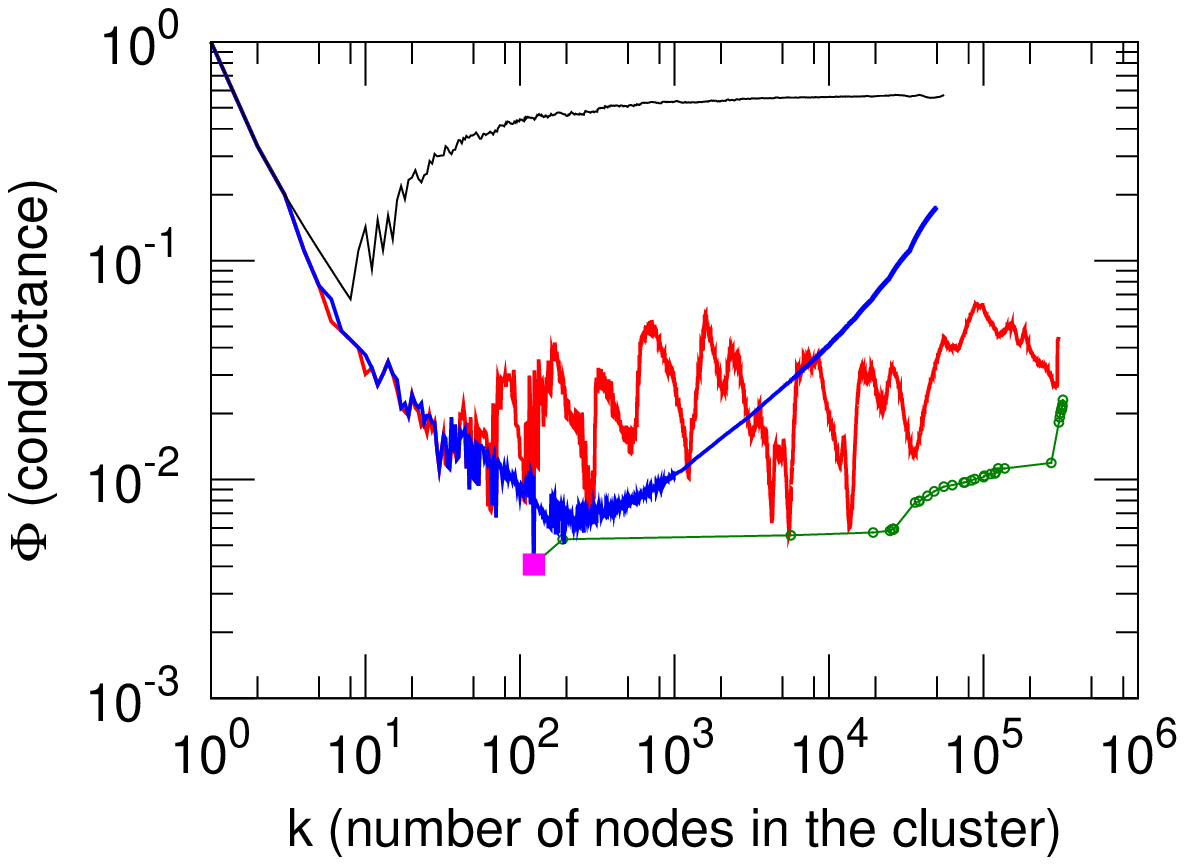} &
  \includegraphics[width=0.3\textwidth]{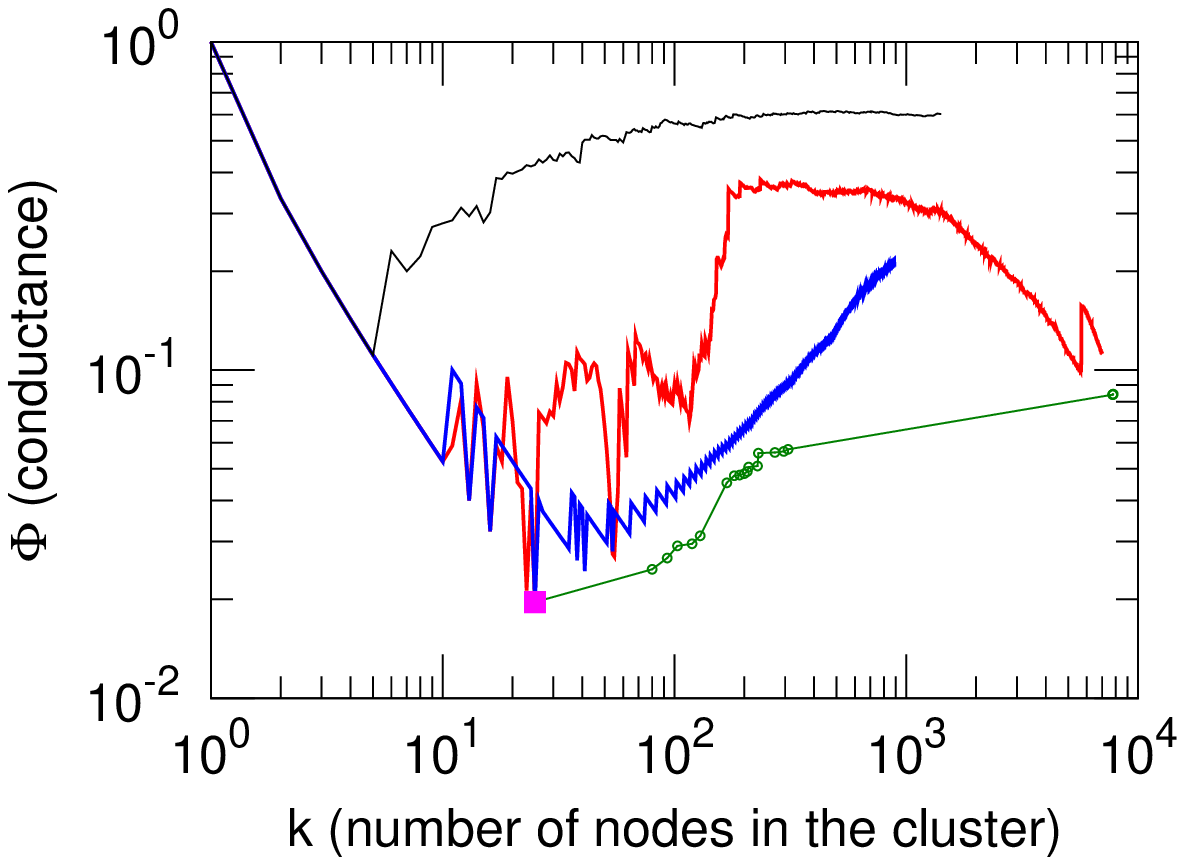} &
	\includegraphics[width=0.3\textwidth]{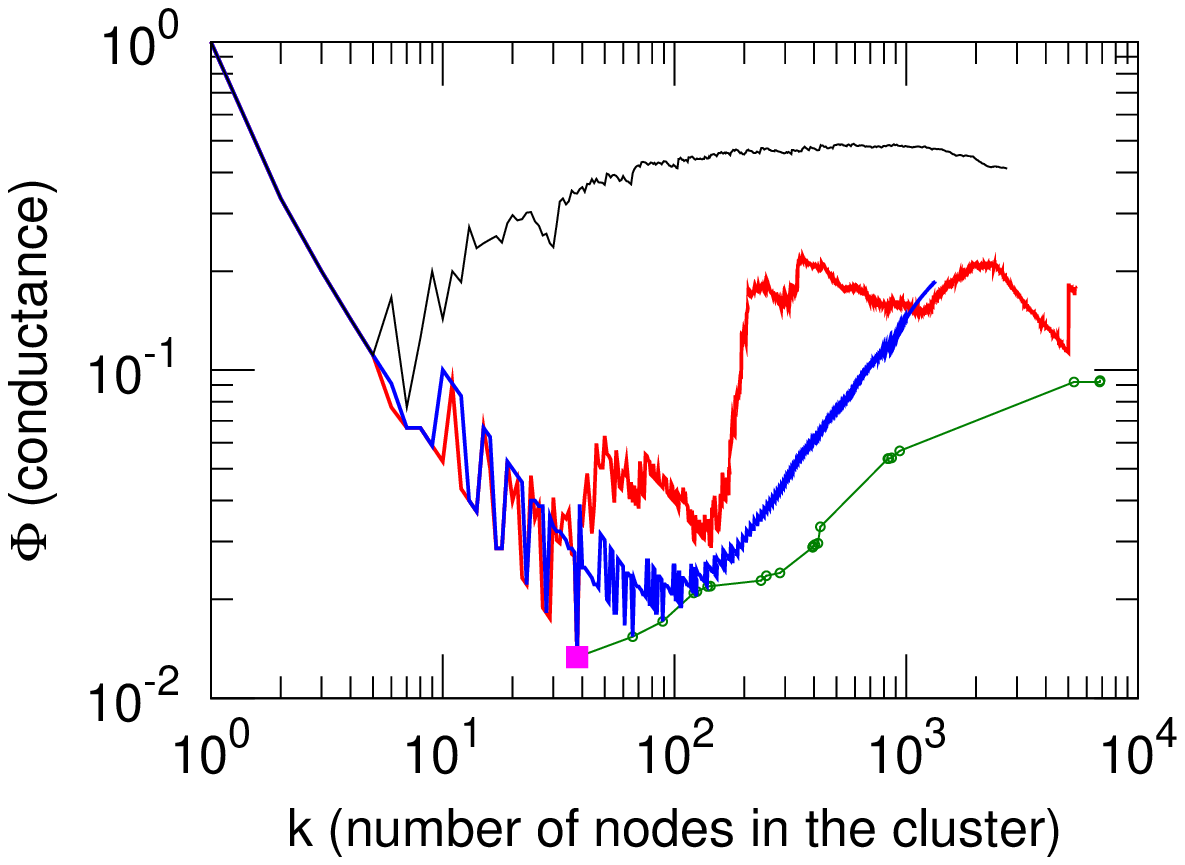} \\
	\net{Imdb-raw07} & \net{Imdb-Mexico}  & \net{Imdb-WGermany}\\
    & & \\
    \multicolumn{3}{c}{\bf Amazon product co-purchasing networks}\\
	\includegraphics[width=0.3\textwidth]{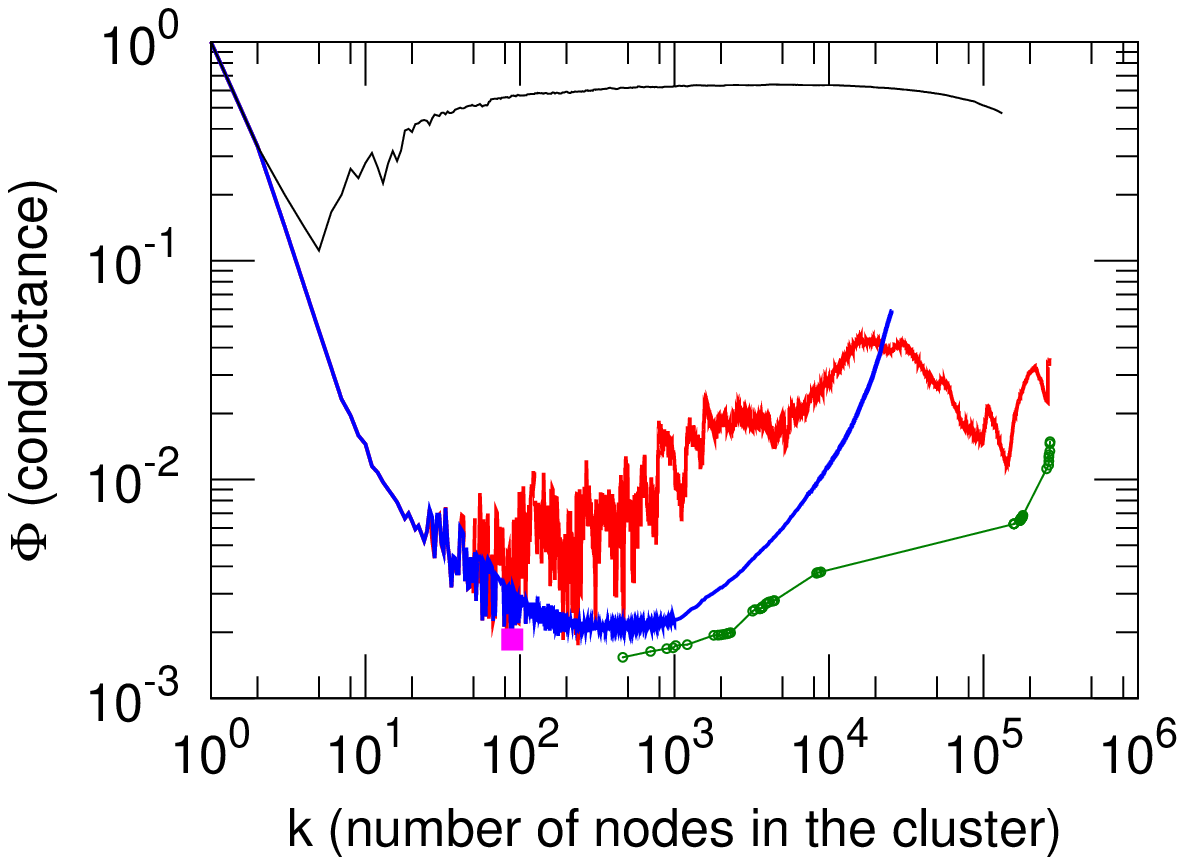} &
  \includegraphics[width=0.3\textwidth]{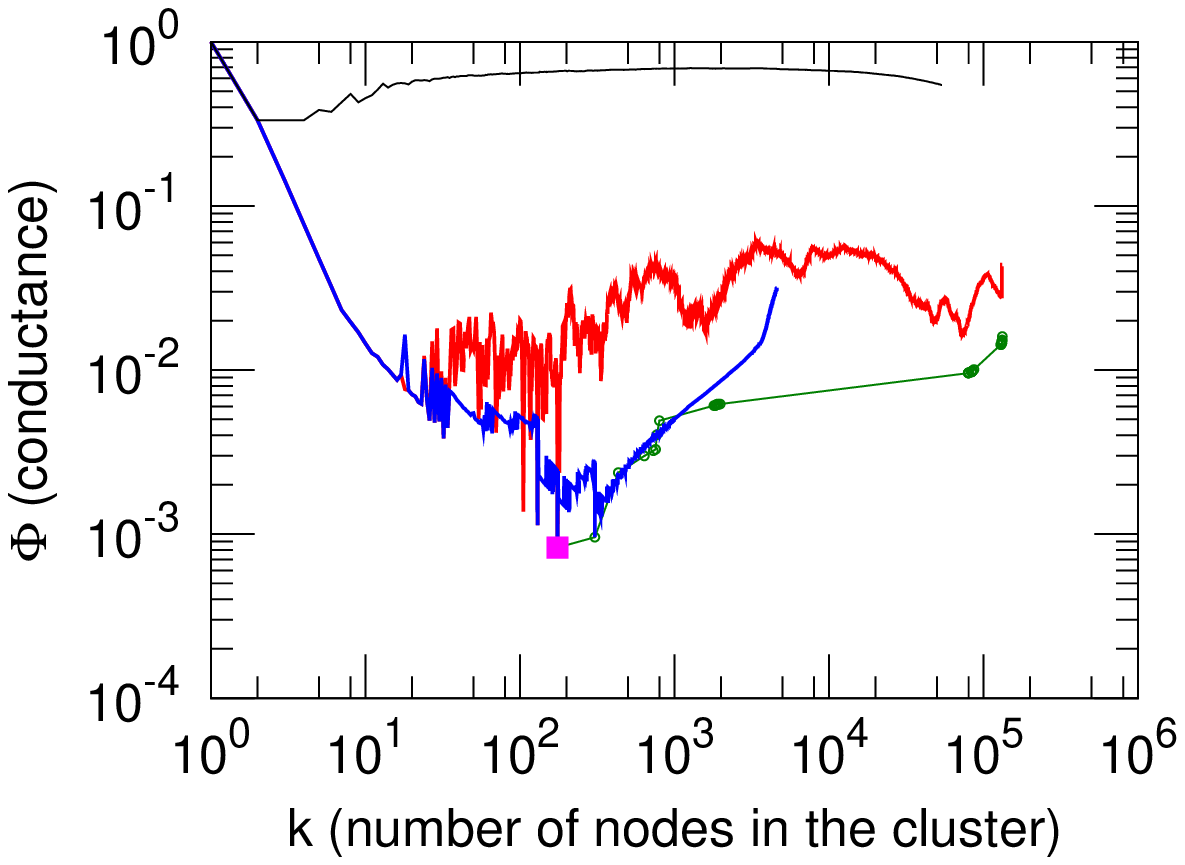} &
	\includegraphics[width=0.3\textwidth]{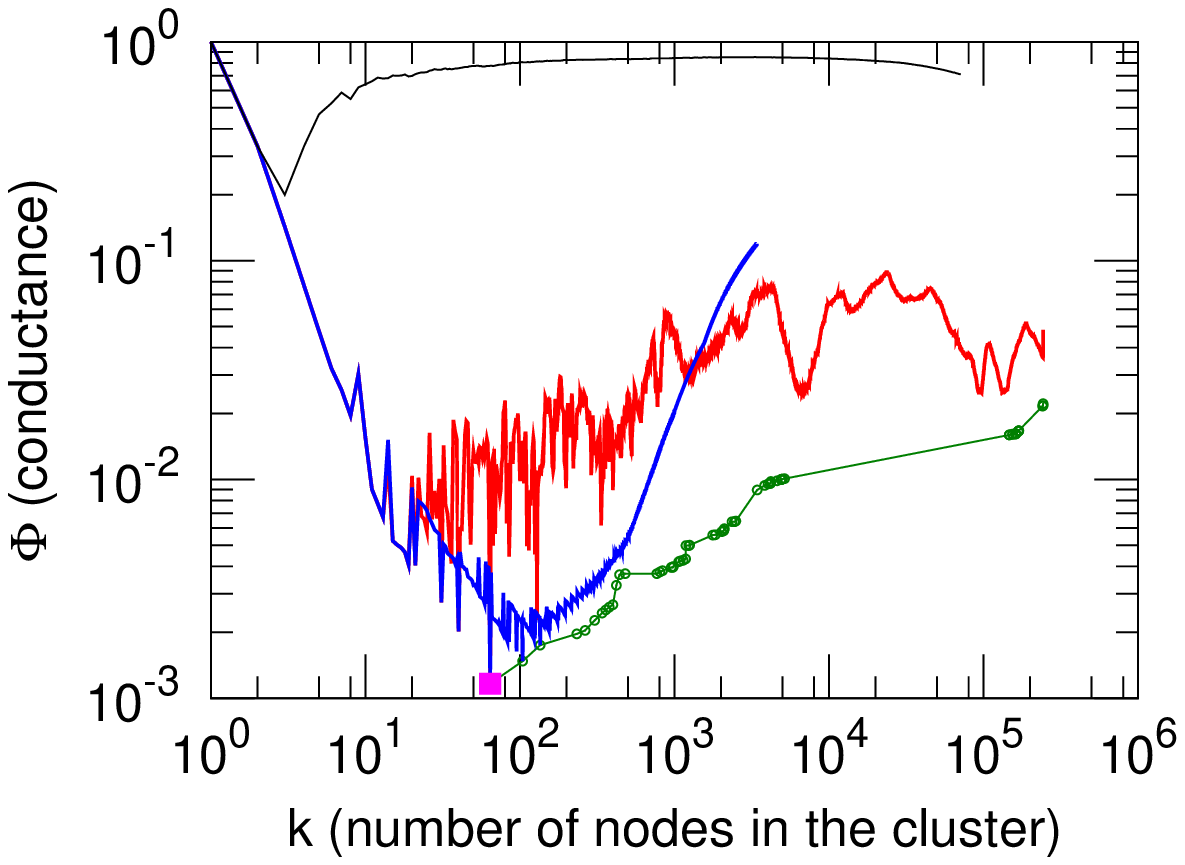} \\
	\net{AmazonAllProd} & \net{Amazon0302}  & \net{AmazonAll}\\
    & & \\
    \multicolumn{3}{c}{\bf Yahoo Answers networks}\\
\includegraphics[width=0.3\textwidth]{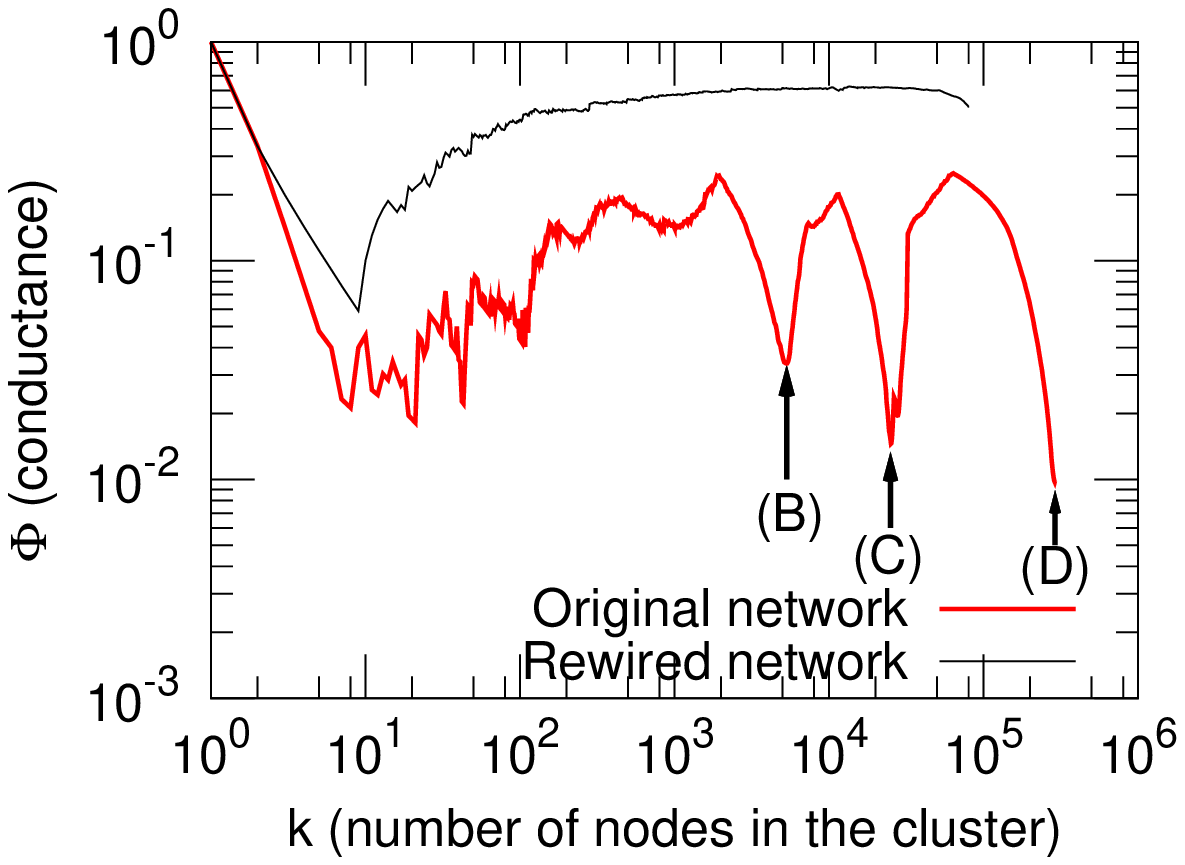} &
\includegraphics[width=0.3\textwidth]{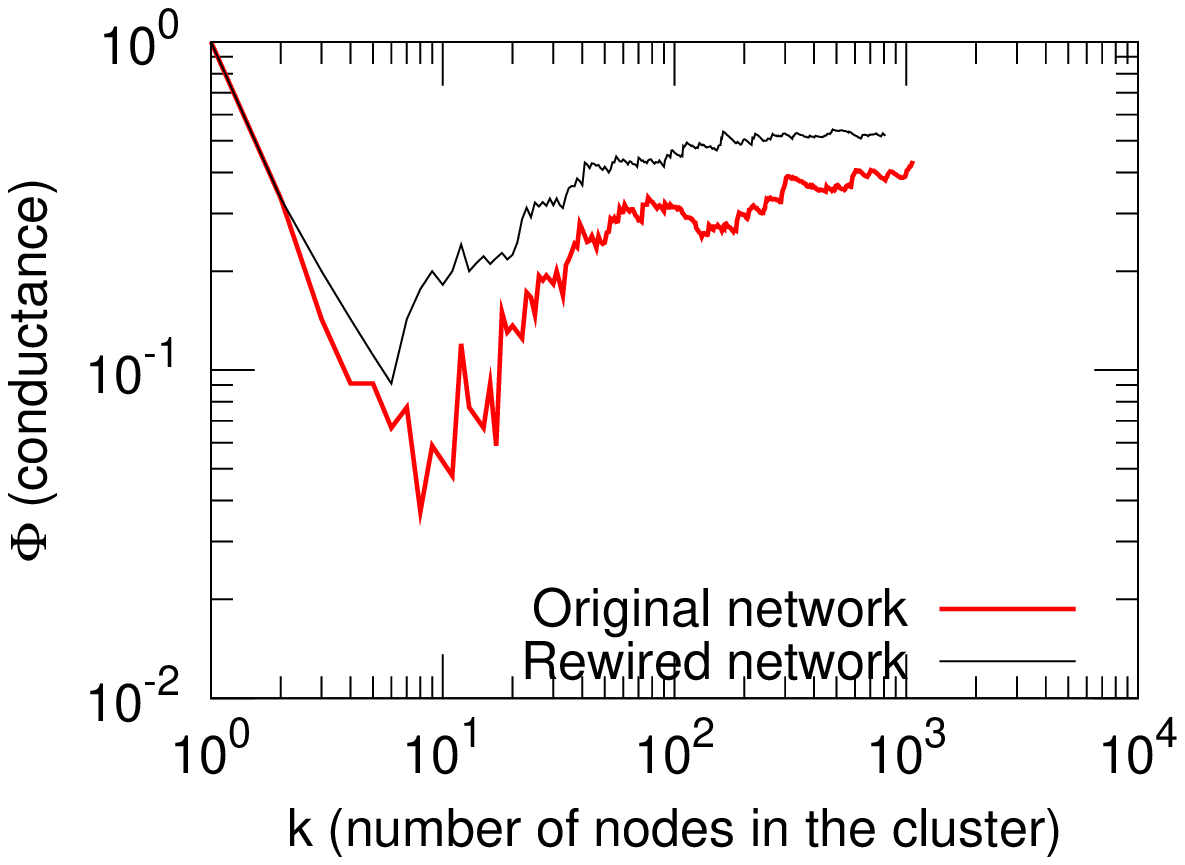} &
\includegraphics[width=0.3\textwidth]{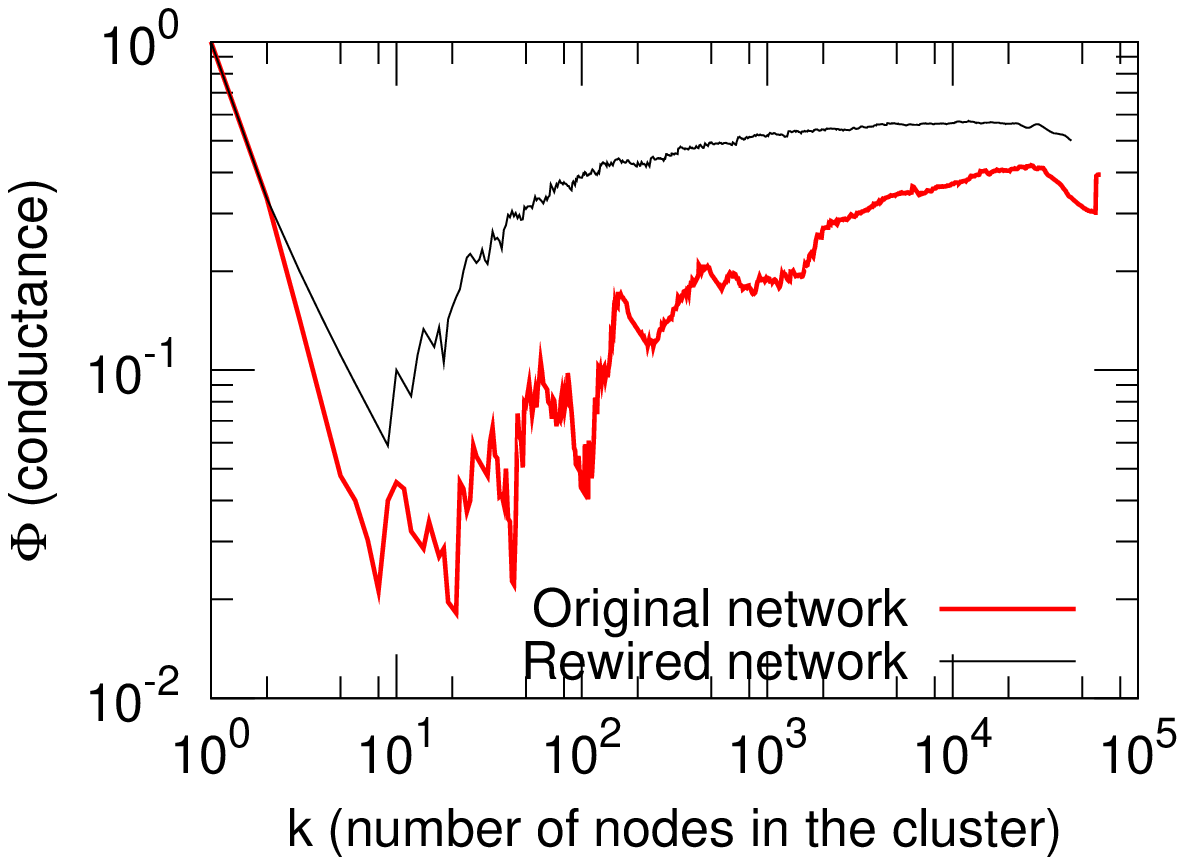} \\
\net{Answers} & \net{Answers-5}  & \net{Answers-6}\\
  \multicolumn{3}{c}{\includegraphics[width=0.6\textwidth]{phiTR-legend2}}   \\
	\end{tabular}
	\end{center} \caption{ [Best viewed in color.] Community profile plots
of networks from Table~\ref{tab:data_StatsDesc_3}, as well as
\net{Answers} and two sub-pieces of \net{Answers}. }
\label{fig:phiDatasets3}
\end{figure*}

Figures~\ref{fig:phiDatasets1}, \ref{fig:phiDatasets2}, and
\ref{fig:phiDatasets3} show additional examples of NCP plots for networks
from Tables~\ref{tab:data_StatsDesc_1}, \ref{tab:data_StatsDesc_2}
and~\ref{tab:data_StatsDesc_3}. In the first two rows of
Figure~\ref{fig:phiDatasets1}, we have several examples of purely Social
networks and two email networks, in the third row we have patent and blog
Information/citation networks, and in the final row we have three examples
of actor and author Collaboration networks. In
Figure~\ref{fig:phiDatasets2}, we see three examples each of Web graphs,
Internet networks, Bipartite affiliation networks, and Biological
networks. Finally, in the first row of Figure~\ref{fig:phiDatasets3}, we
see Low-dimensional networks, including two road and a manifold network;
in the second row, we have an IMDB Actor-to-Movie graphs and two subgraphs
induced by restricting to individual countries; in the third row, we see
three Amazon product co-purchasing networks; and in the final row we see a
Yahoo! Answers networks and two subgraphs that are large good conductance
cuts from the full network.


For most of these networks, the same four versions of the NCP plot are
plotted that were presented in Figure~\ref{fig:netPhiPlot}. Note that, as
before, the scale of the vertical axis in these graphs is not all the
same; the minima range from $10^{-2}$ to $10^{-5}$. These network datasets
are drawn from a wide range of areas, and these graphs contain a wealth of
information, a full analysis of which is well beyond the scope of the
paper. Note, however, that the general trends we discussed in
Section~\ref{sxn:ncpp:large_sparse} still manifest themselves in nearly
every network.

The \net{Imdb-raw07} network is interesting in that its NCP plot does not
increase much (at least not the version computed by the Local Spectral
Algorithm) and we clearly observe large sets with good conductance values.
Upon examination, many of the large good conductance cuts seem to be
associated with different language groups. Two things should be noted.
First, and not surprisingly, in this network and others, we have observed
that there is some sensitivity to how the data are prepared. For example,
we obtain somewhat stronger communities if ambiguous nodes (and there are
a lot of ambiguous nodes in network datasets with millions of nodes) are
removed than if, \emph{e.g.}, they are assigned to a country based on a
voting mechanism of some other heuristic. A full analysis of these data
preparation issues is beyond the scope of this paper, but our overall
conclusions seem to hold independent of the preparation details. Second,
if we examine individual countries---two representative examples are
shown---then we see substantially less structure at large size scales.

The Yahoo! Answers social network (see \net{Answers}) also has several
large cuts with good conductance value---actually, the best cut in the
network has more $10^5$ nodes. (It is likely that exogenous factors are
responsible for these large deep cuts.) Using standard graph partitioning
procedures, we obtained four large disjoint clusters consisting of ca.
$5,300$, $25,400$, $27,000$, and $290,000$ nodes, respectively,
corresponding to the four dips (two of which visually overlap) in the NCP
plot. We then examined the community profile plots for each of these
pieces. The two representative examples of which we show clearly indicate
a NCP plot that is much more like other network datasets we have examined.
\section{More structural observations of our network datasets}
\label{sxn:obs_struct}

We have examined in greater detail our network datasets in order to
understand which structural properties are responsible for the observed
properties of the NCP plot. We first present statistics for our network
datasets in Section~\ref{sxn:obs_struct:stats}. Then, in
Section~\ref{sxn:obs_struct:whiskers} we describe a heuristic to identify
small sets of nodes that have strong connections amongst themselves but
that are connected to the remainder of the network by only a single edge.
In Section~\ref{sxn:obs_struct:bags_of_whiskers}, we show that these
``whiskers'' (or disjoint unions of them) are often the ``best''
conductance communities in the network. Last, in
Section~\ref{sxn:obs_struct:remove_whiskers} we examine NCP plots for
networks in which these whiskers have been removed.

\subsection{General statistics on our network datasets}
\label{sxn:obs_struct:stats}

In Tables~\ref{tab:data_StatsDesc_1}, \ref{tab:data_StatsDesc_2},
and~\ref{tab:data_StatsDesc_3}, we also present the following statistics for
our network datasets:
the number of nodes $N$;
the number of edges $E$;
the fraction of nodes in the largest biconnected component $N_b/N$;
the fraction of edges in the largest biconnected component $E_b/E$;
the average degree $\bar{d}=2E/N$;
the empirical second-order average degree~\cite{ChungLu:2006} $\tilde{d}$;
average clustering coefficient~\cite{watts98collective} $\bar{C}$;
the estimated diameter $D$;
and the estimated average path length $\bar{D}$.
(The diameter was estimated using the following algorithm: pick a random node,
find the farthest node $X$ (via shortest path); move to $X$ and find the
farthest node from $X$; iterate this procedure until the distance to the
farthest node does not increase anymore.
The average path length was estimated based on $10,000$ randomly sampled nodes.)

In nearly every network we have examined, there is a substantial fraction of
nodes that are barely connected to the main part of the network, {\em i.e.},
that are part of a small cluster of ca. $10$ to $100$ nodes that are attached
to the remainder of the network via one or a small number of edges.
In particular, a large fraction of the network is made out of nodes that are
not in the biconnected core.%
\footnote{
In this paper, we are slightly abusing standard terminology by using
the term bi-connectivity to mean 2-edge-connectivity.  We {\em are}
running the classic DFS-based bi-connectivity algorithm, which
identifies both bridge edges and articulation nodes, but then we are
only knocking out the bridge edges, not the articulation nodes,
so we end up with 2-edge-connected pieces.
}

For example, the \net{Epinions} network has $75,877$ nodes and $405,739$
edges, and the core of the network has only $36,111$ ($47\%$) nodes and
$365,253$ ($90\%$) edges. For \net{Delicious}, the core is even smaller:
it contains only $40\%$ of the nodes, and $65\%$ of the edges. Averaging
over our network datasets, we see that the largest biconnected component
contains around only $60\%$ of the nodes and $80\%$ of the edges of the
original network. This is somewhat akin to the so-called ``Jellyfish''
model~\cite{tauro01topology,siganos06_jellyfish} (which was proposed as a
model for the graph of internet topology) and also to the ``Octopus''
model (for random power law graphs~\cite{ChungLu:2006}, which is described
in more detail in Section~\ref{sxn:models:sparse_Gw}). Moreover, the
global minimum of the NCP plot is nearly always one of these pieces that
is connected by only a single edge. Since these small barely-connected
pieces seem to have a disproportionately large influence on the community
structure of our network datasets, we examine them in greater detail in
the next section.

\subsection{Network ``whiskers'' and the ``core''}
\label{sxn:obs_struct:whiskers}

We define \emph{whiskers}, or more precisely \emph{$1$-whiskers}, to be
maximal subgraphs that can be detached from the rest of the network by
removing a {\em single} edge. (Occasionally, we use the term whiskers
informally to refer to barely connected sets of nodes more generally.) To
find $1$-whiskers, we employ the following algorithm. Using a depth-first
search algorithm, we find the largest biconnected component $B$ of the
graph $G$. (A graph is biconnected if the removal of any single edge does
not disconnect the graph.) We then delete all the edges in $G$ that have
one of their end points in $B$. We call the connected components of this
new graph $G'$ $1$-whiskers, since they correspond to largest subgraphs
that can be disconnected from $G$ by removing just a single edge. Recall
that Figure~\ref{fig:intro_graph} contains a schematic picture a network,
including several of its whiskers.

\begin{figure}
   \begin{center}
   \begin{tabular}{cc}
      \subfigure[\net{LiveJournal01}]{
         \includegraphics[width=0.40\textwidth]{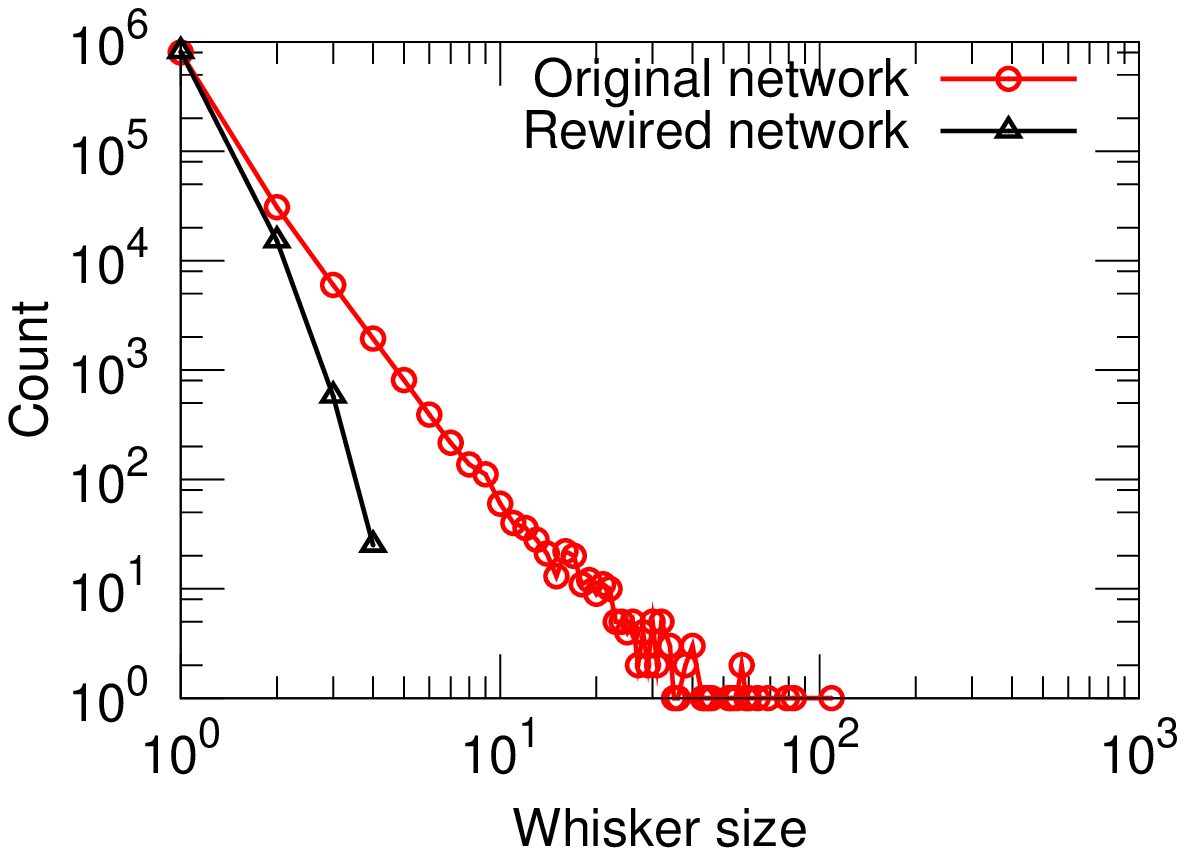}
         \label{fig:whiskerDistr:LiveJournal1}
      } &
      \subfigure[\net{Epinions}]{
         \includegraphics[width=0.40\textwidth]{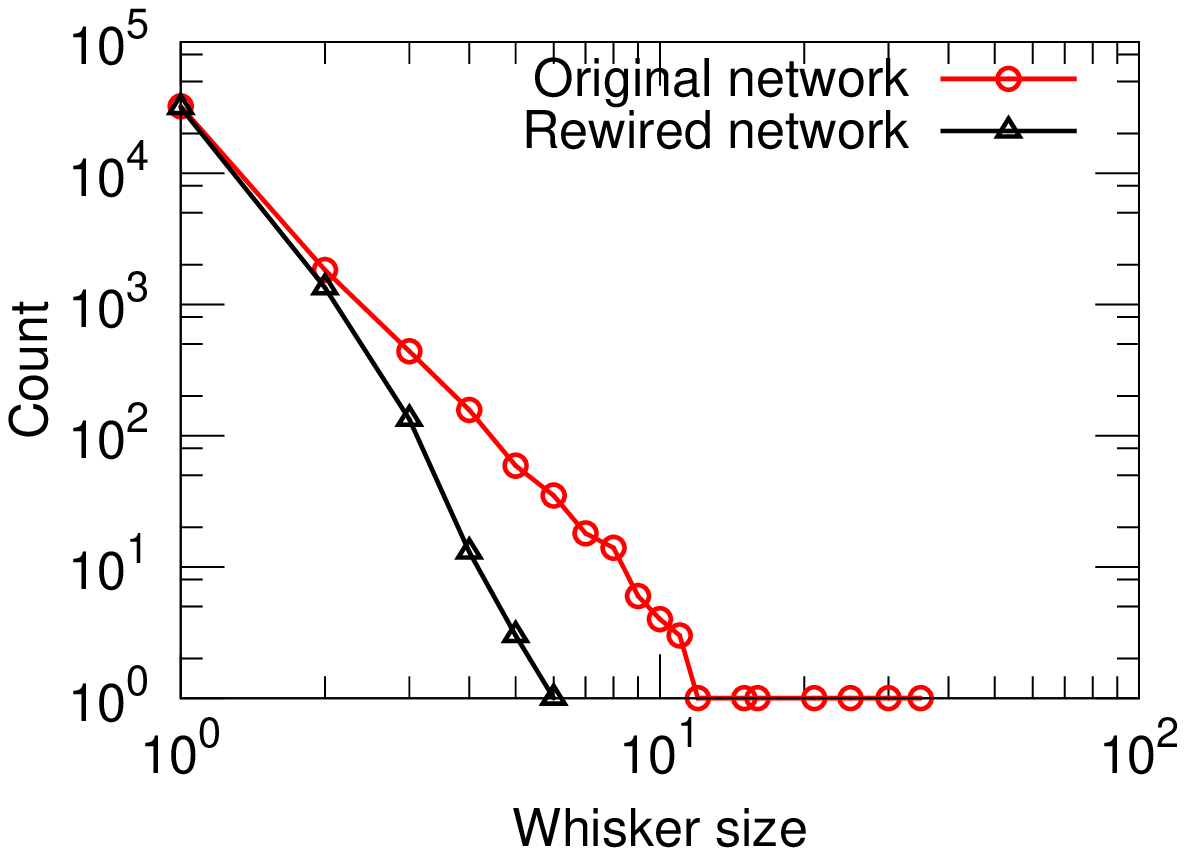}
         \label{fig:whiskerDistr:Epinions}
      } \\
      \subfigure[\net{Cit-hep-th}]{
         \includegraphics[width=0.40\textwidth]{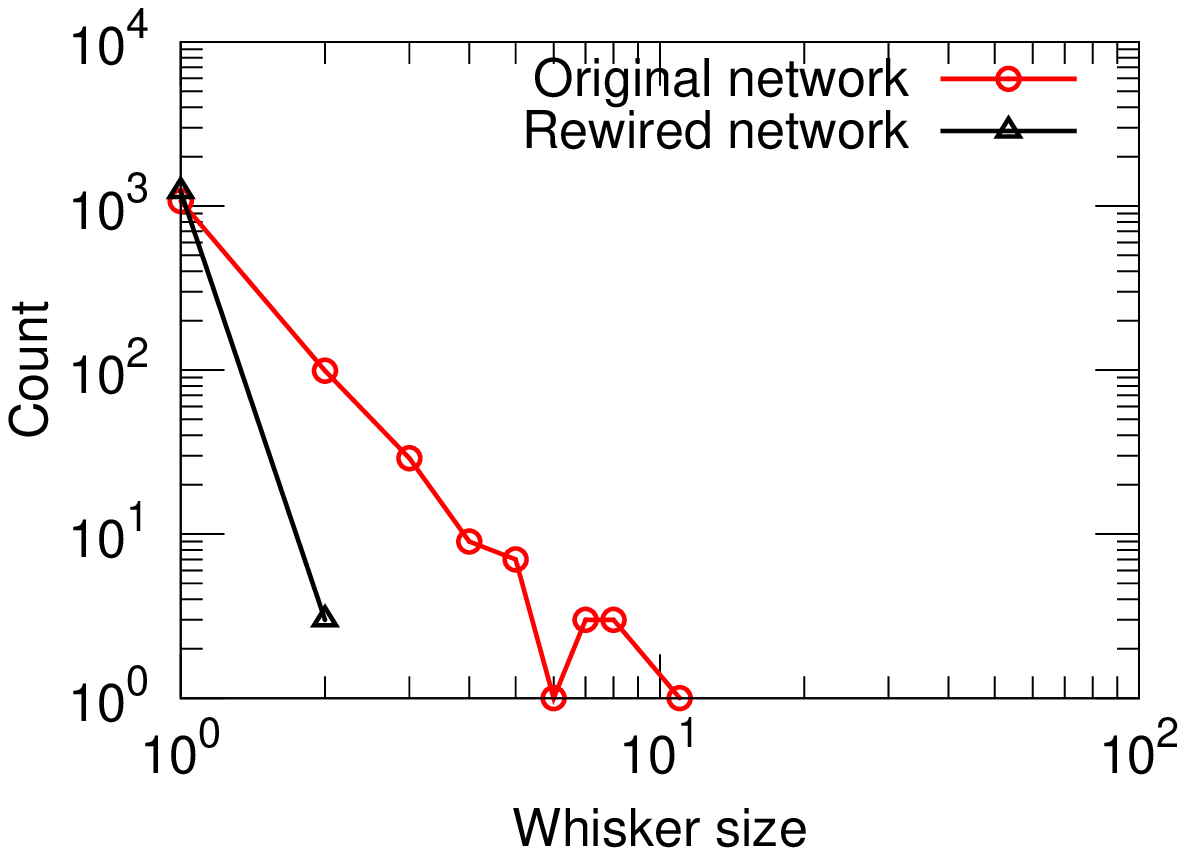}
         \label{fig:whiskerDistr:Cit-hep-th}
      } &
      \subfigure[\net{Web-Google}]{
         \includegraphics[width=0.40\textwidth]{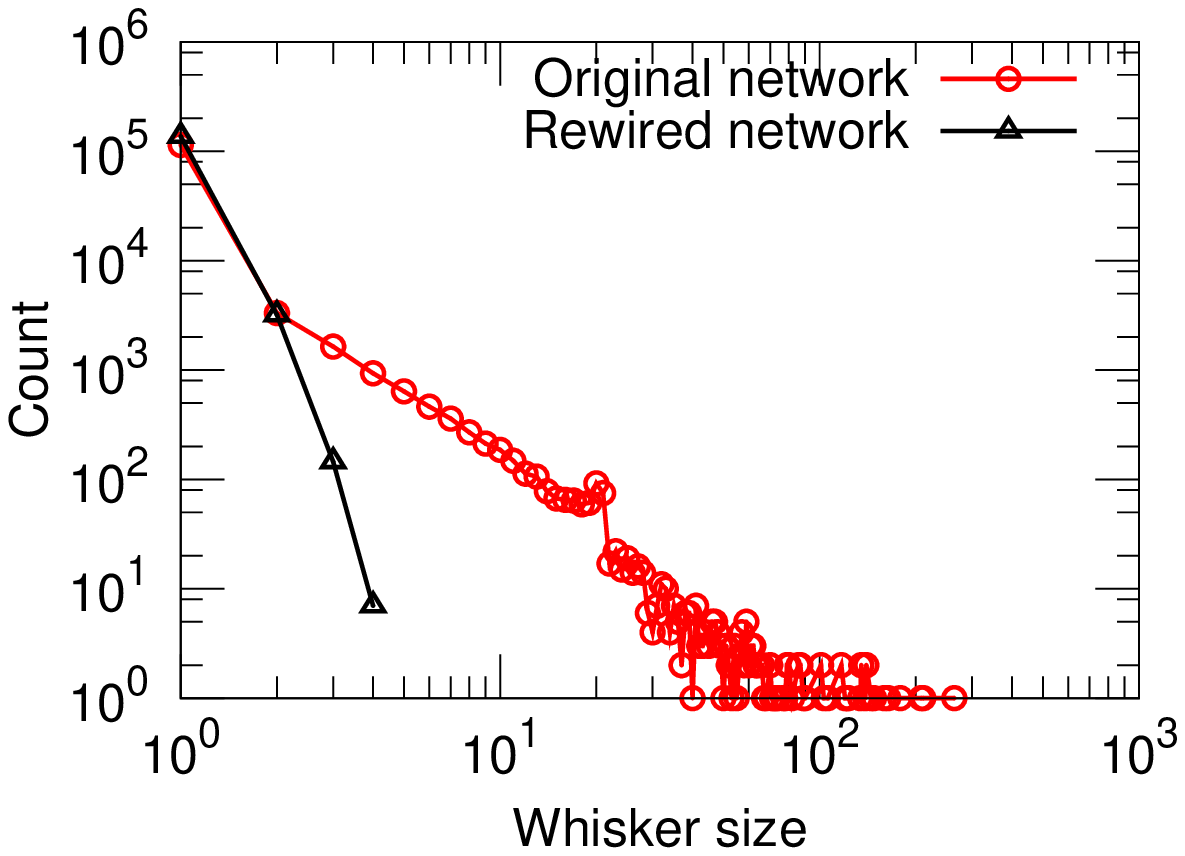}
         \label{fig:whiskerDistr:Web-Google}
      } \\
      \subfigure[\net{Atp-DBLP}]{
         \includegraphics[width=0.40\textwidth]{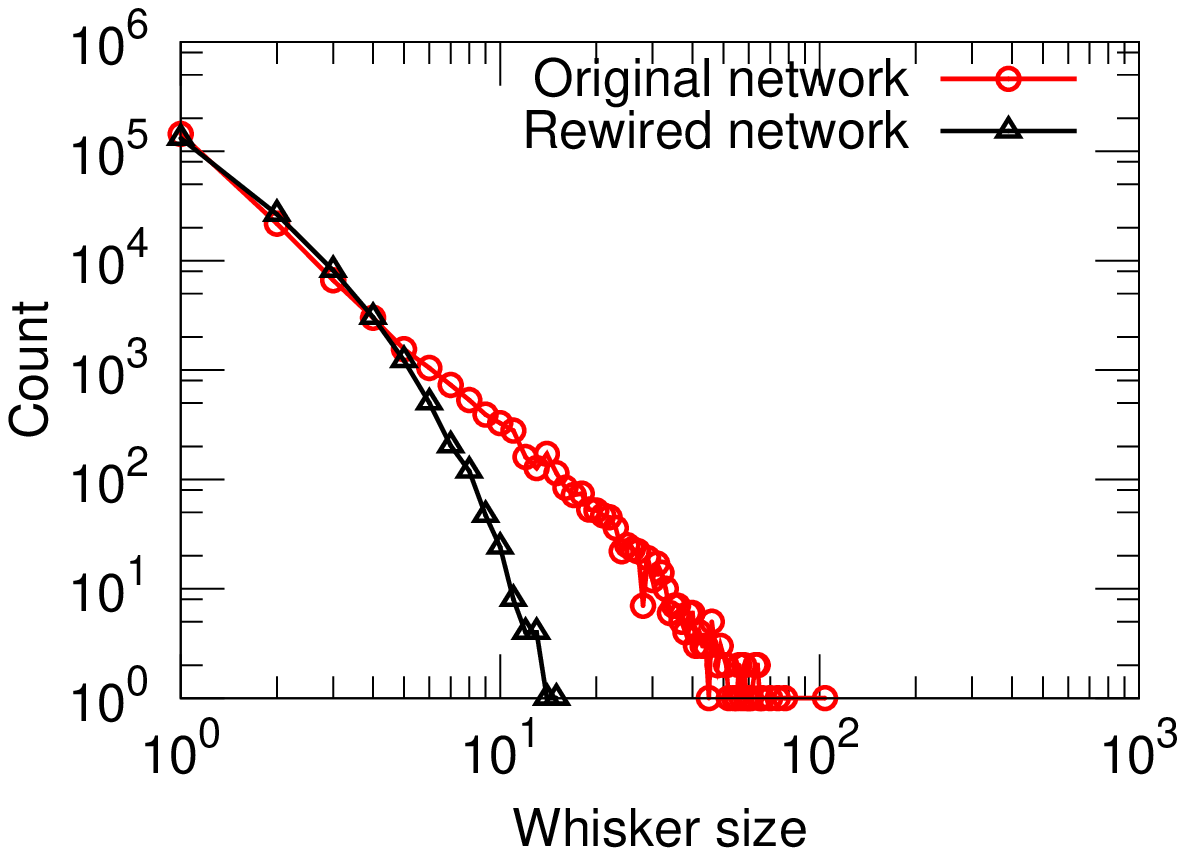}
         \label{fig:whiskerDistr:Atp-DBLP}
      } &
      \subfigure[\net{Gnutella-31}]{
         \includegraphics[width=0.40\textwidth]{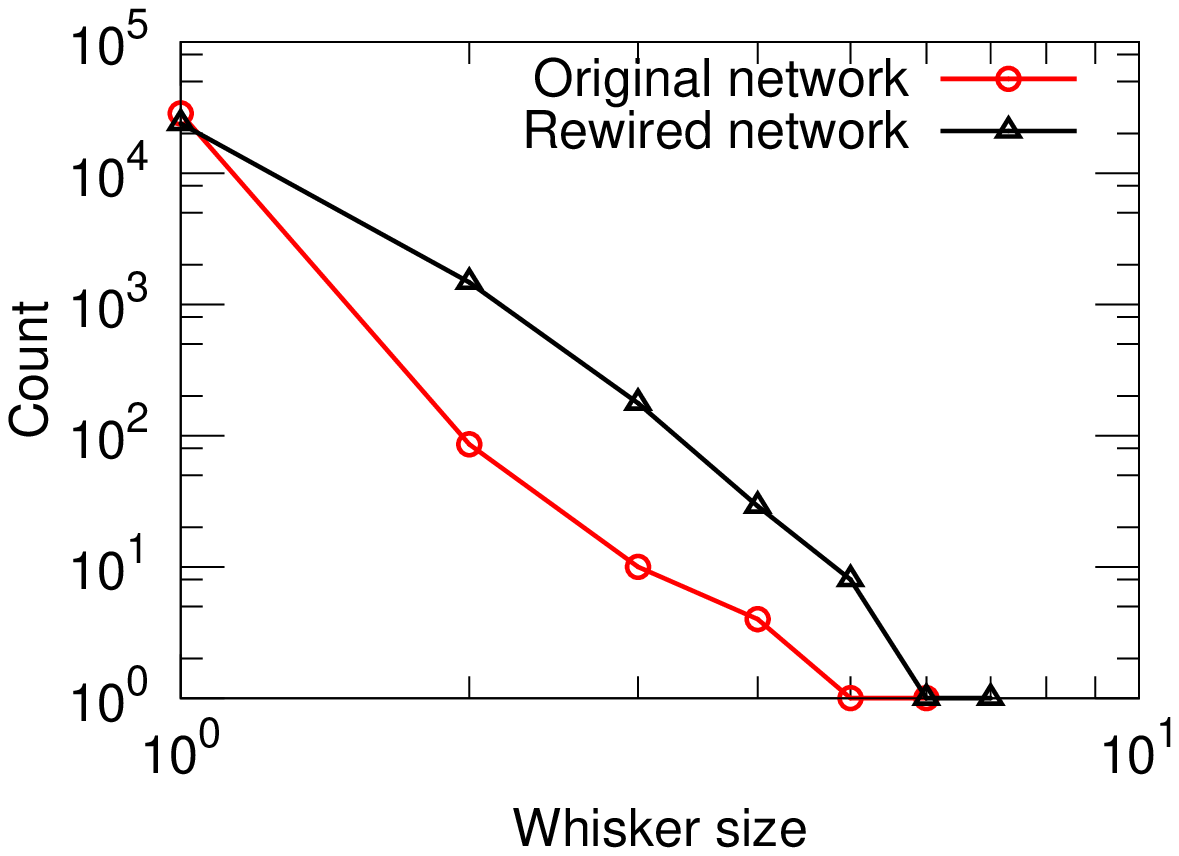}
         \label{fig:whiskerDistr:Gnutella}
      } \\
   \end{tabular}
   \end{center}
\caption{
Distribution of whisker sizes in the true network and the rewired network
(random graph with same degree distribution) for the six networks presented in
Figure~\ref{fig:netPhiPlot}.
The ten largest whiskers for the \net{Epinions} social network (the full
distribution of which is presented here in panel~(b)) are presented in
Figure~\ref{fig:whiskerEpinions}. 
}
\label{fig:whiskerDistr}
\end{figure}

Not surprisingly, there is a wide range of whisker sizes and shapes.
Figure~\ref{fig:whiskerDistr} shows the distribution of $1$-whisker sizes
for a representative selection of our network datasets. Empirically,
$1$-whisker size distribution is heavy-tailed, with the largest whisker
size ranging from around less than $10$ to well above $100$. The largest
whiskers in co-authorship and citation networks have around $10$ nodes,
whiskers in bipartite graphs also tend to be small, and very large
whiskers are found in a web graph. Figure~\ref{fig:whiskerDistr} also
compares the size of the whiskers with the sizes of whiskers in a rewired
version of the same network. (The first thing to note is that due to the
sparsity of the networks, the rewired versions all have whiskers.) In
rewired networks the whiskers tend to be much smaller than in the original
network. A particularly noteworthy exception is found in the Autonomous
systems networks and the \net{Gnutella-31} network. (See
Figure~\ref{fig:whiskerDistr:Gnutella} for an example of the latter.) In
these cases, the whiskers are so small that even the rewired version of
the network has more and larger whiskers. This makes sense, given how
those networks were designed: clearly, many large whiskers would have
negative effects on the Internet connectivity in case of link failures.

\begin{figure}
	\begin{center}
	\begin{tabular}{ccccc}
    \includegraphics[scale=0.3]{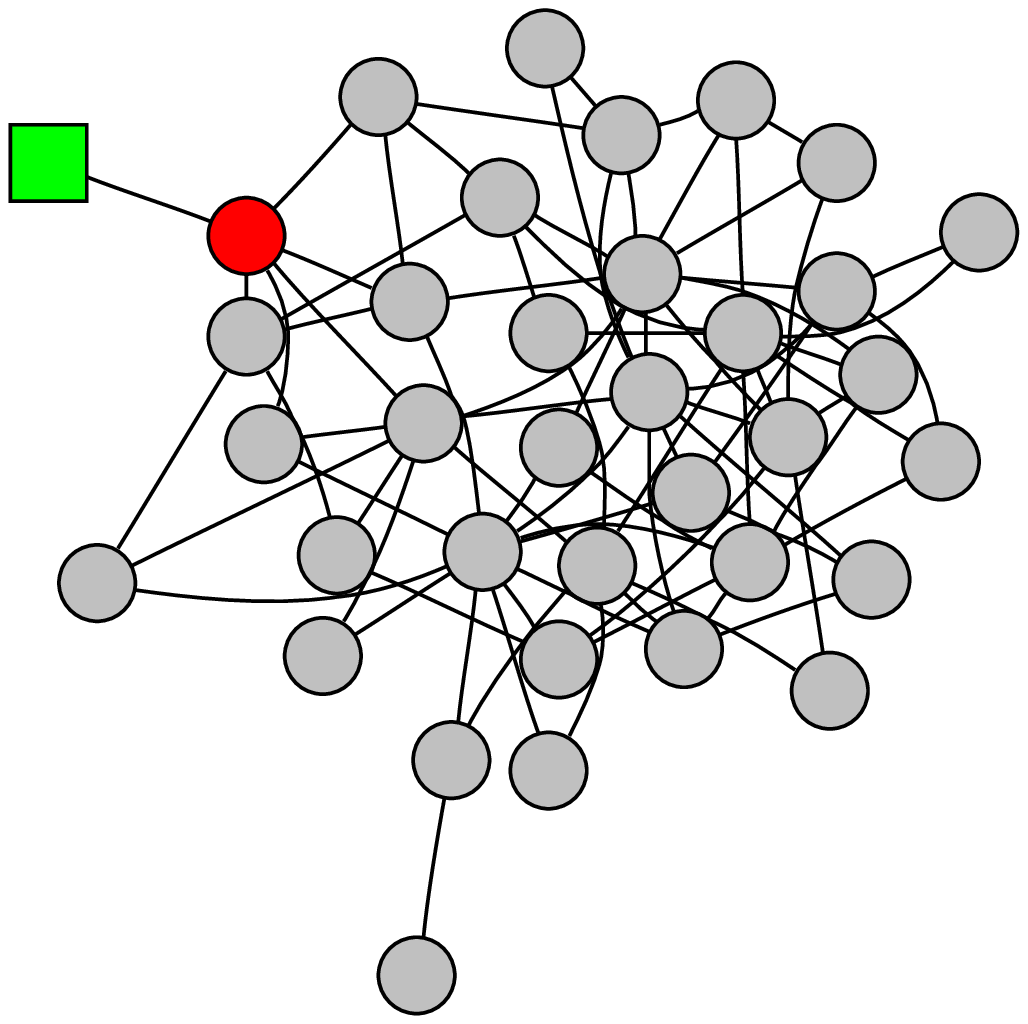} &
    \includegraphics[scale=0.3]{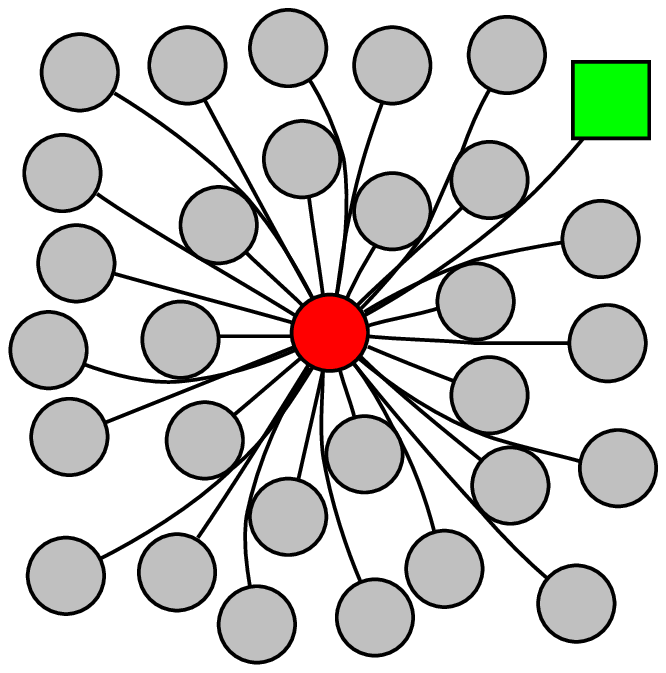} &
    \includegraphics[scale=0.3]{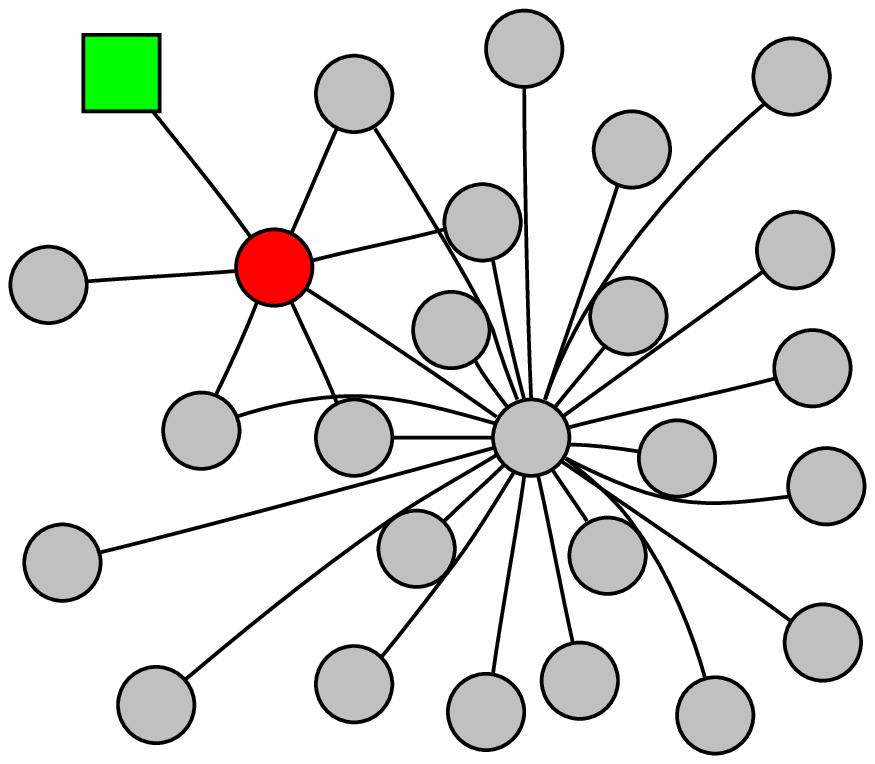} &
    \includegraphics[scale=0.3]{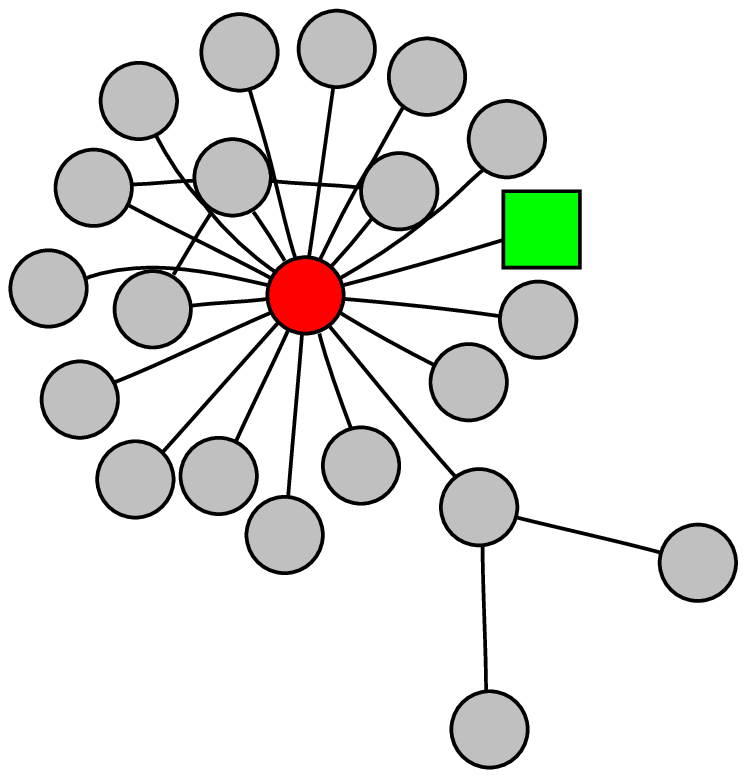} &
    \includegraphics[scale=0.3]{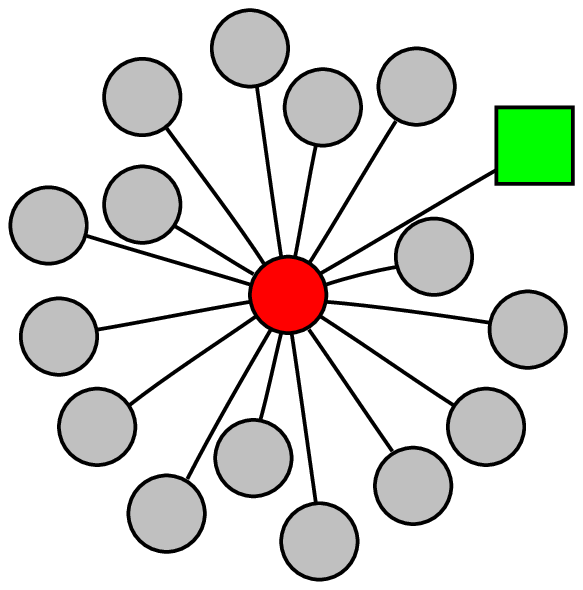} \\
    & & &  & \\
    \includegraphics[scale=0.3]{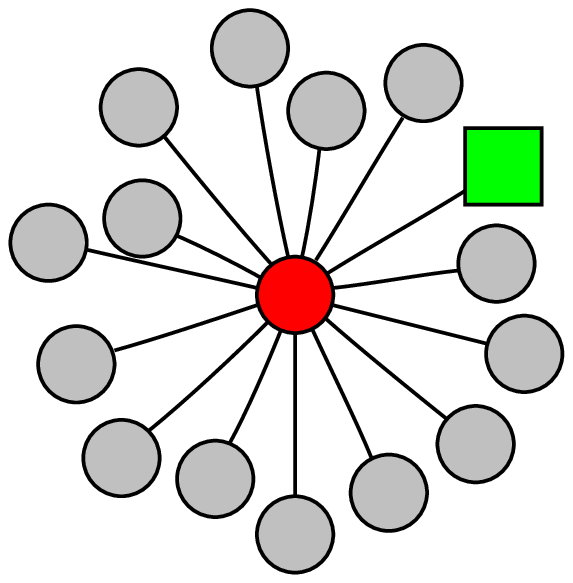} &
    \includegraphics[scale=0.3]{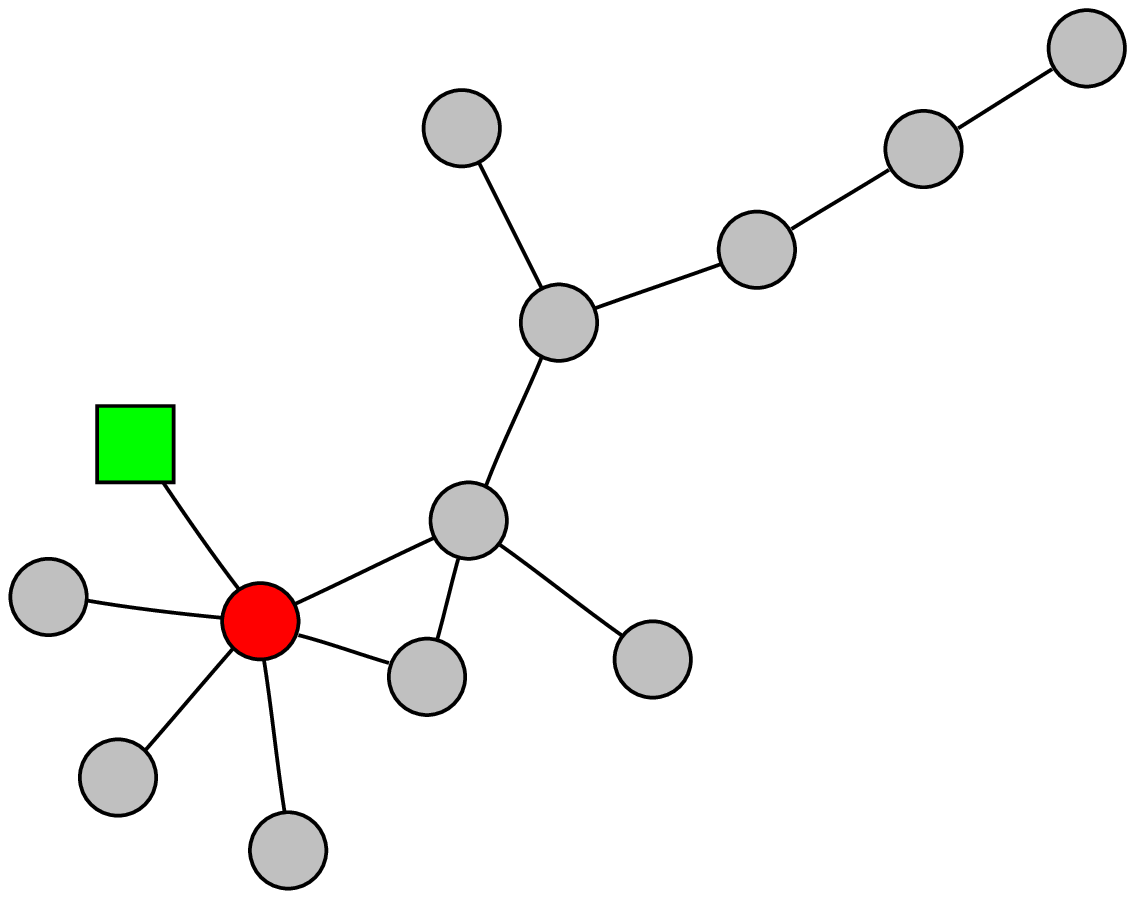} &
    \includegraphics[scale=0.3]{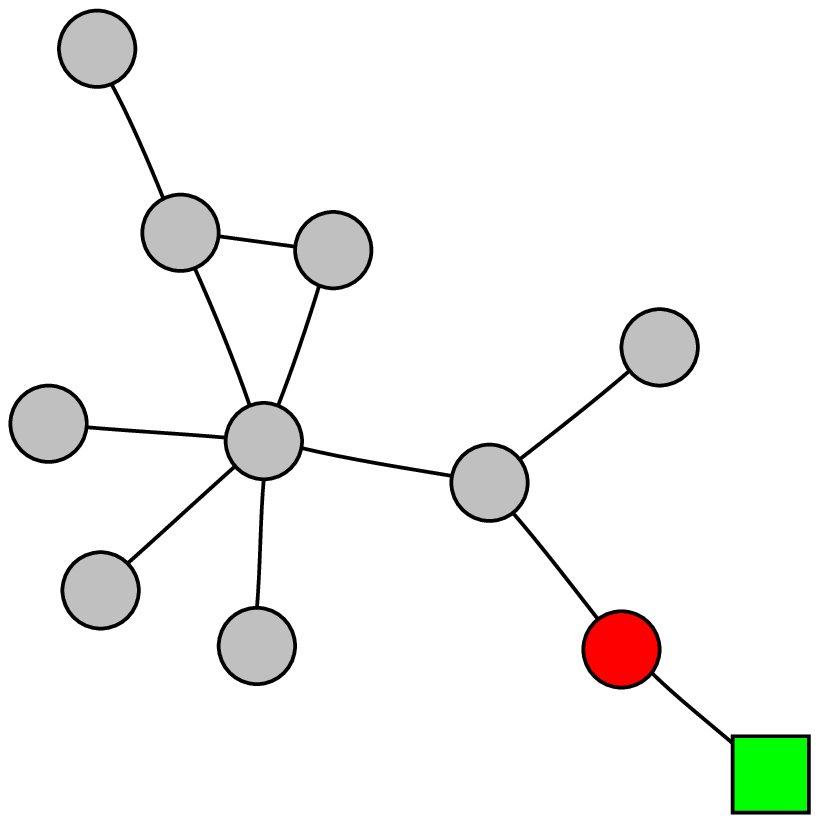} &
    \includegraphics[scale=0.3]{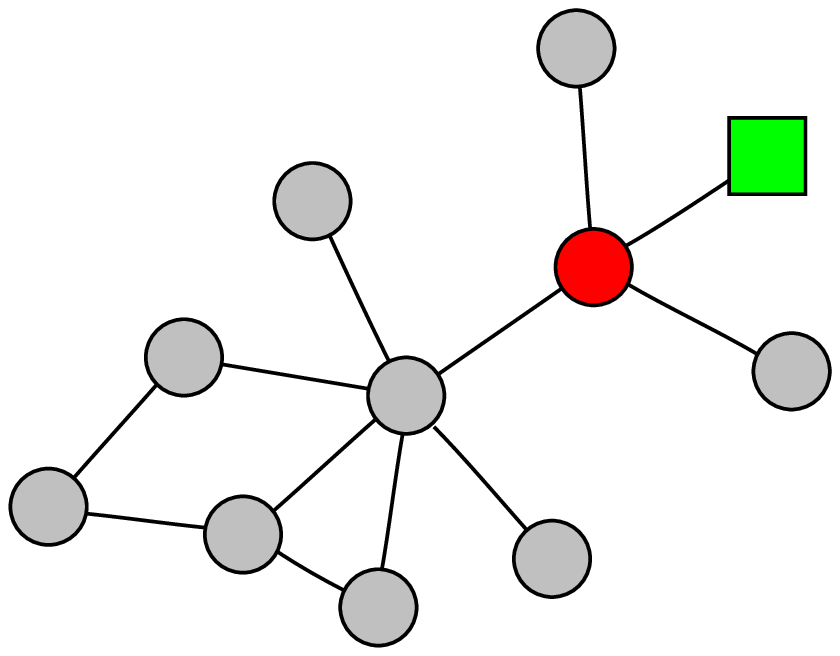} &
    \includegraphics[scale=0.3]{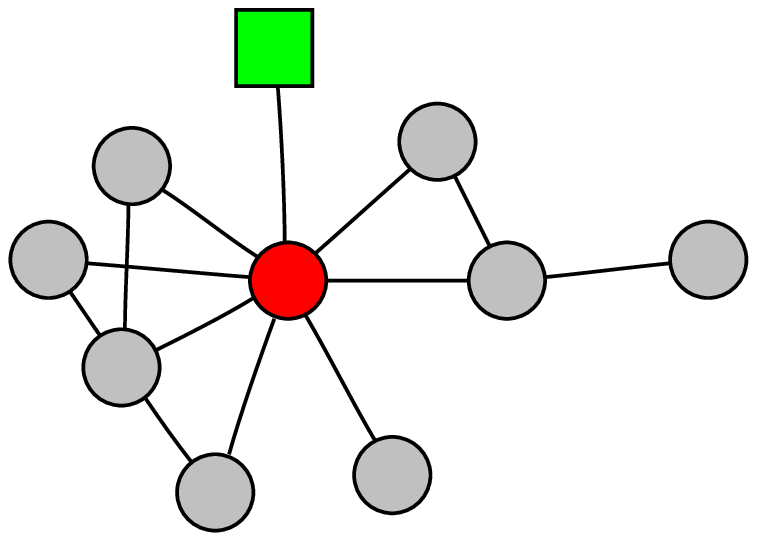} \\
	\end{tabular}
    \end{center}
\caption{
Ten largest whiskers of the \net{Epinions} social network.
The green square node is the node from the bi-connected core of the network to
which the whisker is connected.
For visual clarity, the whisker node that connects to the core of the network
is displayed in red, and thus it is the edge between the red circle and the
green square node that if cut disconnects the whisker from the core.
The distribution of whisker sizes and comparison to rewired network is
plotted in Figure~\ref{fig:whiskerDistr:Epinions}.
}
\label{fig:whiskerEpinions}
\end{figure}

\begin{figure}
	\begin{center}
	\begin{tabular}{ccccc}
    \includegraphics[scale=0.35]{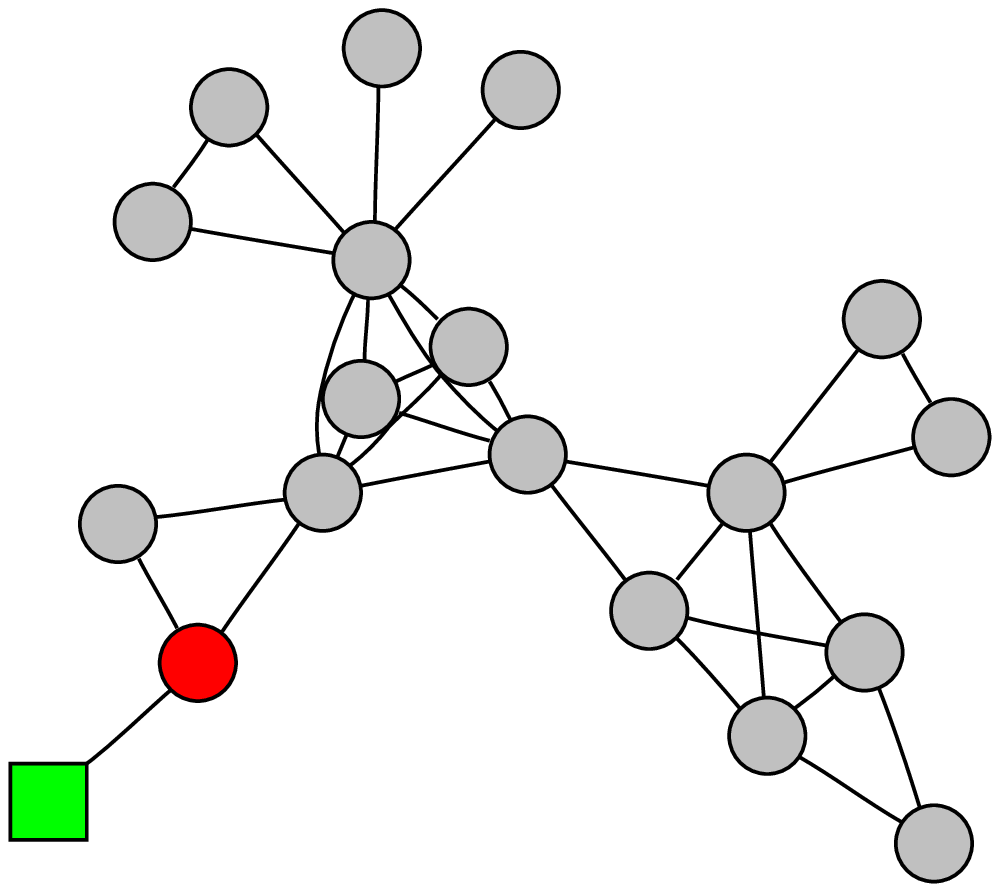} &
    \includegraphics[scale=0.35]{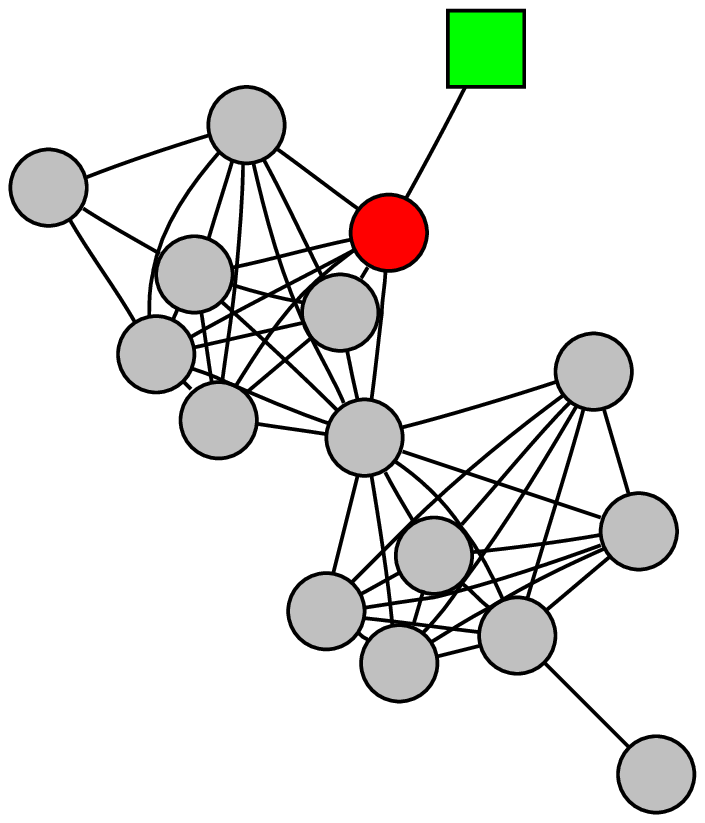} &
    \includegraphics[scale=0.35]{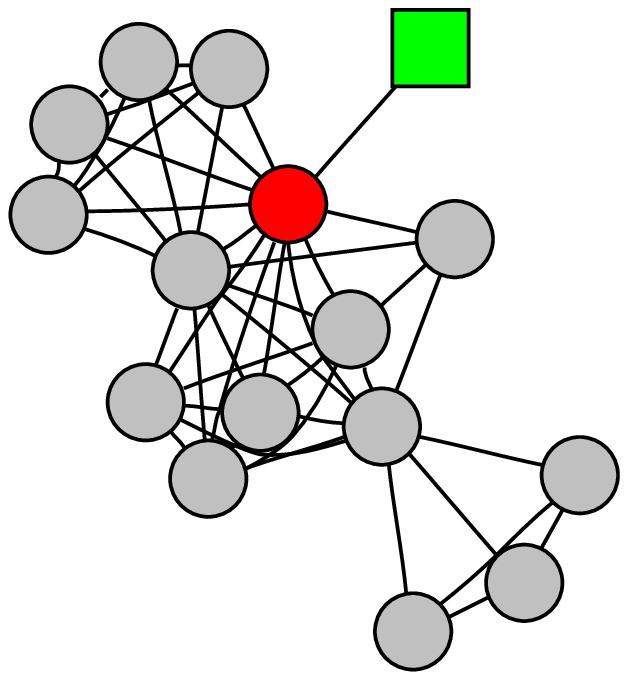} &
    \includegraphics[scale=0.35]{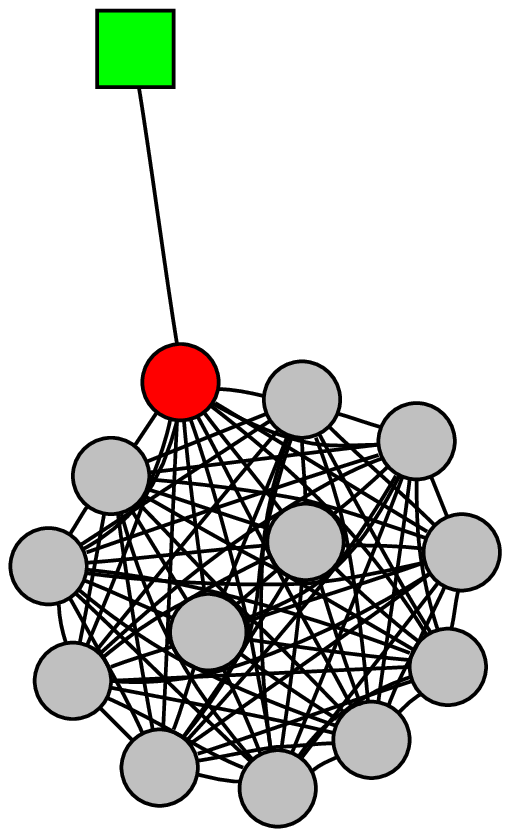} &
    \includegraphics[scale=0.35]{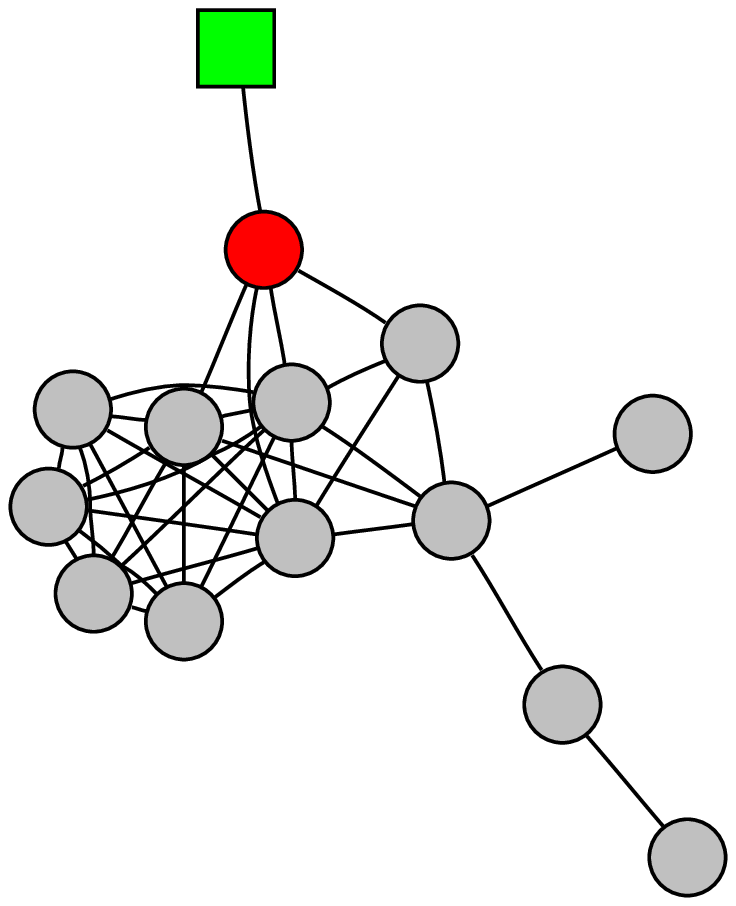} \\
    & & &  & \\
    \includegraphics[scale=0.35]{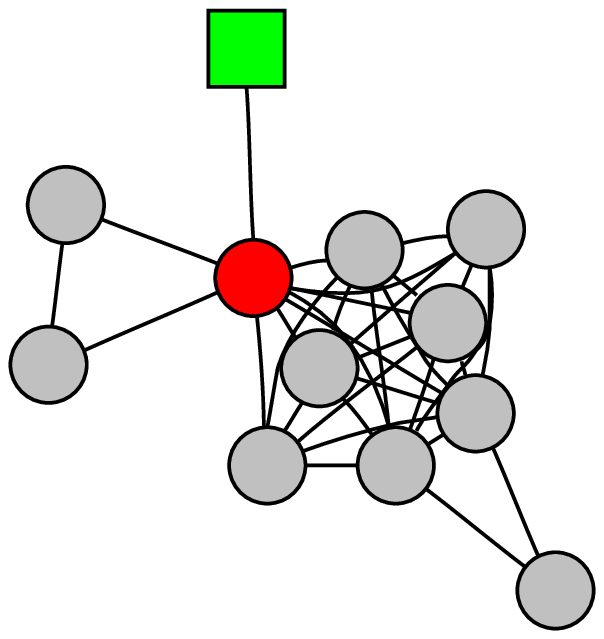} &
    \includegraphics[scale=0.35]{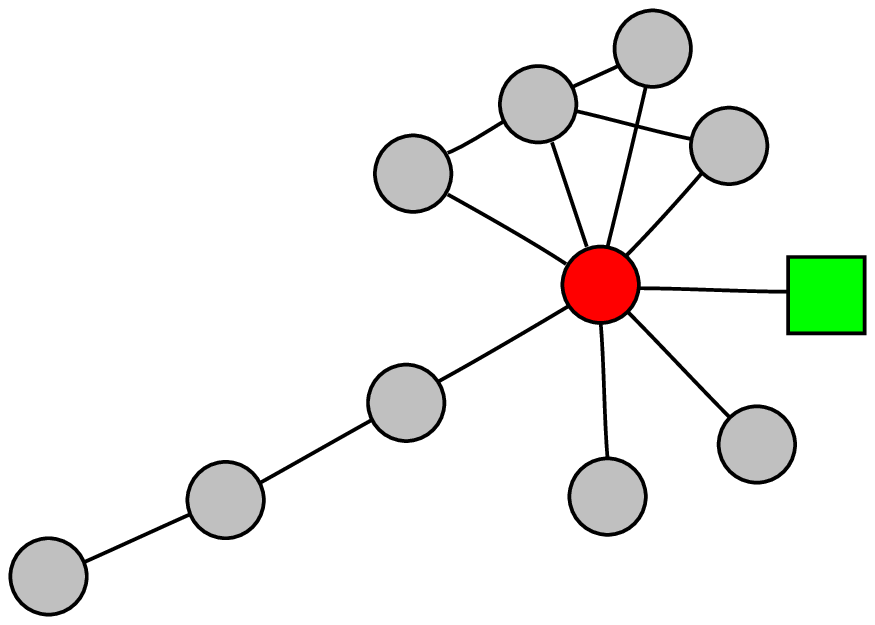} &
    \includegraphics[scale=0.35]{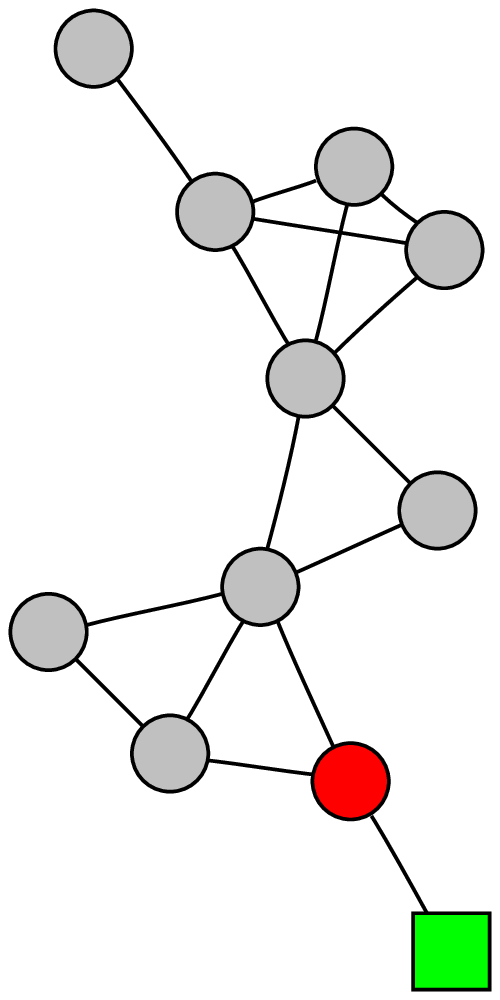} &
    \includegraphics[scale=0.35]{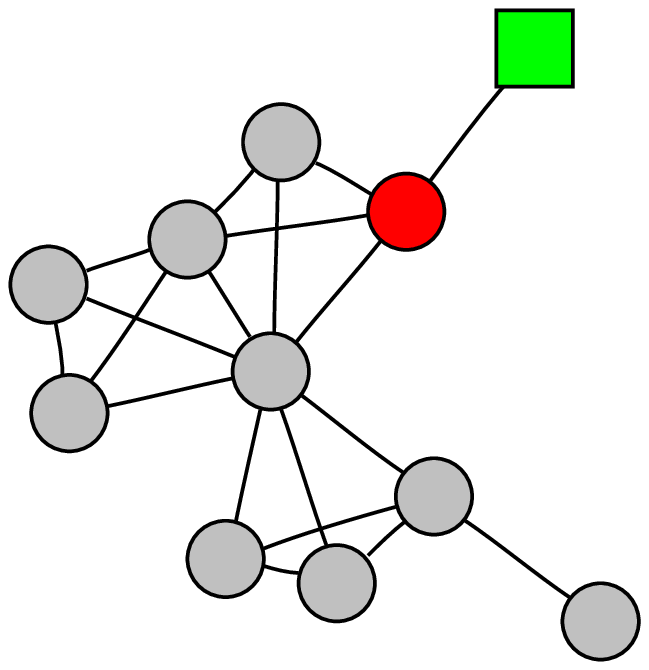} &
    \includegraphics[scale=0.35]{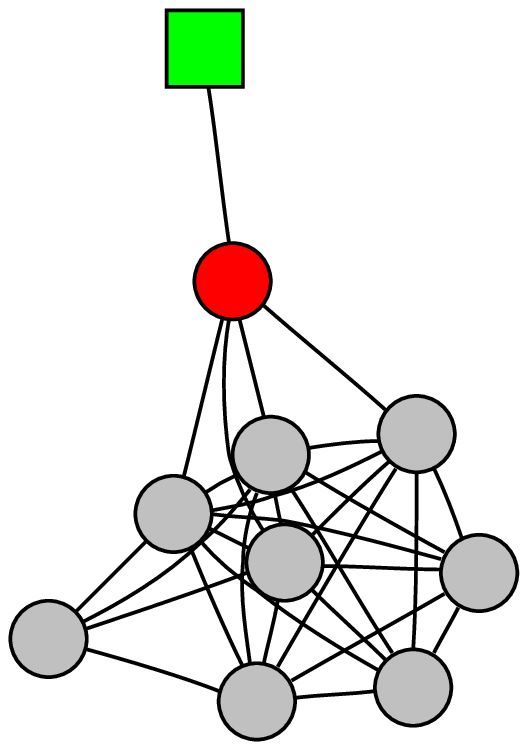} \\
	\end{tabular}
    \end{center}
\caption{
Ten largest whiskers of the \net{CA-cond-mat} co-authorship network.
The green square node belongs to the network core, and by cutting the
edge connecting it with red circular node we separate the community of
circles from the rest of the network (depicted as a green square).
}
\label{fig:whiskerCondMat}
\end{figure}

Figure~\ref{fig:whiskerEpinions} shows the ten largest whiskers of the
\net{Epinions} social network, the full size distribution of which was
plotted in Figure~\ref{fig:whiskerDistr:Epinions}, and
Figure~\ref{fig:whiskerCondMat} shows the ten largest whiskers of the
\net{CA-cond-mat} co-authorship network. In these networks, the whiskers
have on the order of $10$ nodes, and they are seen to have a rich internal
structure. Similar but substantially more complex figures could be
generated for networks with larger whiskers. In general, the results we
observe are consistent with a knowledge of the fields from which the
particular datasets have been drawn. For example, in \net{Web-Google} we
see very large whiskers. This probably represents a well-connected network
between the main categories of a website (\emph{e.g.}, different
projects), while the individual project websites have a main index page
that then points to the rest of the documents.

The discrepancy between the sizes of the whiskers in the original and the
rewired networks gives hints that real networks have much richer structure
than that imposed by their heavy-tailed degree distribution. One might ask
whether the conclusion from this is that real-world graphs should be
thought of as being somewhat like sparse random graphs, since, {\em e.g.},
both have whiskers, or should be thought of as very different than sparse
random graphs, since, {\em e.g.}, the whiskers have much more internal
structure. We will return to this issue in Section~\ref{sxn:models}.

\subsection{Bags of whiskers and communities of composed whiskers}
\label{sxn:obs_struct:bags_of_whiskers}

Empirically, if one looks at the sets of nodes achieving the minimum in
the NCP plot (green Metis+MQI curve), then before the global NCP minimum
communities are whiskers and above that size scale they are often unions
of disjoint whiskers. To understand the extent to which these whiskers and
unions of whiskers are responsible for the ``best'' conductance sets of
different sizes, we have developed the \emph{Bag-of-Whiskers Heuristic}.
We artificially compose ``communities'' from disconnected whiskers and
measure conductance of such clusters. 
Clearly, interpreting and relating such communities to real-world 
communities makes little sense as these communities are in fact 
disconnected. 

In more detail, we performed the following experiment: suppose we have a set $W =
\{w_1,w_2, \ldots\}$ of whiskers. In order to construct the optimal
conductance cluster of size $k$, we need to solve the following problem:
find a set $C$ of whiskers such that $\sum_{i\in C} N(w_i) = k$ and
$\sum_{i\in C}\frac{d(w_i)}{|C|}$ is maximized, where $N(w_i)$ is the
number of nodes in $w_i$ and $d(w_i)$ is its total internal degree. We
then use a dynamic programming to get an approximate solution to this
problem.
This way, for each size $k$, we find a cluster that is composed solely from
(disconnected) whiskers. Figure~\ref{fig:netPhiPlot} as well as
Figures~\ref{fig:phiDatasets1}, \ref{fig:phiDatasets2}
and~\ref{fig:phiDatasets3} show the results of this heuristic applied to
many of our network datasets (blue curve).

There are several observations we can make:
\begin{itemize}
\item The largest whisker (denoted with a red square) is
    the lowest point in nearly all NCP plots. This means that the best
    conductance community is in a sense trivial as it cuts just a
    single edge, and in addition that a very simple heuristic can find
    this set.
\item For community size below the critical size of $\approx 100$ nodes 
    (\emph{i.e.}, of size smaller than the largest whisker), the best 
    community in the network is actually a whisker and can be cut by a 
    single edge (blue and red curve overlap).
\item For community size larger than the critical size of $\approx
    100$, the Bag-of-Whiskers communities have better scores
    than the internally well-connected communities extracted by Local
    Spectral (red curve). 
The shape of this blue curve in that size region depends on the distribution of 
sizes of whiskers, but in nearly every case it is seen to yield better 
conductance sets than the Local Spectral Algorithm.
\end{itemize}
Moreover, the Bag-of-Whiskers Heuristic often agrees, exactly or
approximately, with results from Metis+MQI (green curve).
In particular, the best conductance
sets of a given size are often disconnected, and when they are connected
they are often only tenuously connected. Thus, if one only cares about
finding good cuts then the best cuts in these large sparse graphs are
obtained by composing unrelated disconnected pieces. Intuitively, a
compact cluster is internally well and evenly connected. Possible
measures for cluster compactness include: cluster connectedness, diameter,
conductance of the cut inside the cluster, ratio of conductance of the cut
outside versus the cut inside. We discuss this in more detail in
Section~\ref{algo-notes-section}.

\subsection{Community profile of networks with no $1$-whiskers}
\label{sxn:obs_struct:remove_whiskers}

\begin{figure}
	\begin{center}
    \begin{tabular}{cc}
    	\includegraphics[width=0.45\textwidth]{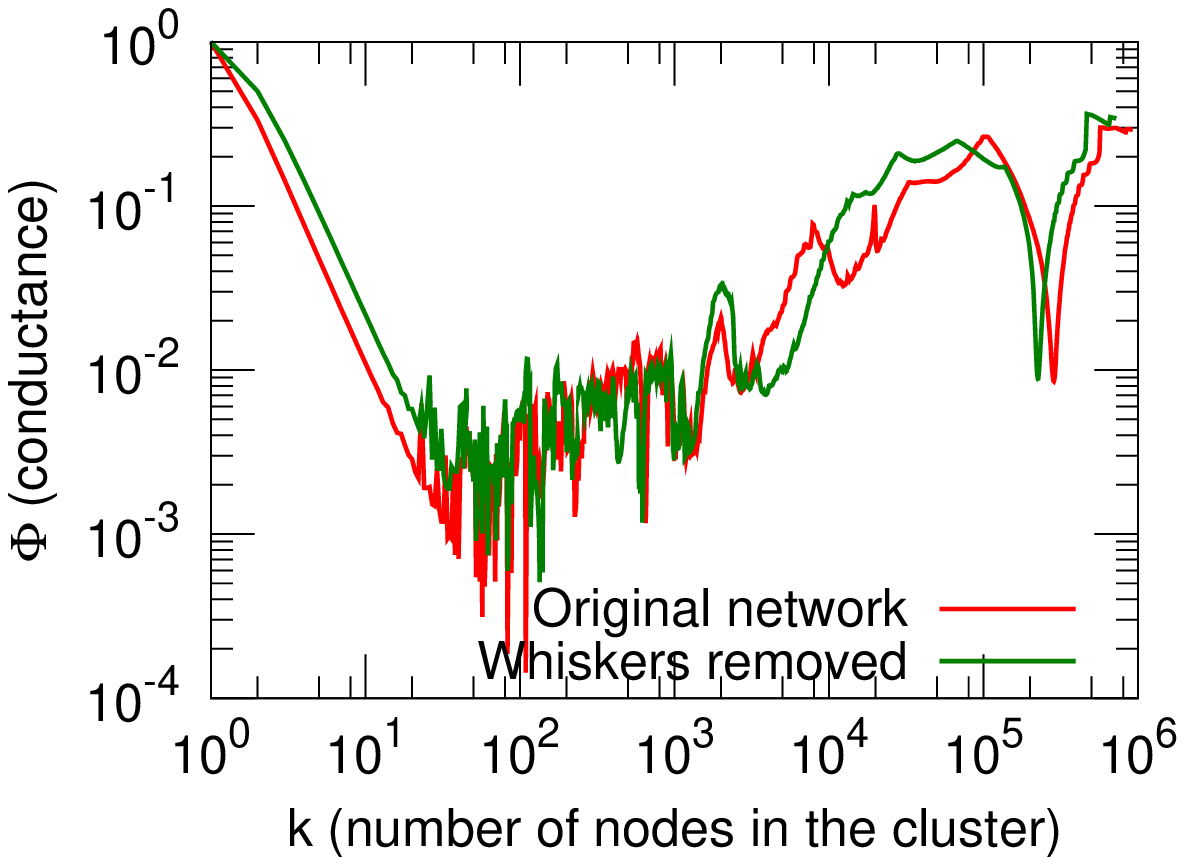} &
  	\includegraphics[width=0.45\textwidth]{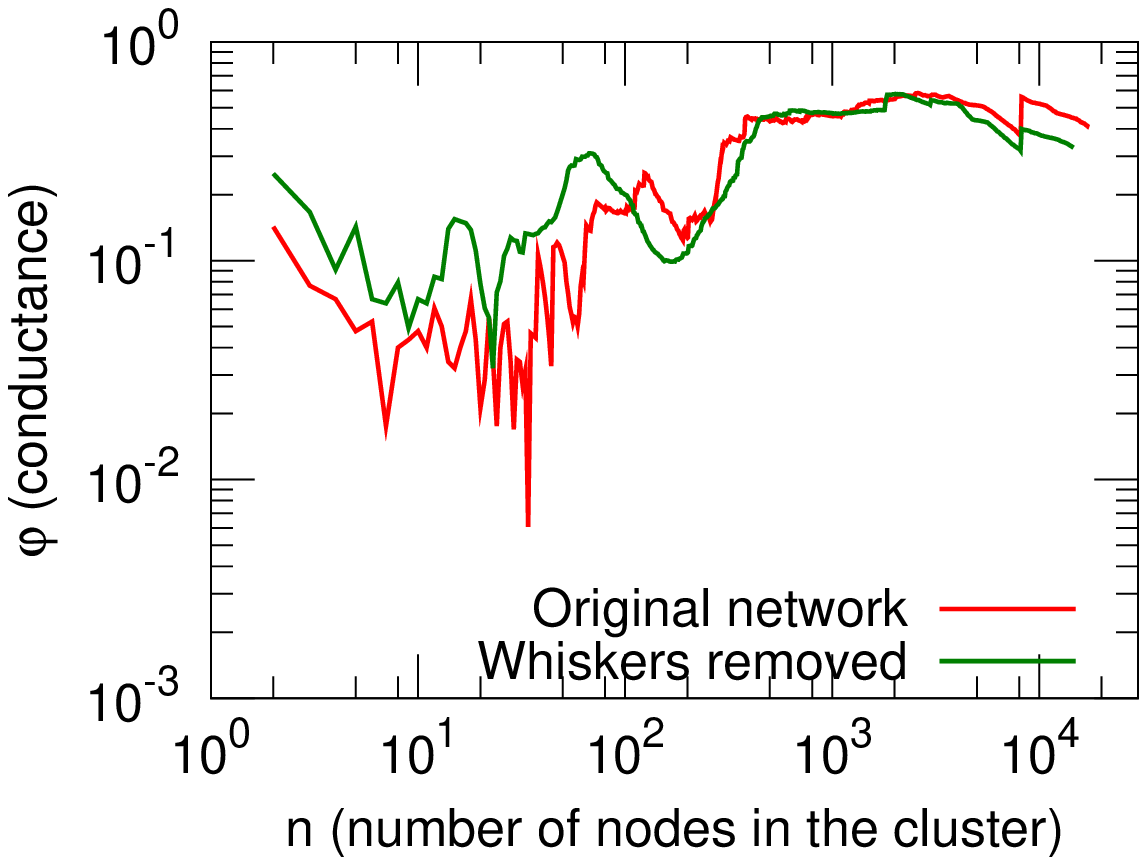} \\
  	(a) \net{LiveJournal01} & (b) \net{Epinions} \\
  	\includegraphics[width=0.45\textwidth]{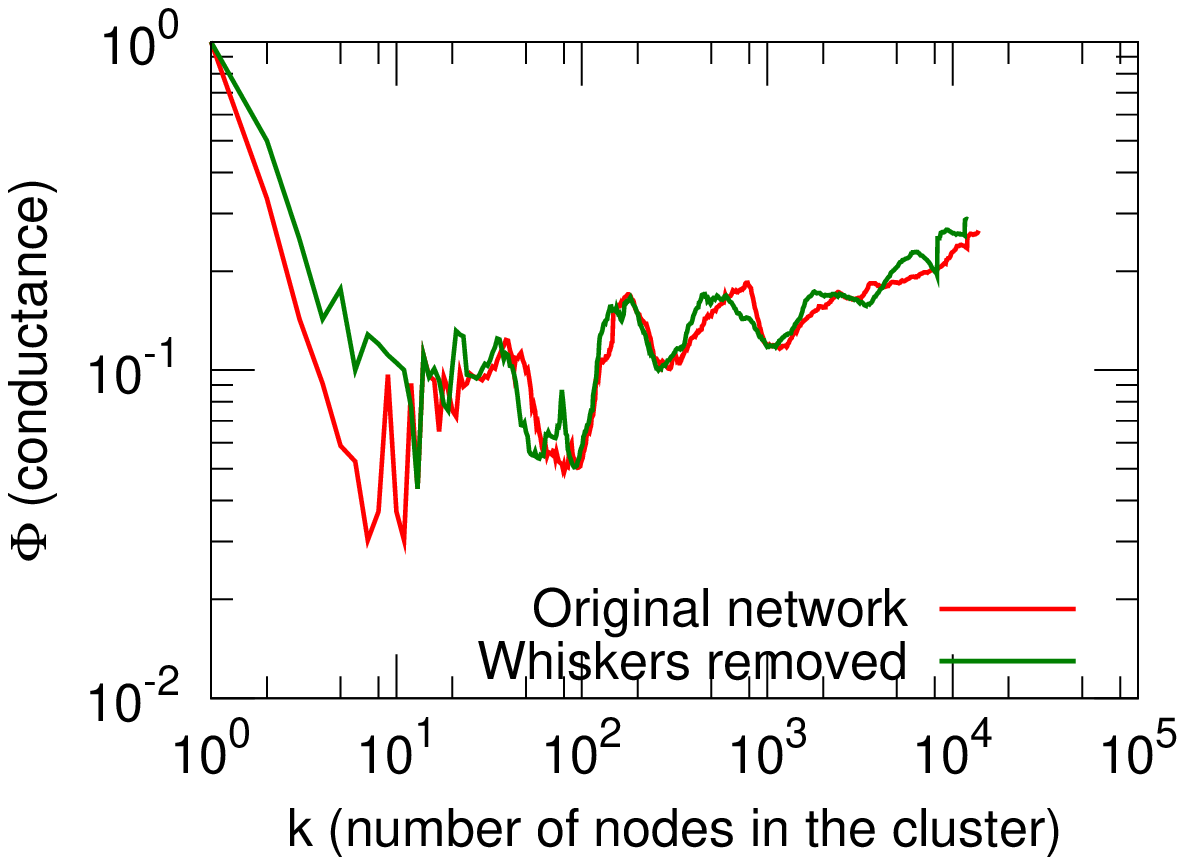} &
  	\includegraphics[width=0.45\textwidth]{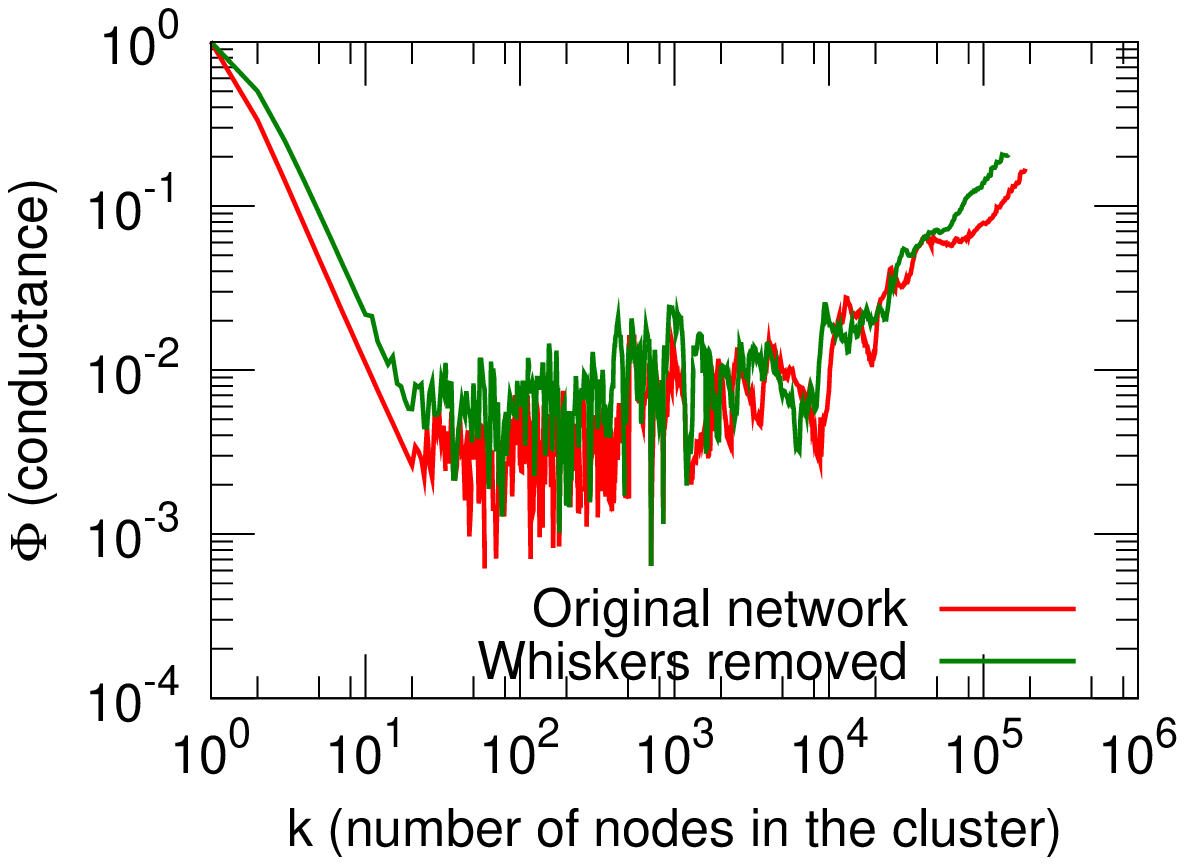} \\
  	(c) \net{Cit-hep-th} & (d) \net{Web-Google} \\
  	\includegraphics[width=0.45\textwidth]{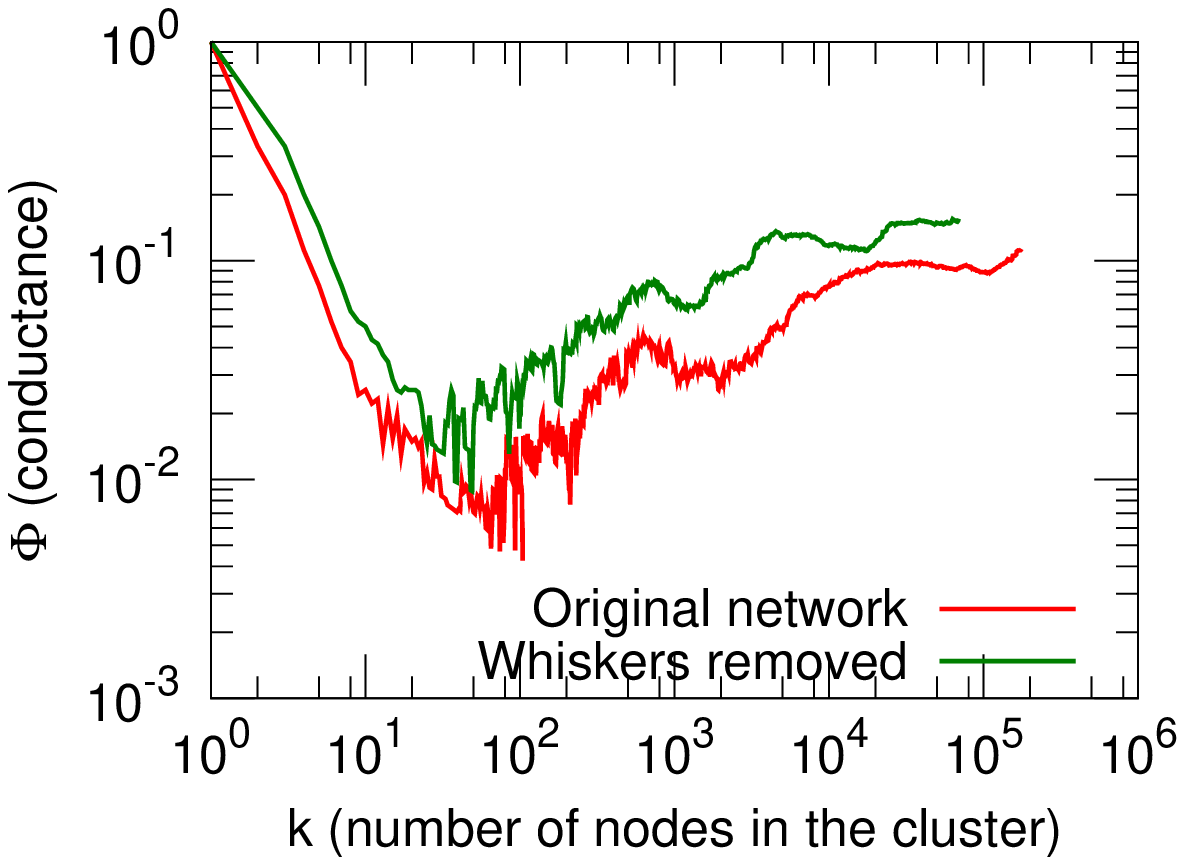} &
  	\includegraphics[width=0.45\textwidth]{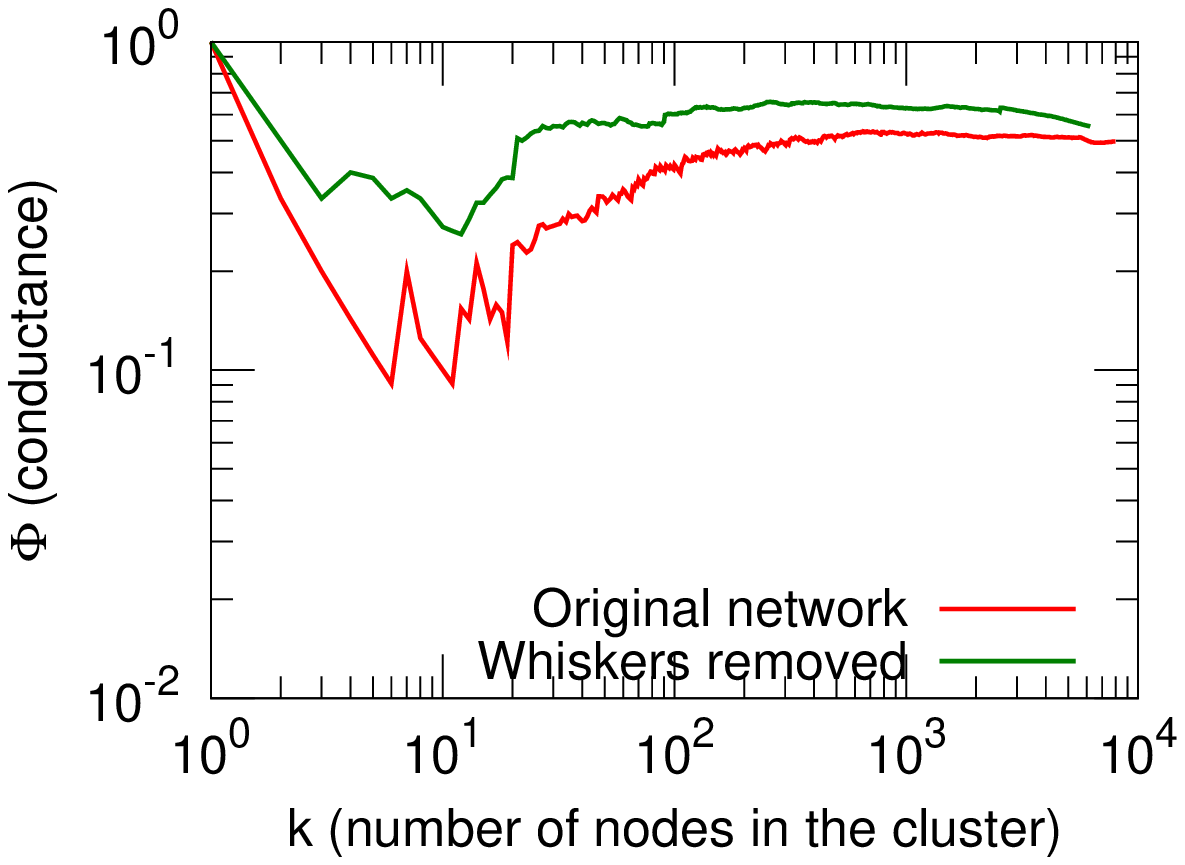} \\
  	(e) \net{Atp-DBLP} & (f) \net{Gnutella-31} \\	
    \end{tabular}
	\end{center}
\caption{
[Best viewed in color.]
Network community profile plots with (in red) and without (in green) 
$1$-whiskers, for each of the six networks shown Figure~\ref{fig:netPhiPlot}.
Whiskers were removed as described in the text.
In the former case, we plot results for the full network, and in the latter 
case, we plot results for the largest bi-connected component.
}
\label{fig:phiCore}
\end{figure}

Given the surprisingly significant effect on the community structure of
real-world networks that whiskers and unions of disjoint whiskers have,
one might wonder whether we see something qualitatively different if we
consider a real-world network in which these barely-connected pieces have
been removed. To study this, we found all $1$-whiskers and removed them
from our networks, using the procedure we described in
Section~\ref{sxn:obs_struct:whiskers}, \emph{i.e.}, we selected the
largest biconnected component for each of our network datasets. 
This way, we kept only the network core, and we then computed the NCP plots for these 
modified networks.
Figure~\ref{fig:phiCore} shows the NCP plots of networks constructed when 
we remove whiskers (\emph{i.e.}, keep only the network core) for the six 
networks we studied in detail before.

Notice that whisker removal does not change the NCP plot much: the plot
shifts slightly upward, but the general trends remain the same. Upon
examination, the global minimum occurs with a ``2-whisker'' that is
connected by two edges to the remainder of the graph. Intuitively, the
largest biconnected core has a large number of barely connected
pieces---connected now by two edges rather than by one edge---and thus the ``core'' itself has a core-periphery structure. Since the
``volume'' for these pieces is similar to that for the original whiskers,
whereas the ``surface area'' is a factor of two larger, the conductance
value is roughly a factor of two worse. Thus, although we have been
discussing $1$-whiskers in this section, one should really view them as
the simplest example of weakly-connected pieces that exert a significant
effect on the community structure in large real-world networks.

\section{Comparison to other algorithms}
\label{algo-notes-section}

So far, we have been primarily relying on two graph partitioning
algorithms: a Local Spectral Algorithm and Metis+MQI. Next, we want to
demonstrate that what we are observing is a true structural property of our
network datasets, rather than properties of our algorithms; and we want to
use the differences between different approximation algorithms to further
highlight structural properties of our network datasets. In this section
we discuss several meta-issues related to this, including whether or not
our algorithms are sufficiently powerful to recover the true shape of the
minimal conductance curves, and whether we should actually be trying to
optimize a slightly different measure that combines conductance of the
separating cut with the piece compactness.

Recall that we defined the NCP plot to be a curve showing the minimum
conductance $\phi$ as a function of piece size $k$.  Finding the points on
this curve is NP-hard. Any cut that we find will only provide an upper
bound on the true minimum at the resulting piece's size.  Given that
fact, how confident can we be that the curve of upper bounds that we
have computed has the same rising or falling shape as the true curve?

One method for finding out whether any given algorithm is doing a good job
of pushing down the upper bounding curve in a non-size-biased way is to
compare its curves for numerous graphs with those produced by other
algorithms. In such experiments, it is good if the algorithms are very
powerful and also independent of each other. We have done extensive
experiments along these lines, and our choice of Local Spectral and
Metis+MQI as the two algorithms for the main body of this paper was based
on the results. In Section~\ref{algo-cross-checking-section} we mention a
few interesting points related to this.

A different method for reducing our uncertainty about the shape of the
true curve would be to also compute lower bounds on the curve.
Ideally, one would compute a complete curve of tight lower bounds,
leaving a thin band between the upper- and lower-bounding curves,
which would make the rising or falling shape of the true curve
obvious.  In Section~\ref{lower-bounds-section} we discuss some
experiments with lower bounds.  Although we only obtained a few
lower bounds rather than a full curve, the results are consistent with
our main results obtained from upper-bounding curves.

Finally, in Section~\ref{algo-compactness-section} we will discuss our
decision to use the Local Spectral algorithm in addition to Metis+MQI in
the main body of the paper, despite the fact that Metis+MQI clearly
dominates Local Spectral at the nominal task of finding the lowest
possible upper bounding curve for the minimal conductance curve.  The
reason for this decision is that Local Spectral often returns ``nicer''
and more ``compact'' pieces because rather than minimizing conductance
alone, it optimizes a slightly different measure that produces a
compromise between the conductance of the bounding cut and the
``compactness'' of the resulting piece.

\subsection{Cross-checking between algorithms}
\label{algo-cross-checking-section}

As just mentioned, one way to gain some confidence in the upper bounding
curves produced by a given algorithm is to compare them with the curves
produced by other algorithms that are as strong as possible, and as
independent as possible.
We have extensively experimented with several variants of the global
spectral method, both the usual eigenvector-based embedding on a line, and
an SDP-based embedding on a hypersphere, both with the usual
hyperplane-sweep rounding method and a fancier flow-based rounding method
which includes MQI as the last step.  In addition, special post-processing
can be done to obtain either connected or disconnected sets.  
After examining the output of those 8 comparatively expensive
algorithms on more than $100$ graphs, we found that our two cheaper main
algorithms did miss an occasional cut on an occasional graph, but
nothing at all serious enough to change our main conclusions.  All of those
detailed results are suppressed in this paper.

We have also done experiments with a practical version of the Leighton-Rao
algorithm~\cite{Leighton:1988,Leighton:1999}, similar to the
implementation described in~\cite{kevin93_finding} and~\cite{kevin04mqi}.
These results are especially interesting because the Leighton-Rao
algorithm, which is based on multi-commodity flow, 
provides a completely independent check on Metis, and on Spectral
Methods generally, and therefore on our two main algorithms, namely
Metis+MQI and Local Spectral.
The Leighton-Rao algorithm has two phases. In the first phase, edge
congestions are produced by routing a large number of commodities
through the network. We adapted our program to optimize conductance
(rather than ordinary ratio cut score) by letting the expected demand
between a pair of nodes be proportional to the product of their
degrees. In the second phase, a rounding algorithm is used to convert
edge congestions into actual cuts.  Our method was to sweep over node
orderings produced by running Prim's MST algorithm on the congestion
graph, starting from a large number of different initial nodes, using
a range of different scales to avoid quadratic run time. We used two
variations of this method, one that only produces connected sets, and
another one that can also produce disconnected sets.

\begin{figure}[t]
\begin{center}
\includegraphics[width=0.32\linewidth]{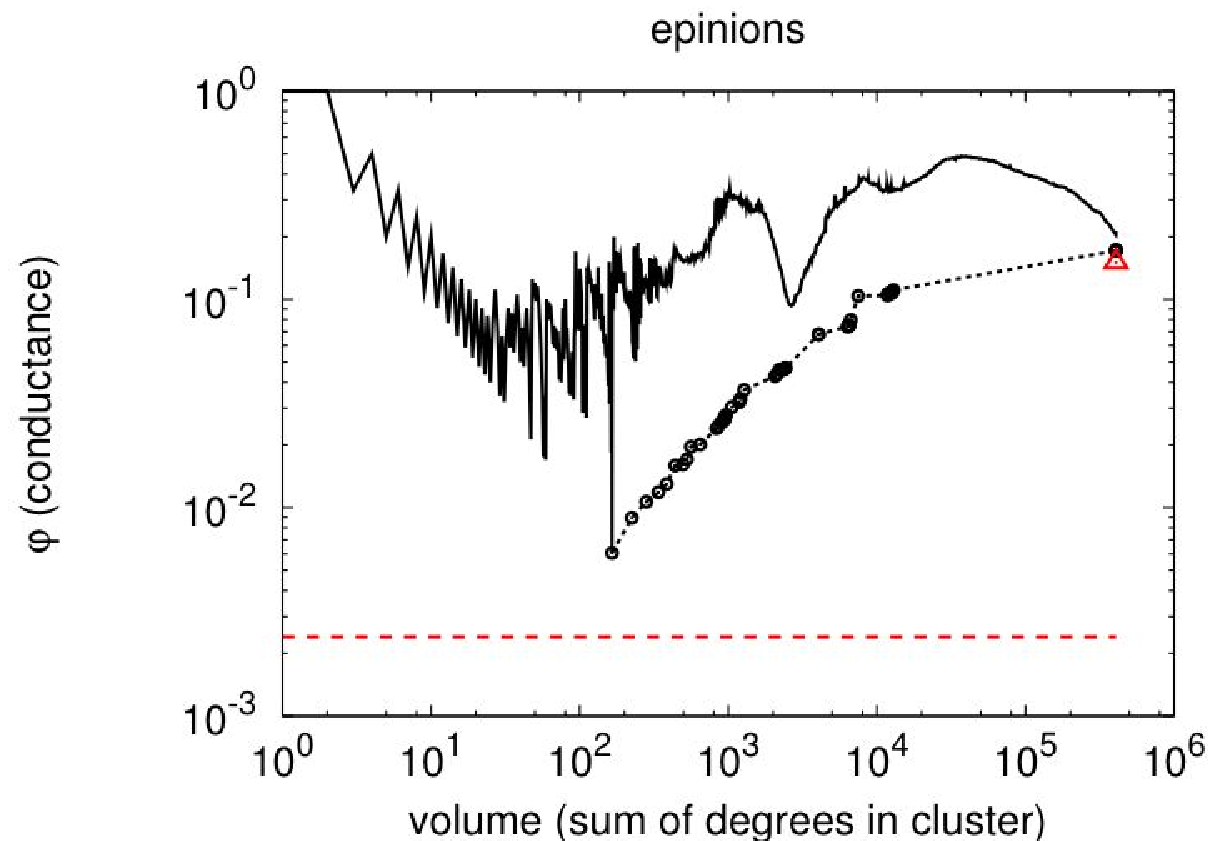}
\includegraphics[width=0.32\linewidth]{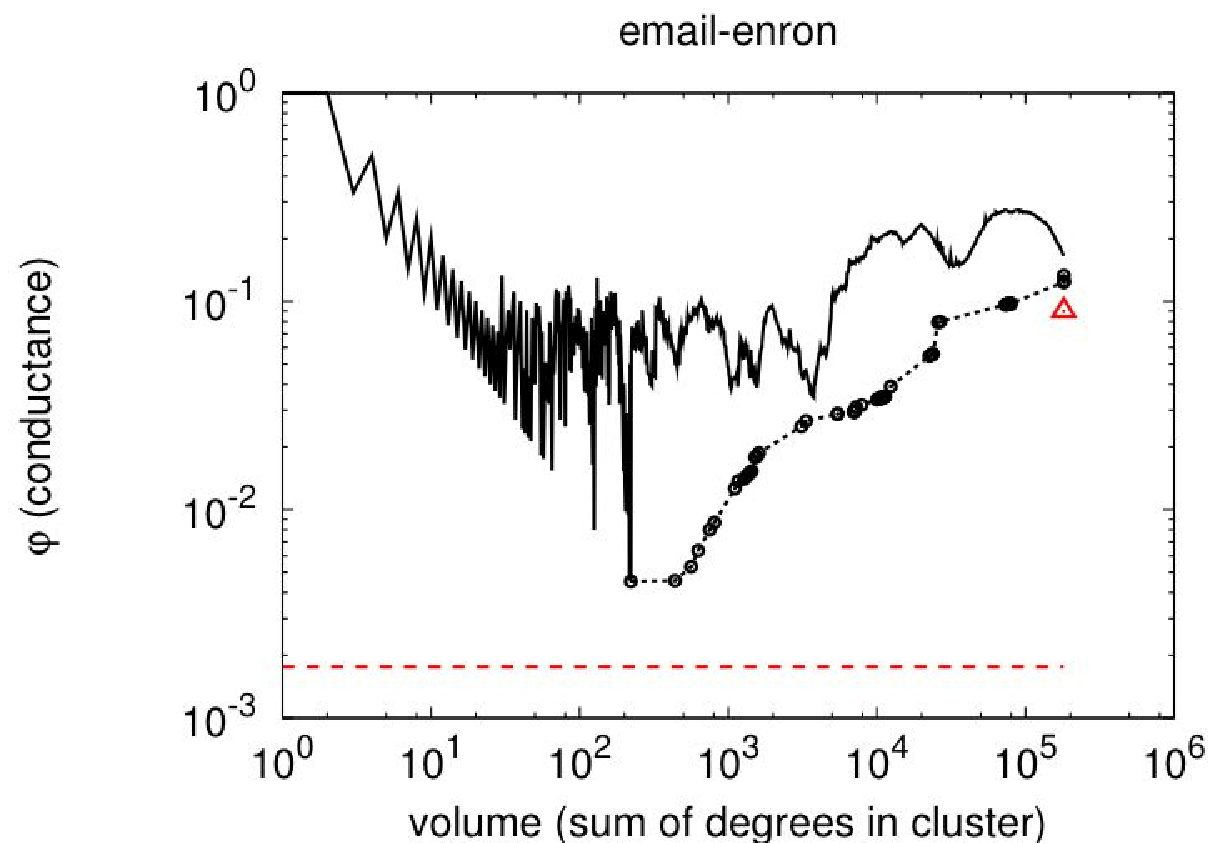}
\includegraphics[width=0.32\linewidth]{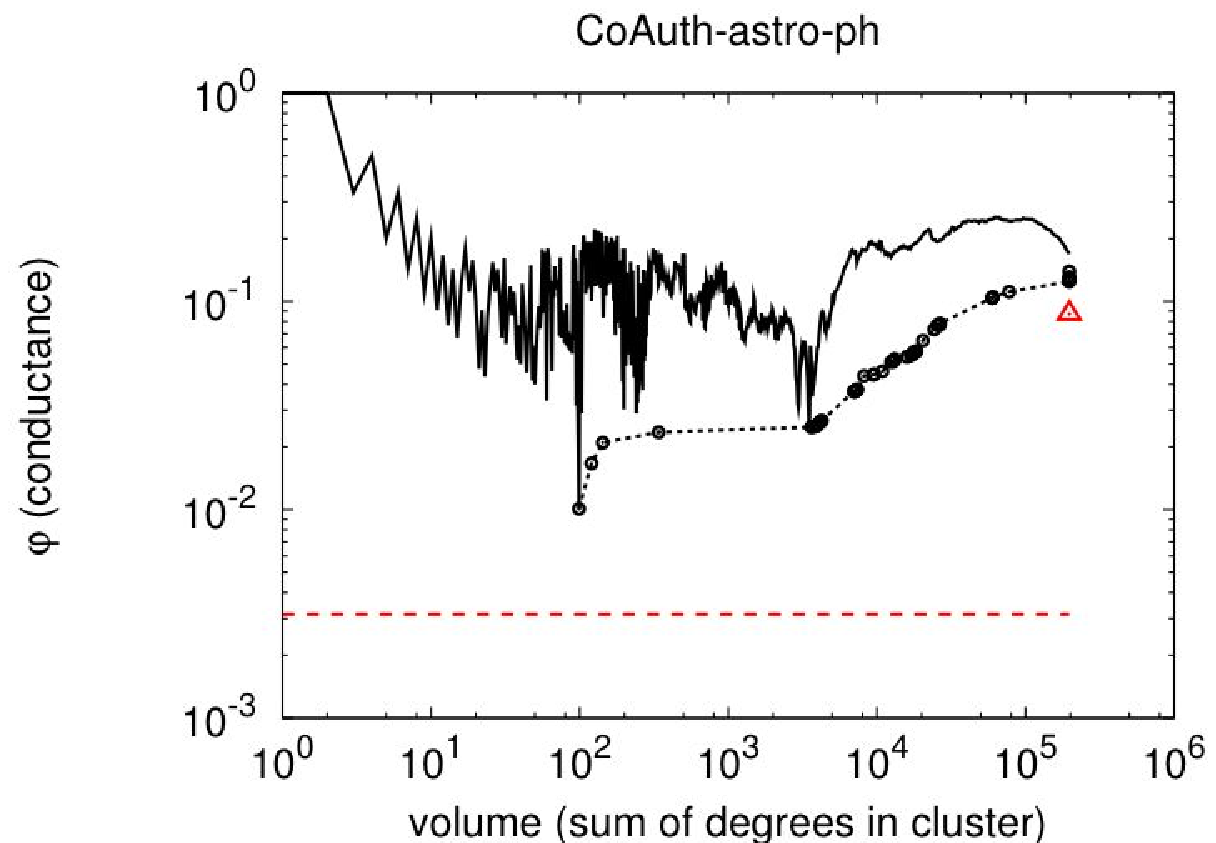}\\
Lower bounds on the conductance of best cut in the network.\\
\includegraphics[width=0.32\linewidth]{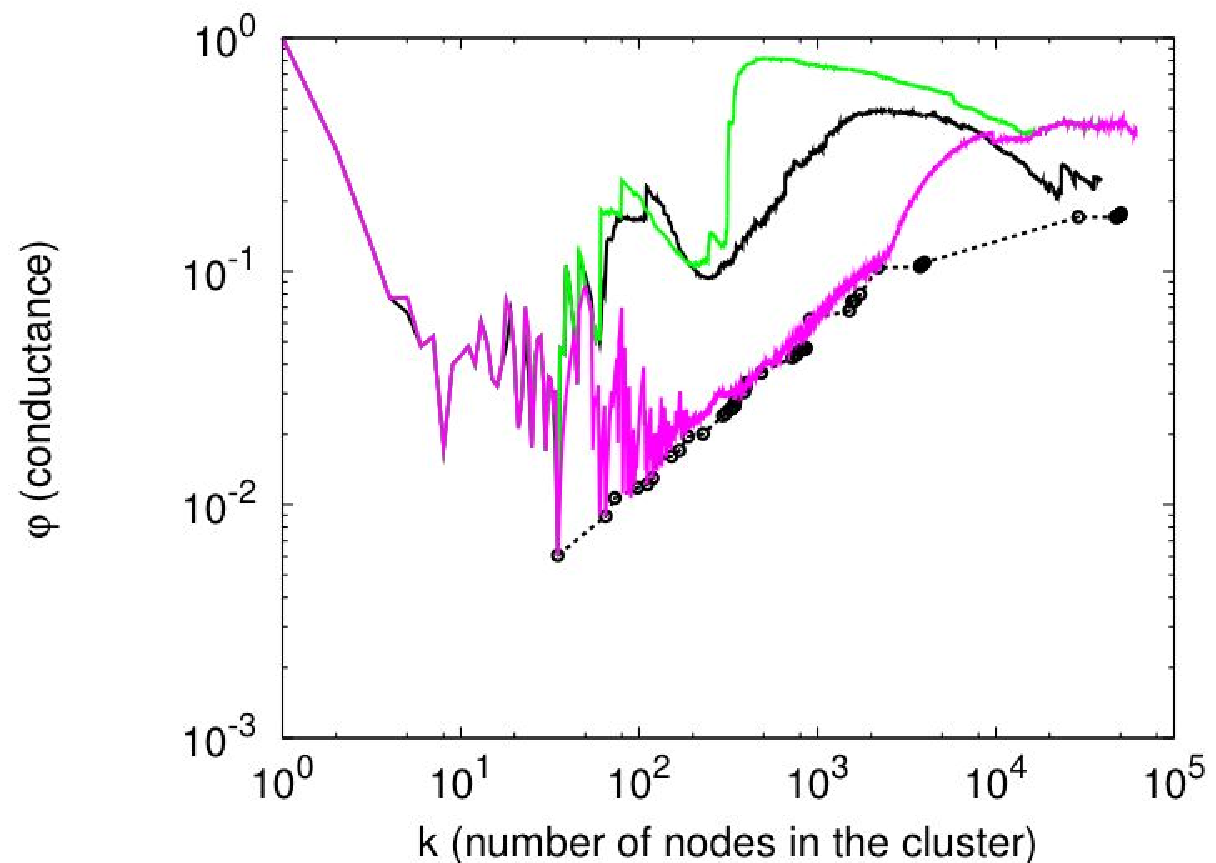}
\includegraphics[width=0.32\linewidth]{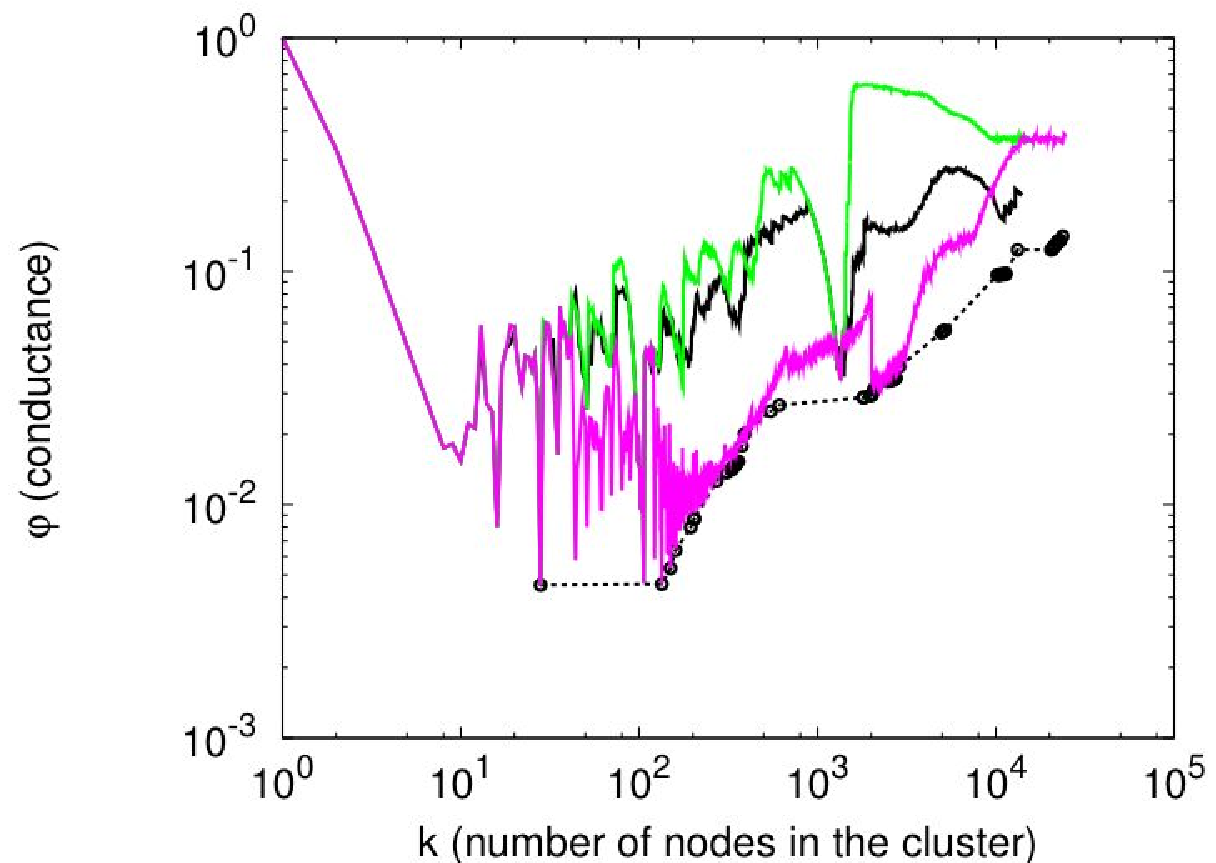}
\includegraphics[width=0.32\linewidth]{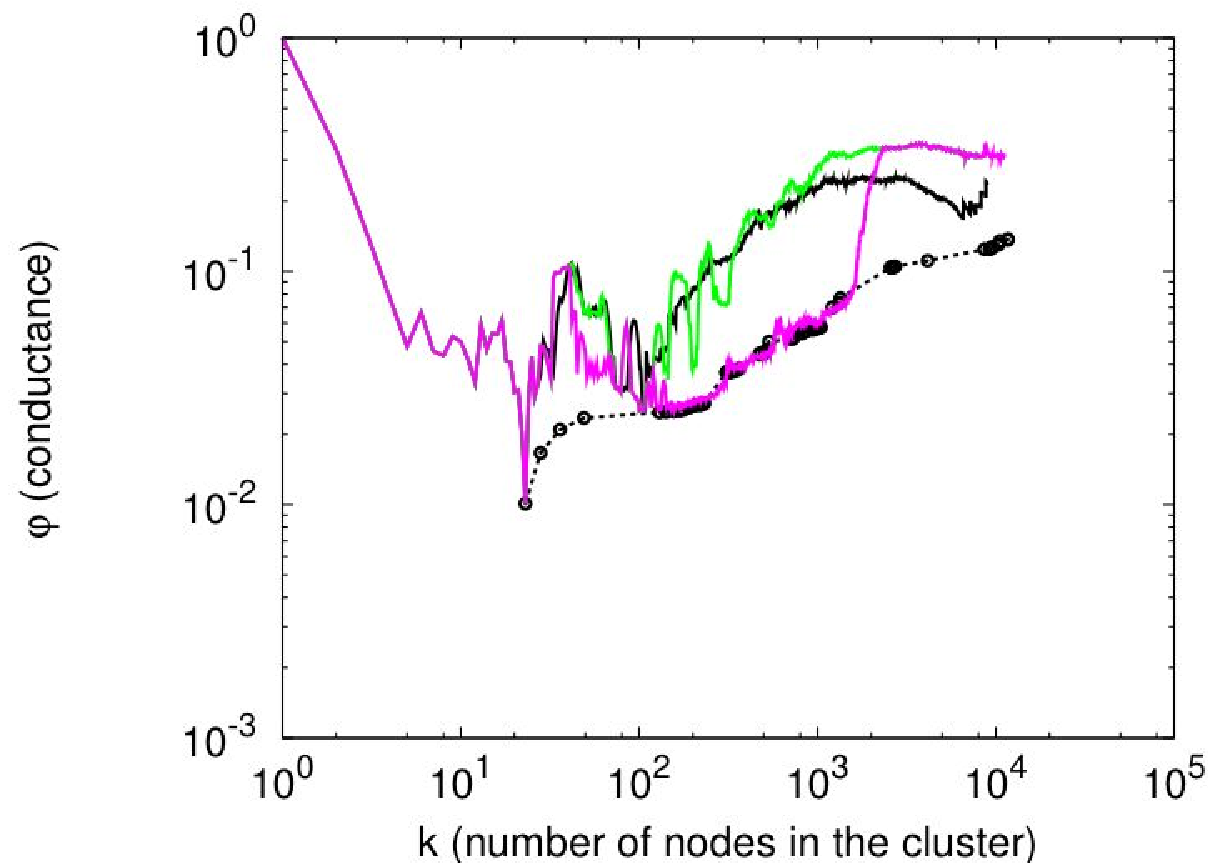}\\
Leighton-Rao: connected clusters (green), disconnected clusters (magenta).\\
\includegraphics[width=0.32\linewidth]{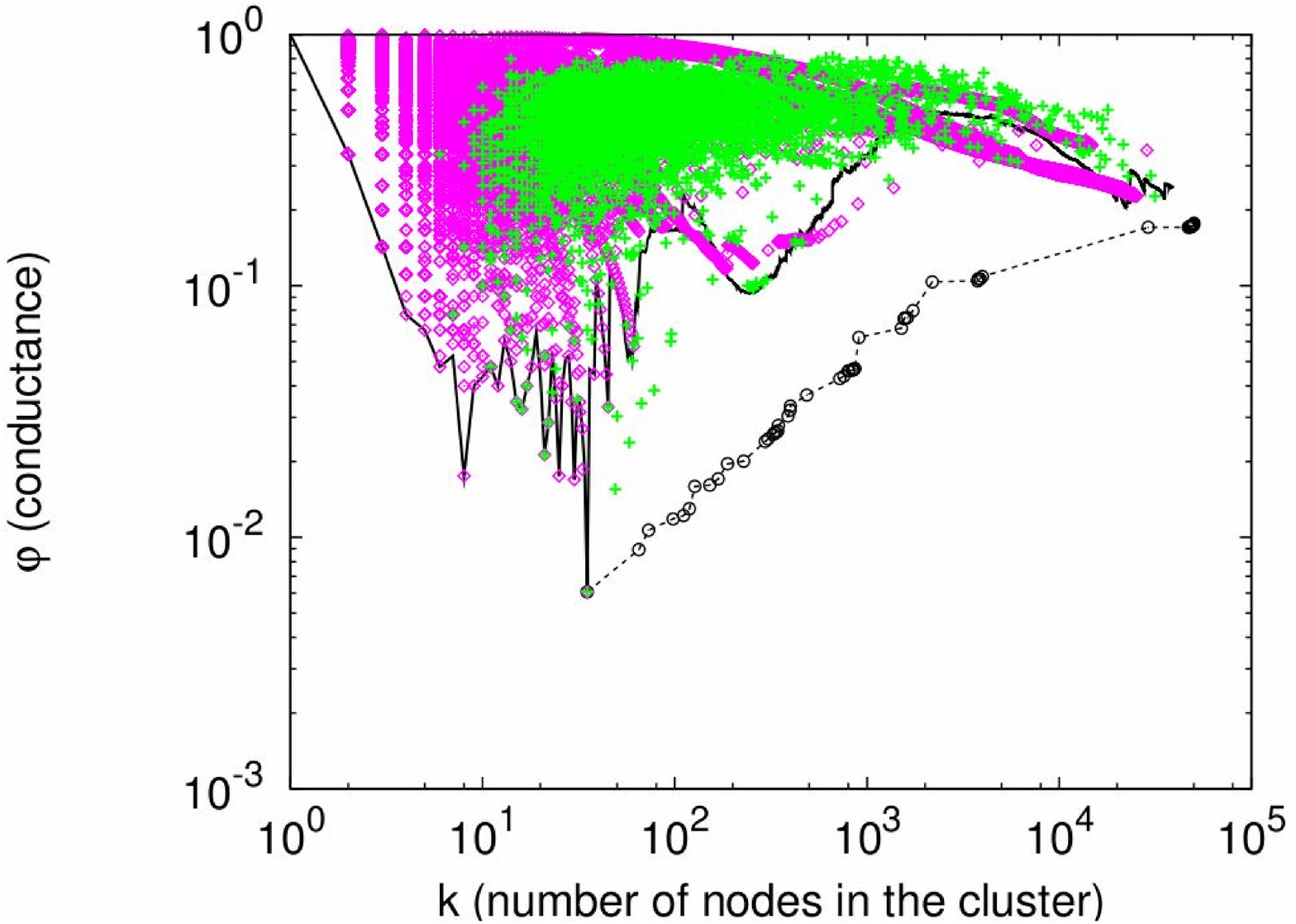}
\includegraphics[width=0.32\linewidth]{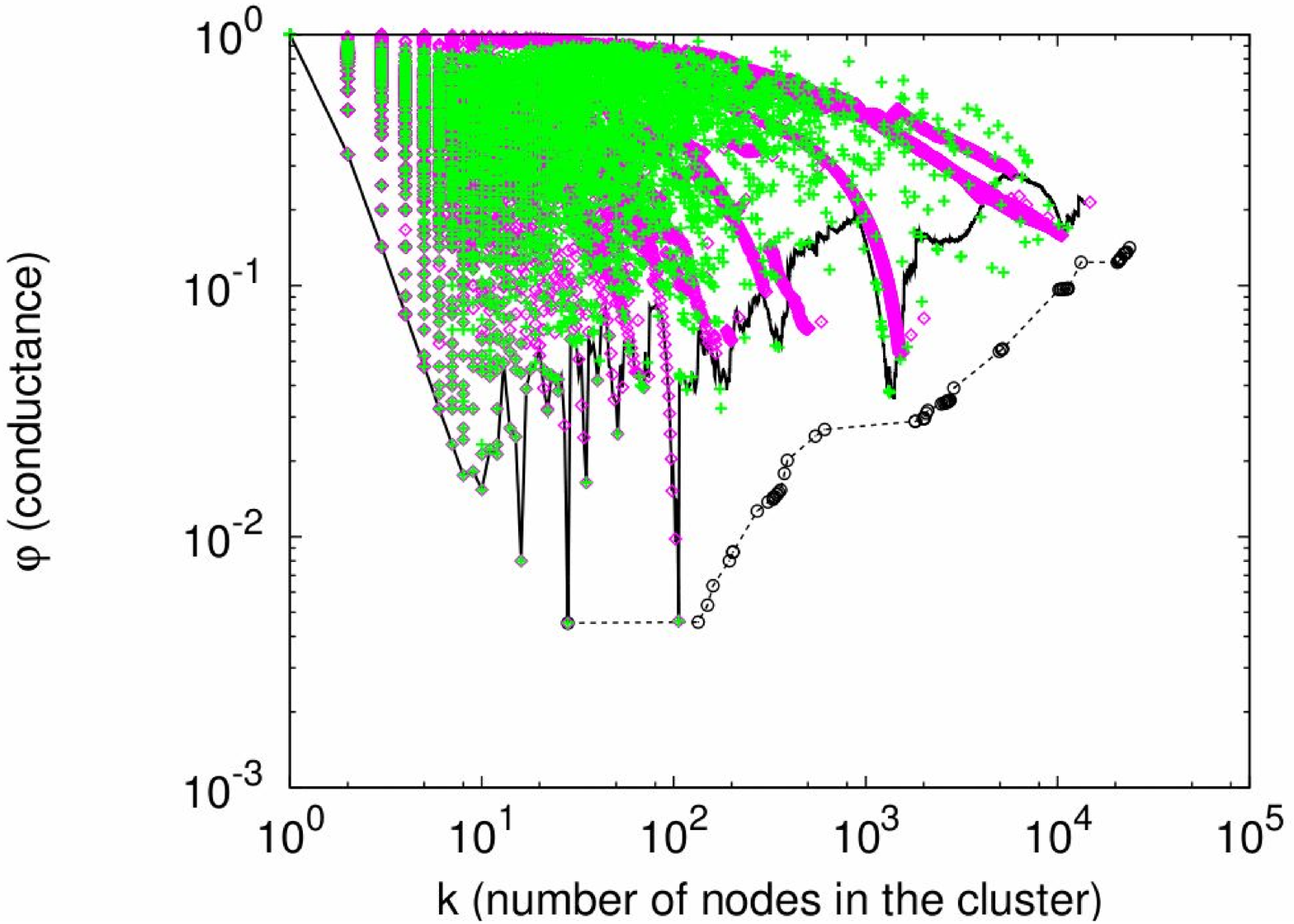}
\includegraphics[width=0.32\linewidth]{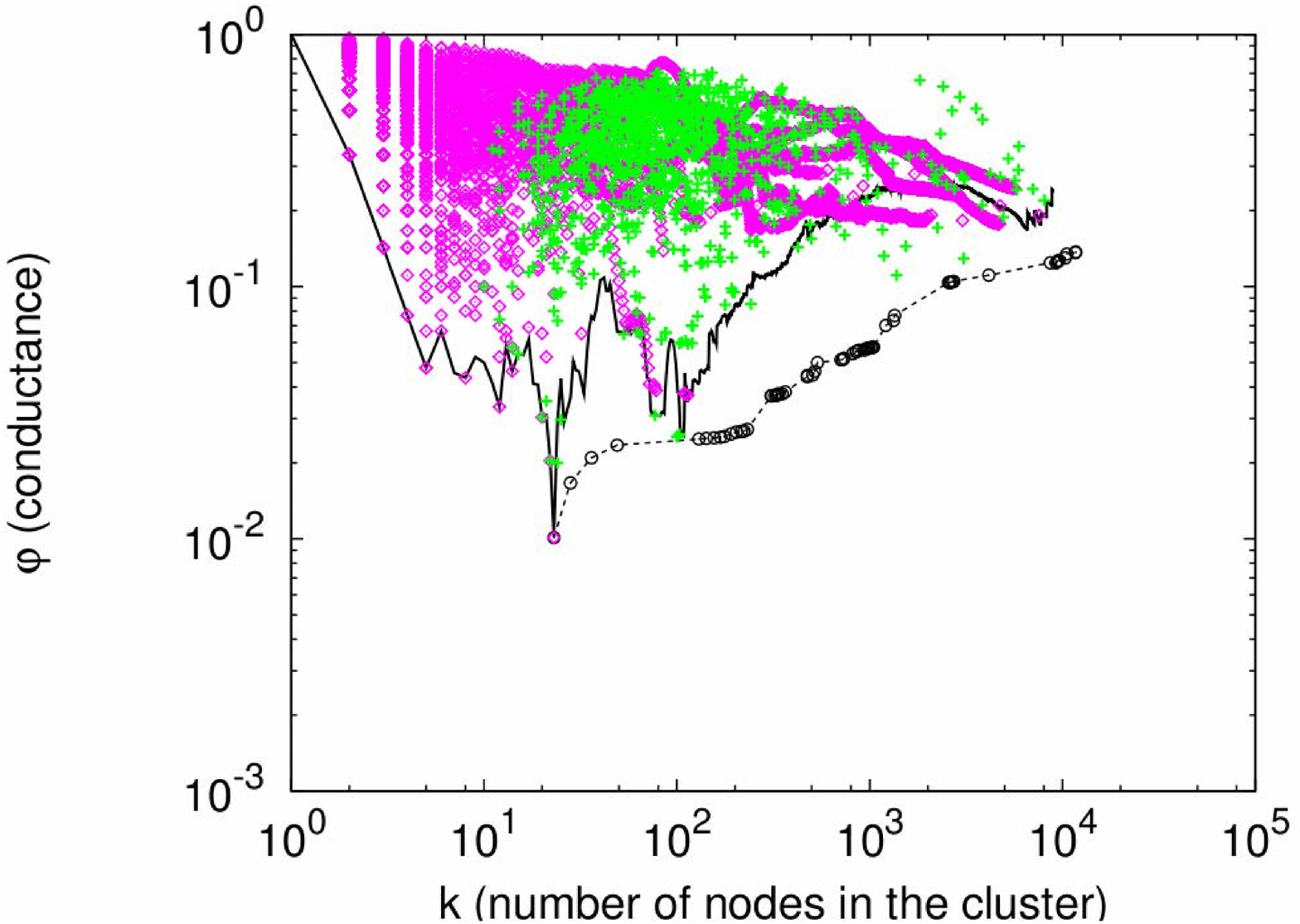}\\
NCP plots obtained by Graclus and Newman's Dendrogram algorithm.\\
\includegraphics[width=0.6\linewidth]{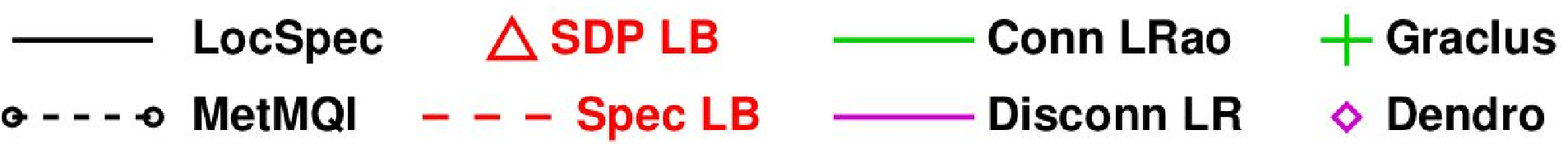}
\end{center}
\caption{
Result of other algorithms for three networks: \net{Epinions},
\net{Email-Enron}, and \net{CA-astro-ph}. Top row plots (in black) conductance
curves as obtained by Local Spectral and Metis+MQI. Top row also shows
lower bounds on conductance of any cut (Spectral lower bound, dashed line)
and the cut separating the graph in half (SDP lower bound, red triangle).
Middle row shows NCP plots for connected (green) and disconnected
(magenta) pieces from our implementation of the Leighton-Rao algorithm. %
Bottom row shows the conductance of some cuts found by Graclus and by
Newman's Dendrogram algorithm.
The overall conclusion is that the qualitative shape of the NCP plots is a 
structural property of large networks and the plot remains practically 
unchanged regardless of what particular community detection algorithm we use.
}
\label{other-algos-fig}
\end{figure}

In the second row of Figure~\ref{other-algos-fig}, we show Leighton-Rao
curves for three example graphs. Our standard Local Spectral and Metis+MQI
curves are drawn in black, while the Leighton-Rao curves for connected and
possibly disconnected sets are drawn in green and magenta respectively. We
note that for small to medium scales, the Leighton-Rao curves for
connected sets resemble the Local Spectral curves, while the Leighton-Rao
curves for possibly disconnected sets resemble the Metis+MQI curves.  This
is big hint about the structure of the sets produced by Local Spectral and
Metis+MQI, that we will discuss further in
Section~\ref{algo-compactness-section}.

At large scales, the Leighton-Rao curves for these example graphs
shoot up and become much worse than our standard curves. This is not
surprising because expander graphs are known to be the worst case
input for the Leighton-Rao approximation guarantee, and we believe
that these graphs contain an expander-like core that is necessarily
encountered at large scales.
We remark that Leighton-Rao does not work poorly at large scales on
every kind of graph. (In fact, for large low-dimensional mesh-like
graphs, Leighton-Rao is a very cheap and effective method for finding
cuts at all scales, while our local spectral method becomes
impractically slow at medium to large scales.  We will not discuss
this point further, except to note that in the main body of the paper we have 
silently substituted Leighton-Rao curves for local spectral curves for the
large road networks and similar graphs.)

We have now covered the main theoretical algorithms that are practical
enough to actually run, which are based on spectral embeddings and on
multi-commodity flow. Starting with~\cite{Arora:2004}, there has been a
recent burst of theoretical activity showing that spectral and flow-based
ideas, which were already known to have complementary strengths and
weaknesses, can in fact be combined to obtain the best ever
approximations. At present none of the resulting algorithms are
sufficiently practical at the sizes that we require, so they were not
included in this study.

Finally, we mention that in addition to the above theoretically-based
practical methods for finding low-conductance cuts, there exist a very
large number of heuristic graph clustering methods. We have tried a number
of them, including Graclus~\cite{dhillon07graclus} and Newman's modularity
optimizing program (we refer to it as
Dendrogram)~\cite{newman02community}. Graclus attempts to find a
partitioning of a graph into pieces bounded by low-conductance cuts using
a kernel k-means algorithm. We ran Graclus repeatedly, asking for
$2,3,\dots,i,\dots,i*\sqrt{2},...$ pieces. Then we measured the size and
conductance of all of the resulting pieces. Newman's Dendrogram program
constructs a recursive partitioning of a graph (that is, a dendrogram)
from the bottom up by repeatedly deleting the surviving edge with the
highest betweenness centrality. A flat partitioning could then be obtained
by cutting at the level which gives the highest modularity score, but
instead of doing that, we measured the size of conductance of every piece
defined by a subtree in the dendrogram.

In the bottom row of Figure~\ref{other-algos-fig}, we present these
results as scatterplots.  Again our two standard curves are drawn in
black. No Graclus or Dendrogram point lies below the Metis+MQI curve. The
lower-envelopes of the points are roughly similar to those produced by
Local Spectral.

Our main point with these experiments is that the lowest points produced
by either Graclus or Dendrogram gradually rise as one moves from small
scales to larger scales, so in principle we could have made the same
observations about the structure of large social and information networks 
by running one of those easily downloadable programs instead of the 
algorithms that we did run.
We chose the algorithms we did due to their speed and power, although they 
may not be as familiar to many readers.

\subsection{Lower bounds on cut conductance}
\label{lower-bounds-section}

As mentioned above, our main arguments are all based on curves which
are actually upper bounds on the true minimum conductance curve. To get a
better idea of how good those upper bounds are, we also compute some lower
bounds. Here we will discuss the spectral lower bound~\cite{Chung:1997} on
the conductance of cuts of arbitrary balance, and we will also discuss a related SDP-based
lower bound~\cite{burer03lowrank} on the conductance of any cut that
divides the graph into two pieces of equal volume.

First, we introduce the following notation: 
$\vec{d}$ is a column vector of the graph's node degrees; 
$D$ is a square matrix whose only nonzero entries are the graph's node 
degrees on the diagonal; 
$A$ is the adjacency matrix of $G$; 
$L=D-A$ is then the non-normalized Laplacian matrix of $G$; 
{\bf 1} is vector of 1's; and 
$A \bullet B = trace (A^T B)$ is the matrix dot-product operator.

Now, consider the following optimization problem (which is well known to be
equivalent to an eigenproblem):
\[
\lambda_G =
  {\rm min}
  \left\{
    \frac{x^T L x}{x^T D x}: x \perp \vec{d}, x \neq 0
  \right\}   .
\]
Let $\hat{x}$ be a vector achieving the minimum value $\lambda_G$. Then
$\frac{\lambda_G}{2}$ is the spectral lower bound on the conductance of
any cut in the graph, regardless of balance, while $\hat{x}$ defines a
spectral embedding of the graph on a line, to which rounding algorithms
can be applied to obtain actual cuts that can serve as upper bounds at
various sizes.

Next, we discuss an SDP-based lower bound on cuts which partition the
graph into two sets of exactly equal volume. Consider:
\[
\mathcal{C}_G =
  {\rm min}
  \left\{
   \frac{1}{4}L\bullet Y: diag(Y) = {\bf 1}, Y\bullet(\vec{d}\,\vec{d}^{\,T}) = 0, Y \succeq 0
  \right\}   ,
\]
and let $\hat{Y}$ be a matrix achieving the minimum value $\mathcal{C}_G$.
Then $\mathcal{C}_G$ is a lower bound on the weight of any cut with
perfect volume balance, and $2 \mathcal{C}_G/{\rm Vol}(G)$ is a lower
bound on the conductance of any cut with perfect volume balance.  We
briefly mention that since $Y \succeq 0$, we can view $Y$ as a Gram matrix
that can be factored can be factored as $RR^T$.  Then the rows of $R$ are
the coordinates of an embedding of the graph on a hypersphere. Again,
rounding algorithms can be applied to the embedding to obtain actual cuts
that can server as upper bounds.

The spectral and SDP embeddings defined here were the basis for the
extensive experiments with global spectral partitioning methods that were
alluded to in Section~\ref{algo-cross-checking-section}. However, in this
section, it is the lower bounds that concern us.
In the top row of Figure~\ref{other-algos-fig}, we present the spectral 
and SDP lower bounds for three example graphs.
The spectral lower bound, which applies to cuts of any balance, is drawn 
as a horizontal line which appears near the bottom of each plot. 
The SDP lower bound, which only applies to cuts separating a specific 
volume, namely ${\rm Vol}(G)/2$, appears as an upwards-pointing triangle 
near the right side of the each plot. 
(Note that plotting this point required us to use volume rather than number 
of nodes for the x-axis of these three plots.)

Clearly, for these graphs, the lower bound at ${\rm Vol}(G)/2$, is higher
than the spectral lower bound which applies at smaller scales. More
importantly, the lower bound at ${\rm Vol}(G)/2$, is higher than our {\em
upper} bounds at many smaller scales, so the true curve must go up, at
least at the very end, as one moves from small to large scales.

Take, for example, the  top left plot of Figure~\ref{other-algos-fig} where in
black we plot the conductance curves obtained by our (Local Spectral and
Metis+MQI) algorithms. With a red dashed line we also plot the lower bound
of best possible cut in the network, and with red triangle we plot the lower bound
for the cut that separates the graph in two equal volume parts. Thus,
the true conductance curve (which is intractable to compute) lies below
black but above red line and red triangle.
This also demonstrates that the conductance curve which starts at upper left corner
of the NCP plot first goes down and reaches the minimum close to the
horizontal dashed line (Spectral lower bound) and then sharply rise and
ends up above the red triangle (SDP lower bound). This verifies that our
conductance curves and obtained NCP plots are not the artifacts of
community detection algorithms we employed.

Finally, in Table~\ref{lower-bound-table} we list for about 40 graphs the
spectral and SDP lower bounds on overall conductance and on volume-bisecting 
conductance, and also the ratio between the two.
It is interesting to see that for these graphs this ratio of lower bounds 
does a fairly good job of discriminating between falling-NCP-plot graphs, 
which have a small ratio, and rising-NCP-plot graphs, which have a large 
ratio.
Small networks (like
\net{CollegeFootball}, \net{ZacharyKarate} and \net{MonksNetwork}) have
downward NCP plot and a small ratio of the SDP and Spectral lower bounds.
On the other hand large networks (\emph{e.g.}, \net{Epinions} or 
\net{Answers-3}) that have downward and then upward NCP plot (as in
Figure~\ref{fig:intro_ncpp}) have large ratio of the two lower bounds.
This is further evidence that small networks have fundamentally different 
community structure from large networks and that one has to examine very large 
networks to observe the gradual absence of communities of size above 
$\approx 100$ nodes.


\begin{table}
{ \footnotesize
\subtable{
\begin{tabular}{|l|r|r|r|}
\hline
                    & Spectral   & SDP              &  ratio   \\
                    & lowerbnd   & lowerbnd         &  of      \\
                    & on $\phi$, & on $\phi$, at    &  lower   \\
            Network & any size.  & ${\rm Vol}(G)/2$ &  bnds    \\
\hline \hline
\net{CollegeFootball}~\cite{newman_netdata}
                    & 0.068402  &  0.091017  &  1.330624  \\
\net{MonksNetwork}~\cite{newman_netdata}
                    & 0.069660  &  0.117117  &  1.681269  \\
\net{ZacharyKarate}~\cite{newman_netdata}
                    & 0.066136  &  0.127625  &  1.929736  \\
\net{PowerGrid}     & 0.000136  &  0.000268  &  1.978484  \\
\net{PoliticalBooks}~\cite{newman_netdata}
                    & 0.018902  &  0.038031  &  2.011991  \\
\net{PoliticalBlogs}~\cite{newman_netdata}
                    & 0.040720  &  0.084052  &  2.064157  \\
\net{RB-Hierarchical}~\cite{ravasz03_hierarchical}
                    & 0.011930  &  0.030335  &  2.542792  \\
\net{Email-InOut}   & 0.038669  &  0.113367  &  2.931752  \\
\net{NetworkScience}~\cite{newman_netdata}
                    & 0.001513  &  0.004502  &  2.974695  \\
\net{As-Oregon}     & 0.012543  &  0.042976  &  3.426417  \\
\net{Blog-nat05-6m} & 0.031604  &  0.108979  &  3.448250  \\
\net{Imdb-India}    & 0.009104  &  0.033318  &  3.659573  \\
\net{Cit-hep-ph}    & 0.007858  &  0.029243  &  3.721553  \\
\net{Bio-Proteins}  & 0.033714  &  0.126137  &  3.741358  \\
\net{As-RouteViews} & 0.018681  &  0.070462  &  3.771821  \\
\net{Gnutella-31}   & 0.029946  &  0.118711  &  3.964127  \\
\net{Imdb-Japan}    & 0.003327  &  0.013396  &  4.026721  \\
\net{Gnutella-30}   & 0.030621  &  0.124929  &  4.079853  \\
\net{DolphinsNetwork}~\cite{newman_netdata}
                    & 0.019762  &  0.103676  &  5.246171  \\
\net{As-Newman}     & 0.009681  &  0.058952  &  6.089191  \\
\net{AtP-gr-qc}     & 0.000846  &  0.006040  &  7.141270  \\
\net{Cit-hep-th}    & 0.009193  &  0.068880  &  7.492522  \\
\net{AtP-cond-mat}  & 0.001703  &  0.013452  &  7.897650  \\
\hline
\end{tabular}
} 
\subtable{
\begin{tabular}{|l|r|r|r|}
\hline
                    & Spectral   & SDP              &  ratio   \\
                    & lowerbnd   & lowerbnd         &  of      \\
                    & on $\phi$, & on $\phi$, at    &  lower   \\
            Network & any size.  & ${\rm Vol}(G)/2$ &  bnds    \\
\hline \hline
\net{Gnutella-25}   & 0.014185  &  0.131032  &  9.237332  \\
\net{Answers-2}     & 0.009660  &  0.107422  &  11.120081  \\
\net{CA-cond-mat}   & 0.003593  &  0.047064  &  13.098027  \\
\net{Answers-1}     & 0.011896  &  0.159251  &  13.386528  \\
\net{Imdb-France}   & 0.003462  &  0.048010  &  13.867591  \\
\net{Answers-5}     & 0.008714  &  0.124703  &  14.311255  \\
\net{Imdb-Mexico}   & 0.003893  &  0.070345  &  18.067513  \\
\net{CA-gr-qc}      & 0.000934  &  0.017421  &  18.659710  \\
\net{AtP-hep-th}    & 0.000514  &  0.009714  &  18.899660  \\
\net{AtP-hep-ph}    & 0.000723  &  0.013770  &  19.040287  \\
\net{Imdb-WGermany} & 0.003025  &  0.065158  &  21.538867  \\
\net{AtP-astro-ph}  & 0.001183  &  0.027256  &  23.036835  \\
\net{CA-hep-th}     & 0.001561  &  0.041125  &  26.350412  \\
\net{CA-astro-ph}   & 0.003143  &  0.086890  &  27.648094  \\
\net{Imdb-UK}       & 0.001283  &  0.036572  &  28.514376  \\
\net{Imdb-Germany}  & 0.000661  &  0.021017  &  31.810460  \\
\net{Blog-nat06all} & 0.002361  &  0.092908  &  39.350874  \\
\net{Imdb-Italy}    & 0.000679  &  0.031954  &  47.077242  \\
\net{Email-Enron}   & 0.001763  &  0.089876  &  50.965424  \\
\net{CA-hep-ph}     & 0.000889  &  0.052249  &  58.755927  \\
\net{Epinions}      & 0.002395  &  0.150242  &  62.739252  \\
\net{Answers-3}     & 0.002636  &  0.185340  &  70.306807  \\
\net{Imdb-Spain}    & 0.000562  &  0.046327  &  82.397702  \\
\hline
\end{tabular}
} 
} 
\caption{
Lower bounds on the conductance for our network datasets.
Recall that the spectral lower bound applies to any cut, while the SDP lower
bound applies to cuts at a specified volume fraction, taken here to be half.
See the top row of Figure~\ref{other-algos-fig} for plots for three of these
networks.
}
\label{lower-bound-table}
\end{table}

\subsection{Local Spectral and Metis+MQI}
\label{algo-compactness-section}

In this section we discuss our rationale for using Local Spectral in 
addition to Metis+MQI as one of our two main algorithms for finding sets 
bounded by low conductance cuts. This choice requires some justification 
because the NCP plots are intended to show the tightest possible upper 
bound on the lowest conductance cut for each piece size, while the 
curve for Local Spectral is generally above that for Metis+MQI.

Our reason for using Local Spectral in addition to Metis+MQI is that Local 
Spectral returns pieces that are internally ``nicer''.  
For graphs with a rising NCP plot, we have
found that many of the low conductance sets returned by Metis+MQI (or
Leighton-Rao, or the Bag-of-Whiskers Heuristic) are actually {\em
disconnected}.  Since internally disconnected sets are not very satisfying
``communities'', it is natural to wonder about NCP plot-style curves with
the additional requirement that pieces must be internally well connected.
In Section~\ref{algo-cross-checking-section}, we generated such a curve
using Leighton-Rao, and found that the curve corresponding to connected
pieces was higher than a curve allowing disconnected sets.

\begin{figure}[t]
\begin{center}
\includegraphics[width=0.32\linewidth]{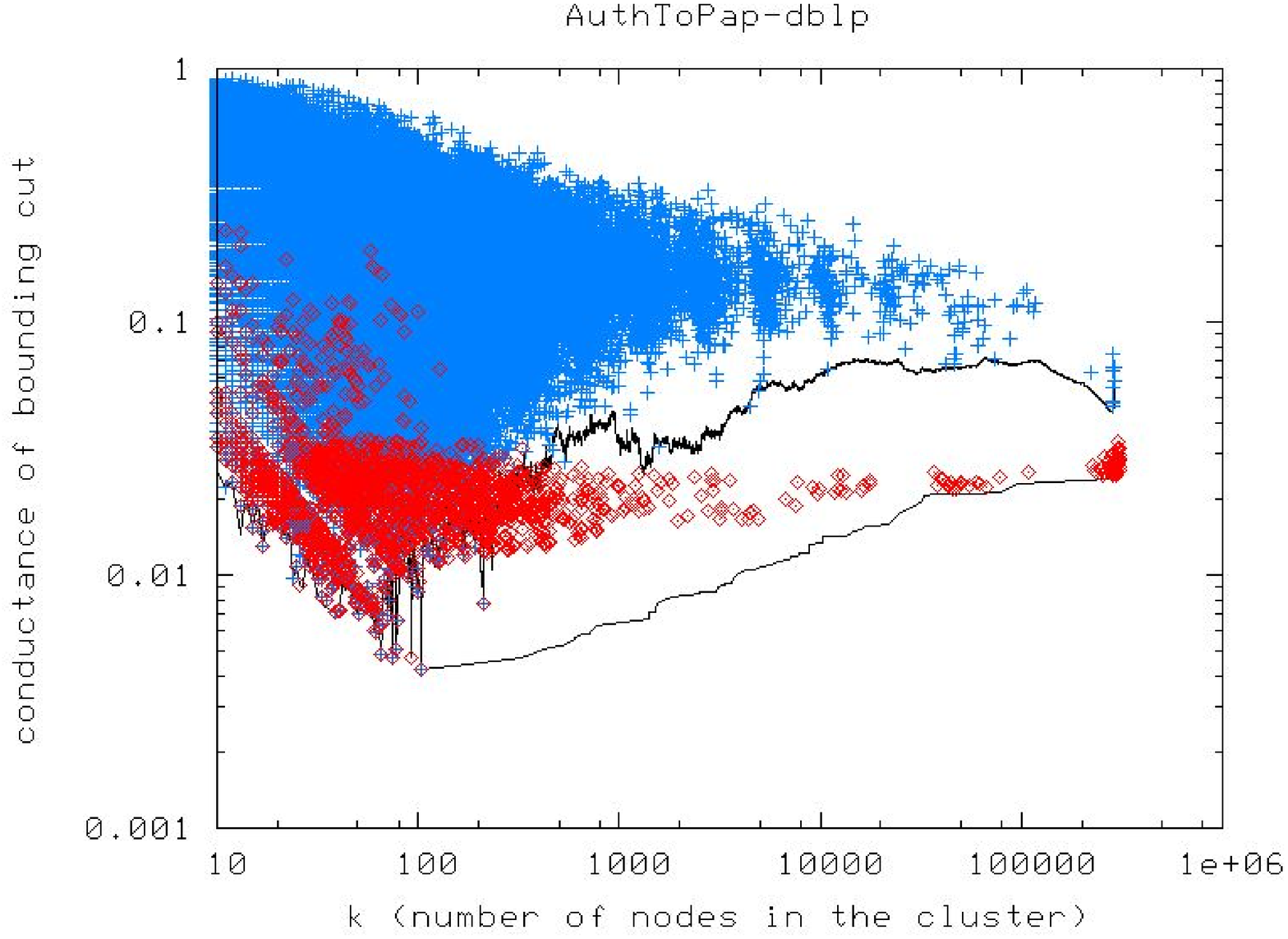}
\includegraphics[width=0.32\linewidth]{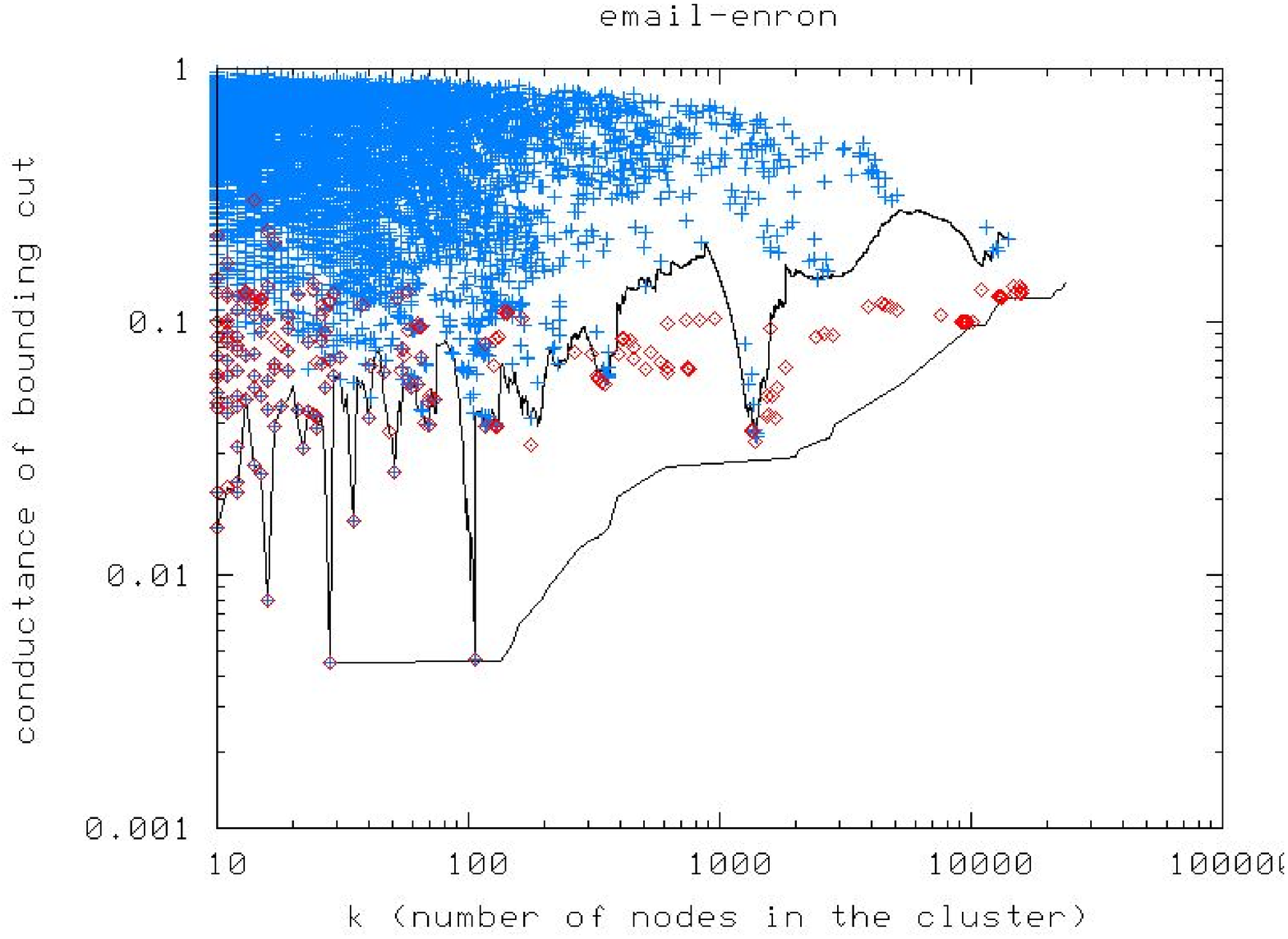}
\includegraphics[width=0.32\linewidth]{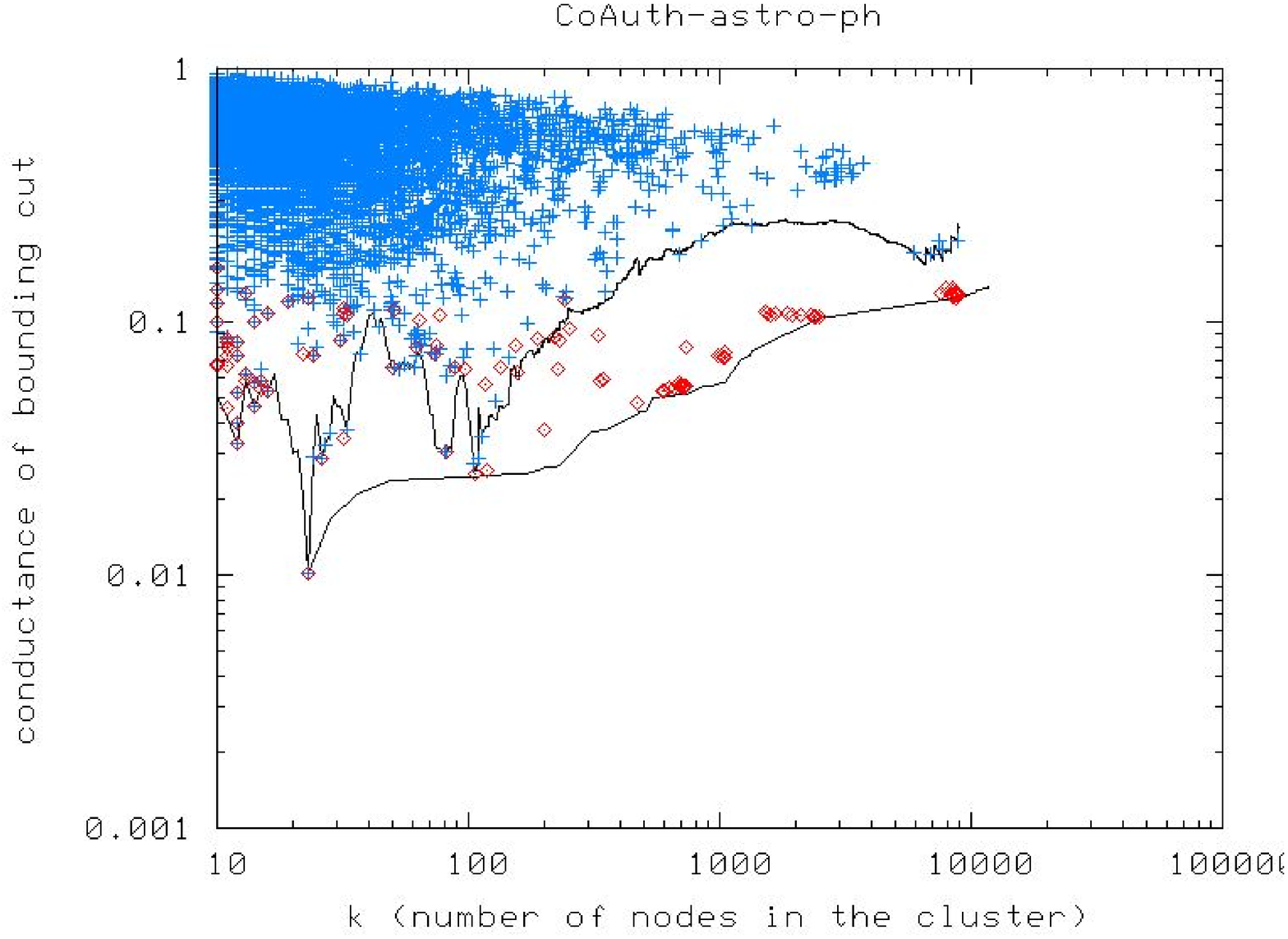}\\
Conductance of connected clusters found by Local Spectral (blue) and Metis+MQI (red)\\
\includegraphics[width=0.32\linewidth]{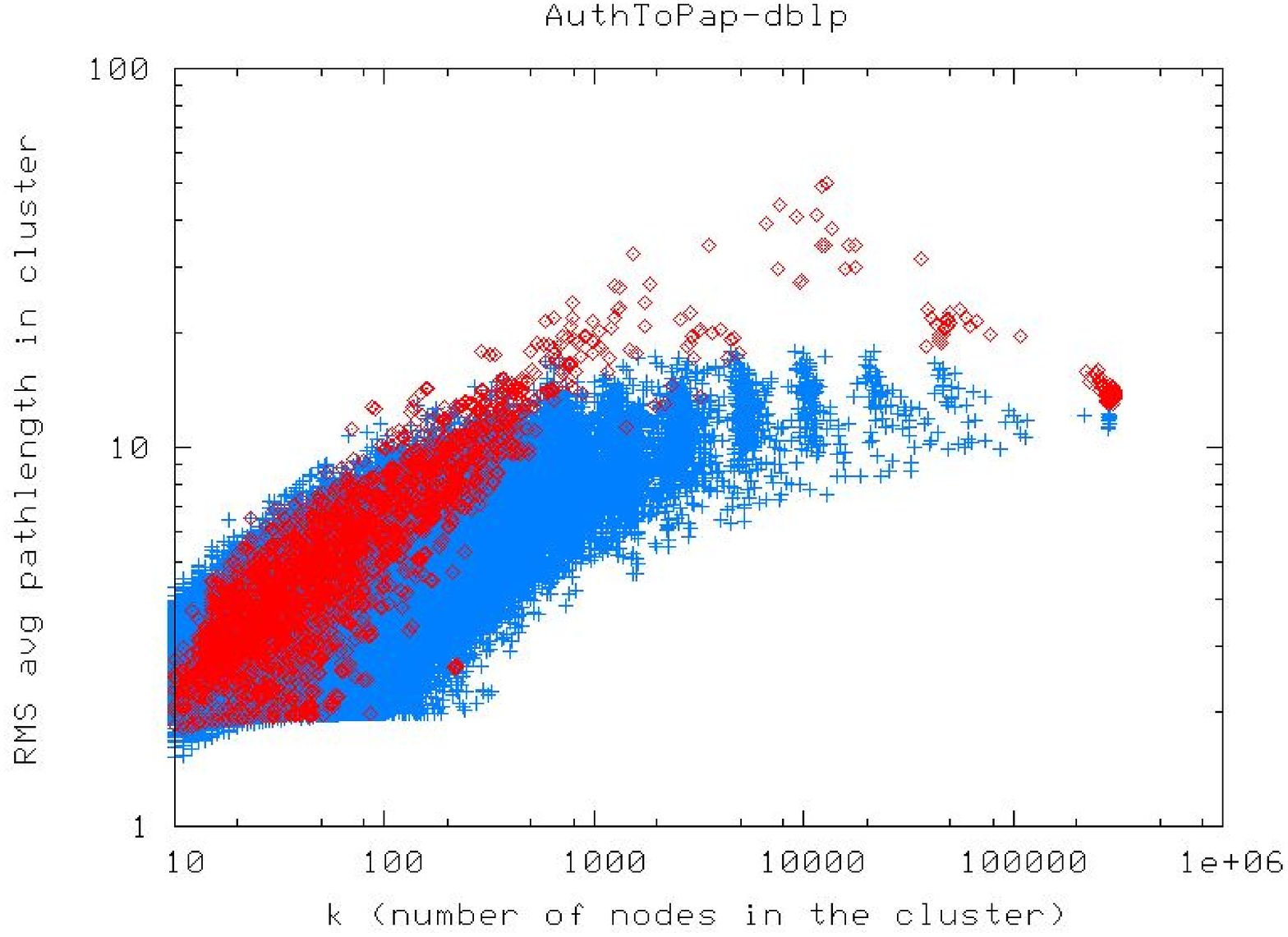}
\includegraphics[width=0.32\linewidth]{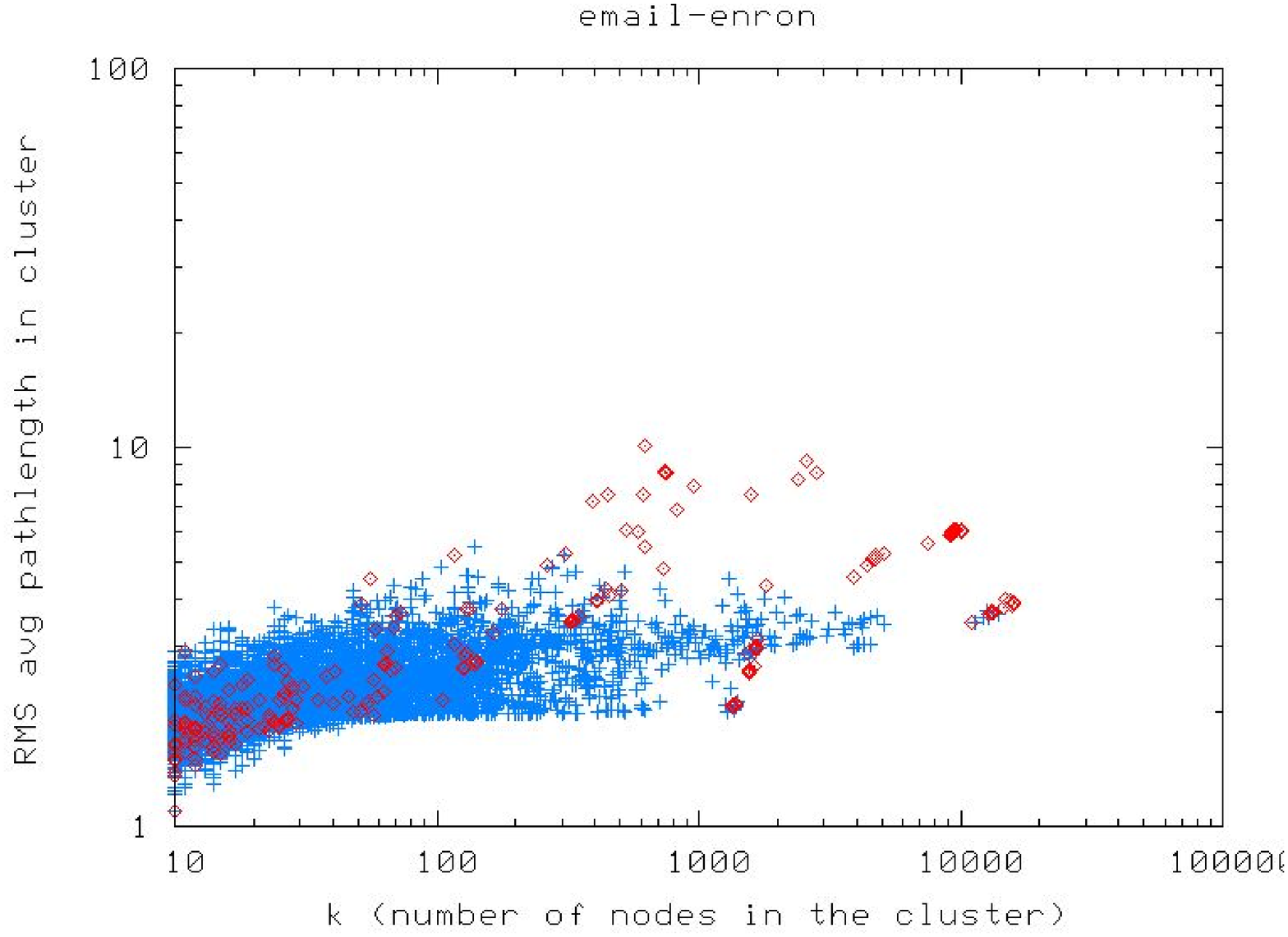}
\includegraphics[width=0.32\linewidth]{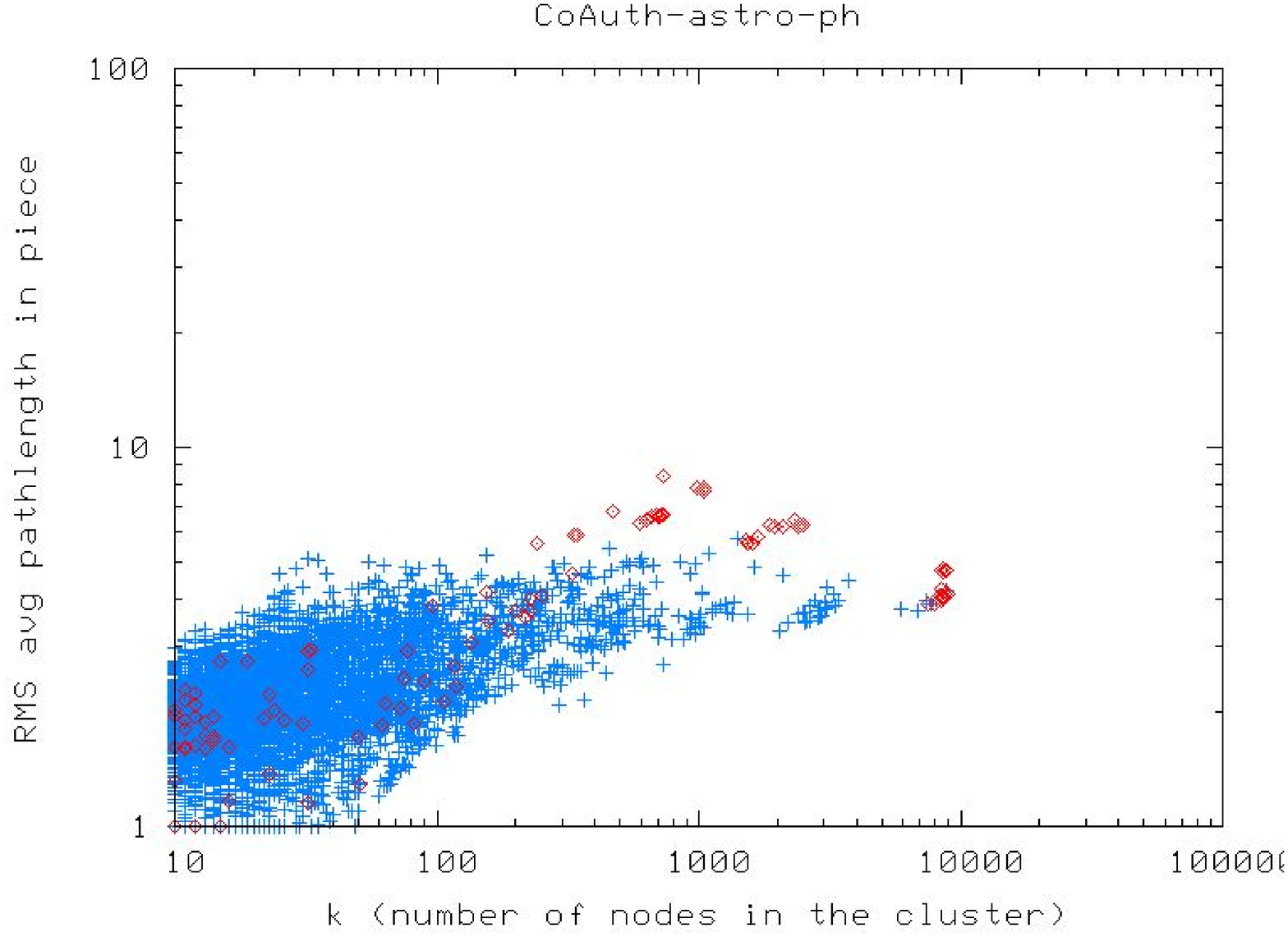}\\
Cluster compactness: average shortest path length\\
\includegraphics[width=0.32\linewidth]{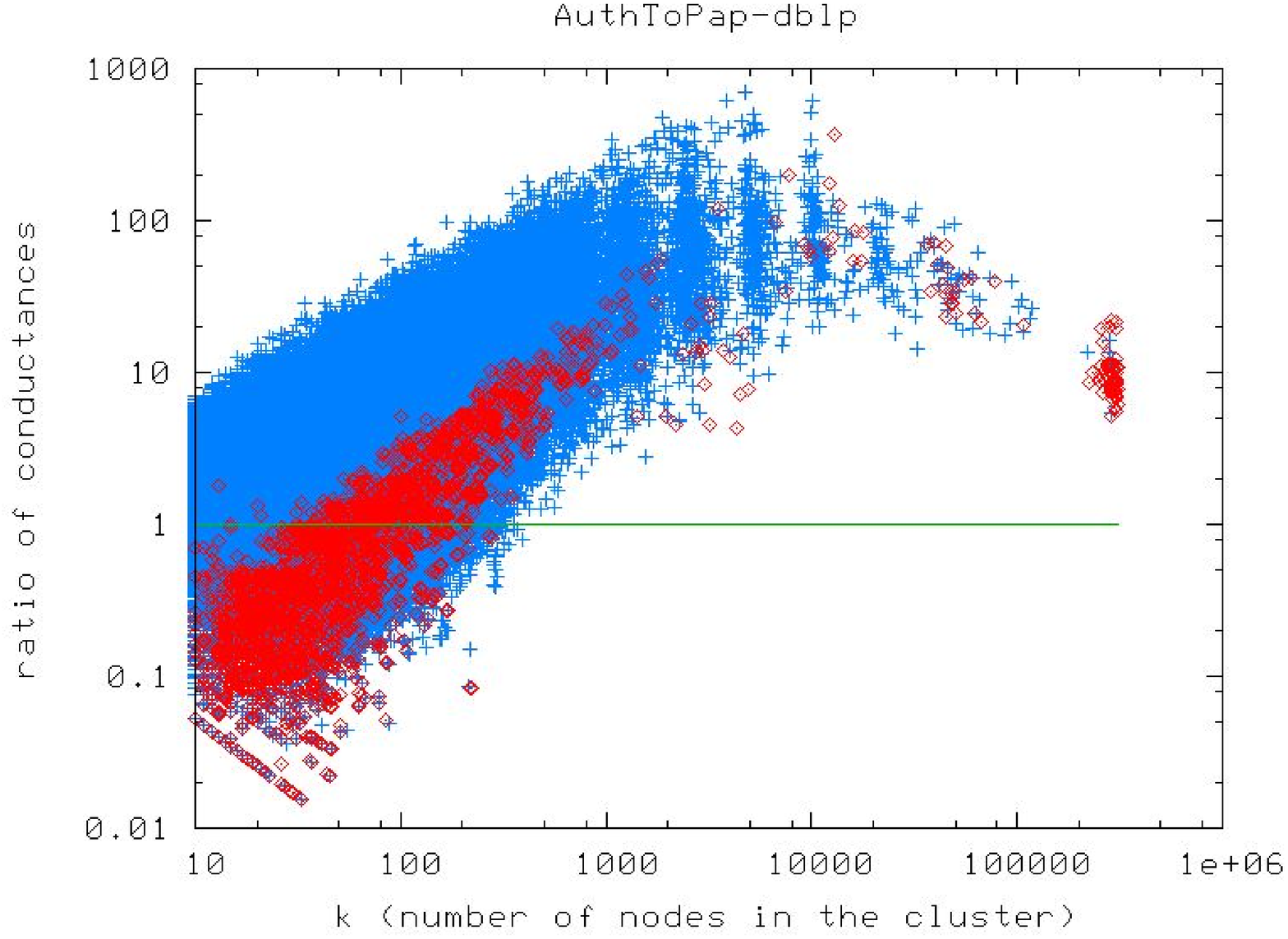}
\includegraphics[width=0.32\linewidth]{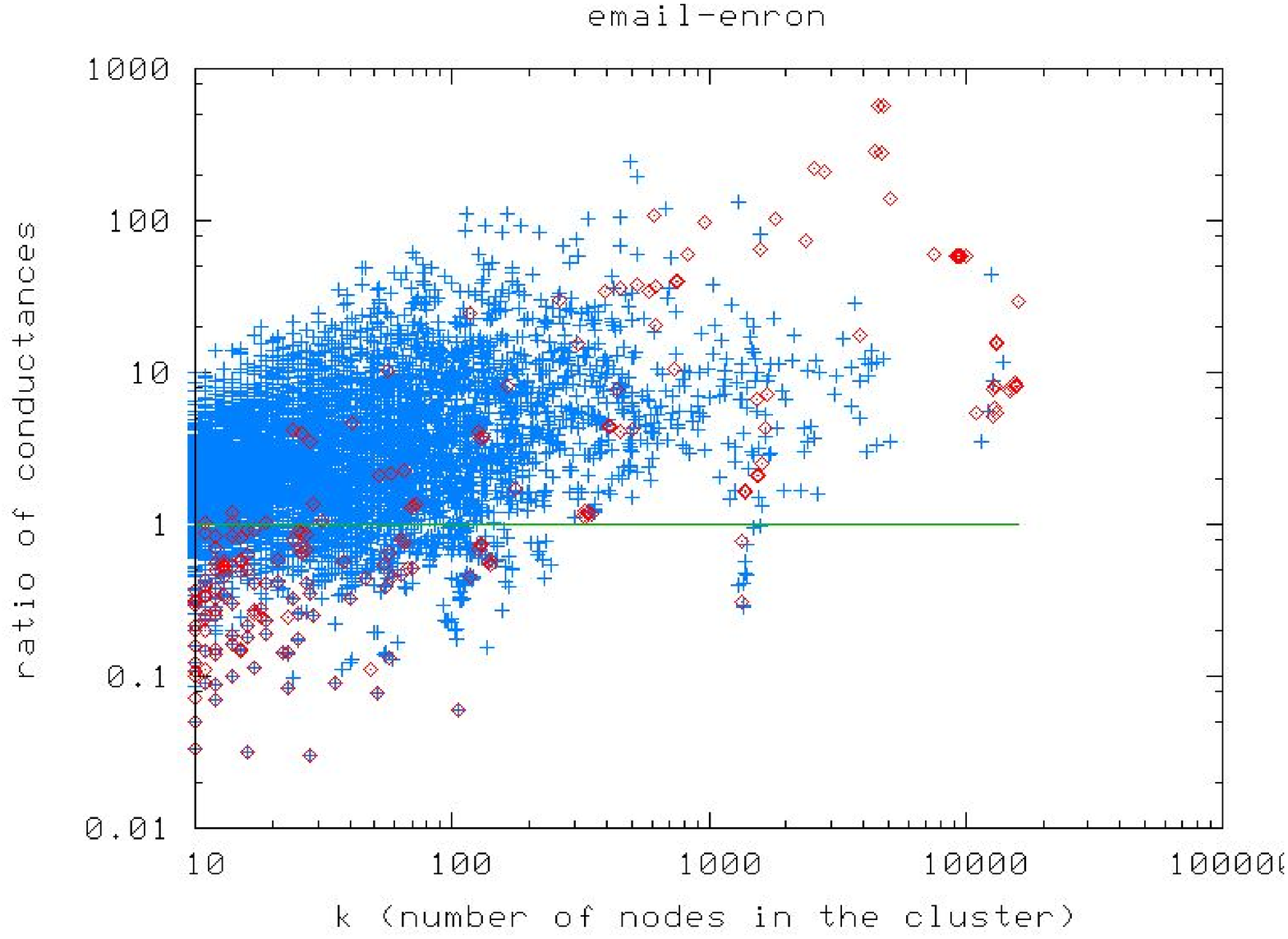}
\includegraphics[width=0.32\linewidth]{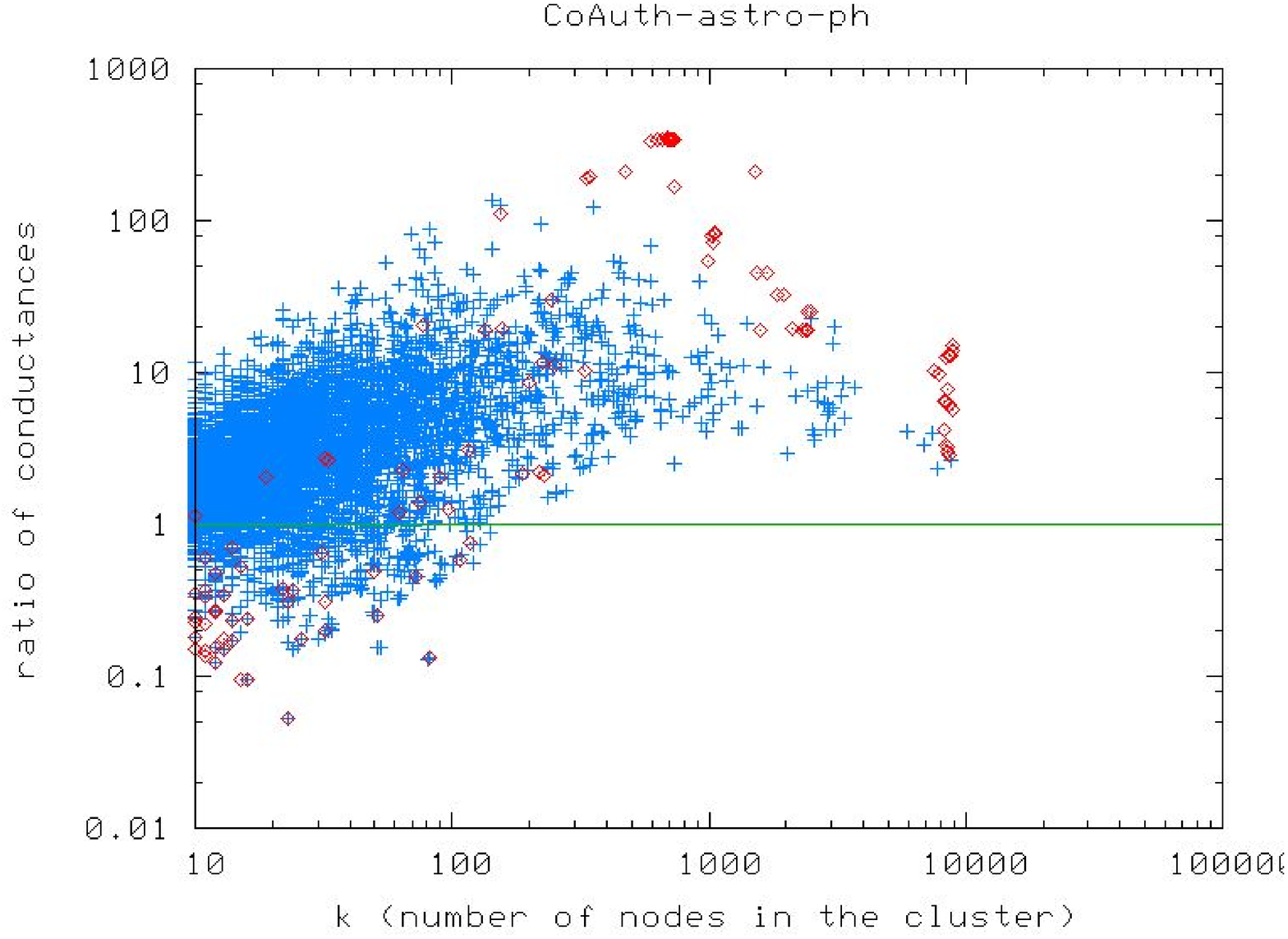}\\
Cluster compactness: external vs. internal conductance\\
\end{center}
\caption{
Result of comparing Local Spectral (blue) and Metis+MQI (red) on connected
clusters for three
networks: \net{AtP-DBLP}, \net{Email-Enron}, and \net{CA-astro-ph}.
In the top row, we plot the conductance of the bounding cut.
In the middle row, we plot the average shortest path length in the
cluster.
In the bottom row, we plot the ratio of the external conductance to the
internal conductance.
Observe that generally Metis+MQI yields better (lower conductance) cuts
while Local Spectral yields pieces that are more
compact: they have shorter path lengths and internal connectivity.
}
\label{compactness-vs-cuts-fig}
\end{figure}

In the top row of Figure~\ref{compactness-vs-cuts-fig}, we show scatter
plots illustrating a similar comparison between the conductance of the
cuts bounding connected pieces generated by Local Spectral and by
Metis+MQI. Our method for getting connected pieces from Metis+MQI here is
simply to separately measure each of the pieces in a disconnected set. The
blue points in the figures show the conductance of some cuts found by
Local Spectral. The red points show the conductance of some cuts found by
Metis+MQI. Apparently, Local Spectral and Metis+MQI find similar pieces at
very small scales, but at slightly larger scales a gap opens up between
the red cloud and the blue cloud. In other words, at those scales
Metis+MQI is finding lower conductance cuts than Local Spectral, even when
the pieces must be internally connected.

However, there is still a measurable sense in which the Local Spectral
pieces are ``nicer'' and more ``compact,'' as shown in the second row of
scatter plots in Figure~\ref{compactness-vs-cuts-fig}.  For each of the
same pieces for which we plotted a conductance in the top row, we are now
plotting the average shortest path length between random node pairs in that
piece. In these plots, we see that in the same size range where Metis+MQI
is generating clearly lower conductance connected sets, we now see that
Local Spectral is generating pieces with clearly shorter internal paths. In other
words, the Local Spectral pieces are more ``compact''.

\begin{figure}[t]
\begin{center}
\includegraphics[width=0.24\linewidth]{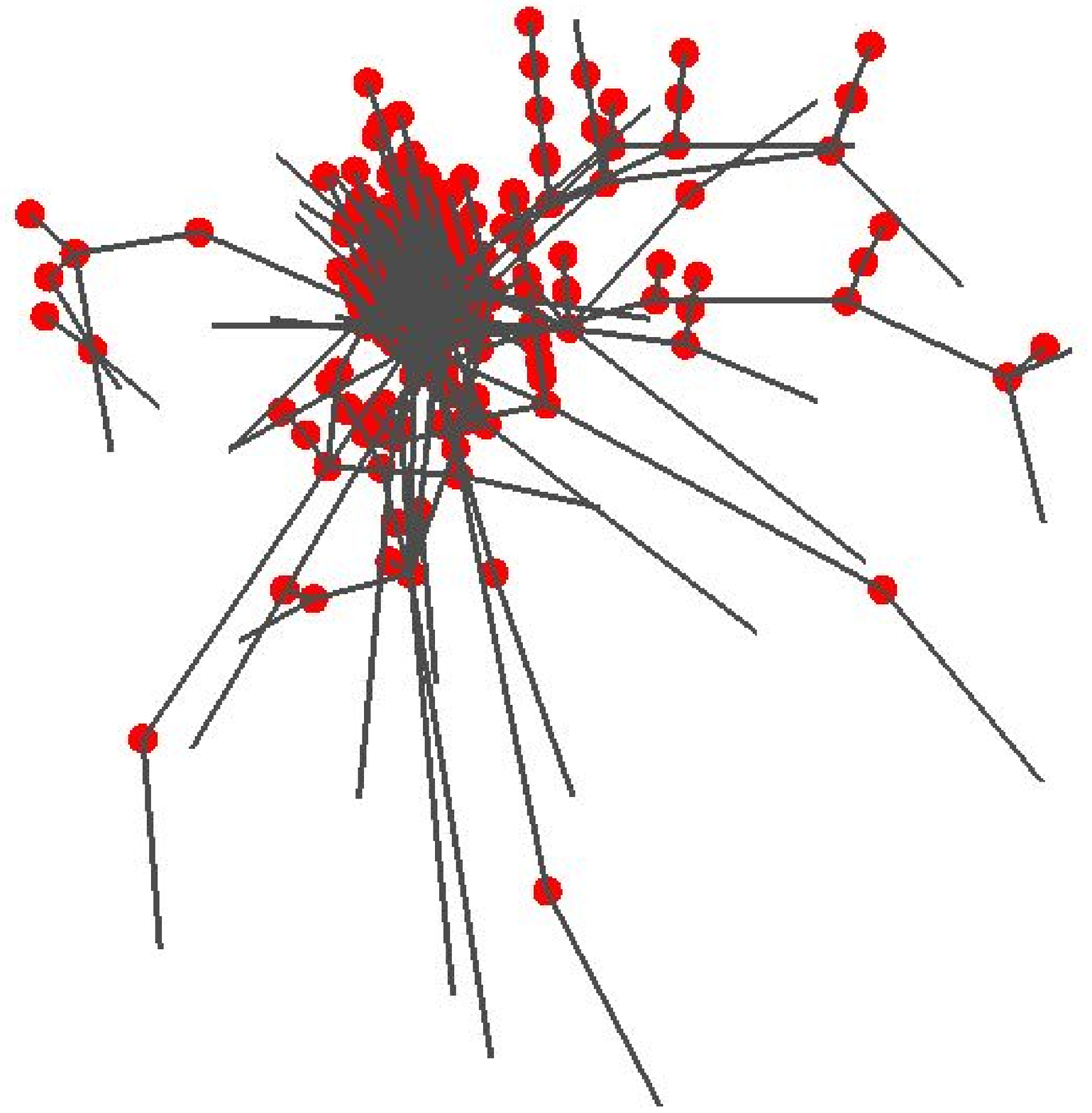}
\includegraphics[width=0.24\linewidth]{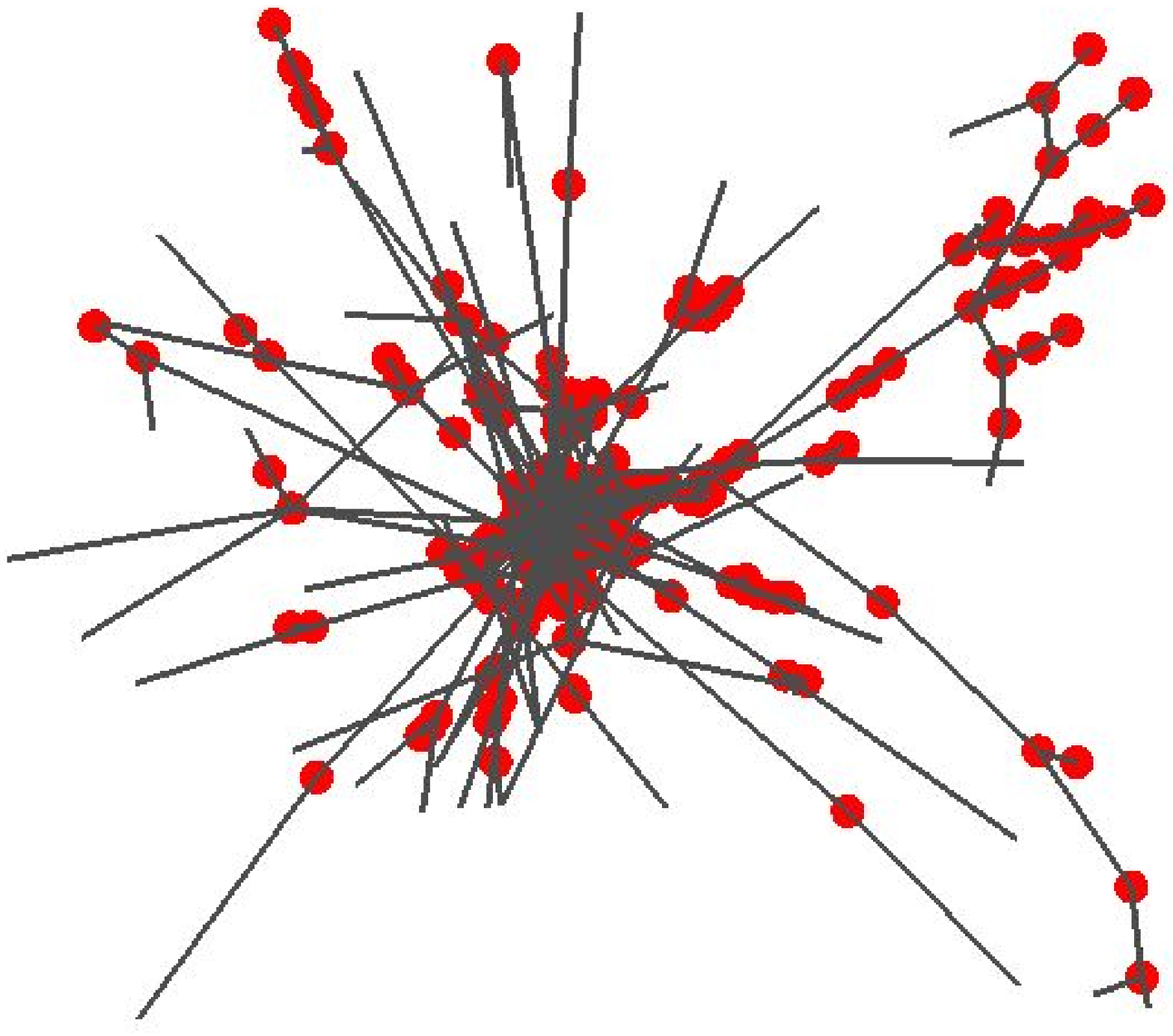}
\includegraphics[width=0.24\linewidth]{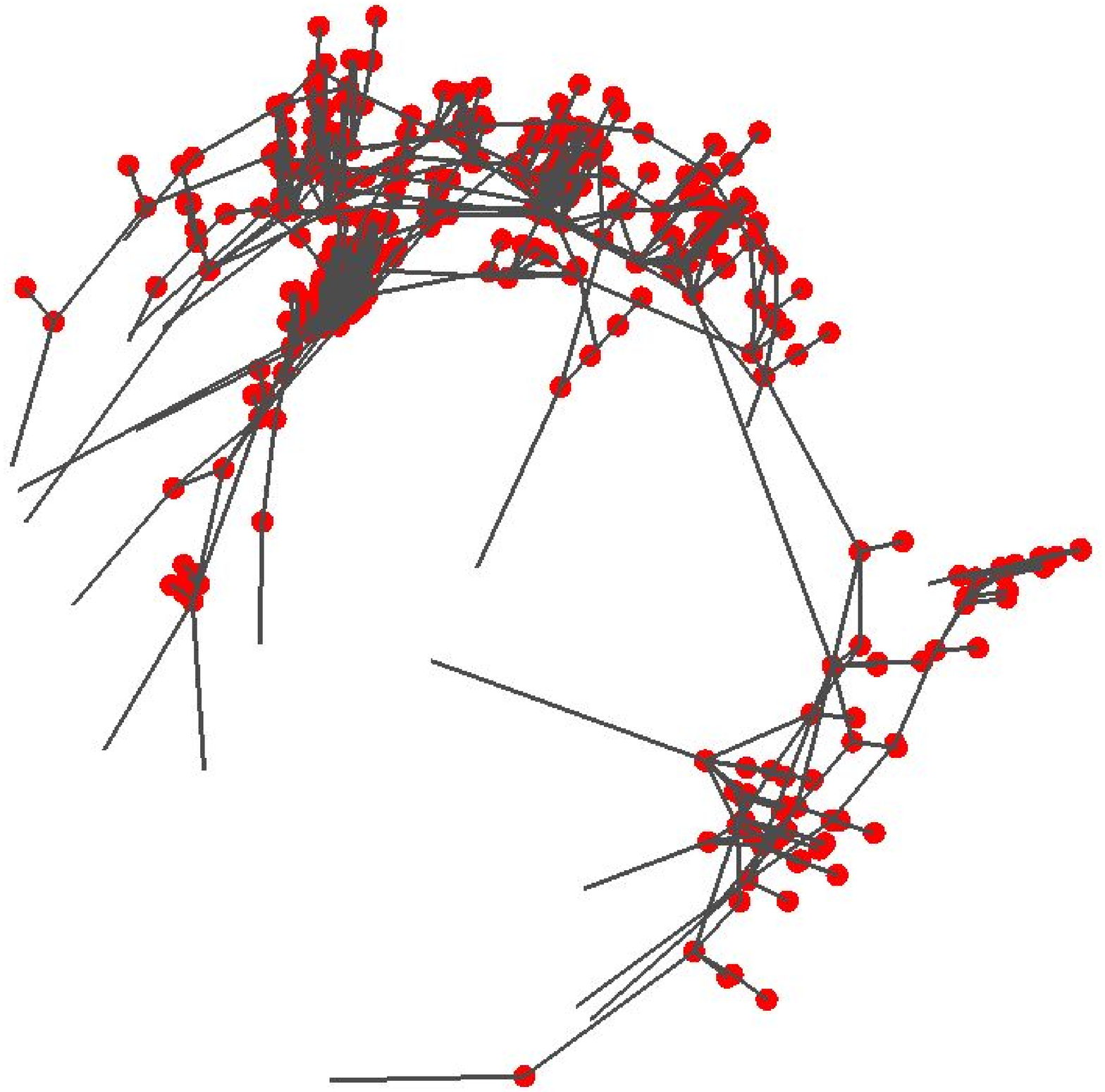}
\includegraphics[width=0.24\linewidth]{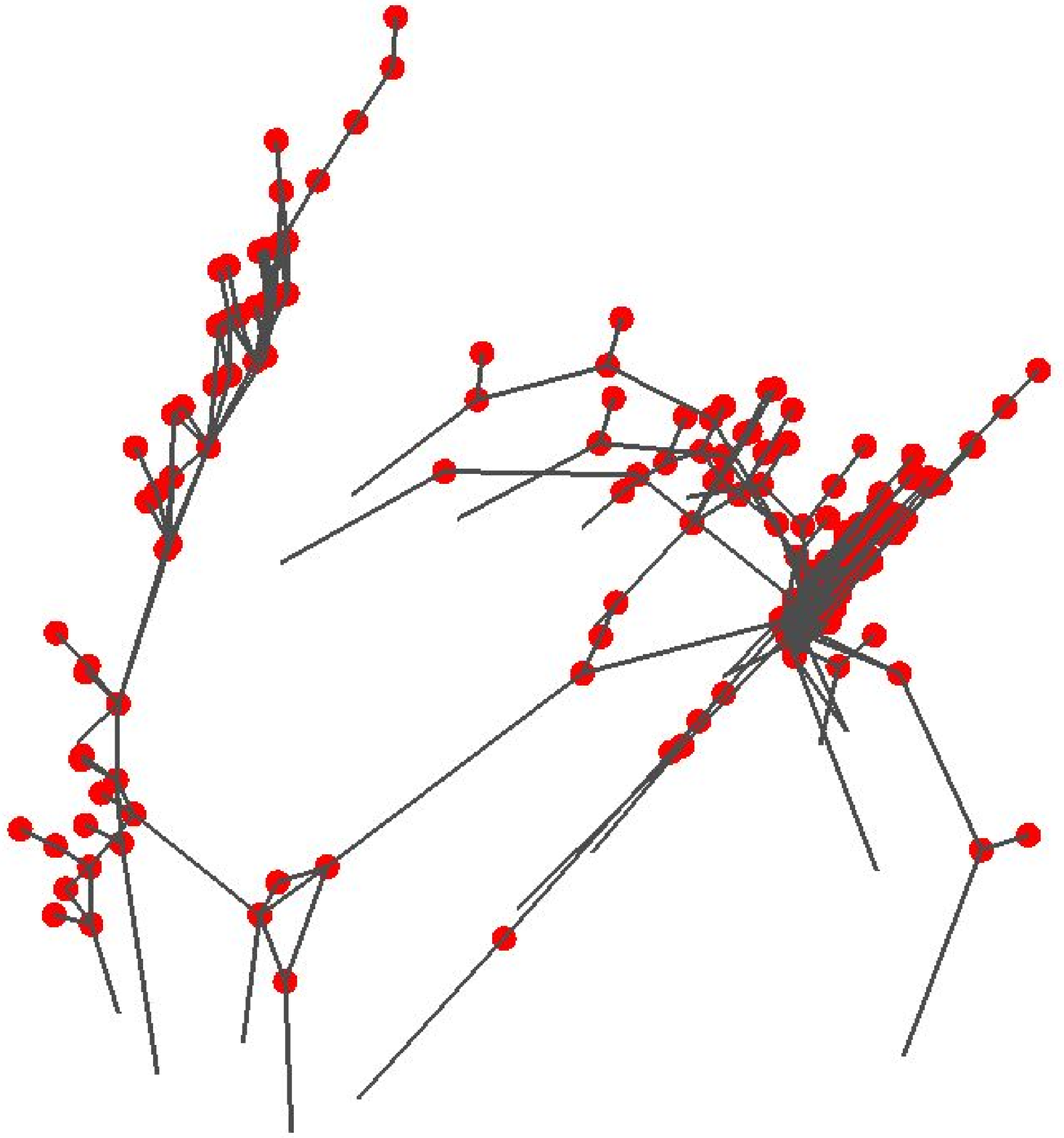}
\end{center}
\caption{
Two examples of ``communities'' found by the LocalSpectral algorithm (on the 
left) and two from the Metis+MQI algorithm (on the right).
Note that the Local Spectral ``communities'' are more compact---they are 
tighter and have smaller diameter since the algorithm has difficulty pushing 
probability mass down long extended paths---while the Metis+MQI ``communities'' 
are more sprawling---they have larger diameter and more diverse internal 
structure, but better conductance scores.
In both cases, we have shown communities with ca. $500$ nodes (many of which
overlap at resolution of this figure), \emph{i.e.}, just above the ``whisker'' 
size scale.
}
\label{sprawl-plots}
\end{figure}

Last, in Figure~\ref{sprawl-plots}, we further illustrate this point with
drawings of some example subgraphs. The two subgraphs shown on the left of
Figure~\ref{sprawl-plots} were found by Local Spectral, while the two
subgraphs shown on the right of Figure~\ref{sprawl-plots} were found by
Metis+MQI.  Clearly, these two pairs of subgraphs have a qualitatively
different appearance, with the Metis+MQI pieces looking longer and
stringier than the Local Spectral pieces. All of these subgraphs contain
roughly 500 nodes, which is a bit more than the natural cluster size for
that graph, and thus the differences between the algorithms start to show
up. In these cases, Local Spectral has grown a cluster out a bit past its
natural boundaries (thus the spokes), while Metis+MQI has strung together
a couple of different sparsely connected clusters.
(We remark that the tendency of Local Spectral to trade off cut quality in
favor of piece compactness isn't just an empirical observation, it is a
well understood consequence of the theoretical analysis of spectral
partitioning methods.)

Finally, in the bottom row of Figure~\ref{compactness-vs-cuts-fig} we
briefly introduce the topic of internal vs. external cuts, which is
something that none of our algorithms are explicitly trying to optimize.
These are again scatter plots showing the same set of Local Spectral and
Metis+MQI pieces as before, but now the y-axis is external conductance
divided by internal conductance. External conductance is the quantity that
we usually plot, namely the conductance of the cut which separates the
piece from the graph.  Internal conductance is the score of a low
conductance cut {\em inside} the piece (that is, in the induced subgraph
on the piece's nodes). Intuitively, good communities should have small
ratios, ideally below 1.0, which would mean that they are well separated
from the rest of the network, but that they are internally well-connected.
However, the three bottom-row plots show that for these three
sample graphs, there are mostly no ratios well below 1.0 except at small 
sizes.
(Of course, any given graph could happen to contain a very distinct piece of 
any size, and the roughly thousand-node piece in the \net{Email-Enron} 
network is a good example.)

This demonstrates another aspect of our findings: small communities of size
below $\approx 100$ nodes are internally compact and well separated from the
remainder of the network, whereas larger pieces are so hard to separate that 
separating them from the network is more expensive than separating them 
internally.
\section{Models for network community structure}
\label{sxn:models}

In this section, we use results from previous sections to devise a model
that explains the shape of NCP plots. In
Section~\ref{sxn:models:common_models}, we examine the NCP plot for a wide
range of existing commonly-used network generation models, and we see that
none of them reproduces the observed properties, at even a qualitative
level. Then, in Section~\ref{sxn:models:sparse_Gw}, we analytically
demonstrate that certain aspects of the NCP plot, {\em e.g.}, the
existence of deep cuts at small size scales, can be explained by very
sparse random graph models. Then, in Section~\ref{sxn:models:toy}, we
present a simple toy model to develop intuition about the effect we must
reproduce with a realistic generative model. Finally, in
Section~\ref{sxn:models:ff}, we will combine these and other ideas to
describe a Forest Fire graph generation model that reproduces quite well
our main observations.

\subsection{Community profile plots for commonly-used network generation models}
\label{sxn:models:common_models}

\begin{figure}
   \begin{center}
   \begin{tabular}{cc}
      \subfigure[Preferential attachment]{
         \includegraphics[width=0.40\textwidth]{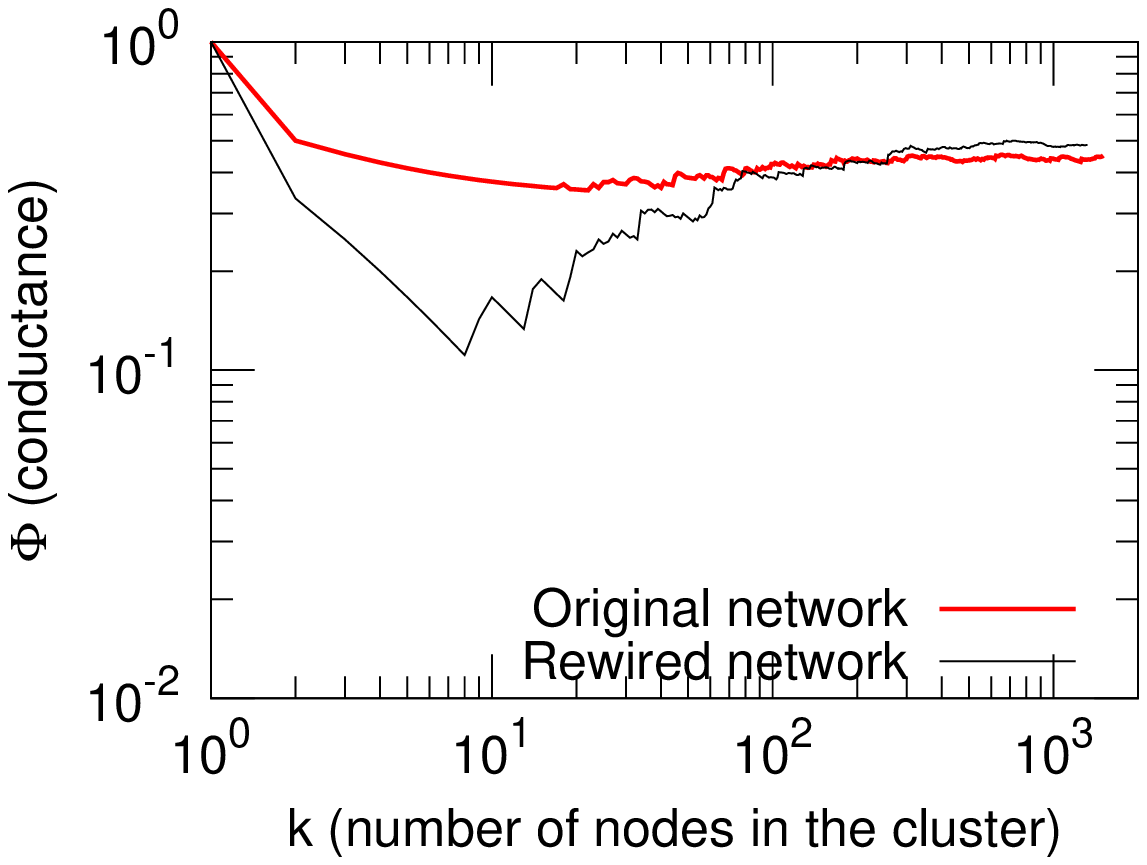}
         \label{fig:phiModels:PA}
      } &
      \subfigure[Copying model]{
         \includegraphics[width=0.40\textwidth]{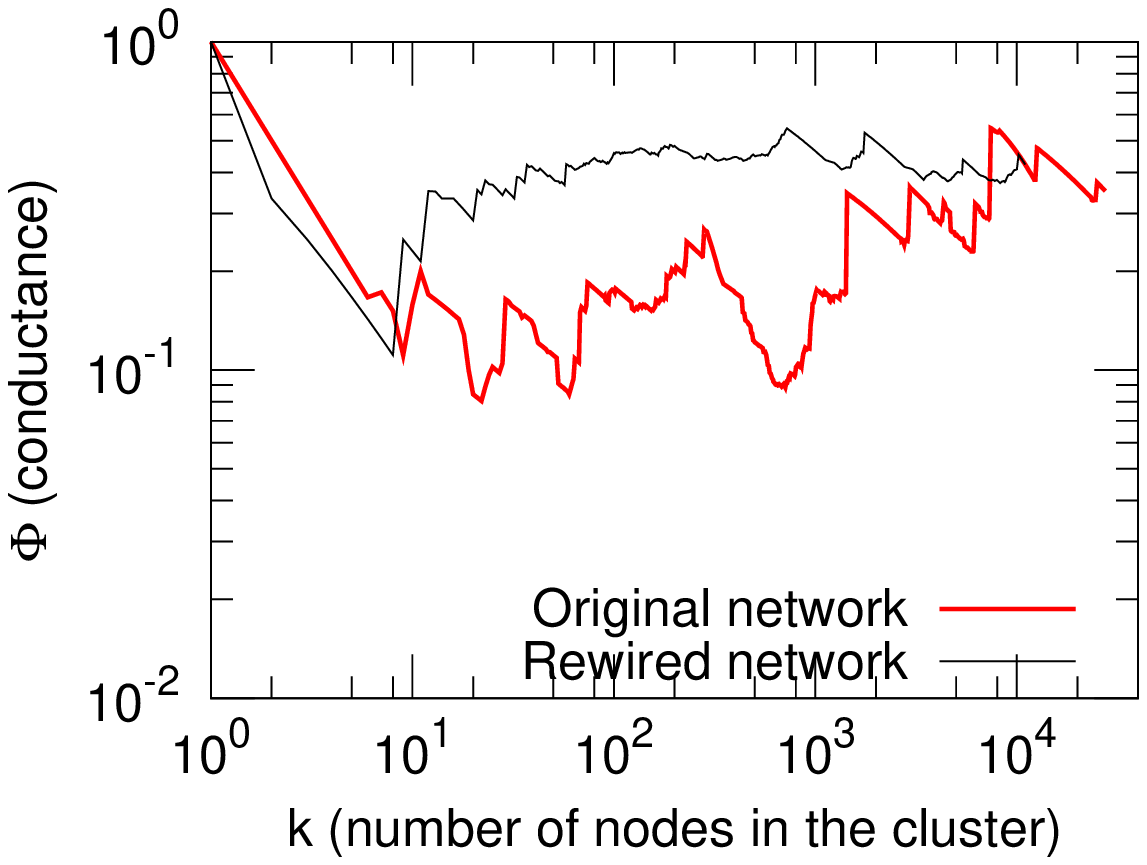}
         \label{fig:phiModels:Copy}
      } \\
      \subfigure[Barabasi Hierarchical]{
         \includegraphics[width=0.40\textwidth]{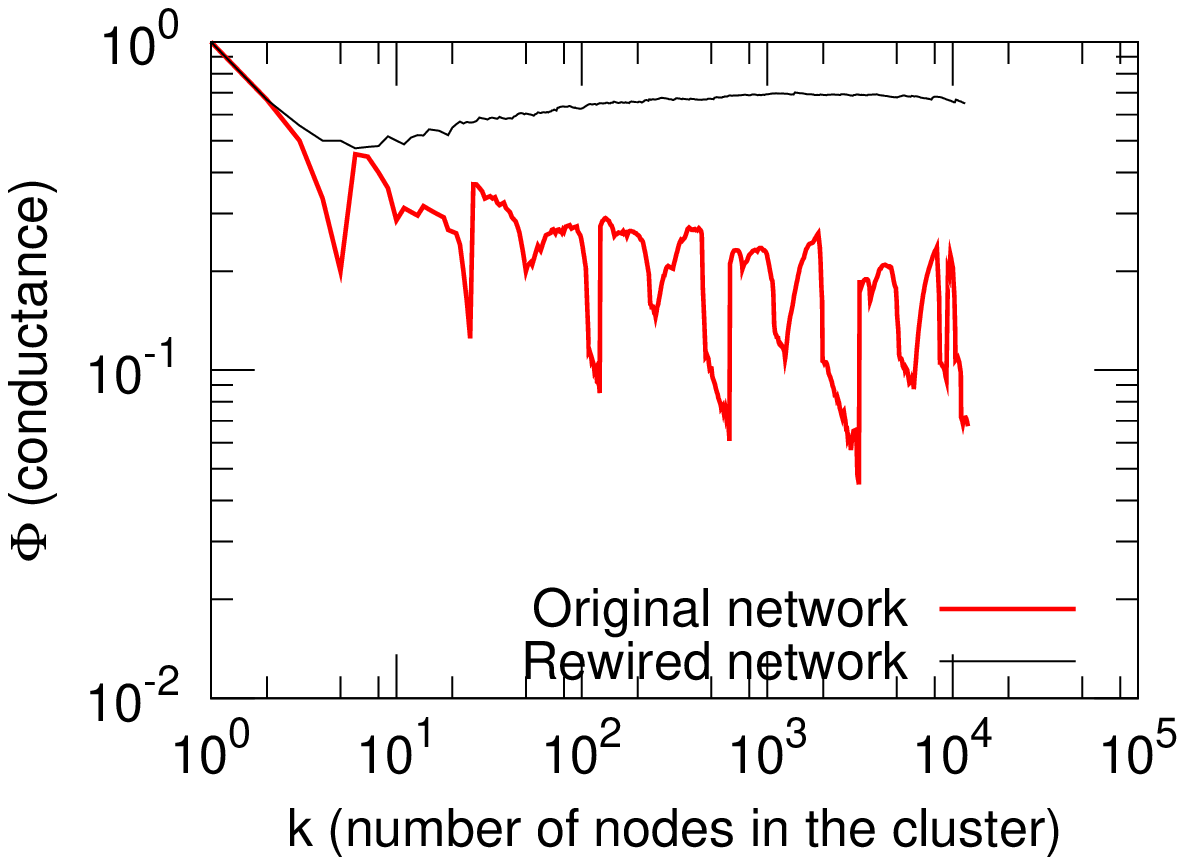}
         \label{fig:phiModels:BH}
      } &
      \subfigure[Community guided attachment]{
         \includegraphics[width=0.40\textwidth]{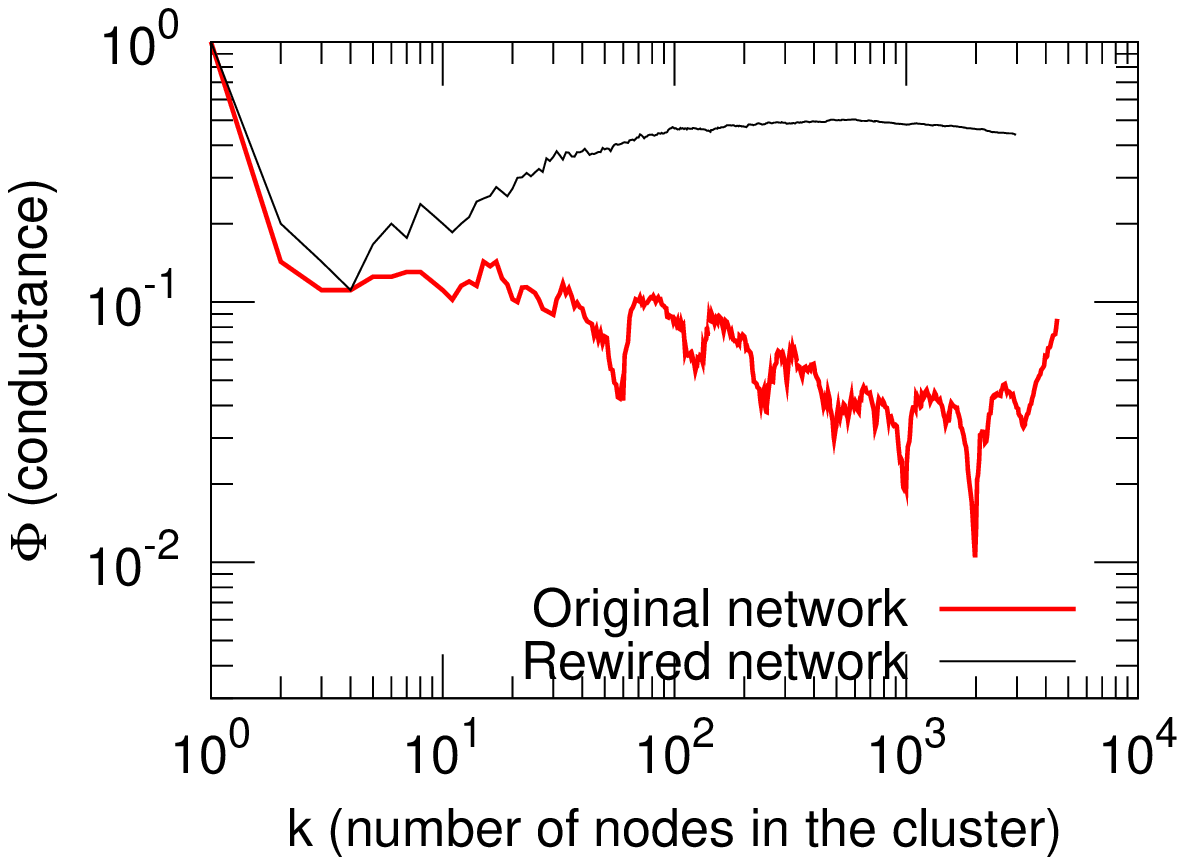}
         \label{fig:phiModels:CGA}
      } \\
      \subfigure[Geometric PA~\cite{FFV04_geometric1}]{
         \includegraphics[width=0.40\textwidth]{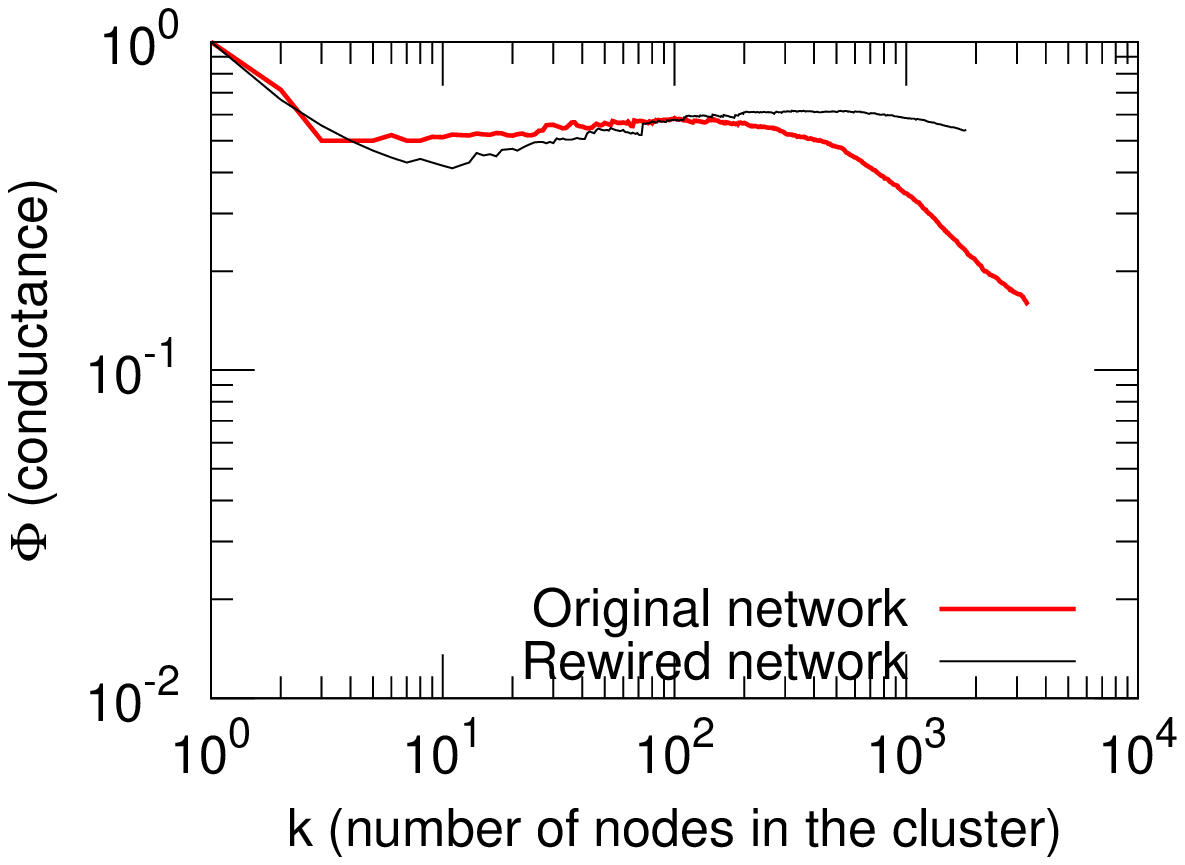} \quad
         \label{fig:phiModels:GPA}
      } &
      \subfigure[Nested Communities]{
         \includegraphics[width=0.40\textwidth]{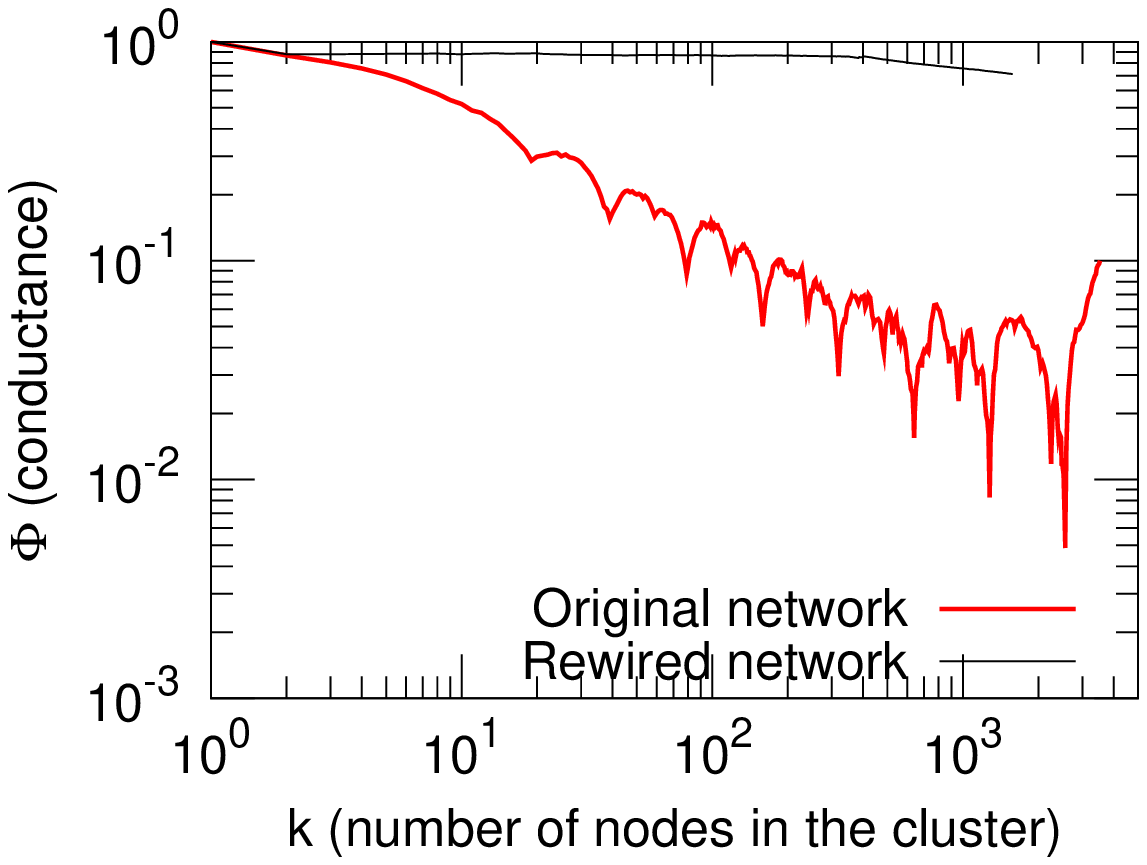}
         \label{fig:phiModels:Nested}
      } \\
   \end{tabular}
   \end{center}
\caption{
[Best viewed in color.]
Network community profile for networks generated from commonly-used procedures
to generate graphs with heavy-tailed degree distributions:
(\ref{fig:phiModels:PA}) Preferential attachment;
(\ref{fig:phiModels:Copy}) Copying model;
(\ref{fig:phiModels:BH}) Hierarchical model;
(\ref{fig:phiModels:CGA}) Community guided attachment;
(\ref{fig:phiModels:GPA}) Geometric preferential attachment; and
(\ref{fig:phiModels:Nested}) Nested community model.
See the text for details.
Red curves plot the results of the Local Spectral Algorithm on the specified 
network, and black curves plot the results of the Local Spectral Algorithm 
applied to a randomly rewired version of the same network. 
}
\label{fig:phiModels}
\end{figure}

We have studied a wide range of commonly-used network generative models in
an effort to reproduce the upward-sloping NCP plots and to understand the
structural properties of the real-world networks that are responsible for
this phenomenon. In each case, we have experimented with a range of
parameters, and in no case have we been able to reproduce our empirical
observations, at even a qualitative level. In Figure~\ref{fig:phiModels},
we summarize these results.

There has been a large body of work subsequent to that of Albert and
Barab\'{a}si~\cite{barabasi99emergence} on models in which edges are added
via a preferential-attachment or rich-gets-richer
mechanism~\cite{newman2003_review,bollobas03_review}. Much of this work
aims at reproducing properties of real-world graphs such as heavy-tailed
degree
distributions~\cite{albert1999dww,broder00_web,faloutsos99_powerlaw}. In
these preferential attachment models, one typically connects each new node
to the existing network by adding exactly $m$ edges to existing nodes with
a probability that depends on the current degree of that existing node.
Figure~\ref{fig:phiModels:PA} shows the NCP plot for a $10,000$ node
network generated according to the original preferential attachment
model~\cite{barabasi99emergence}, where at each time step a node joins the
graph and connects to $m=2$ existing nodes. Note that the NCP plot is very
shallow and flat (more even than the corresponding rewired graph), and
thus the network that is generated is expander-like at all size scales.

A different type of generative model is one in which edges are added
via a copying mechanism~\cite{kumar00stochastic}. In this copying
model, a new node joins the network by attaching exactly $m$ edges
to existing nodes as follows: the new node first selects uniformly
at random a ``seed'' or ``ambassador'' node $u$; then, for each of
its $m$ edges, with probability $\beta$ the new node links to an
existing node chosen randomly, and with probability $1-\beta$ it
links to a random neighbor of node $u$. In
Figure~\ref{fig:phiModels:Copy}, we show the results for a network
with $50,000$ nodes, generated with $m=2$ and $\beta=0.05$. Although
intuitively the copying model aims to produce communities by linking
a new node to neighbors of a existing node, this does not seem to be
the right mechanism  to reproduce the NCP plot since potential
ambassador nodes are all treated similarly and since new nodes
always create the same number of edges.

Next, in Figure~\ref{fig:phiModels:BH}, we consider an example of a
network that was designed to have a recursively hierarchical community
structure~\cite{ravasz02_metabolic,ravasz03_hierarchical}. In this model,
we start with a $5$-node square-like structure with a central node, and
then recursively expand the square and link it to the middle node of the
network. This network has power-law degree distribution, and clustering
coefficient that decays as in a characteristic
manner~\cite{ravasz03_hierarchical}. In this case, however, the NCP plot
is sloping downwards. The local dips in the plot correspond to multiples
of the size of the basic module of the graph. Although the model generates
links such that nodes that are farther apart in the hierarchy link less
frequently, the NCP plot clearly indicates that in aggregate larger
communities are easily separated than smaller communities.

A different way to generate power-law degree distributions is the
Community Guided Attachment model~\cite{jure05dpl}. Here we
decompose the nodes of a graph into a nested groups of nodes, such
that the difficulty of forming links between nodes in different
groups increases exponentially with the distance in the community
hierarchy. Graphs generated by this principle have both power-law
degree distributions and they also obey the Densification Power
Law~\cite{jure05dpl,jure07evolution}. As
Figure~\ref{fig:phiModels:CGA} shows, though, the NCP plot is sloping
downward. Qualitatively this plot from CGA is very similar to the
plot of the recursive hierarchical construction in
Figure~\ref{fig:phiModels:BH}, which is not surprising given the
similarities of the models.

Figure~\ref{fig:phiModels:GPA} shows the NCP plot for a geometric
preferential attachment model~\cite{FFV04_geometric1,FFV07_geometric2}.
This model aims to achieve a heavy-tailed degree distribution as well edge
locality, and it does so by making the connection probabilities depend
both on the two-dimensional geometry and on the preferential attachment
scheme. As we see, the effect of the underlying geometry eventually
dominates the NCP plot since the best bi-partitions are fairly
well-balanced~\cite{FFV04_geometric1}. 
Intuitively, geometric preferential attachment graphs look locally 
expander-like, but at larger size scales the the union of such small 
expander graphs behaves like a geometric mesh.
We also experimented with the small-world model by Watts and
Strogatz~\cite{watts98collective}, in which the NCP plot in some sense behaves exactly the
opposite (plot not shown): first the NCP plot decreases, and then it flattens out. 
Intuitively, a small-world network looks locally like a mesh, but when one 
reaches larger size scales, the randomly rewired edges start to appear and 
the graph looks like an expander.

Finally, we explored in more detail networks with explicitly planted
community structure. For example, we started with $10$ isolated
communities generated using the $G_{n,p}$ model, and then we generated a
random binary tree. For each internal node at height $h$ we link the nodes
in both sides of the tree with probability $p^h$, for a probability
parameter $p$. This and other related networks gives a graph of nested
communities resembling the hierarchical clustering algorithm of Newman and
Girvan~\cite{newman04community}. We see, however, from
Figure~\ref{fig:phiModels:Nested} that the NCP plot slopes steadily
downward, and furthermore we observe that dips correspond to the cuts that
separate the communities.

These experiments demonstrate that hierarchically nested networks and
networks with underlying geometric or expander like structure exhibit very
different NCP plots than observed in real networks. So the question still
remains: what causes NCP plot to decrease and then start to increase?

\subsection{Very sparse random graphs have very unbalanced deep cuts}
\label{sxn:models:sparse_Gw}

In this section, we will analyze a very simple random graph model which
reproduces relatively deep cuts at small size scales and which has a NCP 
plot that then flattens out.  
Understanding why this happens will be instructive as a baseline for understanding the 
community properties we have observed in our real-world networks.

Here we work with the random graph model with given expected degrees, as
described by Chung and
Lu~\cite{ChungLu:2006,ChungLu02_components,Chung03_eigenvalues,Chung02_distancesPNAS,Chung03_distancesIM,Chung03_spectraPNAS,Chung04_spectraIM,chung06_volume}.
Let $n$, the number of nodes in the graph, and a vector
${\bf{w}}=(w_1,\ldots,w_n)$, which will be the expected degree sequence
vector (where we will assume that $\max_i w_i^2 < \sum_k w_k$), be given.
Then, in this random graph model, an edge $e_{ij}$ between nodes $i$ and
$j$ is added, independently, with probability $p_{ij} = w_i w_j /\sum_k
w_k$. Thus, $P(e_{ij} = 1) = p_{ij}$ and $P(e_{ij} = 0) = 1 - p_{ij}$. We
use $G(\bf{w})$ to denote a random graph generated in this manner. (Note
that this model is different than the so-called ``configuration model'' in
which the degree distribution is exactly specified and which was studied
by Molloy and Reed~\cite{molloy95_critical,molloy98_size} and also Aiello,
Chung, and Lu~\cite{aiello00_modelSTOC,aiello01_modelJRNL}. This model is
also different than generative models such as preferential attachment
models~\cite{barabasi99emergence,newman2003_review,bollobas03_review} or
models based on
optimization~\cite{doyle00_hotPRE,doyle02_hotPNAS,fabrikant02_hot},
although common to all of these generative models is that they attempt to
reproduce empirically-observed power-law
behavior~\cite{albert1999dww,faloutsos99_powerlaw,broder00_web,newman2005_zipf,CSN07_powerlaw}.)

In this random graph model, the expected average degree is
$w_{av}=\frac{1}{n}\sum_{i=1}^{n}w_i$ and the expected second-order
average degree is $\tilde{w}=\sum_{i=1}^{n}w_i^2/\sum_k w_k$. Let
$w_G=\sum_i w_i$ denote the expected total degree. Given a subset $S$ of
nodes, we define the volume of $S$ to be $w_S = \sum_{v\in S} w_v$ and we
say that $S$ is $c$-giant if its volume is at least $c w_G$, for some
constant $c>0$. We will denote the actual degrees of the graph $G$ by
$\{d_1,d_2,\ldots,d_n\}$, and will define $d(S)$ to be the sum of the
actual degrees of the vertices in $S$. Clearly, by linearity of
expectation, for any subset $S$, $E(d(S)) = w_S$.

The special case of the $G(\bf{w})$ model in which $\bf{w}$ has a power
law distribution is of interest to us here. (The other interesting special
case, in which all the expected degrees $w_i$ are equal to $np$, for some
$p\in [0,1]$, corresponds to the classical Erd\"{o}s-Renyi $G_{np}$ random
graph model~\cite{bollobas85_rg}.) Given the number of nodes $n$, the
power-law exponent $\beta$, and the parameters $w$ and $w_{\max}$, Chung
and Lu~\cite{ChungLu:2006} give the degree sequence for a power-law graph:
\begin{equation}
\label{eqn:heavy_tail_probs} w_i = c i^{-1/(\beta-1)} \mbox{ for } i
\mbox{ s.t. } i_0 \le i < n+i_0   ,
\end{equation}
where, for the sake of consistency with their notation, we index the
nodes from $i_0$ to $n+i_0-1$, and where $c=c(\beta,w,n)$ and
$i_0=i_0(\beta,w,n,w_{\max})$ are as follows:
\begin{equation}
c=\alpha w n^{1/(\beta-1)} \mbox{ and }
i_0 = n \left(\alpha\frac{w}{w_{\max}}\right)^{\beta-1} ,
\end{equation}
where we have defined $\alpha = \frac{\beta - 2}{\beta - 1}$. It is
easy to verify that: $w_{\max} = \max_i w_i$ is the maximum expected
degree; the average expected degree is given by $w_{av} =
\frac{1}{n}\sum_{i=1}^{n}w_i = w(1 + o(1))$; the minimum expected
degree is given by $w_{\min} = \min_i w_i =  w\alpha(1 - o(1))$; and
the number of vertices that have expected degree in the range $(k-1,
k]$ is proportional to $k^{-\beta}$.

The following theorem characterizes the shape of the NCP plot for this
$G(\bf{w})$ model when the degree distribution follows
Equation~(\ref{eqn:heavy_tail_probs}), with $\beta\in(2,3)$. The theorem
makes two complementary claims. First, there exists at least one (small
but moderately deep) cut in the graph of size $\Theta(\log n)$ and
conductance $\Theta(\frac{1}{\log n})$. Second, for some constants $c'$
and $\epsilon$, there are no cuts in the graph of size greater than $c'
\log n$ having conductance less than $\epsilon$. That is, this model has
clusters of of logarithmic size with logarithmically deep cuts, and once
we get beyond this size scale there do not exist any such deep cuts.

\begin{theorem}
\label{thm:mainGw} Consider the random power-law graph model
$G(\bf{w})$, where $\bf{w}$ is given by
Equation~(\ref{eqn:heavy_tail_probs}), where $w > 5.88$, and the
power-law exponent $\beta$ satisfies $2<\beta<3$. Then, then with
probability $1-o(1)$:
\begin{enumerate}
\item
There exists a cut of size $\Theta(\log n)$ whose conductance is
$\Theta\left(\frac{1}{\log n}\right)$.
\item
There exists $c',\epsilon>0$ such that there are no sets of size
larger than $c' \log n$ having conductance smaller than $\epsilon$.
\end{enumerate}
\end{theorem}
\begin{proof}
Combine the results of Lemma~\ref{lemma:smallcut} and
Lemma~\ref{lemma:nocut}.
\end{proof}

\begin{figure}[t]
   \begin{center}
   \begin{tabular}{cc}
      \subfigure[NCP plot suggested by our main theorem]{
         \includegraphics[width=0.40\textwidth]{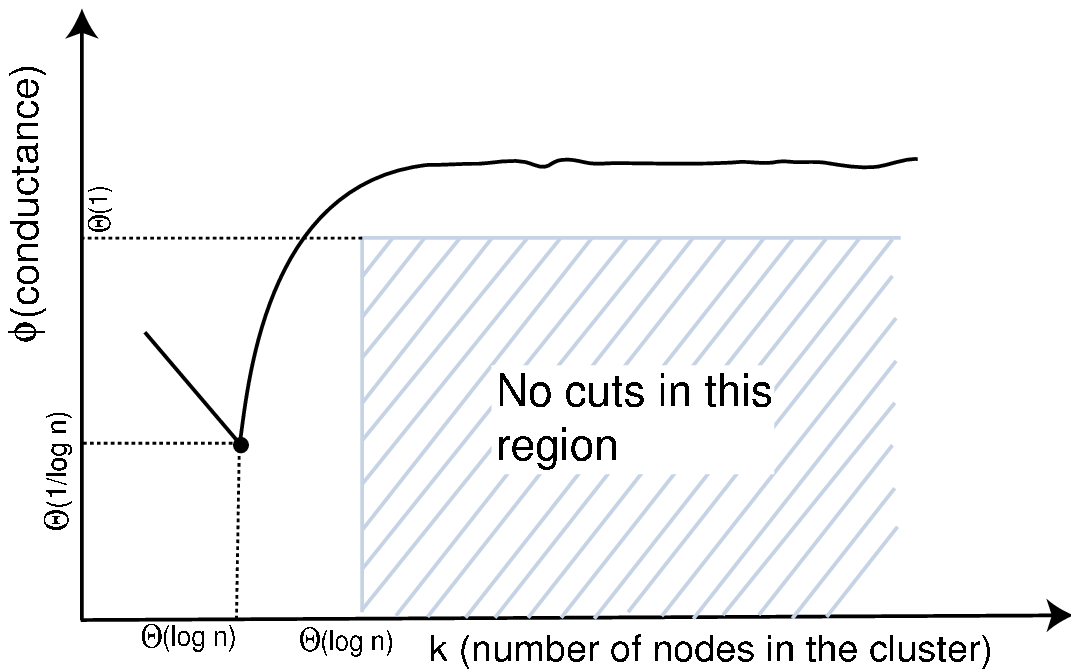}
         \label{fig:models_rand1}
      } &
      \subfigure[Structure of $G(\bf{w})$ model]{
         \includegraphics[width=0.40\textwidth]{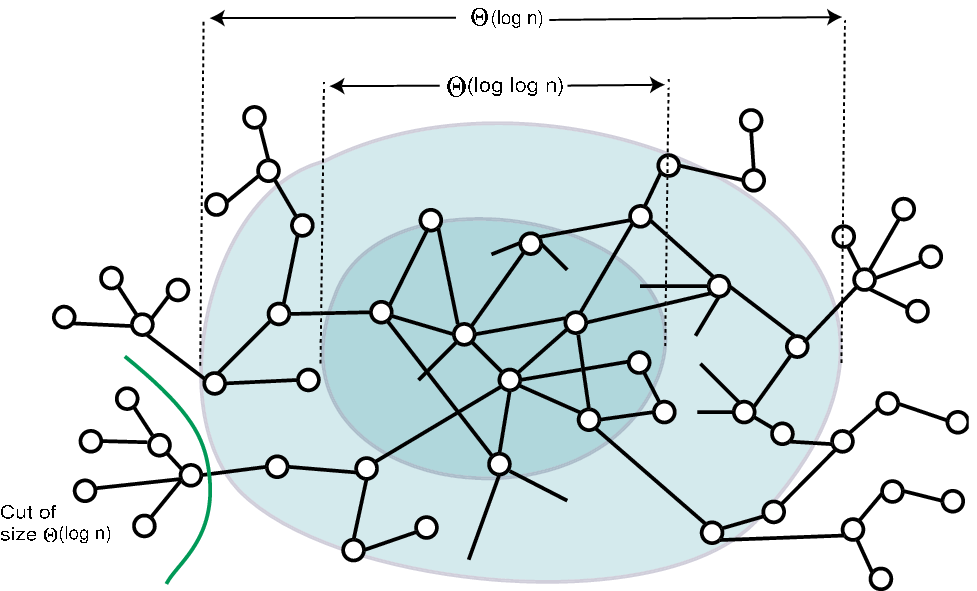}
         \label{fig:models_rand2}
      }
   \end{tabular}
   \end{center}
\caption{
The $G(\bf{w})$ model in the sparse $\beta\in(2,3)$ parameter regime.
(a) Network community profile plot, as suggested by our main theorem.
(b) Caricature of network structure.
}
\label{fig:models_rand}
\end{figure}

The two claims of Theorem~\ref{thm:mainGw} are illustrated in
Figure~\ref{fig:models_rand1}. Note that when $w \ge \frac{4}{e}$
and $\beta\in(2,3)$ then a typical graph in this model is not fully
connected but does have a giant component~\cite{ChungLu:2006}. (The
well-studied $G_{n,p}$ random graph model~\cite{bollobas85_rg} has a
similar regime when $p\in(1/n,\log n/n)$, as will be discussed in
Section~\ref{sxn:discussion:technical}.)

In addition, under certain conditions on the average degree and second
order average degree, the average distance between nodes is in
$O\left(\log \log n\right)$ and yet the diameter of the graph is
$\Theta\left(\log n\right)$. Thus, in this case, the graph has an
``octopus'' structure, with a subgraph containing $n^{c/(\log \log n)}$
nodes constituting a deep core of the graph. The diameter of this core is
$O(\log \log n)$ and almost all vertices are at a distance of $O(\log \log
n)$ from this core. However, the pairwise average distance of nodes in the
entire graph is $O(\log n/\log \tilde{w})$. A schematic picture of the
$G(\bf{w})$ model when $\beta\in(2,3)$ is presented in
Figure~\ref{fig:models_rand2}.

Our first lemma claims that for the $G(\bf{w})$ model, if the degree
distribution $\bf{w}$ follows the above power-law, then there exists
a moderately large cut with small conductance. In order to prove the
existence of a cut of size $\Theta(\log n)$ and conductance
$\Theta(\frac{1}{\log n})$, it is sufficient to concentrate on the
existence of whiskers that are large enough. In particular, to prove
the following lemma, we compute the probability that there exists a
cut of both volume and size $\Theta(\log n)$ and cut-size $1$. (Note
that although we formally state the lemma in terms of the power-law
random graph model, the proof will show that the main claim holds
for a more general representation of the heavy-tailed degree
distribution.)

\begin{lemma}
\label{lemma:smallcut} For the $G(\bf{w})$ model, where $w$ follows
a power-law degree distribution with $2<\beta<3$ then, with
probability $1-o(1)$ there exists a set of size $\Theta(\log n)$
with conductance $\Theta(\frac{1}{\log n})$.
\end{lemma}
\begin{proof}
Let $S$ be a subset with the following description. $S = \{ v_0,
v_1, \ldots, v_k\}$, where $k = c_1\log n$. Let $w_i$ denote the
degree of $v_i$. We have that $w_0\in[c_2\log n, 2c_2\log n]$ and
$w_i \leq w$ for all $i > 0$. Thus the expected volume of $S$ is
$w_S \in [(2\alpha w c_2+c_1) \log n, (2\alpha w c_2+c_1)\log n]$,
and the size of $S$ is $c_1 \log n + 1$. Note that the expected
volume of the graph can be computed as $w_G = wn$, and hence $\rho =
\frac{1}{w_G} = \frac{1}{wn}$.

Now, let $n_1$ denote the number of vertices of expected degree at
most $2\alpha w$. By simple calculation, $n_1 \geq n/2$. The number
of possible choices for the vertex $v_0$ can be computed as follows.
Let $B$ be the set of vertices having degree greater than $2\alpha w
c_2\log n$ and $A$ be the set of vertices with degree at most
$2\alpha w c_2\log n$. Then the number of nodes with degree in $[
c_2\log n, 2 c_2\log n]$ is given by the size of $V\setminus(A\cup
B)$ which is
\begin{align*}
 \alpha w \left(\frac{n}{c_2\log n}\right)^{\beta-1} -
\alpha w \left(\frac{n}{2c_2\log n}\right)^{\beta-1} \geq \alpha w
\left(\frac{n}{c_2\log n}\right)^{\beta-1}  ,
\end{align*}
since $\beta > 2$. Thus the number of possible such subsets $S$ is
given by the number of choices for $v_0$ times the number of
possible choices for the nodes $v-1, \ldots,v_k$. Thus, the number
$N$ of possible such subsets $S$ is at least
\[
N = {n_1 \choose c_1\log n}\times \alpha \left(\frac{n}{2c_2\log
n}\right)^{\beta-1}.
\]
We say that $S$ is good if, after instantiating all the edges, $S$
has a star of size $c_1\log n$ centered at $v_0$, and $v_0$ is
connected to $\bar{S}$ by exactly one edge, and none of the other
vertices in $S$ have any edge to $\bar{S}$. The probability that a
particular set $S$ is good is the product of the following terms:
the probability $p_1$ that there is star of size $c_1\log n$ with
$v_0$ at the center, the probability $p_2$ that none of the nodes
$v_1,\ldots,v_k$ link to any nodes in $\bar{S}$, and the probability
$p_3$ that $v_0$ connects to $\bar{S}$ using exactly one edge. We
now calculate the three probabilities as follows. First,
\[
p_1 = \prod_{i\in [1\ldots k]} w_0w_i\rho \geq (w_0\alpha
w\rho)^{c_1\log n},
\]
since each $w_i\geq w_{\min} \geq \alpha w$. Next,
\[
p_2 = \prod_{i = 1, \ldots k} \prod_{j \notin S} (1-w_j\rho)\geq
 \prod_{i = 1, \ldots k} \prod_{j \notin S}  e^{-w_i\rho/2} = e^{-(c_1 \rho 2\alpha w w_{\bar{S}}\log n)/2 },
\]
obtained by using $1- x \geq e^{-x/2}$ for $0<x<1$, and $w_i \leq
2\alpha w$ for $i\in S, i>1$. Finally, we get $p_3$ as  follows.
First note
\begin{align*} p_3 = &\; \sum_{j \in \bar{S}}w_0w_j \rho \prod_{k
\neq j, k \in
\bar{S}} (1-w_kw_0\rho)\\
\geq&\; \sum_{j \in \bar{S}}w_0w_j \rho e^{-(w_{\bar{S}}\; -
w_j)w_0\rho/2}\\
=&\; w_0 \rho e^{-w_{\bar{S}} w_0 \rho/2}(\sum_{j \in \bar{S}} w_j
e^{w_jw_0\rho/2})   .
\end{align*}
Then, since $w_jw_0\rho \ll 1$ and since $e^{x} \geq 1
+ x$, we have that
\begin{align*}
p_3 \geq&\; w_0 \rho e^{-w_{\bar{S}} w_0
\rho/2}(\sum_{j\in\bar{S}}w_j (1 +
\frac{w_jw_0\rho}{2}))\\
 \geq&\; w_0 \rho e^{-w_{\bar{S}} w_0 \rho/2}(w_{\bar{S}} + w_0\rho
\tilde w_{\bar{S}}/2)   ,
\end{align*} where $\tilde w_{\bar{S}} = \sum_{j \in \bar{S}} w_j^2$. So
the final probability of goodness of $S$ is\begin{eqnarray*}  p =
p_1\times p_2\times p_3 &\geq&  (w_0\alpha w\rho)^{c_1\log n}\times
e^{-(c_1 \rho 2\alpha w w_{\bar{S}}\log n)/2 }\times w_0 \rho
e^{-w_{\bar{S}} w_0
\rho/2}(w_{\bar{S}} + w_0\rho \tilde w_{\bar{S}}/2)\\
& = & (w_0\alpha w\rho)^{c_1\log n}\times e^{-(c_1 \gamma 2\alpha w
\log n)}\times w_0 \rho e^{-\gamma w_0}(w_{\bar{S}} + w_0\rho \tilde
w_{\bar{S}}/2)   ,
 \end{eqnarray*}
using $\gamma = \rho w_{\bar{S}}/2$. So the expected number of such
good subsets $S$ is \begin{align*} Np & \geq  {n_1 \choose c_1\log
n}\times \alpha w \left(\frac{n}{2c_2\log n}\right)^{\beta-1} \times
(w_0\alpha w\rho)^{c_1\log n}\times e^{-(c_1
\gamma 2\alpha w \log n)}\times w_0 \rho e^{-\gamma w_0}(w_{\bar{S}} + w_0\rho \tilde w_{\bar{S}}/2)\\
&\geq\left(\frac{n_1}{c_1\log n}\right)^{c_1\log n}\times
\frac{\alpha w n^{\beta -1}}{(2c_2\log n)^{\beta-1}} \times
(w_0\alpha w\rho)^{c_1\log n}\times e^{-(c_1 \gamma 2\alpha w \log
n)}\times w_0 \rho e^{-\gamma w_0} \times nw/2   ,
\end{align*}
using Stirling's formula and the fact that $w_{\bar{S}} \geq nw/2$.
Using the value of $n_1$ and since $nw\rho =1 $,
\begin{align*}
Np &\geq \left(\frac{n}{2c_1\log n}\right)^{c_1\log n} \times
\frac{\alpha w n^{\beta - 1}}{(2c_2\log n)^{\beta-1}} \times
(w_0\alpha w\rho)^{c_1\log n}\times e^{-(c_1
\gamma 2\alpha w \log n)}\times e^{-\gamma w_0} \times w_0/2\\
&\geq \left(\frac{w_0 \alpha}{2 c_1\log n}\right)^{c_1\log n} \times
\frac{\alpha w n^{\beta - 1}}{(2c_2\log n)^{\beta-1}} \times
(w_0\alpha w\rho)^{c_1\log n}\times e^{-(c_1 \gamma 2\alpha w \log
n)}\times e^{-\gamma w_0} \times w_0/2   .
\end{align*}
Using $w_0 \geq c_2\log n$, we have that have that
\begin{align*}
Np &\geq \left(\frac{c_2\alpha}{2c_1}\right)^{c_1\log n} \times
\frac{\alpha w n^{\beta - 1}}{2(2c_2\log n)^{\beta-2}} \times
e^{-(c_1
\gamma 2\alpha w \log n)}\times e^{-\gamma w_0}\\
&\geq e^{\Theta\log n}\times \frac{\alpha w}{2(2c_2\log
n)^{\beta-2}}   ,
 \end{align*}
where $\Theta = c_1\log(\frac{c_2\alpha}{2c_1}) + (\beta -1)-
\gamma \alpha w c_1 - 2\gamma c_2$. Note that for $2<\beta <3$, we
have that $0<\alpha < \frac{1}{2}$. Also, $\gamma = \frac{1}{2} -
o(1)$. Thus, choosing $c_2 = 2 ec_1/\alpha$ and $c_1 = \frac{\beta
-2 }{2\gamma \alpha w + 4\gamma e/\alpha - 1}$, we get $\Theta = 1$.
So,
\begin{align*}
Np &\geq e^{\log n}\times \frac{\alpha w}{2(2c_2\log n)^{\beta-2}}
=\Omega(\log n)
\end{align*}
\par
 Then, the probability is a particular set $S$ is good is $p \geq \Omega\left( \frac{(\log n)}{N}\right)$. Hence the probability of getting a good set is
\begin{eqnarray*} 1 - (1 - p)^N \geq 1 - \left( 1 - \Omega\left( \frac{(\log
n)^{\beta -2}}{N}\right)\right)^N \geq 1 - o(1) \end{eqnarray*}
\end{proof}

We next state the well-known Chernoff bound~\cite{ChungLu:2006},
which we will use below.

\begin{lemma}
Let $X = \sum_{i} X_i$ where the $X_i$ are independent random
variables with $X_i \geq -M$. Define $\| X\|^2 = \sum_i \Ex(X_i^2)$.
Then,
\begin{align}
\Pr( X \geq \Ex(X) - \lambda )
   \leq \exp\left( - \frac{\lambda^2}{2(\|X\|^2 + M\lambda/3)}\right).
\end{align}
\end{lemma}

Finally, we show that there are no deep cuts with size greater than
$\Theta(\log n)$. To state this lemma, define a connected set $S$ to
be $\epsilon$-deficit set if it has actual volume $d(S) \leq
\frac{1}{2}d(G)$ and if the conductance of the cut $(S, \bar{S})$ is
at most $\epsilon$, \emph{i.e.}, if the number of edges leaving $S$
is at most $\epsilon d(S)$.

\begin{lemma}
\label{lemma:nocut} For the $G(\bf{w})$ model, where $w$ follows a
power-law degree distribution with $2<\beta<3$, if the average
degree $w$ satisfies $w \geq 5.88$, then with probability $1-o(1)$
there exists constants $c', \epsilon$ such that there is no
$\epsilon$-deficit set of size more than $c' \log n$.
\end{lemma}
\begin{proof}
Let $e(S,\bar{S})$ denote the actual number of edges between $S$ and
$\bar{S}$. First we compute the probability that a given set $S$ is
$\epsilon$-deficit, that is, $S$ satisfies $e(S, \bar{S}) < \epsilon
d(S)$. Let $\delta = \frac{2\epsilon}{1 - \epsilon}$. For our case,
define the variables $X_{(i,j)} = e_{ij}$ for $(i,j)\in (S,\bar{S})$
and $X_{(i,j)} = -\delta e_{ij}$ for $(i,j)\in (S,S)$. Then the sum
$X = \sum X_{(i,j)}  = \sum_{(i,j) \in (S,\bar{S})} e_{ij} - \delta
\sum_{(i,j) \in (S,\bar{S})} e_{ij}$. Note that $e(S, \bar{S}) <
\epsilon d(S) \iff X \leq 0$. Using the fact that $\Ex(e_{ij}) =
w_iw_j\rho$, we have $\|X\|^2 = \sum \Ex(X_{ij}^2) = w_S
w_{\bar{S}}\rho  + \delta^2 w_S^2 \rho$. Furthermore, exploiting the
fact that each $X_i \geq - \delta$, we get that \begin{eqnarray*}
\Pr( X \leq 0 ) &=& \Pr( X \leq\Ex(X) - \Ex(X))\\
& \leq & \exp\left( - \frac{\Ex(X)^2}{2(\|X\|^2 +\delta
\Ex(X)/3)}\right) \\
& =& \exp\left( - \frac{\rho^2 w_S^2(w_{\bar{S}} - \delta w_S)^2
}{2(w_S\rho(w_{\bar{S}} + \delta^2w_S) + \delta w_S\rho(w_{\bar{S}}
- \delta w_S)/3 )}\right)  . \end{eqnarray*} Canceling $\rho w_S$ from
both numerator and denominator,

 \begin{eqnarray*}
\Pr( X \leq 0 ) & \leq & \exp\left( - \frac{\rho w_S (w_{\bar{S}} -
\delta w_S)^2 }{2(w_{\bar{S}} + \delta^2w_S  + \delta w_{\bar{S}}/3
- \delta^2 w_S/3 )}\right)\\ &\leq & \exp\left( - \frac{\rho
w_S(w_{\bar{S}} - \delta w_S)^2 }{ 2(1 + \delta/3 +
2\delta^2/3)w_{\bar{S}}}\right) \leq \exp\left( - \frac{\rho w_S
w_{\bar{S}}(1 - 2\delta w_S/w_{\bar{S}}) }{ 2(1
+ \delta/3 + 2\delta^2/3)}\right)\\
&\leq & \exp\left( - \frac{\rho w_S w_{\bar{S}}(1 - 2\delta) }{ 2(1
+ \delta/3 + 2\delta^2/3)}\right)
 \leq \exp\left( -\rho w_S w_{\bar{S}} A_{\delta}/2\right)  ,
  \end{eqnarray*}
where $A_{\delta} = \frac{1 - 2\delta}{(1 + 2\delta/3 +
 2\delta^2/3)}$. So this bounds the probability that a particular
 set $S$ of size $k$ is $\epsilon$-deficit. We will bound the expected number of
 such $\epsilon$-deficit subsets of size $k$. First, let $N_{k,\epsilon, \gamma}$ denote the expected number of
$\epsilon$-deficit sets of size $k$ that have expected volume
 $w_S \leq \gamma w_G$. By linearity of expectation,
 \begin{eqnarray*} N_{k,\epsilon,\gamma} &\leq &
\sum_{\stackrel{S:|S| = k}{w_S\leq \gamma w_G}} w_{i_1}\ldots
w_{i_k} w_S^{k-2}\rho^{k-1}\exp\left( -\rho w_S w_{\bar{S}}
A_{\delta}/2
\right)\\
 &\leq & \sum_{\stackrel{S:|S| = k}{w_S\leq \gamma w_G}} \frac{w_S^{2k-2}}{k^k} \rho^{k-1}\exp(\left( -w_S(1-\gamma)
A_{\delta} \right))   , \end{eqnarray*} where we used the fact that
$\gamma = \rho w_{\bar{S}}/2$ and also the AM-GM inequality to say
that $\prod_{i\in S} w_i \leq \left(\frac{\sum_{i\in S}
w_i}{k}\right)^{k}$. Now, $F(x) =
x^{2k-2}e^{-xA_{\delta}(1-\gamma)}$ is maximized at  $x =
\frac{2k-2}{A_{\delta}(1-\gamma)}$. Thus, the above sum is maximized
when $w_S = \frac{2k-2}{A_{\delta}(1-\gamma)}$. Hence,
 \begin{eqnarray*}
 N_{k,\epsilon,\gamma} &\leq & \frac{n^k}{k!}\frac{\rho^{k-1}}{k^k}\frac{2^{(2k-2)}\cdot
 (k-1)^{(2k-2)}}{(A_{\delta}(1-\gamma))^{(2k-2)}}\exp(-2k+2)\\
 &\leq& \frac{(n\rho)^k}{\rho \sqrt{k}(k/e)^k}\frac{1}{k^k}\frac{2^{(2k-2)}\cdot
 (k-1)^{(2k-2)}}{(A_{\delta}(1-\gamma))^{(2k-2)}}\exp(-2k+2)   .
 \end{eqnarray*}
 Using $(1 - \frac{1}{k})^{2k}\leq e^{-2}$, it follows that
 \begin{eqnarray*} N_{k,\epsilon,\gamma}
&\leq& \frac{1}{4e\sqrt{k}(k-1)^2}\left( \frac{4}{ewA_{\delta}^2
 (1-\gamma)^2}\right)^{k}   .
 \end{eqnarray*}
We would like $\sum_{k=c\log n}^{cn} N_{k,\epsilon,\gamma}$ to be
$o(1)$, for which we need
\[
\frac{4}{ewA_{\delta}^2 (1-\gamma)^2} < 1,
\]
which gives a bound on average degree:
\[
w \geq \frac{4}{A_{\delta}^2(1-\gamma)^2e}.
\]
For sets of volume $w_S \geq \gamma w_G$, we have the following.
From the double-sided Chernoff bound, for any fixed set $S$,
\begin{eqnarray*} |w_S - d(S)| \leq \lambda \mbox{ with probability
} 1 - 2\exp\left( - \frac{\lambda^2}{2(w_S +\lambda/3) }\right)   .
\end{eqnarray*} So if $\lambda = \sqrt{w_S}\log n$, we have the
above statement with probability $1 - 2\exp( - 3\log^2 n/8)$.
Similarly,
\begin{equation*} | e(S,\bar{S}) - \Ex(e(S,\bar{S})) | \leq \lambda
\mbox{ with probability }  1 - 2\exp\left( - \frac{\lambda^2}{2(\rho
w_S w_{\bar{S}} +\lambda/3) }\right)   . \end{equation*}
By having $\lambda = \sqrt{\rho w_S w_{\bar{S}}}\log n$
the above probability becomes $1 - 2\exp( - 3\log^2 n/8)$. Now, if
both these events occur, then the conductance of the set $S$ is at
least $1/3$. So the only way we can get an $\epsilon$-deficit set is
at by having one of these conditions to be invalid. The total number
of sets of expected volume $\gamma w_G$ is bounded by $w_G \choose
\gamma w_G$. So, the expected number of $\epsilon$-deficit sets of
volume at least $\gamma w_G$ is bounded by \begin{eqnarray*}
\sum_{\gamma\leq \theta\leq 1/2}{w_G\choose \theta w_G} 4\exp( -
3\log^2 n/8) \leq \int_{\gamma\leq \theta\leq 1/2}
\frac{1}{\sqrt{\theta w_G}}\left(\frac{1}{\theta}\right)^{\theta
w_G}4\exp( - 3\log^2 n/8) \leq o(1).\end{eqnarray*} Thus, putting
the two bounds together, the expected number of $\epsilon$-deficit
sets of size greater that $c\log n$ is at most $o(1)$. Thus with
probability $1-o(1)$ there does not exist an $\epsilon$-deficit set
of size greater than $c\log n$.
\end{proof}

\subsection{An intuitive toy model for generating an upward-sloping NCP plot}
\label{sxn:models:toy}

We have seen that commonly-studied models, including preferential
attachment models, copying models, simple hierarchical models, and models
in which there is an underlying mesh-like or manifold-like geometry are
not the right way to think out the community structure of large social and
information networks. We have also seen that the extreme sparsity of the
networks, coupled with randomness, can be responsible for the deep cuts at small scales. 

\begin{figure}
    \begin{center}
    \begin{tabular}{cc}
      \subfigure[A toy network model $\ldots$]{
         \includegraphics[width=0.40\textwidth,height=10.5em]{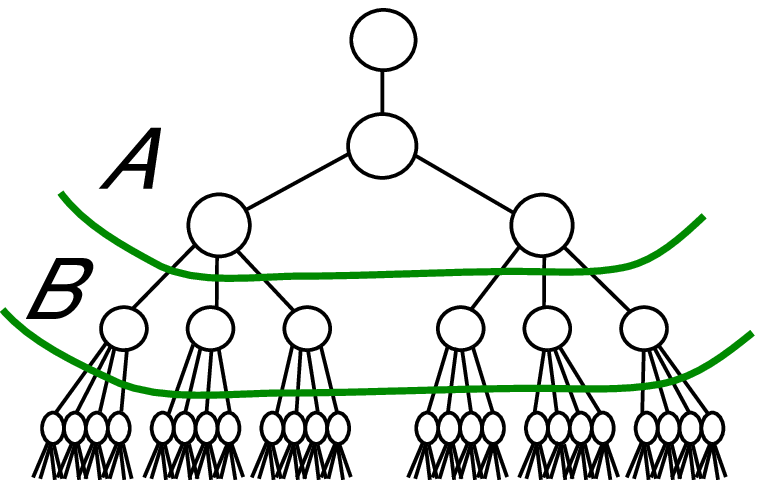} 
         \label{fig:models_clique1}
      } &
      \subfigure[$\ldots$ and its community profile plot.]{
         \includegraphics[width=0.45\textwidth]{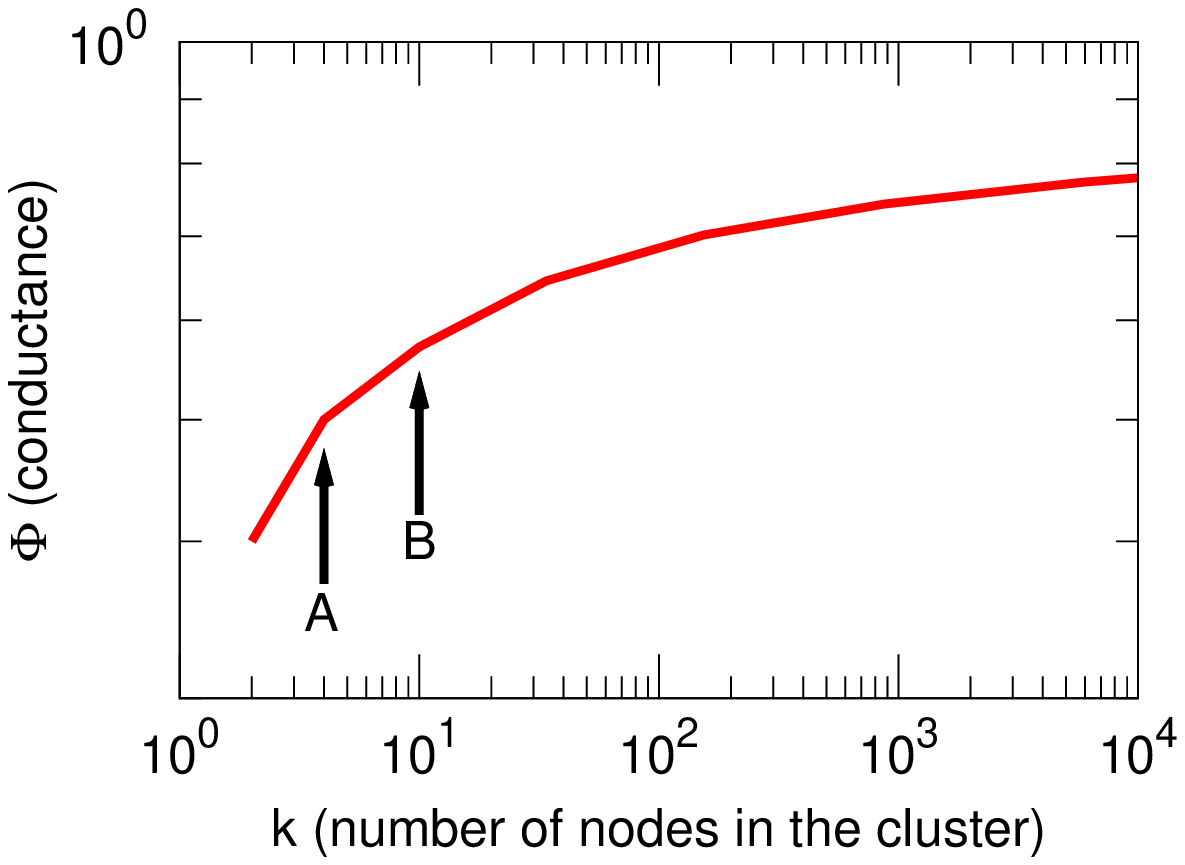} 
         \label{fig:models_clique2}
      } 
    \end{tabular}
    \end{center}
\caption{
Schematic picture of the properties of a network responsible for the 
upward-sloping community profile plot. 
(a) This toy model is designed so that the optimal conductance cuts are 
achieved by cutting nodes from the top of the tree.
(b) The minimum of the NCP plot is achieved by cutting the single top node, and 
then larger and larger cuts have gradually worse and worse conductance values.
}
\label{fig:models_clique}
\end{figure}

To build
intuition as to what the gradually increasing NCP plot might mean,
consider Figure~\ref{fig:models_clique}. This is a toy example of a
network construction in which the NCP plot has a deep dip at a small size
scale and then steadily increases.
The network shown in Figure~\ref{fig:models_clique1} is an infinite tree
that has two parts. The top part, a subtree (with one node in this
example, but more generally consisting of $n_T$ nodes) is indicative of
the whiskers, or the ``small scale'' structure of the graph. The remaining
tree has the property that the number of children increases monotonically
with the level of the node. This property is indicative of the fact that
as the size of a cluster grows, the number of neighbors that it has also
increases.
The key insight in this construction is that the best conductance cuts
first cut at the top of the growing tree and then gradually work their way
``down'' the tree, starting with the small subtrees and moving gradually
down the levels, as depicted in Figure~\ref{fig:models_clique1}.

Thus, intuitively, one can think of small well-separated
communities---those below the $n_T$ size scale that consist of subsets of
the small trees---starting to grow, and as they pass the $n_T$ size scale
and become bigger and bigger, they blend in more and more with the central
part of the network, which (since it exhibits certain expander-like
properties) does not have particularly well-defined communities. Note
(more generally) that if there are $n_T$ nodes in the small tree at the
top of the graph, then the dip in the NCP plot in
Figure~\ref{fig:models_clique2} is of depth $2/(n_T+1)$. In particular, if
$n_T=\Theta(\log n)$ then the depth of this cut is $\Theta(1/\log n)$.

Intuitively, the NCP plot increases since the ``cost'' per edge for every 
additional edge inside a cluster increases with the size of the cluster. 
For example, in cut $A$ in Figure~\ref{fig:models_clique1}, the ``price'' 
for having $3$ internal edges is to cut $6$ edges, \emph{i.e.}, $2$ edges 
cut per edge inside. 
To expand the cluster by just a single edge, one has to move one level down 
in the tree (toward the cut $B$) where now the price for a single edge is 
$4$ edges, and so on.

\subsection{A more realistic model of network community structure}
\label{sxn:models:ff}

The question arises now as to whether we can find a simple generative
model that can explain both the existence of small well-separated
whisker-like clusters and also an expander-like core whose best clusters
get gradually worse as the purported communities increase in size.
Intuitively, a satisfactory network generation model must successfully
take into account the following two mechanisms:

\begin{itemize}
\item[(a)] The model should produce a relatively large number of
    relatively small---but still large when compared to random
    graphs---well connected and distinct whisker-like communities.
    (This should reproduce the downward part of the community profile
    plot and the minimum at small size scales.)
\item[(b)] The model should produce a large expander-like core, which
    may be thought of as consisting of intermingled communities,
    perhaps growing out from the whisker-like communities, the
    boundaries of which get less and less well-defined as the
    communities get larger and larger and as they gradually blend in
    with rest of the network. (This should reproduce the gradual
    upward sloping part of the community profile plot.)
\end{itemize}

The so-called \emph{Forest Fire Model}~\cite{jure05dpl,jure07evolution}
captures exactly these two competing phenomena. The Forest Fire Model is a
model of graph generation (that generates directed graphs---an effect we
will ignore) in which new edges are added via a recursive ``burning''
mechanism in an epidemic-like fashion. Since the details of the recursive
burning process are critical for the model's success, we explain it in
some detail.

To describe the Forest Fire Model
of~\cite{jure05dpl,jure07evolution}, let us fix two parameters, a
\emph{forward burning probability} $p_f$ and a \emph{backward
burning probability} $p_r$. We start the entire process with a
single node, and at each time step $t>1$, we consider a new node $v$
that joins the graph $G_t$ constructed thus far. The node $v$ forms
out-links to nodes in $G_t$ as follows:
\begin{itemize}
\item[(i)] Node $v$ first choose a node $w$, which we will refer to as
    a ``seed'' node or an ``ambassador'' node, uniformly at random and
    forms a link to $w$.
\item[(ii)] Node $v$ selects $x$ out-links and $y$ in-links of $w$
    that have not yet been visited. ($x$ and $y$ are two geometrically
    distributed random numbers with means $p_f/(1-p_f)$ and
    $p_r/(1-p_r)$, respectively. If not enough in-links or out-links
    are available, then $v$ selects as many as possible.) Let
    $w_1,w_2,\ldots,w_{x+y}$ denote the nodes at the other ends of
    these selected links.
\item[(iii)] Node $v$ forms out-links to $w_1,w_2,\ldots,w_{x+y}$, and
    then applied step~(ii) recursively to each of the
    $w_1,w_2,\ldots,w_{x+y}$, except that nodes cannot be visited a
    second time during the process.
\end{itemize}

Thus, burning of links in the Forest Fire Model begins at node $w$,
spreads to $w_1,w_2,\ldots,w_{x+y}$, and proceeds recursively until
the process dies out. One can view such a process intuitively as
corresponding to a model in which a person comes to the party and
first meets an ambassador who then introduces him or her around. If
the person creates a small number of friendships these will likely
be from the ambassadors ``community,'' but if the person happens to
create many friendships then these will likely go outside the
ambassador's circle of friends. This way, the ambassador's community
might gradually get intermingled with the rest of the network.

\begin{figure}
   \begin{center}
         \includegraphics[width=0.15\textwidth]{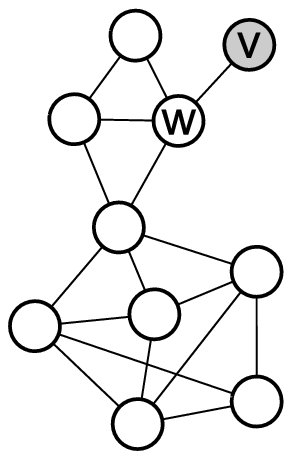}
         \hspace{1.5cm}
         \includegraphics[width=0.15\textwidth]{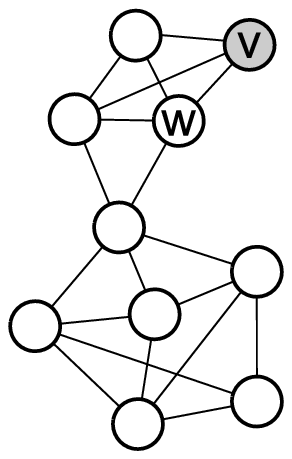}
         \hspace{1.5cm}
         \includegraphics[width=0.17\textwidth]{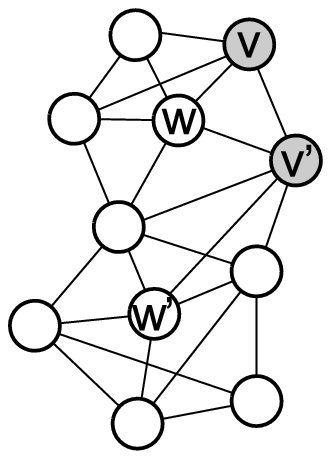}
   \end{center}
\caption{The Forest Fire burning process. 
Left: a new node $v$ joins the network and selects a seed node $w$. 
Middle: $v$ then attaches itself by recursively linking to $w$'s 
neighbors, $w$'s neighbor-neighbors, and so on, according to the 
``forest fire'' burning mechanism described in the text.  
Right: a new node $v'$ joins the network, selects seed $w'$, and 
recursively adds links using the same ``forest fire'' burning mechanism. 
Notice that if $v'$ causes a large ``fire,'' it links to a large number 
of existing nodes.  
In this way, as potential communities around node $w$ grow, the NCP plot 
is initially decreasing, but then larger communities around $w$ gradually 
blend-in with the rest of the network, which leads the NCP plot to 
increase.}
\vspace{-3mm}
\label{fig:ForestFire}
\end{figure}

Two properties of this model are particularly significant. First, although
many nodes might form one or a small number of links, certain nodes can
produce large conflagrations, burning many edges and thus forming a large
number of out-links before the process ends. Such nodes will help generate
a skewed out-degree distribution, and they will also serve as ``bridges''
that connect formerly disparate parts of the network. This will help make
the NCP plot gradually increase. Second, there is a locality structure in
that as each new node $v$  arrives over time, it is assigned a ``center of
gravity'' in some part of the network, {\em i.e.}, at the ambassador node
$w$, and the manner in which new links are added depends sensitively on
the local graph structure around node $w$. Not only does the probability
of linking to other nodes decrease rapidly with distance to the current
ambassador, but because of the recursive process regions with a higher
density of links tend to attract new links.

Figure~\ref{fig:ForestFire} illustrates this. 
Initially, there is a small community around node $w$. 
Then, node $v$ joins and using the Forest Fire mechanism locally attaches 
to nodes in the neighborhood of seed node $w$. 
The growth of the community around $w$ corresponds to downward part of the 
NCP plot.
However, if a node $v'$ then joins and causes a large fire, this has the 
effect of larger and larger communities around $w$ blending into and 
merging with the rest of the network.

\begin{figure}
   \begin{center}
      \begin{tabular}{cc}
         \includegraphics[width=0.45\textwidth]{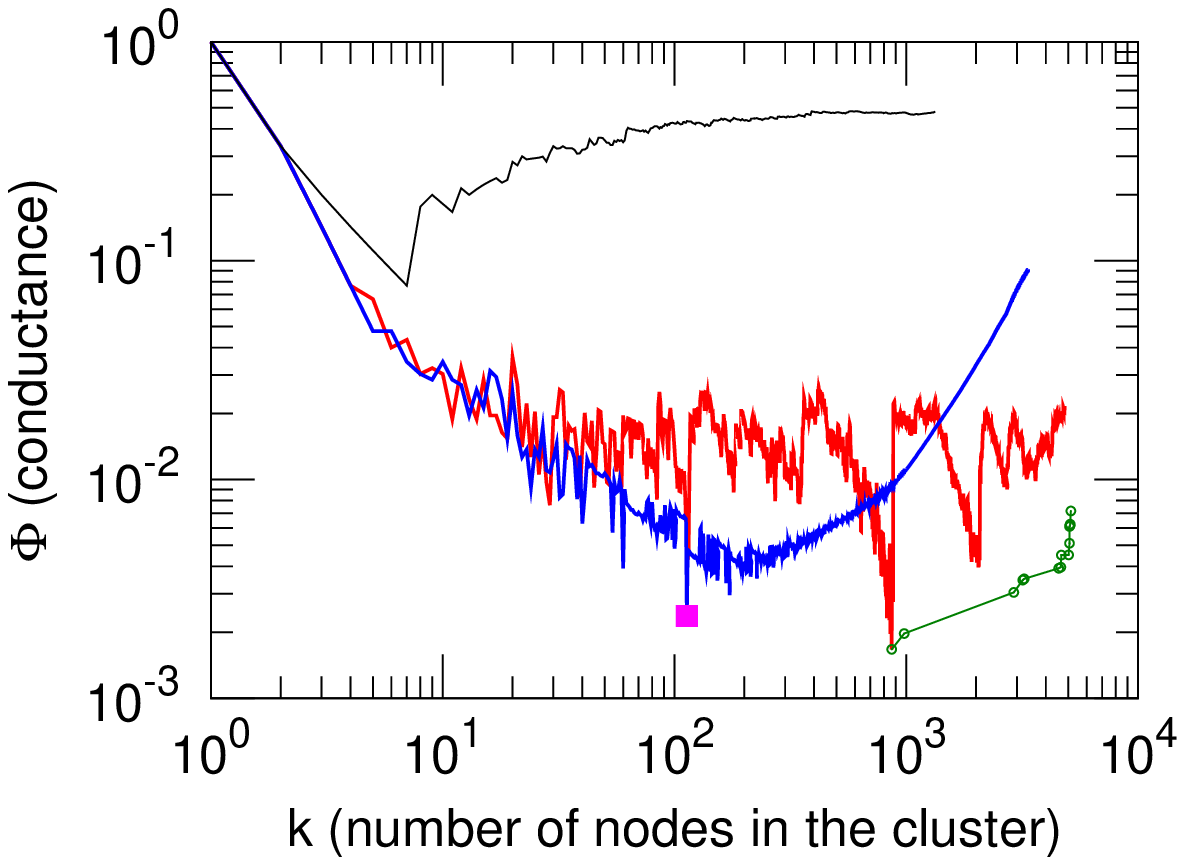} &
         \includegraphics[width=0.45\textwidth]{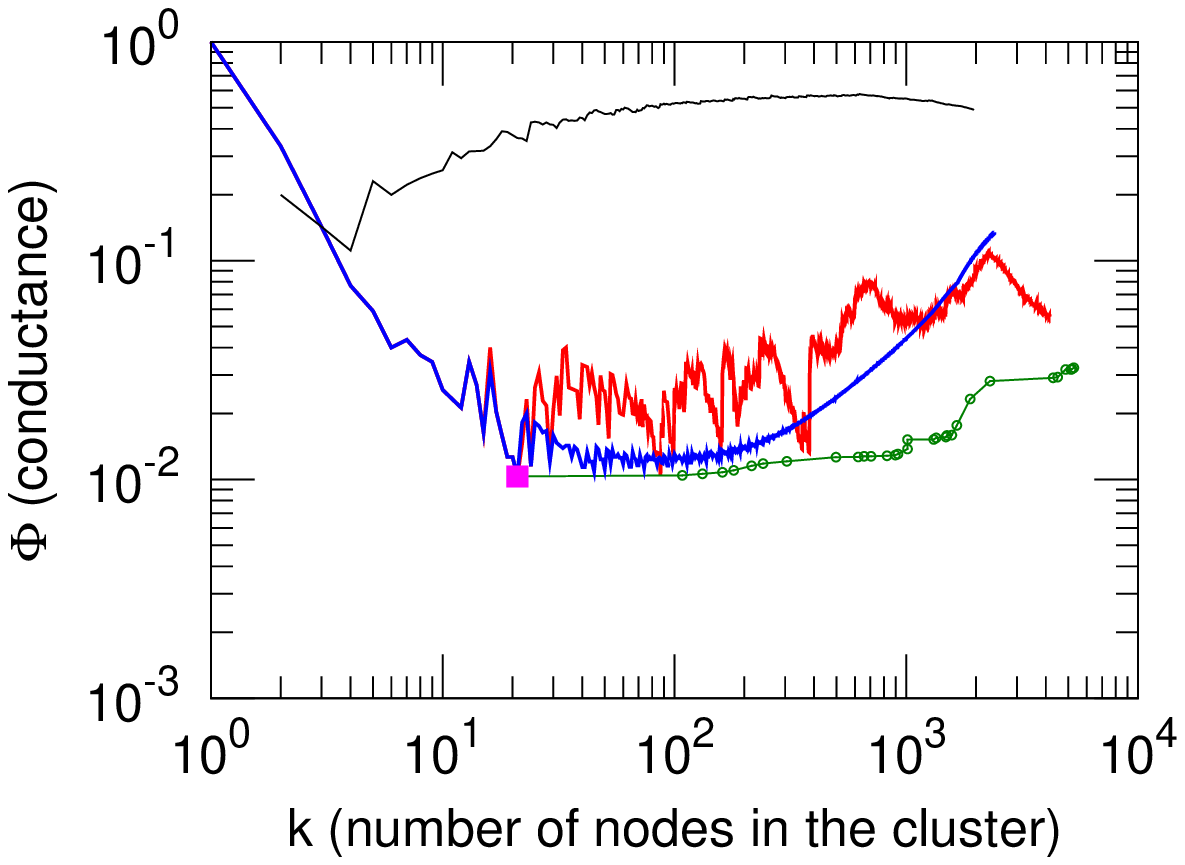} \\
         \includegraphics[width=0.45\textwidth]{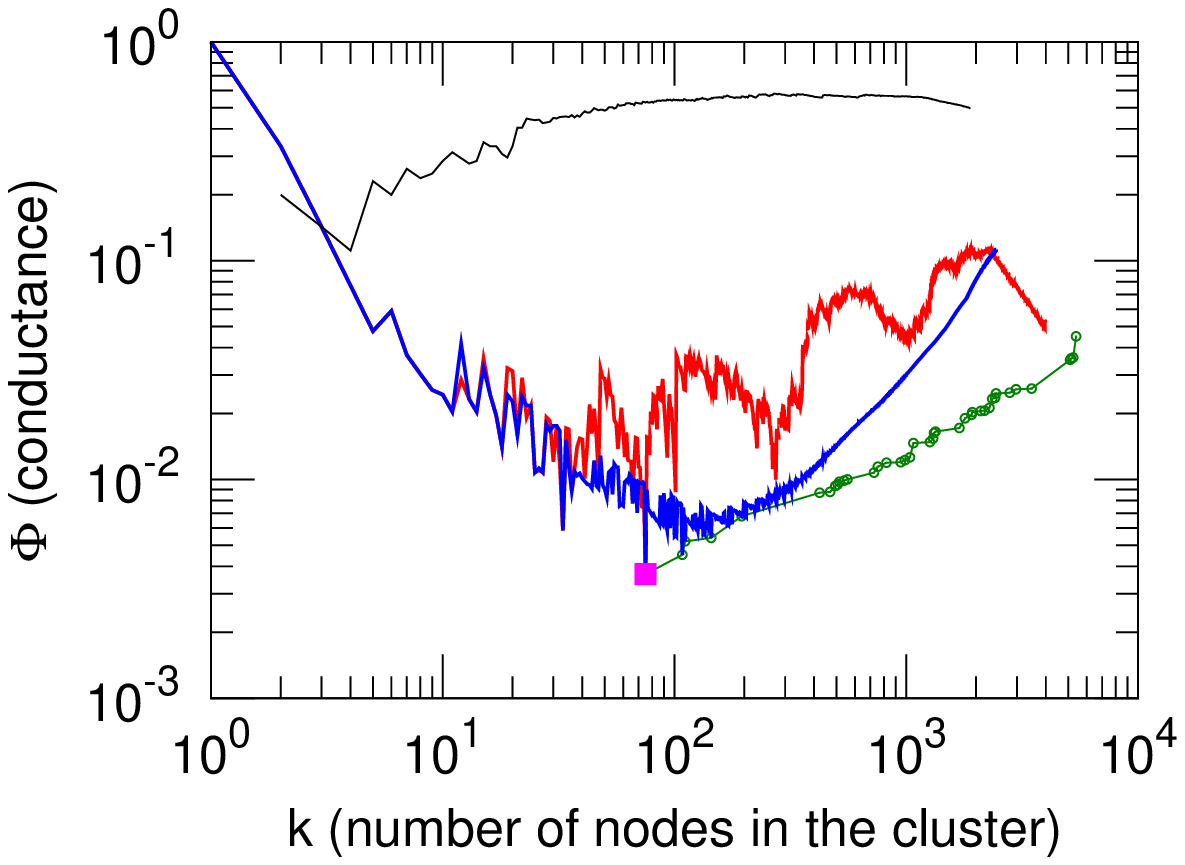} &
         \includegraphics[width=0.45\textwidth]{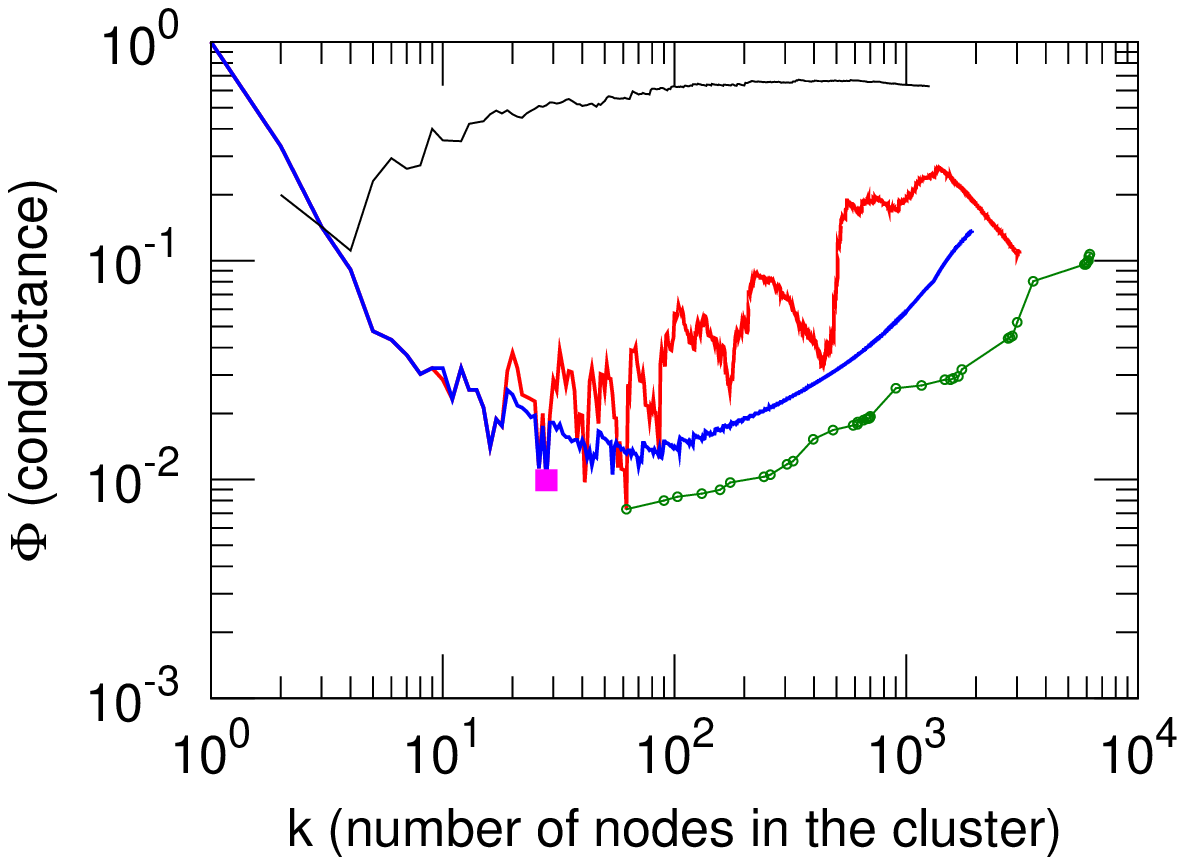} \\
         \includegraphics[width=0.45\textwidth]{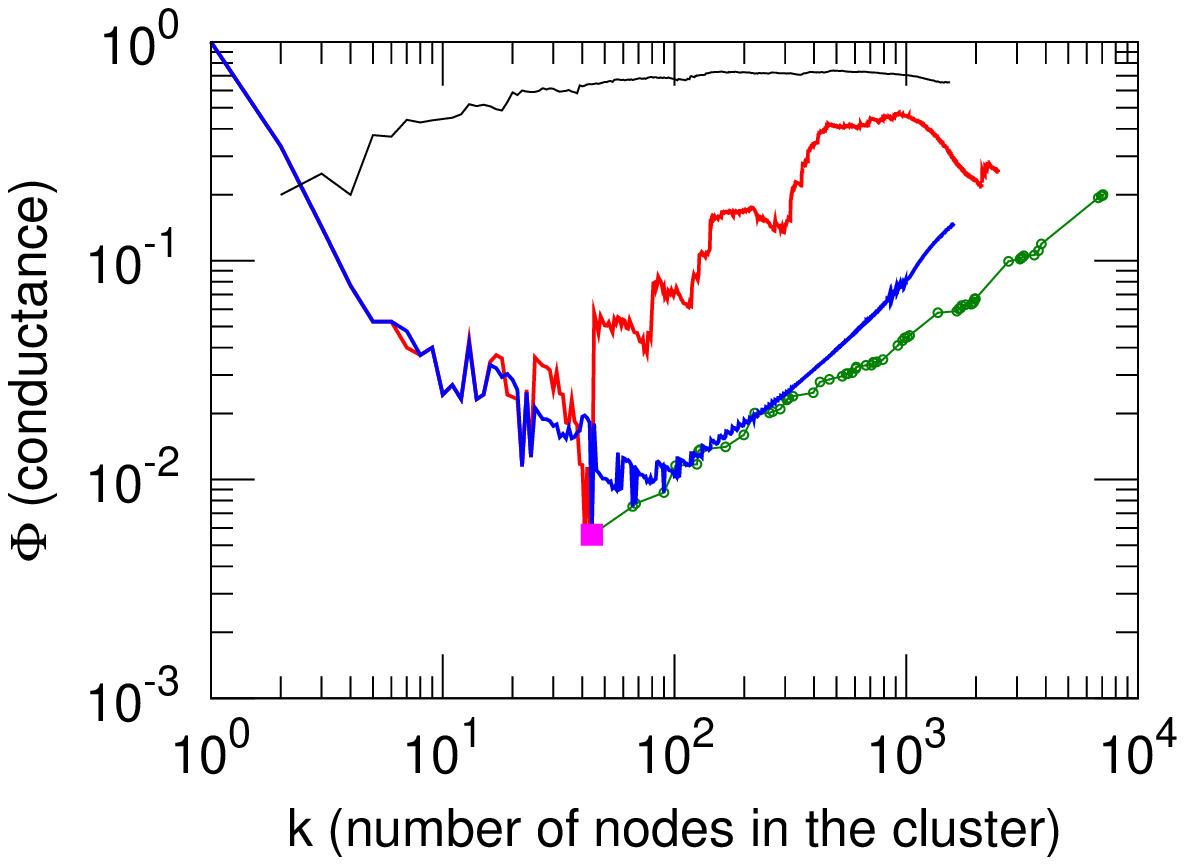} &
         \includegraphics[width=0.45\textwidth]{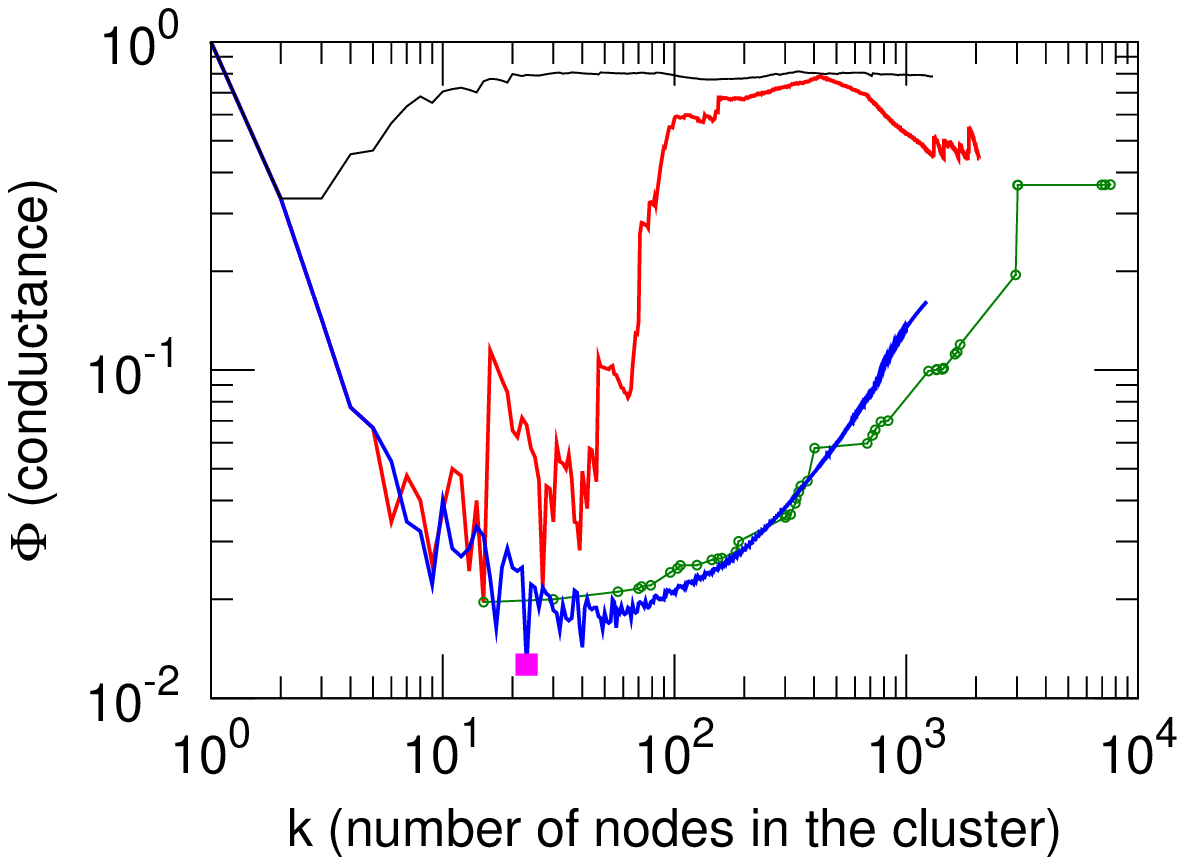} \\
         \multicolumn{2}{c}{\includegraphics[width=0.6\textwidth]{phiTR-legend2}}
      \end{tabular}
   \end{center}
\caption{
[Best viewed in color.]
Community profile plots for the Forest Fire Model at various parameter
settings.
The backward burning probability is $p_b=0.3$, and we increase (left to right,
top to bottom) the forward burning probability
$p_f=\{0.26, 0.31, 0.33, 0.35, 0.37, 0.40\}$.
Note that the largest and smallest values for $p_f$ lead to less realistic
community profile plots, as discussed in the text.
}
\label{fig:phiForestFire}
\end{figure}

Not surprisingly, however, the Forest Fire Model is sensitive to the
choice of the burning probabilities $p_f$ and $p_b$. We have experimented
with a wide range of network sizes and values for these parameters, and in
Figure~\ref{fig:phiForestFire}, we show the community profile plots of
several $10,000$ node Forest Fire networks generated with $p_b=0.3$ and
several different values of $p_f$. The first thing to note is that since
we are varying $p_f$ the six plots in Figure~\ref{fig:phiForestFire}, we
are viewing networks with very different densities. Next, notice that if,
{\em e.g.}, $p_f=0.33$ or $p_f=0.35$ then we observe a very natural
behavior: the conductance nicely decreases, reaches the minimum somewhere
between $10$ and $100$ nodes, and then slowly but not too smoothly
increases. Not surprisingly, it is in this parameter region where the
Forest Fire Model has been shown to exhibit realistic time evolving graph
properties such as densification and shrinking
diameters~\cite{jure05dpl,jure07evolution}.

Next, also notice that if $p_f$ is too low or too high, then we obtain
qualitatively different results. For example, if $p_f=0.26$, then the
community profile plot gradually decreases for nearly the entire plot. For
this choice of parameters, the Forest Fire does not spread well since the
forward burning probability is too small, the network is extremely sparse
and is tree-like with just a few extra edges, and so we get large well
separated ``communities'' that get better as they get larger. On the other
hand, when burning probability is too high, {\em e.g.}, $p_f=0.40$, then
the NCP plot has a minimum and then rises extremely rapidly. For this
choice of parameters, if a node which initially attached to a whisker
successfully burns into the core, then it quickly establishes many
successful connections to other nodes in the core. Thus, the network has
relatively large whiskers that failed to establish such a connection and a
very expander-like core, with no intermediate region, and the increase in
the community profile plot is quite abrupt.

\begin{figure}
\begin{center}
   \includegraphics[width=2.2cm]{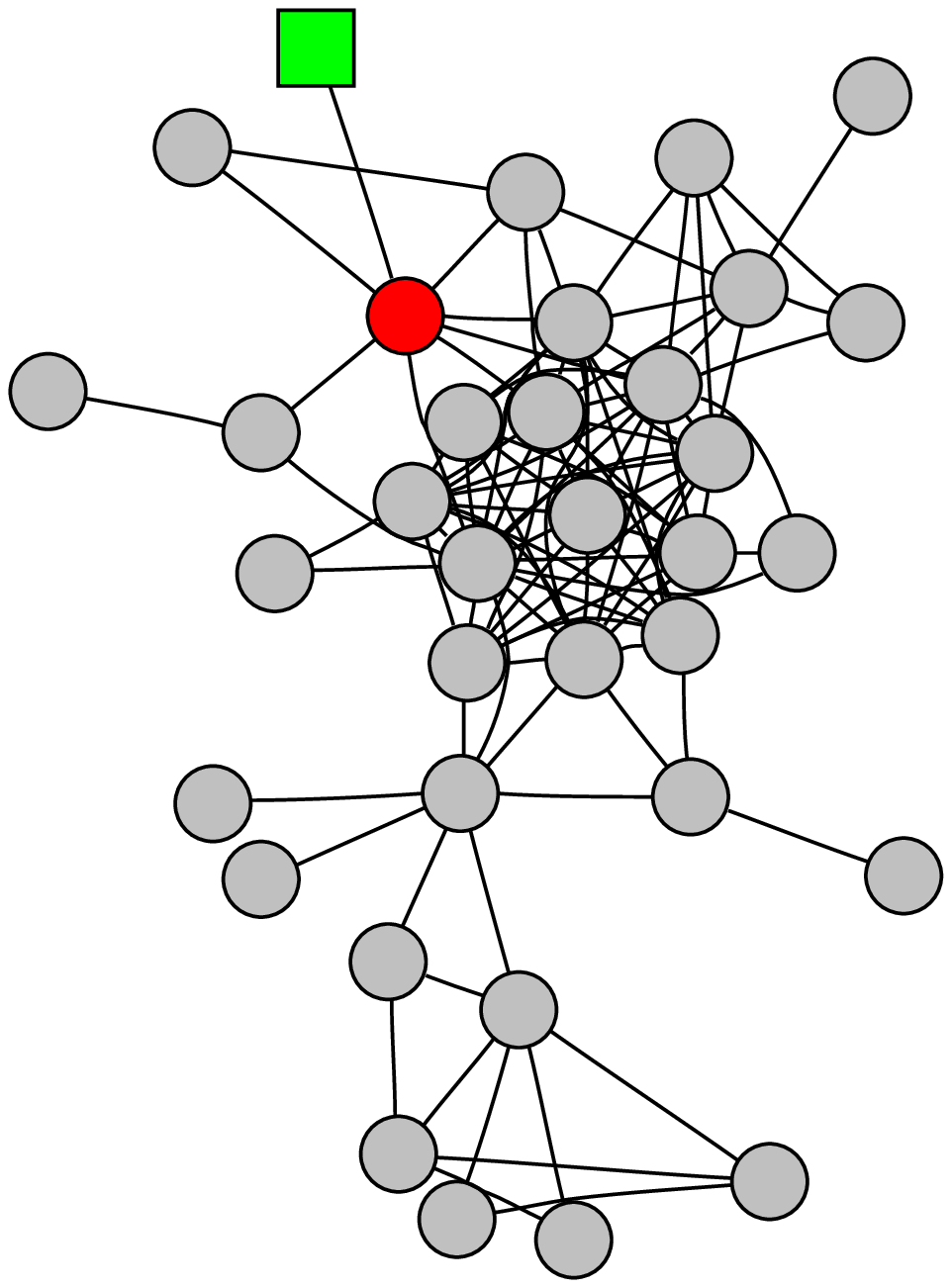}
   \includegraphics[width=2.2cm]{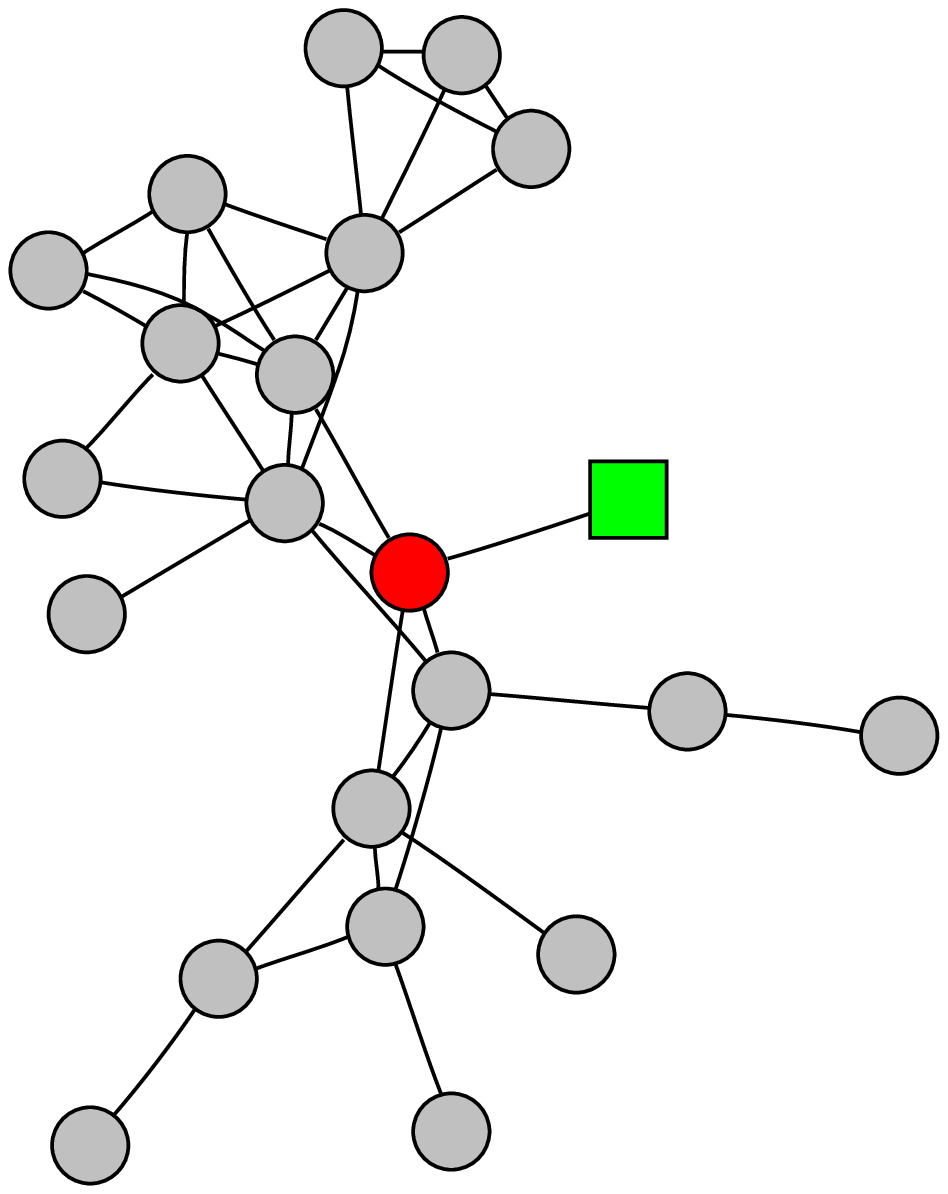}
   \includegraphics[width=2cm]{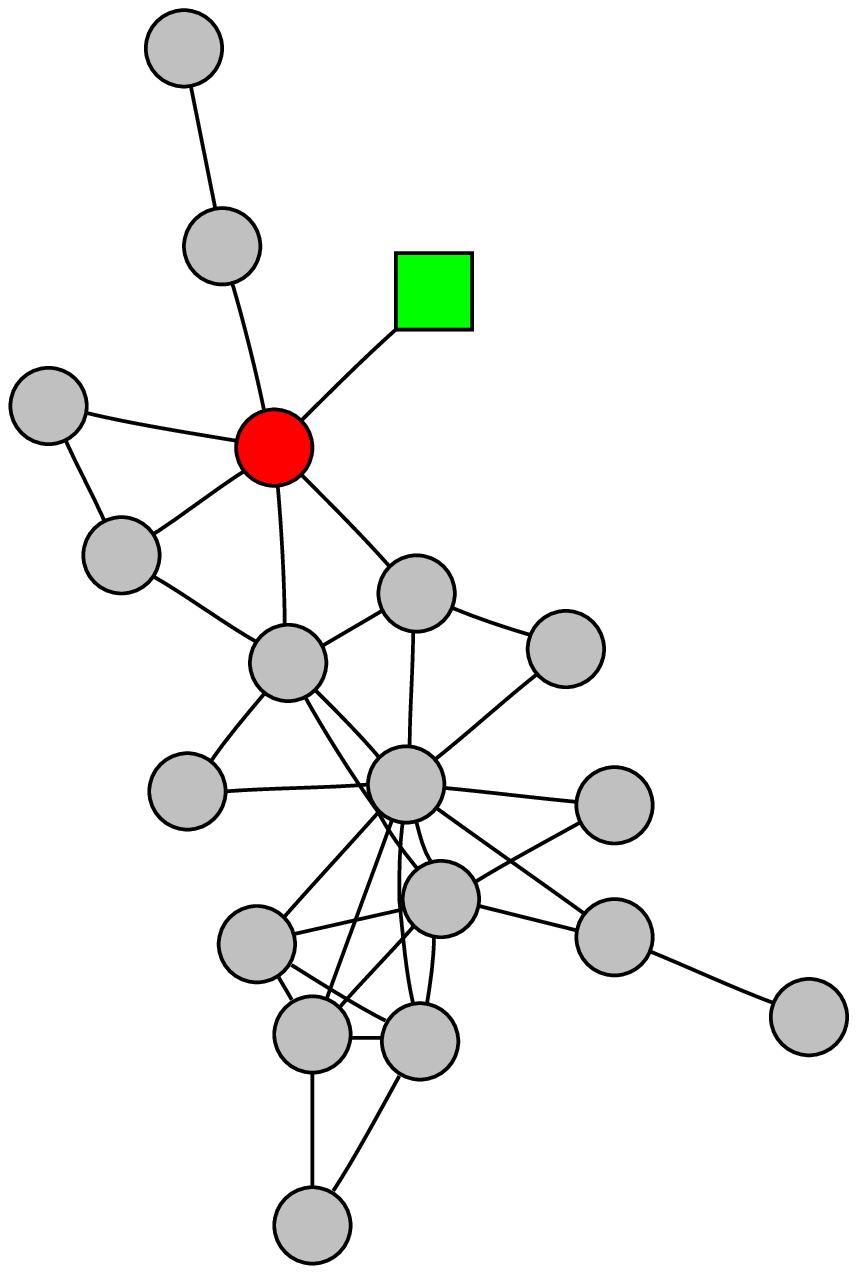}
   \includegraphics[width=2cm]{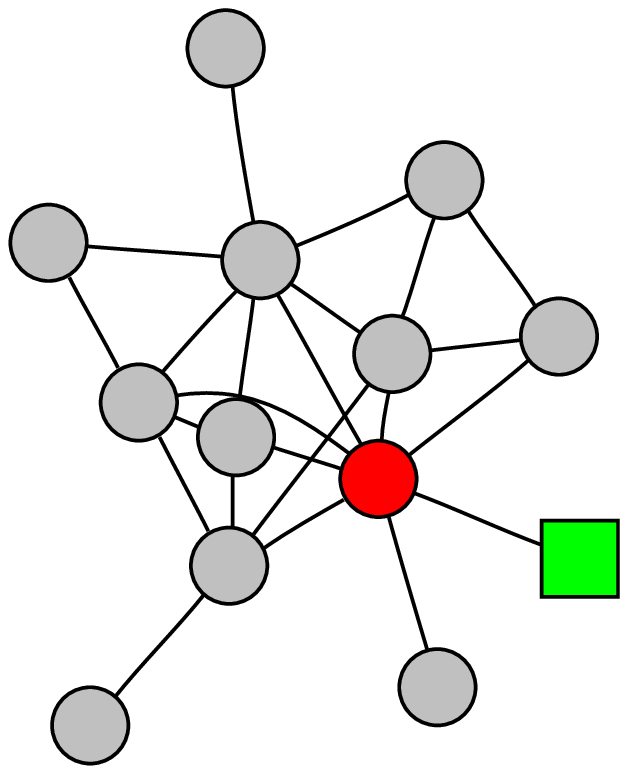}
   \includegraphics[width=2.3cm]{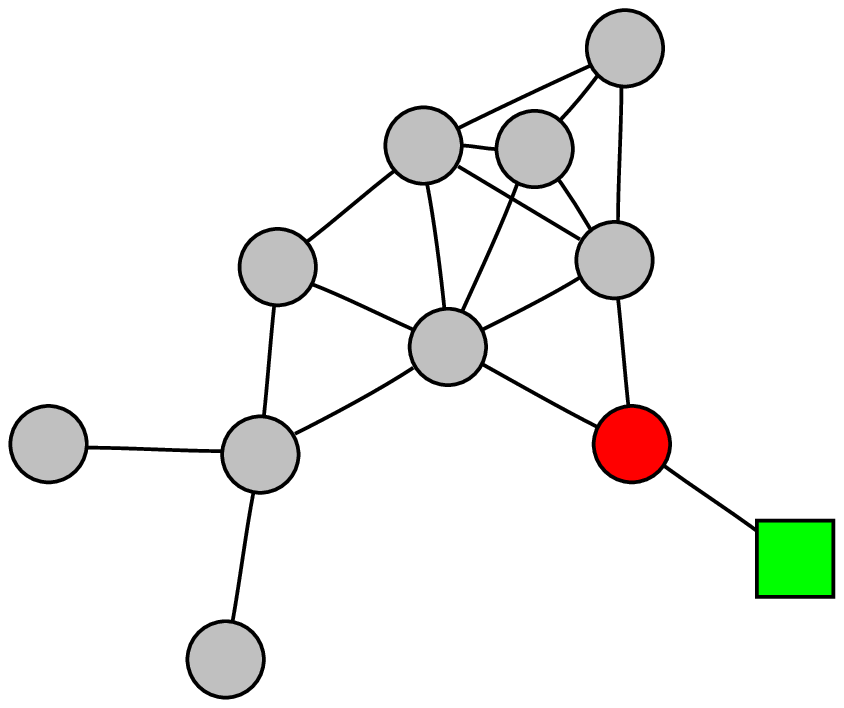}
   \includegraphics[width=1.5cm]{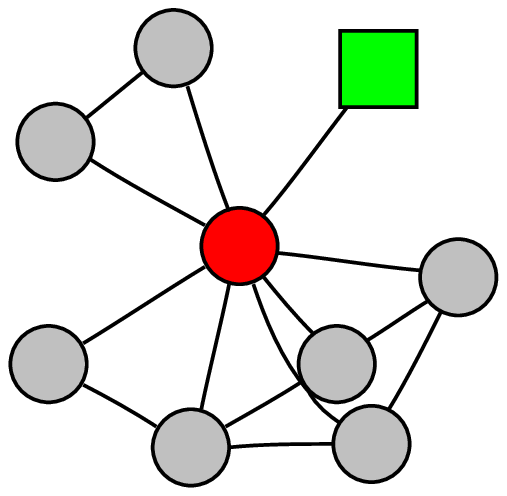}
   \includegraphics[width=1.5cm]{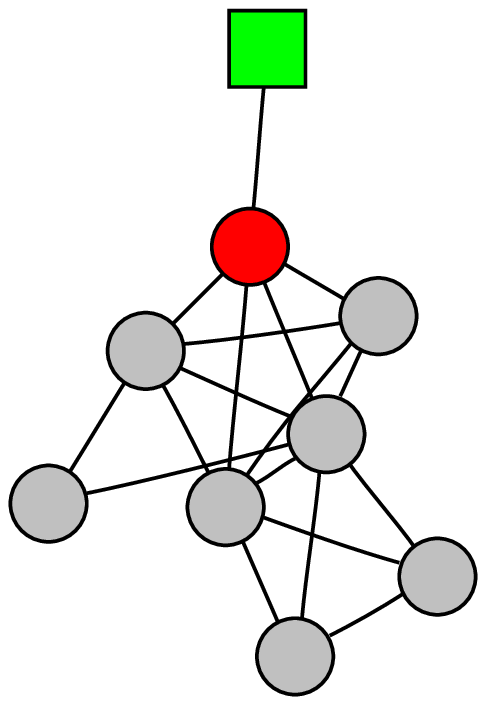}
\end{center}
\caption{
Examples of whiskers from a simulation of the Forest Fire Model with
parameter settings $p_f=0.37$ and $p_b=0.3$. The green square node belongs
to the network core, and by cutting the edge connecting it with red
circular node we separate the community of circles from the rest
of the network (depicted as a green square).}
\label{fig:whiskFF}
\end{figure}

We have examined numerous other properties of the graphs generated by the
Forest Fire Model and have found them to be broadly consistent with the
social and information networks we have examined. One property, however,
that is of particular interest is what the whiskers look like.
Figure~\ref{fig:whiskFF} shows an example of several whiskers generated by
the Forest Fire Model if we choose $p_b=0.30$ and $p_f=0.37$. They are
larger and more well-structured than the tree-like whiskers from the
random graph model of Section~\ref{sxn:models:sparse_Gw}. Also notice that
they all look plausibly community-like with a core of the nodes densely
linked among themselves and the bridge edge then connects the whisker to
the rest of the network.

We conclude by noting that there has also been interest in developing
hierarchical graph generation models, {\em i.e.}, models in which a
hierarchy is given and the linkage probability between pairs of nodes
decreases as a function of their distance in the
hierarchy~\cite{ravasz02_metabolic,ravasz03_hierarchical,chakrabarti04_rmat,abello04_hierarchical,jure05_realistic,clauset06structural,xuan06_growth,jure07_scalable}.
The motivation for this comes largely from the intuition that nodes in
social networks and are joined in to small relatively tight groups that
are then further join into larger groups, and so on. As
Figures~\ref{fig:phiModels:BH} and~\ref{fig:phiModels:CGA} make clear,
however, such models do not immediately lead to community structure
similar to what we have observed and which has been reproduced by the
Forest Fire Model. On the other hand, although there are significant
differences between hierarchical models and the Forest Fire
Model,~\cite{jure05dpl,jure07evolution} notes that there are similarities.
In particular, in the Forest Fire Model a new node $v$ is assigned an
ambassador $w$ as an entry point to the network. This is analogous to a
child having a parent in the hierarchy which helps to determine how that
node links to the remainder of the network. Similarly, many hierarchical
models have a connection probability that decreases exponentially in the
hierarchical tree distance. In the Forest Fire Model, the probability that
a node $v$ will burn along a particular path to another node $u'$ is
exponentially small in the path length, although the analogy is not
perfect since there may exist many possible paths.
\section{Discussion}
\label{sxn:discussion}

In this section, we discuss several aspects of our main results in a
broader context. In particular, in
Section~\ref{sxn:discussion:ground_truth}, we compare to several data sets
in which there is some notion of ``ground truth'' community and we also
describe several broader non-technical implications of our results. Then,
in Section~\ref{sxn:discussion:community}, we describe recent work on
community detection and identification. Finally, in
Section~\ref{sxn:discussion:technical}, we discuss several technical and
algorithmic issues and questions raised by our work.

\subsection{Comparison with ``ground truth'' and sociological communities}
\label{sxn:discussion:ground_truth}

In this subsection, we examine the relationship between network communities 
of the sort we have been discussing so far and some notion of ``ground 
truth.'' 
When considering a real network, one hopes that the output of a community finding 
algorithms will be ``real'' communities that exist in some meaningful sense 
in the real world. 
For example, in the Karate club network in 
Figure~\ref{fig:ncpp_small:karate_graph}, the cut found by the algorithm 
corresponds in some sense to a true community, in that it splits the nodes 
almost precisely as they split into two newly formed karate clubs.
In this section, we take a different approach: we take networks in which 
there are explicitly defined communities, and we examine how well these 
communities are separated from the rest of the network. 
In particular, we examine a minimum conductance profile of several network 
datasets, where we can associate with each node one or more community labels 
which are exogenously specified. 
Note that we are overloading the term ``community'' here, as in this context 
the term might mean one of two things: first, it can refer to groups of 
nodes with good conductance properties; and second, it can refer to groups 
of nodes that belong to the same self-defined or exogenously-specified group.

We consider the following five datasets:

\begin{itemize}
\item \net{LiveJournal12}~\cite{lars06groups}: LiveJournal is an
    on-line blogging site where users can create friendship links to
    other users. In addition, users can create groups which other
    users can then join. In LiveJournal, there are $385,959$ such
    groups, and a node belongs to $3.5$ groups on the average. Thus,
    in addition to the information in the interaction graph, we have
    labels specifying those groups with which a user is associated,
    and thus we may view each such group as determining a ``ground
    truth'' community.
\item \net{CA-DBLP}~\cite{lars06groups}: We considered a co-authorship
    network in which nodes are authors and there is an edge if authors
    co-authored at least one paper. Here, publication venues ({\em
    e.g.}, journals and conferences) can play the role of ``ground
    truth'' communities. That is, an author is a member of a
    particular group or community if he or she published at a
    particular conference or in a particular journal. In our DBLP
    network, there are $2,547$ such groups, with a node belonging to
    $2.6$ on the average.
\item \net{AmazonAllProd}~\cite{clauset04large}: This is a network of
    products that are commonly purchased together at \url{amazon.com}.
    (Intuitively one might expect that, {\em e.g.}, gardening books
    are frequently purchased together, so the network structure might
    reflect a well-connected cluster of gardening books.) Here, each
    item belongs to one or more hierarchically organized categories
    (book, movie genres, product types, etc.), and products from the
    same category define a group which we will view as a ``ground
    truth'' community. Items can belong to $49,732$ different groups,
    and each item belongs to $14.3$ groups on the average.
\item \net{AtM-IMDB}: This network is a bipartite actors-to-movies
    network composed from IMDB data, and an actor $A$ is connected to
    movie $B$ if $A$ appeared in $B$. For each movie we also know the
    language and the country where it was produced. Countries and
    languages may be taken as ``ground truth'' communities or groups,
    where every movie belongs to exactly one group and actors belongs
    to all groups to which movies that they appeared in belong. In our
    dataset, we have $393$ language groups and $181$ country groups.
\item \net{Email-Inside} and \net{Email-InOut}~\cite{jure07evolution}:
    This is an email communication network from a large European
    research organization conducting research in natural sciences:
    physics, chemistry, biology and computer science. Each of $986$
    members of the organization belongs to exactly one of $45$
    departments, and we use the department memberships to define
    ``ground truth'' communities.
\end{itemize}

Although none of these notions of ``ground truth'' is perfect, many
community finding algorithms use precisely this form of anecdotal
evaluation: a network is taken, network communities are found, and then
the correspondence of network communities to ``ground truth'' communities
is evaluated. Note, in contrast, we are evaluating how ``ground truth''
communities behave at different size scales with respect to our
methodology, rather than examining how the groups we find relate to
``ground truth'' communities. Furthermore, note that the notions of
``ground truth'' are not all the same---we might expect that people
publish papers across several different venues in a very different way
than actors appear in movies from different countries. More detailed
statistics for each of these networks may be found in
Tables~\ref{tab:data_StatsDesc_1}, \ref{tab:data_StatsDesc_2}
and~\ref{tab:data_StatsDesc_3}.

\begin{figure}
\begin{center}
	\mbox{
		\subfigure[\net{LiveJournal12}]{
			\includegraphics[width=0.4\textwidth]{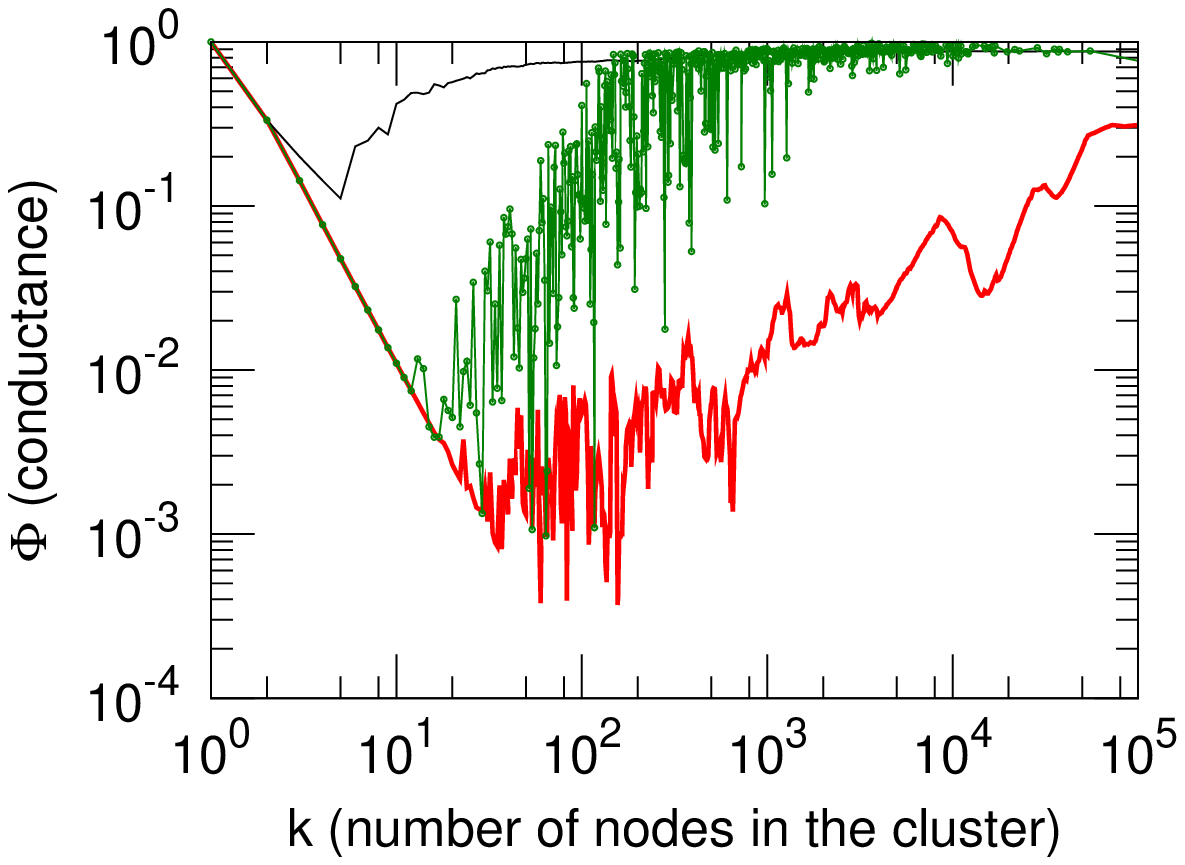}
			\label{fig:explicitCmty:livejournal}
		}   \quad
		\subfigure[\net{CA-DBLP}]{
			\includegraphics[width=0.4\textwidth]{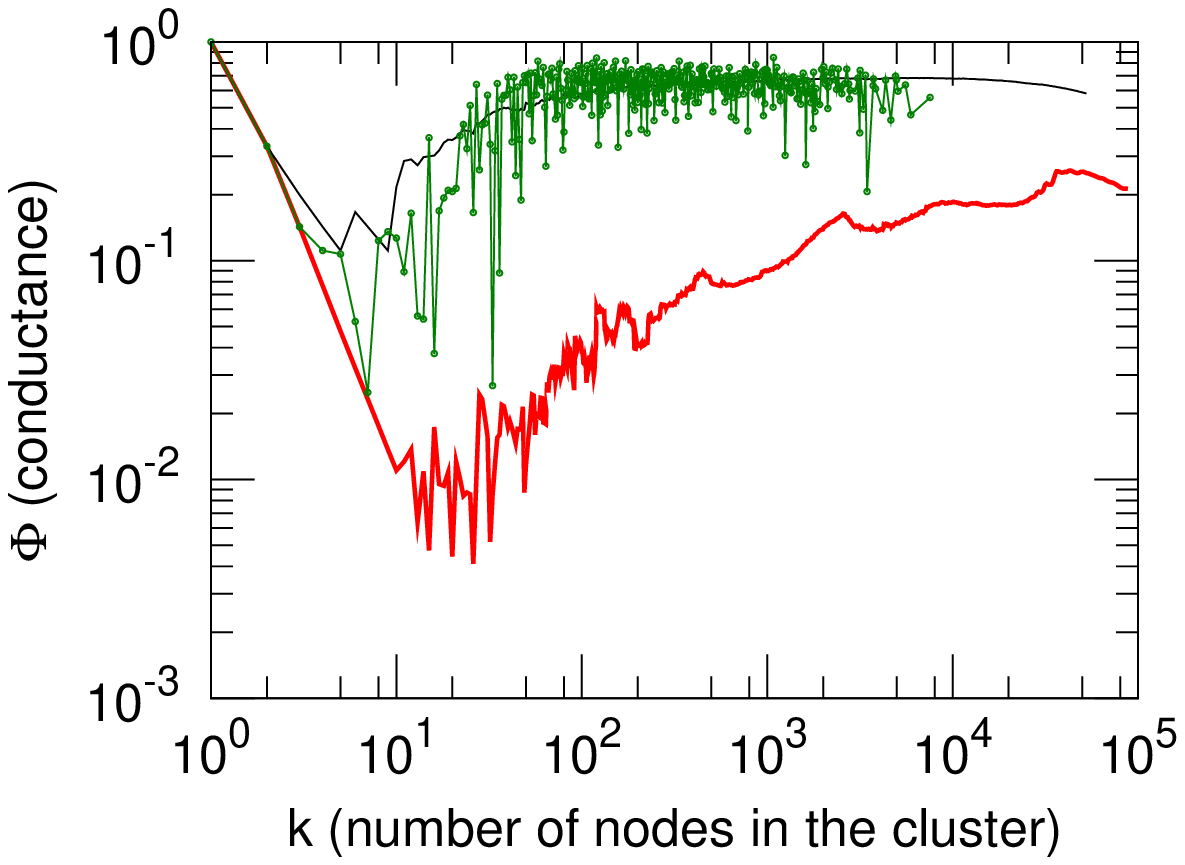}
			\label{fig:explicitCmty:dblp}
		}   \quad
	}
	\mbox{
		\subfigure[\net{AmazonAllProd}]{
			\includegraphics[width=0.4\textwidth]{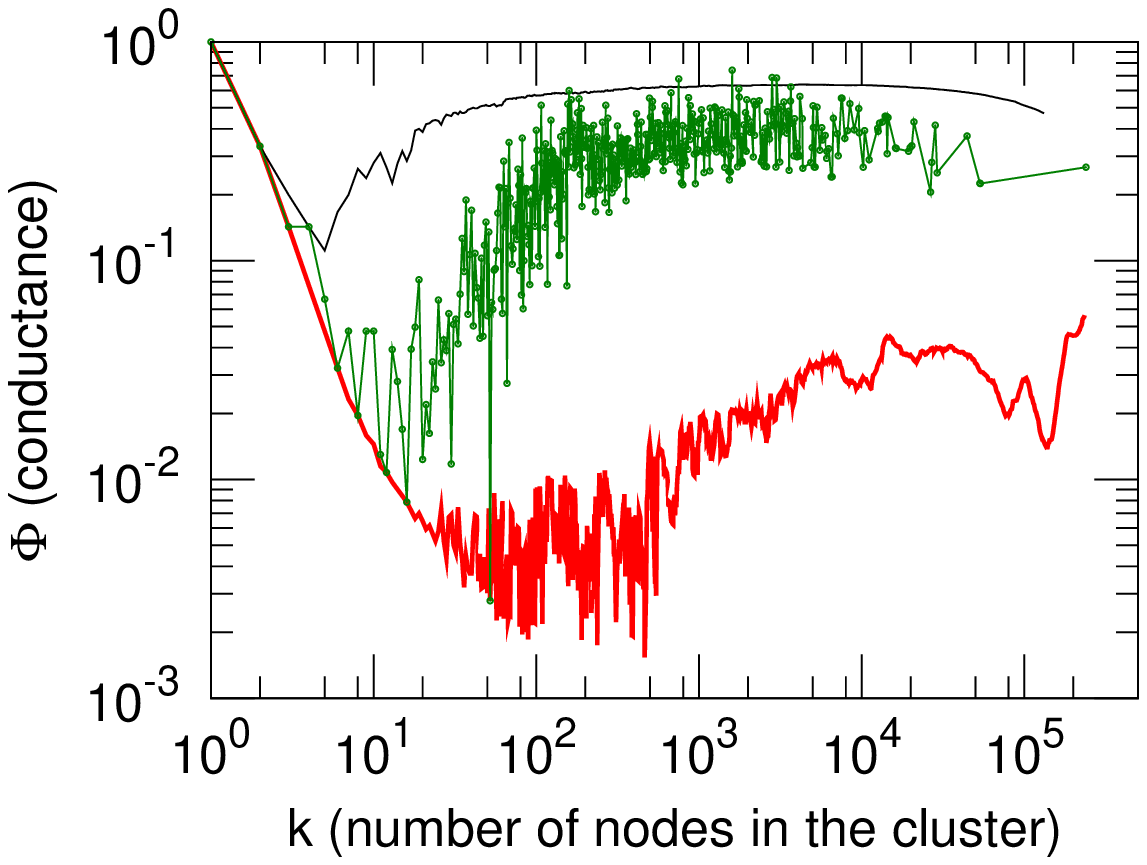}
			\label{fig:explicitCmty:cmty3}
		}   \quad
		\subfigure[\net{Email-Inside}]{
			\includegraphics[width=0.4\textwidth]{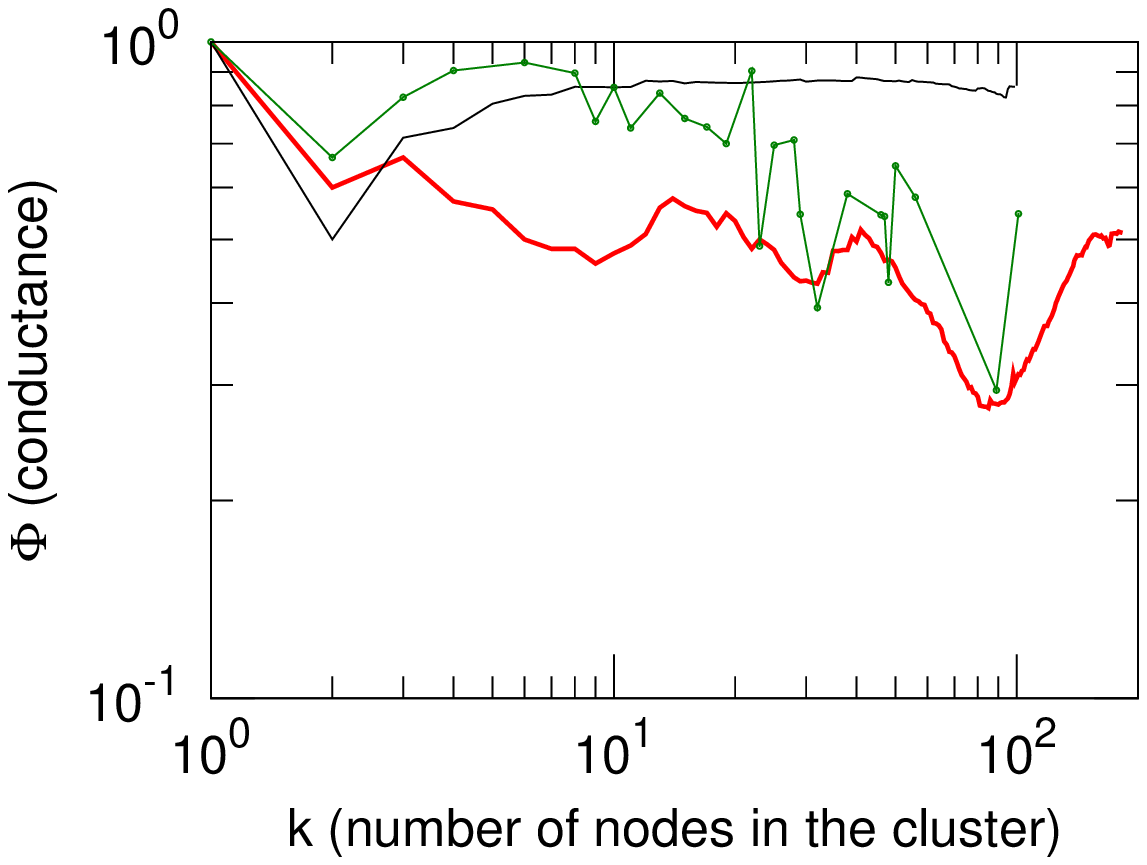}
			\label{fig:explicitCmty:cmty4}
		}   \quad
	}
	\mbox{
		\subfigure[\net{AtM-IMDB} Country]{
			\includegraphics[width=0.4\textwidth]{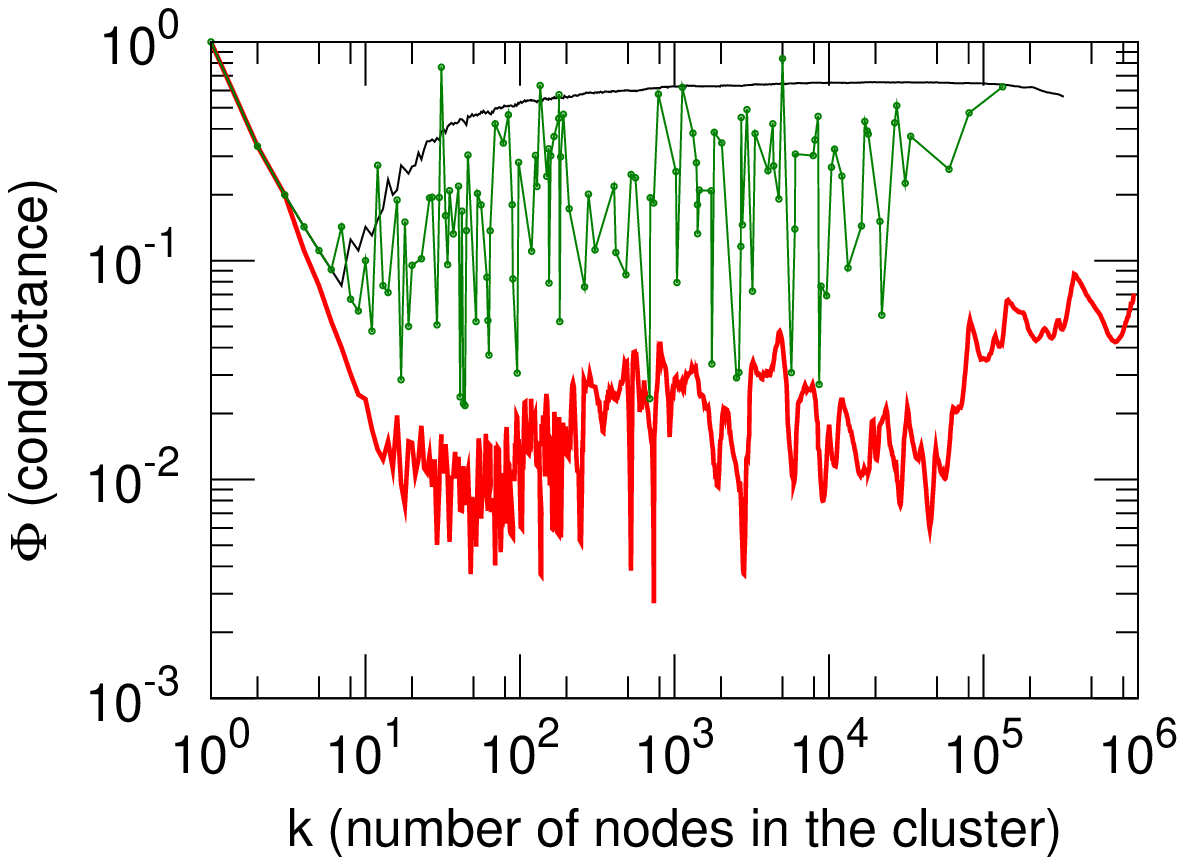}
			\label{fig:explicitCmty:cmty5}
		}   \quad
		\subfigure[\net{AtM-IMDB} Language]{
			\includegraphics[width=0.4\textwidth]{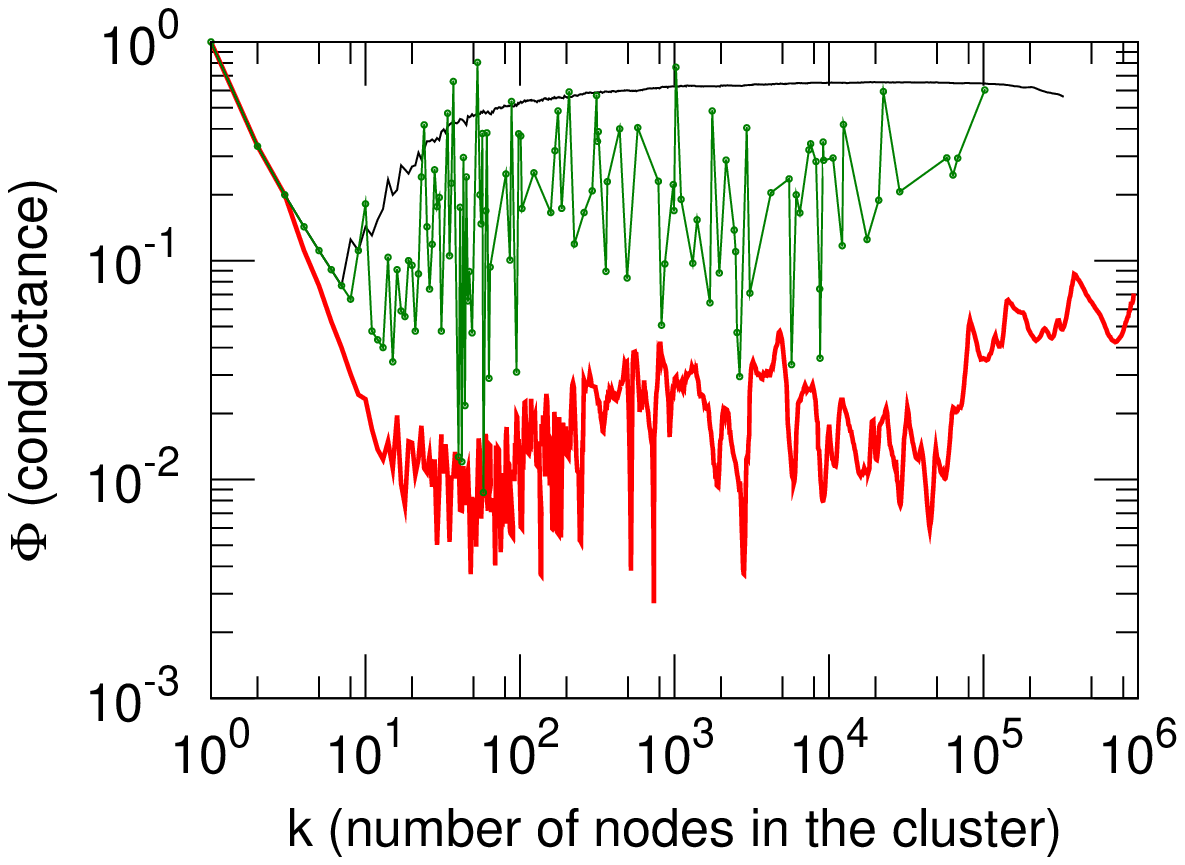}
			\label{fig:explicitCmty:cmty6}
		}   \quad
	}\end{center}
\caption{
[Best viewed in color.]
Network community profile plots for explicitly ``ground truth'' communities 
(green), compared with that for the original network (red) and a rewired version 
of the network (black):
(a) \net{LiveJournal12};
(b) \net{CA-DBLP};
(c) \net{AmazonAllProd};
(d) \net{Email-Inside}; and
(e-f) \net{AtM-IMDB}.
}
\label{fig:explicitCmty}
\end{figure}

To examine the quality of ``ground truth'' communities in the these
network datasets, we take all groups and measure the conductance of the
cut that separates that group from the rest of the network. Thus, we
generated NCP plots in the following way. For every ``ground truth''
community, we measure the conductance of the cut separating it from the
rest of the graph, from which we obtain a scatter plot of community size
versus conductance. Then, we take the lower-envelope of this plot, {\em
i.e.}, for every integer $k$ we find the conductance value of the
community of size $k$ that has the lowest conductance.
Figure~\ref{fig:explicitCmty} shows the results for these network
datasets; the figure also shows the NCP plot obtained from using the Local
Spectral Algorithm on both the original network (plotted in red) and on
the rewired network (plotted in black).

Several observations can be made:
\begin{itemize}
\item The conductance of ``ground truth'' communities follows that for
    the network communities up to until size $10$-$100$ nodes, {\em
    i.e.}, larger communities get successively more community-like. As
    ``ground truth'' communities get larger, their conductance values
    tend to get worse and worse, in agreement with network communities
    discovered with graph partitioning approximation algorithms. Thus,
    the qualitative trend we observed in nearly every large sparse
    real-world network (of the best communities blending in with the
    rest of the network as they grow in size) is seen to hold for
    small ``ground truth'' communities.
\item One might expect that the NCP plot for the ``ground truth''
    communities (the green curves) will be somewhere between the NCP
    plot of the original network (red curves) and that for the rewired
    network (black curves), and this is seen to be the case in
    general. The NCP plot for network communities goes much deeper and
    rises more gradually than for ``ground truth'' communities. This
    is also consistent with our general observation that only small
    communities tend to be dense and well separated, and to separate
    large groups one has to cut disproportionately many edges.
\item For the two social networks we studied (\net{LiveJournal12} and
    \net{CA-DBLP}), larger ``ground truth'' communities have
    conductance scores that get quite ``random'', {\em i.e.}, they are
    as well separated as they would be in a randomly rewired network
    (green and black curves overlap). This is likely associated with
    the relatively weak and overlapping notion of ``ground truth'' we
    associated with those two network datasets. On the other hand, for
    \net{AmazonAllProd} and \net{AtM-IMDB} networks, the general trend
    still remains but large ``ground truth'' communities have
    conductance scores that lie well below the rewired network curve.
\end{itemize}

Our email network illustrates a somewhat different point. The NCP plot for
\net{Email-Inside} should be compared with that for \net{Email-InOut},
which is displayed in Figure~\ref{fig:phiDatasets1}. The
\net{Email-Inside} email network is rather small, and so it has a
decreasing community profile plot, in agreement with the results for small
social networks. Since communication is mainly focused between the members
of the same department, both network and ``ground truth'' communities are
well expressed. Next, compare the NCP plot of \net{Email-Inside} with that
of \net{Email-InOut} (Figure~\ref{fig:phiDatasets1}). 
We see that the NCP plot of \net{Email-Inside} slopes downwards (as we 
consider only the communication inside the organization), but as soon as we 
consider the communication inside the organization and to the outside world 
(\net{Email-InOut}, or alternatively, see \net{Email-Enron}) then we see a
completely different and more familiar picture---the NCP plot drops and then 
slowly increases. 
This suggests that the organizational
structure, (\emph{e.g.}, departments) manifest themselves in the internal
communication network, but as soon as we put the organization into the
broader context ({\em i.e.}, how it communicates to the rest of the world)
then the internal department structure seems to disappear.

\subsection{Connections and broader implications}

In contrast to numerous studies of community structure, we find that there
is a natural size scale to communities. Communities are relatively
small, with sizes only up to about $100$ nodes. We also find that above
size of about $100$, the ``quality'' of communities gets worse and worse
and communities more and more ``blend into'' the network. Eventually, even
the existence of communities (at least when viewed as sets with stronger
internal than external connectivity) is rather questionable. 
We show that large social and information networks can be decomposed into a 
large number of small communities and a large dense and intermingled network 
``core''---we empirically establish that the ``core'' contains on average 
$60\%$ of the nodes and $80\%$ of all edges.
But, as demonstrated by Figure~\ref{fig:phiCore}, the ``core'' itself has a 
nontrivial structure---in particular, it has a core-whisker structure that is analogous to 
the original complete network.

{\bf The Dunbar number:} Our observation on the limit of community size
agrees with Dunbar~\cite{dunbar98gossip} who predicted that roughly $150$ is the
upper limit on the size of a well-functioning human community. Moreover,
Allen~\cite{allen04dunbar} gives evidence that on-line communities have
around $60$ members, and on-line discussion forums start to break down at
about $80$ active contributors. Church congregations, military companies,
divisions of corporations, all are close to the sum of
$150$~\cite{allen04dunbar}. We are thus led to ask: Why, above this size,
is community quality inversely proportional to its size? And why are NCP
plots of small and large networks so different?

Previous studies mainly focused on small networks ({\em e.g.},
see~\cite{danon05community}), which are simply not large enough for the
clusters to gradually blend into one another as one looks at larger size
scales. Our results do not disagree with literature at small sizes. But it
seems that in order to make our observations one needs to look at large
networks. It is only when Dunbar's limit is passed that we find
large communities blurring and eventually vanishing. A second reason is
that previous work did not measure and examine the {\em network community
profile} of cluster size vs. cluster quality.

{\bf Common bond vs. common identity communities:} Dunbar's explanation
aligns well with the common bond vs. common identity theory of group
attachment~\cite{ren07bond} from social psychology. Common identity theory
makes predictions about people's attachment to the group as a whole, while
common bond theory predicts people's attachment to individual group
members. The distinction between the two refers to people's different
reasons for being in a group. Because they like the group as a whole we
get identity-based attachment, or because they like individuals in the
group we get bond-based attachment. Anecdotally, bond-based groups are
based on social interaction with others, personal knowledge of them, and
interpersonal attraction to them. On the other hand, identity-based groups
are based on common identity of its members, \emph{e.g.}, liking to play a
particular on-line game, contributing to Wikipedia, etc. It has been noted
that bond communities tend to be smaller and more
cohesive~\cite{back51influence}, as they are based on interpersonal ties,
while identity communities are focused around common theme or interest.
See~\cite{ren07bond} for a very good review of the topic.

Translating this to our context, the bond vs. identity communities mean
that small, cohesive and well-separated communities are probably based on
common bonds, while bigger groups may be based on common identity, and it
is hard to expect such big communities to be well-separated or
well-expressed in a network sense. This further means the transition
between common bond (\emph{i.e.}, maintaining close personal ties) and
common identity (\emph{i.e.}, sharing a common interest or theme) occurs
at around one hundred nodes. It seems that at this size the cost of
maintaining bond ties becomes too large and the group either dies or
transitions into a common identity community. It would be very interesting
as a future research topic to explore differences in community network
structure as the community grows and transitions from common bond to
common identity community.

{\bf Edge semantics:} Another explanation could be that in small,
carefully collected networks, the semantics of edges is very precise while
in large networks we know much less about each particular edge, {\em
e.g.}, especially when online people have very different criteria for
calling someone a friend. Traditionally social scientists through
questionnaires ``normalized'' the links by making sure each link has the
same semantics/strength.

{\bf Evidence in previous work:} There has also been some evidence that
hints towards the findings we make here. For example, Clauset {\em et
al.}~\cite{clauset04large} analyzed community structure of a graph related
to the \net{AmazonAllProd}, and they found that around $50\%$ of the nodes
belonged to the largest ``miscellaneous'' community. This agrees with the
typical size of the network core, and one could conclude that the largest
community they found likely corresponds to the intermingled core of the network,
and most of the rest of the communities are whisker-like.

In addition, recently there have been several works hinting that the network
communities subject is more complex than it seems at the first sight. For
example, it has been found that even random graphs can have good
modularity scores~\cite{guimera04_fluctuations}. Intuitively, random
graphs have no community structure, but there can still exist sets of nodes
with good community scores, at least as measured by modularity (due to random fluctuations about the mean). Moreover,
very recently a study of robustness of community structure showed that the
canonical example of presence of community structure in
networks~\cite{zachary77karate} may have no significant community
structure~\cite{KLN07_robustness}.


{\bf More general thoughts:}
Our work also raises an important question of what is a natural community
size and whether larger communities (in a network sense) even exist. 
It seems that when community size surpasses some threshold, the community 
becomes so diverse that it stops existing as a traditionally understood 
``network community.'' 
Instead, it blends in with the network, and intuitions based on connectivity
and cuts seem to fail to identify it. 
Approaches that consider both the network structure and node attribute data 
might help to detect communities in these cases.

Also, conductance seems like a very reasonable measure that satisfies
intuition about community quality, but we have seen that if one only worries
about conductance, then bags of whiskers and other internally disconnected
sets have the best scores. This raises interesting questions about cluster
compactness, regularization, and smoothness: what is a good definition of
compactness, what is the best way to regularize these noisy networks, and 
how should this be connected to the notion of community separability?

A common assumption is that each node belongs to exactly one community.
Our approach does not make such an assumption.
Instead, for each given size, we independently find best set of nodes, and 
``communities'' of different sizes often overlap. 
As long there is a boundary between communities (even if boundaries overlap), 
cut- and edge-density- based techniques (like modularity and conductance) 
may have the opportunity to find those communities. 
However, it is the absence of clear community boundaries that makes the NCP 
plot go upwards.

\subsection{Relationship with community identification methods}
\label{sxn:discussion:community}

A great deal of work has been devoted to finding communities in large
networks, and much of this has been devoted to formalizing the intuition
that a community is a set of nodes that has more and/or better
intra-linkages between its members than inter-linkages with the remainder
of the network. Very relevant to our work is that of Kannan, Vempala, and
Vetta~\cite{kannan04_gbs}, who analyze spectral algorithms and describe a
community concept in terms of a bicriterion depending on the conductance
of the communities and the relative weight of inter-community edges.
Flake, Tarjan, and Tsioutsiouliklis~\cite{FTT03_graph} introduce a similar
bicriterion that is based on network flow ideas, and Flake {\em et
al.}~\cite{flake00_efficient,flake02_selforganize} defined a community as
a set of nodes that has more intra-edges than inter-edges. Similar
edge-counting ideas were used by Radicchi {\em et al.}~\cite{RCCLP04_PNAS}
to define and apply the notions of a strong community and a weak
community.

Within the ``complex networks'' community, Girvan and
Newman~\cite{newman02community} proposed an algorithm that used
``centrality'' indices to find community boundaries. Following this,
Newman and Girvan~\cite{newman04community} introduced \emph{modularity} as
\emph{a posteriori} measure of the strength of community structure.
Modularity measures inter- (and not intra-) connectivity, but it does so
with reference to a randomized null model. Modularity has been very
influential in the recent community detection
literature~\cite{newman2004_detect,danon05community}, and one can use
spectral techniques to approximate
it~\cite{WS05_spectralSDM,newman2006_ModularityPNAS}. On the other hand,
Guimer\`{a}, Sales-Pardo, and Amaral~\cite{guimera04_fluctuations} and
Fortunato and Barth\'{e}lemy~\cite{Fortunato07_ResolutionPNAS} showed that
random graphs have high-modularity subsets and that there exists a size
scale below which communities cannot be identified. In part as a response
to this, some recent work has had a more statistical
flavor~\cite{hastings06_inference,reichardt07_partitioning,rosvall07_informationPNAS,KLN07_robustness,newman07_exploratoryPNAS}.
In light of our results, this work seems promising, both due to potential
``overfitting'' issues arising from the extreme sparsity of the networks,
and also due to the empirically-promising regularization properties
exhibited by local spectral methods.

We have made extensive use of the Local Spectral Algorithm of Andersen,
Chung, and Lang~\cite{andersen06local}. Similar results were originally
proven by Spielman and Teng~\cite{Spielman:2004,spielman04_nearlyARXIV},
who analyzed local random walks on a graph; see
Chung~\cite{chung07_fourproofs,Chung07_localcutsLAA,Chung07_heatkernelPNAS}
for an exposition of the relationship between these methods. Andersen and
Lang~\cite{andersen06seed} showed that these techniques can find (in a
scalable manner) medium-sized communities in very large social graphs in
which there exist reasonably well-defined communities. In light of our
results, such methods seem promising more generally. Other recent work
that has focused on developing local and/or near-linear time heuristics
for community detection
include~\cite{clauset04large,wu04_linear,clauset05_local,BB05_local,albert07_linear}.

In addition to this work we have cited, there exists work which views
communities from a very different perspective. For example, Kumar {\em et
al.}~\cite{KRRT99_trawling} view communities as a dense bipartite subgraph
of the Web; Gibson, Kleinberg, and Raghavan~\cite{GKR98_linktopoogy} view
communities as consisting of a core of central authoritative pages linked
together by hub pages; Hopcroft {\em et
al.}~\cite{hopcroft03_communitiesKDD,hopcroft03_communitiesPNAS} are
interested in the temporal evolution of communities that are robust when
the input data to clustering algorithms that identify them are moderately
perturbed; and Palla {\em et al.}~\cite{palla05_OveralpNature} view
communities as a chain of adjacent cliques and focus on the extent to
which they are nested and overlap. The implications of our results for
this body of work remain to be explored.

\subsection{Relationship with other theoretical work}
\label{sxn:discussion:technical}

In this subsection, we describe the relationship between our work and
recent work with similar flavor in graph partitioning, algorithms, and
graph theory.

Recent work has focused on the expansion properties of power law graphs
and the real-world networks they model. For example, Mihail,
Papadimitriou, and Saberi~\cite{MPS03_connectivity}, as well as
Gkantsidis, Mihail, and Saberi~\cite{GMS03_conductance}, studied Internet
routing at the level of Autonomous Systems (AS), and showed that the
preferential attachment model and a random graph model with power law
degree distributions each have good expansion properties if the minimum
degree is greater than $2$ or $3$, respectively. This is consistent with
the empirical results, but as we have seen the AS graphs are quite
unusual, when compared with nearly every other social and information
network we have studied. On the other hand, Estrada has made the
observation that although certain communication, information, and
biological networks have good expansion properties, social networks do
not~\cite{estrada06_expansion}. This is interpreted as evidence that such
social networks have good small highly-cohesive groups, a property which
is not attributed to the biological networks that were considered. From
the perspective of our analysis, these results are interesting since it is
likely that these small highly-cohesive groups correspond to sets near the
global minimum of the network community profile plot. Reproducing deep
cuts was also a motivation for the development of the geometric
preferential attachment models of Flaxman, Frieze, and
Vera~\cite{FFV04_geometric1,FFV07_geometric2}. Note, however, that the
deep cuts they obtain arise from the underlying geometry of the model and
thus are nearly bisections.

Consider also recent results on the structural and spectral properties of
very sparse random graphs. Recall that the $G_{np}$ random graph
model~\cite{bollobas85_rg} consists of those graphs on $n$ nodes, in which
there is an edge between every pair vertices with a probability $p$,
independently. Recall also that if $p\in(1/n,\log n/n)$, then a typical
graph in $G_{np}$ has a giant component, {\em i.e.}, connected subgraph
consisting of a constant fraction of the nodes, but the graph is not fully
connected~\cite{bollobas85_rg}. (If $p<1/n$, the a typical graph is
disconnected and there does not exist a giant component, while if $p >
\log n/n$, then a typical graph is fully connected.) As noted, {\em e.g.},
by Feige and Ofek~\cite{FO05_sparse}, this latter regime is particularly
difficult to analyze since with fairly high probability there exist
vertices with degrees that are much larger than their expected degree. As
reviewed in Section~\ref{sxn:models:sparse_Gw}, however, this regime is
not unlike that in a power law random graph in which the power law
exponent
$\beta\in(2,3)$~\cite{ChungLu01_sparse,Lu01_diameter,ChungLu:2006}.

Chakrabarti {\em et al.}~\cite{chakrabarti07aplots} defined the
``min-cut'' plot which has similarities with our NCP plot.
They used a different approach in which a network was recursively bisected
and then the quality of the obtained clusters was plotted against as a 
function of size; 
and the ``min-cut'' plots were only used as yet-another statistic to 
test when assessing how realistic are synthetically generated graphs. 
Note, however, that the ``min-cut'' plots have qualitatively similar 
behavior to our NCP plots, \emph{i.e.}, they initially decrease, reach a 
minimum, and then increase.

Of particular interest to us are recent results on the mixing time of
random walks in this $p\in(1/n,\log n/n)$ regime of the $G_{np}$ (and the
related $G_{nm}$) random graph model. Benjamini, Kozma, and
Wormald~\cite{BKW06_mixing} and Fountoulakis and
Reed~\cite{FR07_bottlenecks,FR07_evolution}  have established rapid mixing
results by proving structural results about these very sparse graphs. In
particular, they proved that these graphs may be viewed as a ``core''
expander subgraph, whose deletion leaves a large number of
``decorations,'' {\em i.e.}, small components such that a bounded number
are attached to any vertex in the core. The particular constructions in
their proofs is complicated, but they have a similar flavor to the
core-and-whiskers structure we have empirically observed. Similar results
were observed by Fernholz and Ramachandran~\cite{fernholz07_diameter},
whose analysis separately considered the $2$-core of these graphs and then
the residual pieces. They show that a typical longest shortest path
between two vertices $u$ and $v$ consists of a path of length $O(\log n)$
from $u$ to the $2$-core, then a path of length $O(\log n)$ across the
$2$-core, and finally a path of length $O(\log n)$ from the $2$-core to
$v$. Again, this is reminiscent of the core-and-whiskers properties we
have observed. In all these cases, the structure is very different than
traditional expanders~\cite{HLW06_expanders}, which we also empirically
observe. Eigenvalues of power law graphs have also been studied by Mihail
and Papadimitriou~\cite{MP02_eval_powerlaw}, Chung, Lu,
Vu~\cite{Chung03_eigenvalues,Chung03_spectraPNAS,Chung04_spectraIM}, and
Flaxman, Frieze, and Fenner~\cite{FFF04_eigenvalues}.
\section{Conclusion}
\label{sxn:conclusion}

We investigated statistical properties of community-like sets of nodes in
large real-world social and information networks. 
We discovered that community structure in these networks is very different 
than what we expected from the experience with small networks and from what
commonly-used models would suggest.

In particular, we defined a {\em network community profile plot (NCP plot)}, 
and we observed that good network communities exist only up to a size scale 
of $\approx 100$ nodes.
This agrees well with the observations of Dunbar. 
For size scales above $\approx 100$ nodes, the NCP plot slopes upwards as 
the conductance score of the best possible set of nodes gets gradually worse 
and worse as those sets increase in size. 
Thus, if the world is modeled by a sparse ``interaction graph'' and if a 
density-based notion such as conductance is an appropriate measure of 
community quality, then the ``best'' possible ``communities'' in nearly 
every real-world network we examined gradually gets less and less 
community-like and instead gradually ``blends in'' with the rest of the 
network, as the purported communities steadily grow in size. 
Although this suggests that large networks have a {\em core-periphery} or 
{\em jellyfish} type of structure, where small ``whiskers'' connect 
themselves into a large dense intermingled network ``core,'' we also 
observed that the ``core'' itself has an analogous core-periphery structure.

None of the commonly-used network generation models, including 
preferential-attachment, copying, and hierarchical models, generates networks 
that even qualitatively reproduce this community structure property.
We found, however, that a model in which edges are added recursively, via an 
iterative ``forest fire'' burning mechanism, produces remarkably good results.
Our work opens several new questions about the structure of large social and 
information networks in general, and it has implications for the use of graph 
partitioning algorithms on real-world networks and for detecting communities in 
them.


\section*{Acknowledgement}

We thank Reid Andersen, Christos Faloutsos and Jon Kleinberg for discussions,
Lars Backstrom for data, and Arpita Ghosh for assistance with the proof of
Theorem~\ref{thm:mainGw}.

{\footnotesize

}


\end{document}